\documentclass[aip,jap,reprint,floatfix,citeautoscript,longbibliography,superscriptaddress]{revtex4-2}
\usepackage{graphicx,float}
\usepackage{bm,amsmath,amssymb,mathrsfs,dcolumn}
\usepackage{gensymb}
\usepackage[usenames]{color}
\usepackage{subfigure}
\usepackage{ulem}
\usepackage{soul,xcolor}
\usepackage{hyperref}
\hypersetup{colorlinks=true, linkcolor=blue, citecolor=blue, urlcolor=blue}

\begin{document}
\title{Tutorial: Nonlinear  magnonics}

\author{Shasha Zheng}
\affiliation{Intelligent Science $\&$ Technology Academy of CASIC, Beijing 100041, China}
\affiliation{Scientific Research Key Laboratory of Aerospace Defence
Intelligent Systems and Technology, Beijing 100041, China}
\author{Zhenyu Wang}
\affiliation{School of Physics and State Key Laboratory of Electronic Thin Films and Integrated Devices, University of Electronic Science and Technology of China, Chengdu 610054, China}
\author{Yipu Wang}
\affiliation{Interdisciplinary Center of Quantum Information and Zhejiang Province Key Laboratory of Quantum Technology and Device, School of Physics, Zhejiang University, Hangzhou 310027, China}
\author{Fengxiao Sun}
\affiliation{State Key Laboratory for Mesoscopic Physics, School of Physics, Frontiers Science Center for Nano-optoelectronics \& Collaborative Innovation Center of Quantum Matter, Peking University, Beijing 100871, China}
\author{Qiongyi He}
\affiliation{State Key Laboratory for Mesoscopic Physics, School of Physics, Frontiers Science Center for Nano-optoelectronics \& Collaborative Innovation Center of Quantum Matter, Peking University, Beijing 100871, China}
\affiliation{Collaborative Innovation Center of Extreme Optics, Shanxi University, Taiyuan, Shanxi 030006, China}
\affiliation{Peking University Yangtze Delta Institute of Optoelectronics, Nantong, Jiangsu, China}
\affiliation{Hefei National Laboratory, Hefei 230088, China}
\author{Peng Yan}
\email{yan@uestc.edu.cn}
\affiliation{School of Physics and State Key Laboratory of Electronic Thin Films and Integrated Devices, University of Electronic Science and Technology of China, Chengdu 610054, China}
\author{H. Y. Yuan}
\email{huaiyangyuan@gmail.com}
\affiliation{Institute for Theoretical Physics, 3584 CC Utrecht, The Netherlands}
\date{\today}

\begin{abstract}
Nonlinear magnonics studies the nonlinear interaction between magnons and other physical platforms (phonon, photon, qubit, spin texture) to generate novel magnon states for information processing. In this tutorial, we first introduce the nonlinear interactions of magnons in pure magnetic systems and hybrid magnon-phonon and magnon-photon systems. Then we show how these nonlinear interactions can generate exotic magnonic phenomena. In the classical regime, we will cover the parametric excitation of magnons, bistability and multistability, and the magnonic frequency comb. In the quantum regime, we will discuss the single magnon state, Schr\"{o}dinger cat state and the entanglement and quantum steering among magnons, photons and phonons. The applications of the hybrid magnonics systems in quantum transducer and sensing will also be presented. Finally, we outlook the future development direction of nonlinear magnonics.
\end{abstract}

\maketitle

\section{Introduction}
Finding efficient means to store, manipulate and process information is a central topic in the modern science and technology. Differing from the traditional electronics based on transistors, spintronics is a newly-rising field that utilizes the spin degree of freedom of electrons, besides its charge degree of freedoms. Ever since its birth in 1980s, there have emerged many interesting physical concepts including giant magnetoresistance, spin-transfer torque, spin pumping etc. and fruitful applications including magnetic random access memory, magnetic logic gates, mangetic storage, etc \cite{DasmaRMP2004, FertRMP2008, HirohataReview2020}. Among these developments, a branch of spintronics so-called magnon spintronics or magnonics attracted much attentions in the last decade \cite{ChumakNP2015}. Magnons, as quanta of spin-wave excitations in ordered magnets can carry information in their amplitude, phase and angular momentum. Magnonics aims to manipulate the generation, propagation and read-out of magnon current for information processing. As information carriers, magnons can propagate in magnetic insulators and thus the energy consumption accompanied by the Joule heating can be strongly reduced. Furthermore, magnonic system can be integrated with other quantum platforms, for example, photons, qubits, and phonons to achieve hybrid quantum systems. These hybrid platforms benefit the advantages of magnonic system and can be utilized to achieve multifunctional tasks of information processing in both the classical and quantum levels. In particular, the interplay of magnonics and quantum information science have witnessed the birth and development of quantum magnonics, which addresses the quantum states of magnons and opens a new avenue of magnonics \cite{YuanReview2022,BabakReview2022}.

One critical question in magnonics is to generate desirable magnon states for long distance transport. In the traditional magnonics, one usually applies a microwave to excite the coherent magnons in both magnetic nanoparticles and magnetic thin films. Later, it was found that spin-orbit effects can generate spin accumulation at the interface of hybrid magnet-nonmagnet structure and then inject magnon current into the magnetic layer. The magnon transport can well be understood within the theory of linear response. Recently, people found that the nonlinear interaction of magnons can generate exotic magnon states, for example magnon Bose-Einstein condensation and spin superfluidity, which can mediate the long-distance transport of magnon currents.  Magnon splitting and confluence are important for realizing frequency down- and up-conversion, respectively \cite{WangPRL2021, ShengPRL2023}. On the other hand, the nonlinear interaction between magnons and other quasiparticles (photons, qubits, and phonons) is the
cornerstone to generate and stabilize the quantum states of magnons and to entangle magnons with other quasiparticles. Therefore, the nonlinear magnon process has received much attention recently, we dub this subbranch of magnonics as nonlinear magnonics.

In this tutorial, we first give a detailed introduction to the nonlinear magnon interactions including the three- and four-magnon processes, and the nonlinear interaction among magnons, phonons, and photons. This preliminary knowledge provides a basis to study various nonlinear magnonic phenomena. Then we proceed to show how these nonlinear interactions give rise to the novel magnon states.  In the classical regime, we shall focus on the parametric excitation of magnons, bistability and multistability, magnon frequency comb and magnonic Penrose superradiance induced by magnetic nonlinearities. In the quantum regime, we shall first elaborate the single magnon state and Schr\"{o}dinger cat state and then discuss the entanglement and quantum steering among magnons, photons and phonons in hybrid quantum systems. The applications of the hybrid magnonic systems as quantum transducers and sensors will also be introduced in this section. Finally, we give an outlook of the future developments of nonlinear magnonics.
To minimize the overlap with the existing reviews on cavity spintronics and quantum magnonics, we shall focus on the magnonic phenomena merely induced by nonlinear effects with detailed introduction to the underlying physical mechanisms. Unless stated otherwise, we set the reduced Planck constant $\hbar$ as 1.

\section{Nonlinear magnon process}\label{preliminary}
\subsection{Magnon-magnon interaction} \label{mm_interaction}
\begin{figure*}[ht!]
	\includegraphics[width=0.8\textwidth]{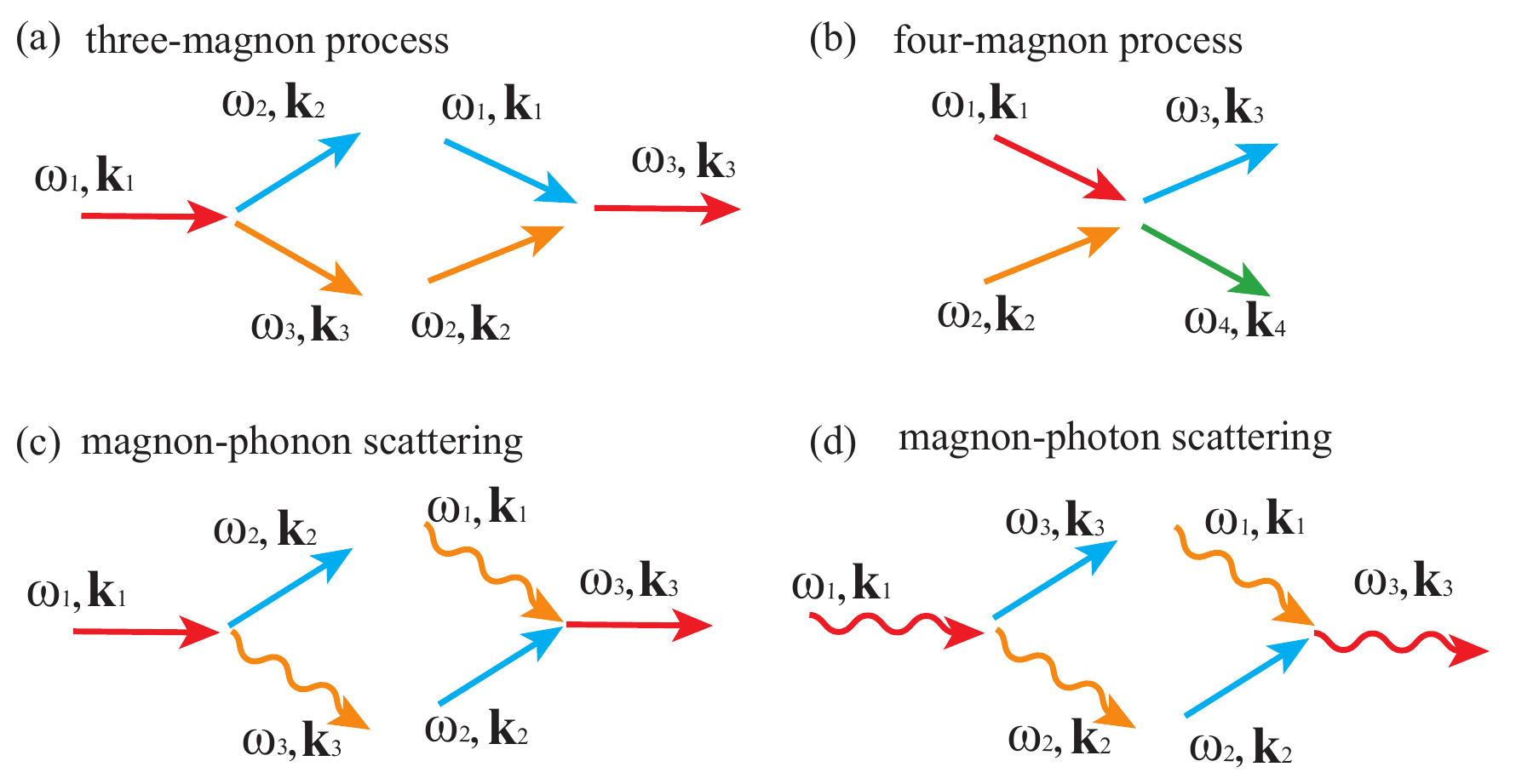}
	\caption{Schematic of three-magnon and four-magnon scattering processes (a-b), pressure-like magnon-phonon interaction (c) and nonlinear magnon-photon process (d). The straight (wavy) arrows represent magnon (phonon/photon) modes.}
	\label{three_four_magnon}
\end{figure*}

In general, a magnet contains huge number of spins coupled with each other through the exchange interaction and the Hamiltonian of the system may be written as
\begin{widetext}
\begin{equation}\label{generalHam}
\hat{\mathcal{H}}=-J\sum_{\langle i,j\rangle} \hat{\mathbf{S}}_i\cdot \hat{\mathbf{S}}_j + \hat{\mathcal{H}}_\mathrm{an}-  \mu_0 g_e \mu_B\sum_i \mathbf{H}_i \cdot \hat{\mathbf{S}}_i- \frac{\mu_0 (g_e\mu_B)^2}{4\pi}\sum_{i<j} \frac{3(\hat{\mathbf{S}}_i\cdot \mathbf{e}_{ij})(\hat{\mathbf{S}}_j\cdot \mathbf{e}_{ij})-\hat{\mathbf{S}}_i\cdot \hat{\mathbf{S}}_j}{|\mathbf{r}_{ij}|^3}.
\end{equation}
\end{widetext}
Here $\hat{\mathbf{S}}_i$ is the spin operator at the $i$-th site and the first term is exchange energy with $J$ the exchange coefficient. $J>0$ ($J<0$) represents a ferromagnetic (antiferromagnetic) coupling between neighboring spins. The second term is crystalline anisotropy that prefers spins to align along particular directions and it is rooted in the spin-orbit coupling. The third term is Zeeman energy with $\mathbf{H}_i$ the external magnetic field on the $i$-th spin, $\mu_0$ the vacuum magnetic permeability, $g_e$ the Land\'{e} $g$-factor, and $\mu_B$ being the Bohr magneton. The last term is dipolar interaction between two arbitrary spins with $\mathbf{r}_{ij}$ the relative position between the $i$-th and $j$-th spins, and $\mathbf{e}_{ij}$ being the unit direction vector connecting sites $i$ and $j$. The competition of the exchange interaction, anisotropy energy, Zeeman energy and dipolar interaction can generate various ground states of the magnetic systems, including uniform domains, domain walls, vortices, skyrmions, etc. To study the magnon excitation above a particular ground state, we perform the standard Holstein-Promakoff (HP) transformation \cite{HP_PR_1940} on the ground state and derive the effective Hamiltonian for the magnons as
\begin{equation}
\hat{\mathcal{H}}=\hat{\mathcal{H}}^{(0)} + \hat{\mathcal{H}}^{(2)} + \hat{\mathcal{H}}^{(3)} + \hat{\mathcal{H}}^{(4)} +...,
\end{equation}
where $\hat{\mathcal{H}}_0$ is ground state energy, $\hat{\mathcal{H}}^{(2)}$ denotes the two-magnon process contributed by exchange field, external field, anisotropy field and dipolar interactions, and it dominates the magnon excitation at low temperature and under weak external drivings. By diagonalizing this quadratic Hamiltonian $\hat{\mathcal{H}}^{(2)}$, one can obtain the eigenspectrum of magnons, which works for both collinear and noncollinear magnetic structures, regardless of magnon number conservation. $\hat{\mathcal{H}}^{(3)}$ represents the three-magnon processes from dipolar interactions and $\hat{\mathcal{H}}^{(4)}$  stands for the four-magnon processes from exchange and dipolar interactions, respectively, as illustrated in Fig. \ref{three_four_magnon}(a-b).

As an example, we consider a uniaxial ferromagnetic system ($J>0$) magnetized along an external field $\mathbf{H}=H\mathbf{e}_z$. The Hamiltonian of the system is
\begin{equation}\label{uniaxialHam}
\hat{\mathcal{H}}=-J\sum_{\langle i,j\rangle} \hat{\mathbf{S}}_i\cdot \hat{\mathbf{S}}_j -K_z \sum_i\left (\hat{S}_{i,z} \right )^2-\mu_0 g_e \mu_B H\sum_i \hat{S}_{i,z},
\end{equation}
where $K_z$ is anisotropy along the $z$ directions with $K_z>0$. The classical ground state of the system is $\langle \hat{\mathbf{S}} \rangle=S\mathbf{e}_z$. To study magnon-magnon interaction, we have to go beyond the linear form of HP transformation and include the higher-order terms. Here we expand the spin operators up to the third-order terms of magnons as below
\begin{subequations}\label{HPtransform}
\begin{align}
&\hat{S}_{i,x}=\frac{\sqrt{2S}}{2}(\hat{a}_i + \hat{a}^\dagger_i-\frac{\hat{a}_i^\dagger \hat{a}_i \hat{a}_i+\hat{a}_i^\dagger \hat{a}_i^\dagger \hat{a}_i}{4S}),\\ &\hat{S}_{i,y}=\frac{\sqrt{2S}}{2i}(\hat{a}_i-\hat{a}_i^\dagger-\frac{\hat{a}_i^\dagger \hat{a}_i \hat{a}_i-\hat{a}_i^\dagger \hat{a}_i^\dagger \hat{a}_i}{4S}),\\
&\hat{S}_{i,z}=S-\hat{a}_i^\dagger \hat{a}_i,
\end{align}
\end{subequations}
where $\hat{a}_i$ ($\hat{a}_i^\dagger$) is the magnon annihilation (creation) operator at the $i$-th spin site, By substituting Eqs. \eqref{HPtransform} into the Hamiltonian \eqref{uniaxialHam} and further making a Fourier transformation $\hat{a}_i = 1/\sqrt{N}\sum_\mathbf{k} \hat{a}_\mathbf{k} e^{i\mathbf{k}\cdot \mathbf{R}_i}$, the quadratic term of the effective Hamiltonian reads
\begin{equation} \label{Ham2}
\mathcal{\hat{H}}^{(2)}= \sum_k \left [\omega_k-\frac{ZJ}{2N} V(\mathbf{k}_1,\mathbf{k}_2,0) \right] \hat{a}^{\dagger}_\mathbf{k} \hat{a}_\mathbf{k},
\end{equation}
where $V(\mathbf{k}_1,\mathbf{k}_2,\mathbf{q}) = \gamma_{\mathbf{k}_1 - \mathbf{q}} + \gamma_{\mathbf{k}_2} - 2\gamma_{\mathbf{k}_1 - \mathbf{k}_2-\mathbf{q}}$ represents a renormalization of the magnon frequency from the magnon-magnon interaction with structure factor of spin-wave $\gamma_\mathbf{k} = 1/Z\sum_\mathbf{\delta} e^{i \mathbf{k} \cdot \mathbf{\delta} }$. At low temperature, the magnon excitation is very low, i.e., $\langle \hat{a}^{\dagger}_\mathbf{k} \hat{a}_\mathbf{k} \rangle \ll N$, the renormalization term can be neglected. For long-wavelength magnons ($kd \ll 1$ with $d$ lattice constant), we can expand the structure factor $\gamma_\mathbf{k}$ up to the quadratic terms of $(kd)^2$ and derive the eigenfrequency as $\omega_k = 2JS(kd)^2 + \mu_0 g_e \mu_B H+2K_zS$. For uniform precession or ferromagnetic resonance mode, i.e., $\mathbf{k}=0$, the magnon frequency is $\omega_0=\mu_0 g_e \mu_B H+2K_zS$. Note that the resonance frequency is tunable by an external field, which is one main advantage of magnonic system for information processing.
%

At the same time, we also derive the quartic terms representing the four-magnon process as
\begin{equation}\label{four_magnon}
\mathcal{\hat{H}}^{(4)}= -\sum_{\mathbf{k}_1,\mathbf{k}_2,\mathbf{q}} \left [ \frac{ZJ}{2N} V(\mathbf{k}_1,\mathbf{k}_2,\mathbf{q}) +\frac{K}{N}\right ] \hat{a}_{\mathbf{k}_1 - \mathbf{q}}^\dagger \hat{a}_{\mathbf{k}_2}\hat{a}_{\mathbf{k}_2 +  \mathbf{q}}^\dagger \hat{a}_{\mathbf{k}_1}.
\end{equation}
It describes the scattering of two magnons with momentum $\mathbf{k}_1$ and $\mathbf{k}_2$ to another two magnons with momentum $\mathbf{k}_1 - \mathbf{q}$ and $\mathbf{k}_2 + \mathbf{q}$, as shown in Fig. \ref{three_four_magnon}(b). Both momentum and energy of magnons are conserved during this scattering process.
In the long wavelength limit ($\mathbf{k}_1,\mathbf{k}_2,\mathbf{q} \rightarrow 0$), $V(\mathbf{k}_1,\mathbf{k}_2,\mathbf{q}) \rightarrow 0$, the anisotropy will dominate the fourth order terms as
$-K/N (\hat{a}^\dagger \hat{a})^2$ with $\hat{a} \equiv \hat{a}_{\mathbf{k} \rightarrow 0}$, which is known as the Kerr nonlinearity.

We note that the three-magnon process, i.e., terms involving three creation/annihilation operators of magnons do not appear in our simple model. This is because we do not include the dipolar interaction yet. When the dipolar interaction is taken into account, it will contribute to both  magnon splitting and confluence process (see Fig. \ref{three_four_magnon}(a)) as

\begin{equation}
\mathcal{\hat{H}}^{(3)}= \sum_{\mathbf{k},\mathbf{q}} \left [ V'(\mathbf{k},\mathbf{q}) \hat{a}_{\mathbf{k}} \hat{a}_{\mathbf{k} - \mathbf{q}}^\dagger \hat{a}_{\mathbf{k} +  \mathbf{q}}^\dagger + V'^\dagger(\mathbf{k},\mathbf{q}) \hat{a}_{\mathbf{k} +  \mathbf{q}} \hat{a}_{\mathbf{k} - \mathbf{q}}\hat{a}_{\mathbf{k}}^\dagger \right ],
\end{equation}
where the first and second terms represent the magnon splitting and confluence process, respectively. Again, the momentum and energy of magnons are conserved in the three-magnon process, however, the magnon number is not.

The strength of the three-magnon process is inversely proportional to $\sqrt{N}$ and also depends on the magnon density, and it is usually very weak at low temperature without any drivings. Under strong drivings, the magnon number will become considerable large and the three-magnon process has to be considered as shown in Sec. \ref{parametric_magnon} below.  On the other hand, such three-magnon process also depends on the gradient of magnetization in noncollinear magnetic textures. Thus, it can be enhanced for a magnetic texture with a very sharp magnetization change at small spatial scale, for example, small skyrmions, narrow domain walls, etc.  To make this point clear, we shall consider the magnon excitation above a magnetic texture in detail. For convenience, we transfer to a continuous fields $\mathbf{m}(r)$ where its gradient can be defined naturally. Let us consider a uniaxial magnetic film with chiral interaction described by the following Hamiltonian
\begin{equation}
\hat{\mathcal{H}}=\int A(\nabla \mathbf{m})^2 - Km_z^2 +D[ m_z \nabla \cdot \mathbf{m} - (\mathbf{m} \cdot \nabla )m_z] dxdy,
\label{ham_nonclinear}
\end{equation}
where $\mathbf{m}$ is the normalized magnetization, $A$ is the exchange stiffness, $K$ is the easy-axis anisotropy, and $D$ is the strength of Dzyaloshinskii-Moriya
interaction (DMI). The competition of these interactions can generate various magnetic states, such as domain walls, skyrmions, vortices, etc. To consider the magnon-magnon interaction, we transfer to a local coordinate system with $\mathbf{e}_3$ pointing along the local magnetization direction of the magnetic texture, and the other two orthogonal axes $\mathbf{e}_1$ and $\mathbf{e}_2$ are determined by
\begin{equation}
\left (\begin{array}{c}
  \mathbf{e}_1 \\
  \mathbf{e}_2 \\
  \mathbf{e}_3
\end{array} \right )= \left (
\begin{array}{ccc}
  \sin \phi & -\cos \phi & 0 \\
  \cos \theta \cos \phi & \cos \theta \sin \phi & -\sin \theta \\
   \sin \theta \cos \phi &  \sin \theta \sin \phi & \cos \theta
\end{array} \right )
\left (\begin{array}{c}
  \mathbf{e}_x \\
  \mathbf{e}_y \\
  \mathbf{e}_z
\end{array} \right ).
\end{equation}
Then the magnetization in this local coordinate and the original $xyz$ coordinate are connected by
\begin{equation}
\left (\begin{array}{c}
   m_x \\
   m_y \\
   m_z
\end{array} \right )= \left (
\begin{array}{ccc}
  \sin \phi & \cos \theta \cos \phi & \sin \theta \cos \phi \\
  -\cos \phi & \cos \theta \sin \phi & \sin \theta \sin \phi \\
   0 &  -\sin \theta & \cos \theta
\end{array} \right )
\left (\begin{array}{c}
  m_1 \\
  m_2 \\
  m_3
\end{array} \right ).
\label{transform}
\end{equation}
By substituting these relations back into Hamiltonian (\ref{ham_nonclinear}), we can reformulate the exchange, anisotropy and DMI energy in the local coordinate as,
\begin{widetext}
\begin{equation}
\begin{aligned}
\hat{\mathcal{H}}_\mathrm{ex}/A&=(\nabla m_1)^2 + (\nabla m_2)^2 + (\nabla m_3)^2\\
&~~+(\nabla \theta )^2(m_2^2 + m_3^2) + (\nabla \phi )^2[m_1^2 + (m_2 \cos \theta + m_3 \sin \theta)^2]
+2 (\nabla \theta)(\nabla \phi)m_1(m_3 \cos \theta -m_2 \sin \theta)\\
&~~+2 (\nabla \phi)[\sin \theta (m_1 \nabla m_3-m_3\nabla m_1) + \cos \theta (m_1 \nabla m_2-m_2 \nabla m_1)]
+2(\nabla \theta) (m_3 \nabla m_2-m_2 \nabla m_3),\\
\hat{\mathcal{H}}_\mathrm{an}/K&=-(-m_2 \sin \theta + m_3 \cos \theta)^2, \\
\hat{\mathcal{H}}_\mathrm{DM}/D&=(- m_2 \sin \theta + m_3 \cos \theta)  \left[ \partial_x (m_1 \sin \phi + m_2 \cos \theta \cos \phi + m_3 \sin \theta \cos \phi)\right] \\
&~~+(- m_2 \sin \theta + m_3 \cos \theta)  \left[ \partial_y(-m_1 \cos \phi + m_2 \cos \theta \sin \phi + m_3 \sin \theta \sin \phi) \right]\\
&~~-(m_1 \sin \phi + m_2 \cos \theta \cos \phi + m_3 \sin \theta \cos \phi)\partial_x(- m_2 \sin \theta + m_3 \cos \theta)\\
&~~-(-m_1 \cos \phi + m_2 \cos \theta \sin \phi + m_3 \sin \theta \sin \phi)\partial_y(- m_2 \sin \theta + m_3 \cos \theta).
\end{aligned}
\label{transform}
\end{equation}
\end{widetext}
Similar to the discrete case, the magnon creation ($\hat{a}^\dagger$) and annihilation ($\hat{a}$) can be reformulated through Holstein-Primakoff transformation \cite{HP_PR_1940} up to the terms with three magnon operators as
\begin{subequations}
\begin{align}
&m_1=\frac{1}{\sqrt{2S}} \left[ \hat{a} + \hat{a}^\dagger - \frac{\hat{a}^\dagger \hat{a}\hat{a} +\hat{a}^\dagger \hat{a}^\dagger \hat{a}}{4S}\right ], \\
&m_2= -\frac{i}{\sqrt{2S}} \left[ \hat{a} - \hat{a}^\dagger - \frac{\hat{a}^\dagger \hat{a}\hat{a} -\hat{a}^\dagger \hat{a}^\dagger \hat{a}}{4S}\right ], \\
&m_3 = 1 - \hat{a}^\dagger \hat{a}/S,
\end{align}
\label{HP_mfield}
\end{subequations}
where $\hat{a}$ ($\hat{a}$) is short for $\hat{a}(\mathbf{r})$ ($\hat{a}^\dagger(\mathbf{r})$).
By substituting Eq. (\ref{HP_mfield}) into Eq. (\ref{transform}), we can derive the leading terms governing the three-magnon process as
\begin{widetext}
\begin{equation}
\begin{aligned}
\hat{\mathcal{H}}^{(3)}&=C_1\hat{a}^\dagger \hat{a} \hat{a} + \hat{a}^\dagger \hat{a} \nabla (C_2\hat{a} + C_3 \hat{a}^\dagger) + (C_4\hat{a} + C_5 \hat{a}^\dagger) \nabla \hat{a}^\dagger \hat{a} + \mathrm{h.c.},  \\
\end{aligned}
\end{equation}
\end{widetext}
where the coefficients $C_i=C_i(A,D,K,\theta,\phi,\nabla \theta,\nabla \phi)$ ($i=1,2,3,4,5$) depends on the magnetic profile $(\theta,\phi)$ as well as the gradient of magnetization $(\nabla \theta,\nabla \phi)$.
Note that all magnon creation and annihilation operators are spatially dependent. Once we transfer to the momentum space through a Fourier transform, $\hat{a}(\mathbf{r})=1/\sqrt{N}\sum_k \hat{a}_\mathbf{k}e^{i\mathbf{k}\cdot \mathbf{r}}$, we will only have the terms,
\begin{widetext}
\begin{equation}
\begin{aligned}
\hat{\mathcal{H}}^{(3)}&=\sum_{\mathbf{k}_1,\mathbf{k}_2,\mathbf{k}_3} \left [ g_p(\mathbf{k}_1,\mathbf{k}_2,\mathbf{k}_3) \hat{a}_{\mathbf{k}_1}\hat{a}_{\mathbf{k}_2}^\dagger \hat{a}_{\mathbf{k}_3}^\dagger +  g_n(\mathbf{k}_1,\mathbf{k}_2,\mathbf{k}_3)\hat{a}_{\mathbf{k}_1}\hat{a}_{\mathbf{k}_2}\hat{a}_{\mathbf{k}_3}^\dagger + \mathrm{h.c.} \right ].\\
\end{aligned}
\label{h3k}
\end{equation}
\end{widetext}
Here the strength of three-magnon process depends strongly on the magnetization gradient inside the textures and it thus could be enhanced in magnetic structures with magnetization variation over small spatial scale, such as domain walls, skyrmions, vortex. In Sec. \ref{magnon_comb}, we will see that such three-magnon process can generate a magnonic frequency comb with a set of equal-spacing frequency lines.

\subsection{Magnon-phonon interaction} \label{magnon_phonon_interaction}
Besides magnon-magnon interaction, magnon can also be coupled to lattice vibrations of the magnetic system, i.e., phonons, through magnetoelastic interaction.
In general, the magnetoelastic interaction reads~\cite{KittelRMP1949,DreherPRB2012,RuckriegelPRB2014,KamraSSC2014}
\begin{equation}\label{ham_me}
\hat{\mathcal{H}}_{\mathrm{int}}= \sum_\mathbf{r} \sum_{pq} [b_{pq} \hat{S}_p(\mathbf{r}) \hat{S}_q(\mathbf{r}) + b_{pq}^{'}\partial_p \hat{\mathbf{S}}(\mathbf{r}) \cdot \partial_q \hat{\mathbf{S}}(\mathbf{r})]\hat{\epsilon}_{pq},
\end{equation}
where $p,q=x,y,z$, the strain tensor $\hat{\epsilon}_{pq}$ is defined in terms of the lattice displacements $\hat{\mathbf{u}}$ as $\hat{\epsilon}_{pq} \equiv (\partial_q \hat{u}_p+\partial_p \hat{u}_q)/2$, $b_{pq}$ and $b_{pq}^{'}$ are the phenomenological magnetoelastic coupling coefficients. Here the first term in $\hat{\mathcal{H}}_{\mathrm{int}}$ is generated by the spin-orbit coupling while the second term is due to the spatial dependence of exchange interaction. The quantization of Eq. \eqref{ham_me} may include both linear and nonlinear interactions between magnons and phonons. Here the linear part (quadratic terms of magnon and phonon operators) contributes to the hybridization of magnon and phonon modes, while the nonlinear part (cubic and even higher-order terms of magnon/phonon operators) describes the higher-order scattering process. The higher-order term that received most attention is the pressure-like interaction between magnons and phonons, i.e., $\hat{a}^\dagger \hat{a} \hat{b}$ with $\hat{a}$ and $\hat{b}$ being the annihilation operators of magnons and phonons, respectively. This is because it resembles the interaction between photons and mechanical oscillations in cavity optomechanics, which has shown its importance to study the entanglement properties of continuous variables and applications in the high-precision measurements \cite{AspelmeyerRMP2014}.

To illustrate how the pressure-like interaction appears, we focus on the long-wavelength magnon mode in a one-dimensional (1D) spin chain with atoms patterned along the $z-$axis. Then the gradient terms in Eq. \eqref{ham_me} proportional to the wavevector of magnons can be safely neglected because they contribute to the higher-order terms on the magnitude of $|\mathbf{k}|^2$ with $\mathbf{k}$ being the wavevector of phonons. The effective Hamiltonian describing the interaction of magnons and phonons \cite{RuckriegelPRB2014} is thus $\hat{\mathcal{H}}_\mathrm{int}=\sum_j b_{zz} (\hat{S}_j^z)^2 \hat{\epsilon}_{zz}$, where the strain tensor $\hat{\epsilon}_{zz}=\partial_{z} \hat{u}_z$ with the atom displacement $\hat{u}_z=\sum_k (2\rho \omega_kV)^{-1/2}(\hat{b}_k + \hat{b}_k^\dagger)e^{ikz}$. Here $\rho$ is the mass density of the magnet, $V$ is the volume, and $\hat{b}_k$ ($\hat{b}_k^\dagger$) is the annihilation (creation) operator of phonon mode with wavevector $k$. The phonon follows the linear dispersion relation $\omega_k = c|k|$ with $c$ being the speed of sound. By employing the HP transformation of magnons ($\hat{S}_j^z= S-\hat{a}_j^\dagger \hat{a}_j$) and the quantization of phonon modes $u_z$, the interaction Hamiltonian can be expressed as
\begin{equation} \label{Hint_real_space}
\hat{\mathcal{H}}_{\mathrm{int}}= \sum_{j} g(\omega_k)\hat{a}_j^\dagger \hat{a}_j \left [ \hat{b}_ke^{ikR_j} -\hat{b}_k^\dagger e^{-ikR_j}\right],
\end{equation}
where we have transformed to a discrete space ($z \rightarrow R_j$) and the parameter $g(\omega_k)=-2iSb_{zz}k/\sqrt{2\rho \omega_kV}$ is the coupling coefficient between magnons and phonons, which depends on the spin number, magetoelastic coupling strength, phonon wavevector and frequency, as well as mass density and volume of the material.

After a Fourier transform of magnon operator $\hat{a}_j =1/\sqrt{N} \sum_k \hat{a}_k e^{ikR_j}$, we can simplify the interaction Hamiltonian \eqref{Hint_real_space} as
\begin{equation}
\hat{\mathcal{H}}_{\mathrm{int}}=  \sum_{k,k'} g(\omega_k)(\hat{a}^\dagger_{k'} \hat{a}_{k'-k} \hat{b}_k - \hat{a}^\dagger_{k'} \hat{a}_{k'+k}\hat{b}_{k}^\dagger).
\end{equation}
This Hamiltonian describes the scattering of a magnon mode $k'-k$ and a phonon mode $k$ to form a new magnon mode $k'$ and vice versa, as shown in Fig. \ref{three_four_magnon}(c). Here the requirement of momentum conservation can be broken by taking into account of impurities, grain boundaries and other inhomogeneities in the magnetic system \cite{MichaelJAP2002, Safonov2003}. In the experiments, one can select a particular phonon mode of magnetic sphere to couple to the magnon such that the sum may be removed \cite{ZhangSA2016}. Then we may simplify the interaction as $\hat{\mathcal{H}}_{\mathrm{int}} \propto \hat{a}^\dagger \hat{a} (\hat{b}-\hat{b}^\dagger)$. Such pressure-like interaction has been widely employed recently to study the entanglement between magnons and phonons, as shall be introduced in Sec. \ref{mpp_entanglement}.

\subsection{Magnon-photon interaction} \label{magnon_photon_interaction}
Magnons can be coupled to photons ranging from microwave, terahertz wave to the optical wave regime. When the magnon frequency matches the frequency of photon, they can couple directly through Zeeman interaction due to the mode overlap between spin-wave and magnetic component of electromagnetic wave inside the magnet. This type of coupling is linear in both magnon and photon operator, and it is the basis of the rising field of cavity magnonics \cite{BabakReview2022}. Depending on the setups to generate photons, both coherent and dissipative couplings can be realized, which correspond to the energy level repulsion and attraction between the two modes, respectively \cite{HarderPRL2018,YuPRL2019}. When the photon frequency is much higher than the magnon frequency, the Zeeman coupling between the spin and magnetic component of electromagnetic wave will be suppressed and the indirect coupling between spin and electric components via spin-orbit interaction becomes important. Following the standard approach \cite{BorovikPR1982}, the magnetic ordering will change the dielectric permittivity of the material as
\begin{equation}
\Delta \epsilon_{jk} = f_{jkp}m_p + g_{jkpq}m_pm_q + ...
\end{equation}
In principle, one can expand the dielectric permittivity to arbitrary orders of the magnetization. Here we focus on the first two terms for their sufficiency to consider the widely studied Faraday effect and Brillouin light scattering.

In a non-absorbing media, the tensor $\epsilon_{jk}$ is Hermitian, i.e., $\epsilon_{jk} =\epsilon^*_{kj}$. According to the Onsager principle, $\epsilon_{jk}(\mathbf{H})=\epsilon_{jk}(-\mathbf{H})$. These two constraints readily imply that $\epsilon_{jk}$ is a purely imaginary anti-symmetric tensor with three independent variables. By further imposing the constraint of crystal symmetry, the number of independent variables can be further reduced. Taking the example of a cubic crystal \cite{BorovikPR1982}, $\Delta \epsilon$ is of the form
\begin{equation}\label{dielectric}
\Delta \epsilon=\left(\begin{array}{ccc}
  0 & ifm_z & -ifm_y \\
  -ifm_z & 0 & ifm_x \\
  ifm_y & -ifm_x & 0
\end{array}\right ).
\end{equation}

Now the magnet-optical energy of the system can be written as
\begin{equation}\label{Ham-magneto-optical}
\hat{\mathcal{H}} = \frac{1}{2}\int \left [ \hat{\mathbf{E}}\cdot (\epsilon_0 + \Delta \epsilon) \cdot \hat{\mathbf{E}}^* + \mu_0 \hat{\mathbf{H}} \cdot \hat{\mathbf{H}}^* \right ]d \mathbf{r},
\end{equation}
where the electric component of electromagnetic wave reads
\begin{equation} \label{emEfield}
\mathbf{E} = i \sum_\mathbf{k} \sqrt{\frac{\omega_\mathbf{k}}{2\epsilon_0}} [\hat{c}_\mathbf{k} \mathbf{u}_k(\mathbf{r}) e^{-i\omega_\mathbf{k} t} - \hat{c}_\mathbf{k}^\dagger \mathbf{u}_k^*(\mathbf{r}) e^{i\omega_\mathbf{k} t}],
\end{equation}
where $\hat{c}_\mathbf{k}$ and $\hat{c}_\mathbf{k}^\dagger$ are the annihilation and creation operators of photons and $\mathbf{u}_\mathbf{k}$ depends the polarization vector and propagation factor of photons. By substituting Eqs. \eqref{HP_mfield}, \eqref{dielectric} and \eqref{emEfield} into the Hamiltonian \eqref{Ham-magneto-optical} and keeping the linear terms of magnon operator, we obtain $\hat{\mathcal{H}} = \sum_\mathbf{k} \omega_\mathbf{k}  \hat{c}_\mathbf{k}^\dagger  \hat{c}_\mathbf{k} +  \hat{\mathcal{H}}_1+  \hat{\mathcal{H}}_2$. Here $\hat{\mathcal{H}}_1$ reads
\begin{equation}
\hat{\mathcal{H}}_1 = \sum_{\mathbf{k},\mathbf{k}'} ( g_{\mathbf{k},\mathbf{k}'} \hat{c}_\mathbf{k}\hat{c}_{\mathbf{k}'}^\dagger + \mathrm{h.c.} ),
\end{equation}
where $g_{\mathbf{k},\mathbf{k}'}= if\sqrt{\omega_k \omega_{k'}}/(4\epsilon_0) \int  (\mathbf{u}_k \times \mathbf{u}_{k'}^*)_zd\mathbf{r}$. This term describes the polarization change of incident photons when they pass through a magnetic media, which is known as Faraday effects. It is independent of magnon excitation and only depends on the strength and direction of static magnetization.

The higher-order term reads
\begin{equation}
\hat{\mathcal{H}}_2 = \sum_{\mathbf{k},\mathbf{k}',\mathbf{k}''} \hat{c}_{\mathbf{k}'}\hat{c}_{\mathbf{k}''}^\dagger \left (G_{\mathbf{k}',\mathbf{k}''}^-\hat{a}_\mathbf{k}+G_{\mathbf{k}',\mathbf{k}''}^+\hat{a}_\mathbf{k}^\dagger \right),
\end{equation}
where
\begin{equation}
G_{\mathbf{k}',\mathbf{k}''}^\pm = \frac{i f \sqrt{\omega_{k'} \omega_{k''}}}{4\epsilon_0\sqrt{2NS}} \int u^\pm (\mathbf{k}',\mathbf{k}'') e^{\mp i\mathbf{k}\cdot \mathbf{r}} d\mathbf{r},
\end{equation}
with $u^\pm (\mathbf{k}',\mathbf{k}'') = (\mathbf{u}_{\mathbf{k}'} \times \mathbf{u}_{\mathbf{k}''}^*)_x \pm i(\mathbf{u}_{\mathbf{k}'} \times \mathbf{u}_{\mathbf{k}''}^*)_y$. $\hat{\mathcal{H}}_2$ describes the splitting of an incident photon into a magnon and a lower-energy photon (Stokes process), and the confluence of a photon and a magnon into a higher-energy photon (anti-Stokes process), as shown in Fig. \ref{three_four_magnon}(d). This process conserves the total momentum and energy of magnons and photons similar to the three-magnon process. Through this interaction, one can know the energy and momentum of magnons by analyzing the incident and scattering photons, which is the basis of the powerful technique called Brillouin Light Scattering Microscopy\cite{SebastianFP2015}. Note that the Raman scattering of photons by the magnons share a similar principle, and allows the detection of magnons in the terahertz regime \cite{FleuryPR1968}, while Brillouin light scattering Microscopy usually works in the gigahertz regime. Furthermore, such nonlinear interaction is important to generate the quantum states of magnons, as we shall see in Sec. \ref{magnon_qm_state}.




\section{Classical region}

\subsection{Parametric excitation of magnons}\label{parametric_magnon}

The most common method of spin-wave excitations is the magnetization oscillation driven by a microwave field created by a microstripe antenna. At a low driving power, this method only excites spin waves with the wavelength larger than the width of the antenna and frequency equal to that of the microwave field, and it is unable to excite short-wavelength spin waves, which limits its applicability in magnonic devices based on short-wavelength spin waves.

The inherent nonlinearity of magnetic dynamics (or the Landau-Lifshitz-Gilbert equation) leads to the spin-wave interactions, resulting in a number of nonlinear phenomena \cite{RezendePIEEE1990}. Of particular interest is the parametric excitation of magnons, which can excite spin waves with a wide range of wavelength including the exchange spin waves with short wavelength \cite{SandwegPRL2011,KurebayashiAPL2011}. Parametric pumping of magnons was first observed in experiments on ferromagnetic resonance (FMR) under a strong microwave field \cite{BloembergenPRB1952,BloembergenPR1954} and then explained by the theory of the spin-wave instability \cite{AndersonPR1955}. Using various spin-wave instability processes, magnons with selected frequency and wave vector can be excited.

\begin{figure}
  \centering
  \includegraphics[width=0.45\textwidth]{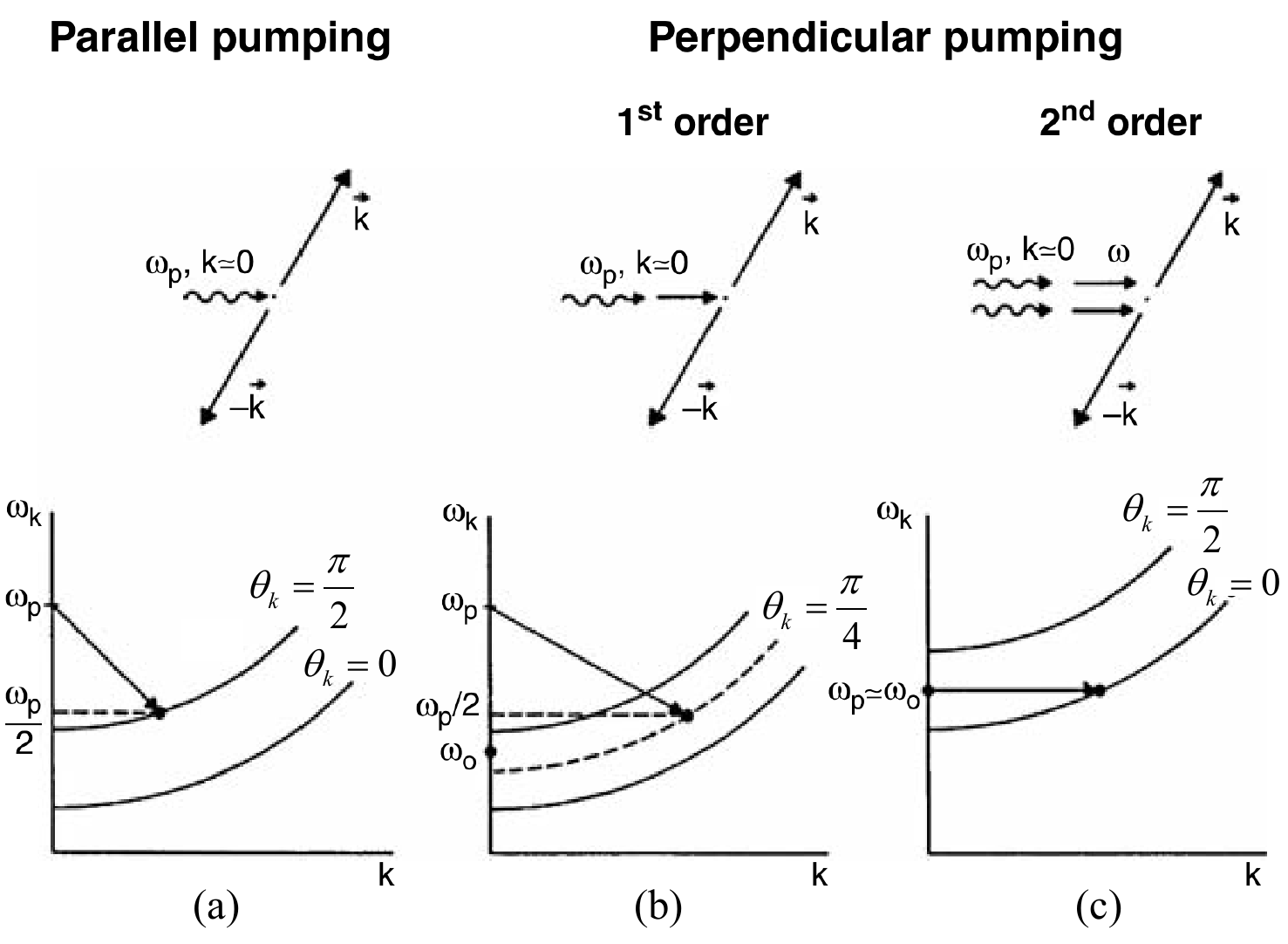}\\
  \caption{Illustration of spin-wave pumping processes \cite{RezendePIEEE1990}. (a) Parallel pumping. (b) Perpendicular pumping, first-order Suhl processes. (c) Perpendicular pumping, second-order Suhl processes. $\theta_{k}$ is the angle between the wave vector $\mathbf{k}$ and static magnetization $\mathbf{m}_{0}$.}\label{fig1}
\end{figure}

In the parallel pumping process \cite{SchlomannJAP1960,BracherPR2017}, a microwave field at a frequency $\omega_p$ is applied parallel to the static magnetization. Due to the ellipticity of magnetization precession, the pumping field directly couples to spin waves with frequency $\omega_k=\omega_p/2$. When the microwave field amplitude exceeds a certain threshold, the pumping energy compensates the spin-wave losses leading to the exponential growth of the spin-wave amplitude. In other words, a microwave photon with the frequency $\omega_p$ splits into a pair of magnons with opposite wavevectors $\pm\mathbf{k}$ and half the frequency of the pumping field $\omega_p/2$, following the conservation of energy and momentum, as illustrated in Fig. \ref{fig1}(a). Beside the critical amplitude of the pumping field, another condition of parallel pumping is that the half of the pumping frequency should be larger than the minimum frequency of the spin-wave band \cite{OkanoPRB2019}.

In the perpendicular pumping, the microwave field is perpendicular to the static magnetization and excite the uniform magnetization oscillation (FMR mode, $k=0$). Due to the spin-wave interactions, magnons with $k\neq0$ are coupled to the FMR mode excited by the pumping field. When the amplitude of the FMR mode is larger than a threshold value, the energy acquired by spin waves from the FMR mode exceeds the spin-wave losses and the spin-wave instability occurs \cite{SuhlJPCS1957}. Two main processes can be distinguished. In the first-order Suhl process, spin waves with $\omega_k=\omega_p/2$ and opposite wave vectors $\pm k$ can be excited parametrically by the FMR mode via three-magnon interactions, as shown in Fig. \ref{fig1}(b). In the second-order Suhl process, a pair of spin waves with the frequency $\omega_k=\omega_p$ and opposite wave vectors $\pm\mathbf{k}$ are driven by two FMR mode magnons via four-magnon interaction, as illustration in Fig. \ref{fig1}(c).

From the spin-wave excitation efficiency point of view, parallel pumping is superior to perpendicular pumping. This is because magnons are directly excited by the microwave photons in the parallel pumping process, while magnons are indirectly generated from the pumped FMR modes via magnon-magnon interactions in the perpendicular pumping process. Recently, it has been demonstrated that parametric excitation of magnons can also be realized by electric field \cite{VerbaPRA2014,ChenNL2017} and acoustic waves \cite{AlekseevAPL2020}.

\subsection{Bistability and multi-stability}
There is intrinsic magnetocrystalline anisotropy in magnetic materials~\cite{Gurevich,spaldin2010magnetic}. The corresponding anisotropic energy, i.e., the potential energy that needs to be overcome when the magnetization is deflected from the easy axis to the hard axis, is nonlinearly dependent on the magnetization. Usually, the leading contribution of the anisotropic energy can be expressed as the quadratic term of the magnetization~\cite{Gurevich}. This anisotropy naturally becomes the source of nonlinearity in magnetic materials. Further, these nonlinearities can serve as the basis for diverse magnetic control and functionalities.  In this section, we discuss the magnon Kerr nonlinear effect induced by magnetocrystalline anisotropy, and the bistability/multi-stability of the magnon polariton in a driven-dissipative nonlinear cavity magnonic system. In addition, we will present a selection of recent theoretical proposals exploiting the magnon Kerr nonlinear effect, such as non-reciprocity~\cite{PhysRevApplied.12.034001}, higher-order sidebands~\cite{Liu:18,PhysRevApplied.18.044074,PhysRevA.104.033708}, and chaos~\cite{Liu:19,Wang_2019} in the nonlinear cavity magnonics, as well as some other quantum proposals.

The spin-orbit interaction is the primary cause of magnetocrystalline anisotropy in magnetic materials~\cite{PhysRev.110.1341,doi:10.1063/1.2000425}. Magnetocrystalline anisotropy can be expressed as the magnon Kerr effect, which gives rise to the bistability and multi-stability of magnon modes. Bistability and multi-stability in this context indicate that when magnetic materials are strongly driven by microwaves, the excitation number of magnons in magnetic materials can exist in two or more stable states under the same driving condition. The formalism of the magnon Kerr effect is derived as follows. Without loss of generality, we concentrate on the magnon mode in a ferrimagnetic sphere, such as yttrium iron garnet (YIG).

Under the bias magnetic field $\mathbf{B}_{0}$, the Hamiltonian of the magnon mode $\hat{\mathcal{H}}_{\rm{m}}$ reads~\cite{Blundell01, Wang16}
\begin{equation}\label{YP-1}
	\hat{\mathcal{H}}_{\rm{m}}=-\int_{V_{\rm{m}}}\mathbf{M}\cdot\mathbf{B}_{0}d\tau
	-\frac{\mu_{0}}{2}\int_{V_{\rm{m}}}\mathbf{M}\cdot\mathbf{H}_{\rm{an}}d\tau,
\end{equation}
where the first term represents the Zeeman energy and the second term is the magnetocrystalline anisotropy energy. In Eq.~(\ref{YP-1}), $V_{\rm{m}}$ is the volume of the YIG sphere, $\mathbf{M}=(M_{x},M_{y},M_{z})$ is the magnetization of the YIG sphere, $\mu_{0}$ is the vacuum permeability, and $\mathbf{H}_{\rm{an}}$ is the anisotropic field due to the magnetocrystalline anisotropy in the YIG crystal.

We adopt the direction of the bias magnetic $\mathbf{B}_{0}$ as the $z$ direction  ($\mathbf{B}_{0}=B_{0}\mathbf{e}_{z}$). When the [100] crystal axis of the YIG sphere is aligned along the bias magnetic field, the anisotropic field is given by~\cite{stancil2009spin}
\begin{equation}\label{YP-2}
	\mathbf{H}_{\rm{an}}=\frac{2K_{\rm{an}}M_{z}}{\mu_{0}M_s^{2}}\mathbf{e}_{z},
\end{equation}
where $K_{\rm{an}}$ is the first-order magnetocrystalline anisotropy constant and $M_s$ is the saturation magnetization. The second-order magnetocrystalline anisotropy is small, which can be approximately ignored even at high drive power before entering the Suhl instability regime~\cite{PhysRev.100.1788,stancil2009spin}. Then, the Hamiltonian of the magnon mode is
\begin{equation}\label{YP-3}
	\hat{\mathcal{H}}_{\rm{m}}=-B_{0}M_{z}V_{\rm{m}}-\frac{K_{\rm{an}}M_{z}^2}{M_s^{2}}V_{\rm{m}}.
\end{equation}
Using the macrospin operator $\mathbf{S}=\mathbf{M}V_{\rm{m}}/\gamma\equiv(S_{x},S_{y},S_{z})$~\cite{Soykal10}, where $\gamma$ is the gyromagnetic ratio, the Hamiltonian $\hat{\mathcal{H}}_{\rm{m}}$ can be written as
\begin{equation}\label{YP-4}
	\begin{aligned}
		\hat{\mathcal{H}}_{\rm{m}}=-\gamma B_{0}S_{z}-\frac{ K_{\rm{an}}\gamma^2}{M_s^{2}V_{\rm{m}}}S_{z}^2.
	\end{aligned}
\end{equation}
It has been found that the magnetocrystalline anisotropy energy can be expressed as a quadratic term of magnetic moment, indicating that it may be a nonlinear source.

To define the magnon Kerr effect, we set the Kerr coefficient as
\begin{equation}\label{sup6}
	\begin{aligned}
		K=-\frac{K_{\rm{an}}\gamma^{2}}{M_s^{2}V_{\rm{m}}}.
	\end{aligned}
\end{equation}
With the parameters $M_s=143.2~\rm{kA/m}$, $K_{\rm{an}}=-610~\rm{J/m^3}$ at the room temperature~\cite{stancil2009spin}, $\gamma/2\pi=28~\rm{GHz/T}$, and $V_{\rm{m}}=4.19\times{10}^{-9}~\rm{m^{3}}$ for the YIG sphere of 1 mm in diameter, we obtain the Kerr coefficient $K/2 \pi=0.0295~{\rm{nHz}}$ for the [100] crystal axis of the YIG sphere aligned along the bias magnetic field.

The magnitude of the nonlinear coefficient reflects the effect of a single magnon excitation. Considering that the eigenfrequency of the magnon mode is around several gigahertz, this is a relatively weak nonlinear effect. Therefore, a significant magnon excitation is required for the magnon Kerr effect to manifest visibly. This modest Kerr nonlinear coefficient restricts its application in the quantum domain, as the nonlinear effect at low excitation numbers is nearly unobservable unless the size of the magnets falls into nanometer scale. Superconducting qubits~\cite{GU20171,doi:10.1063/1.5089550}, as a quantum device with strong nonlinearity, may supply the essential nonlinear source when coupled with magnon~\cite{doi:10.1126/science.aaa3693}, and serve as a tool for demonstrating diverse quantum effects~\cite{doi:10.1126/sciadv.1603150,doi:10.1126/science.aaz9236,xu2022quantum}.

In contrast to other nonlinear systems, the magnon Kerr nonlinear effect is distinguished by the fact that the magnon Kerr coefficient can be changed from positive to negative by rotating the YIG sphere to adjust the angle between the crystal axis and the bias magnetic field~\cite{Gurevich}. This characteristic may have specific applications.

Equation (\ref{YP-4}) does not make apparent the effect generated by the Kerr nonlinearity when pumping the YIG sphere.  In fact, when a large number of magnon excitations are created, the magnon Kerr nonlinearity causes the frequency shift of the magnon modes. It will make the physical picture clearer if the relevant Hamiltonian is described by the magnon creation ($\hat{m}^{\dagger}$) and annihilation ($\hat{m}$) operators. To this end, we employ the Holstein-Primakoff transformation~\cite{HP_PR_1940}: $\hat{S}^{+}=\sqrt{2 S-\hat{m}^{\dagger} \hat{m}}\hat{m}$, $\hat{S}^{-}=\hat{m}^{\dagger} \sqrt{2 S-\hat{m}^{\dagger} \hat{m}}$, and $\hat{S}_{z}=S-\hat{m}^{\dag}\hat{m}$, where $S$ is the total spin of the sample. The Hamiltonian of the nonlinear magnonic system becomes
\begin{eqnarray}\label{YP-6}
	\begin{aligned}
		\hat{\mathcal{H}}_{\rm{m}}=\omega_{m} \hat{m}^{\dagger} \hat{m}+K \hat{m}^{\dagger} \hat{m} \hat{m}^{\dagger} \hat{m},
	\end{aligned}
\end{eqnarray}
where $\omega_{m}$ is the frequency of the magnon mode and $K$ is the Kerr nonlinear coefficient. Under the mean-field approximation, the mean excitation number of magnons in the YIG sphere can be determined using the frequency shift $\chi_{m}$ and Kerr coefficient $K$ via the relation $\chi_{m}=2K\langle \hat{m}^{\dagger} \hat{m}\rangle$. The experimentally detectable frequency shift acts as the method for observing the bistability of magnon modes.

\begin{figure*}[ht!]
	\includegraphics[width=0.88\textwidth]{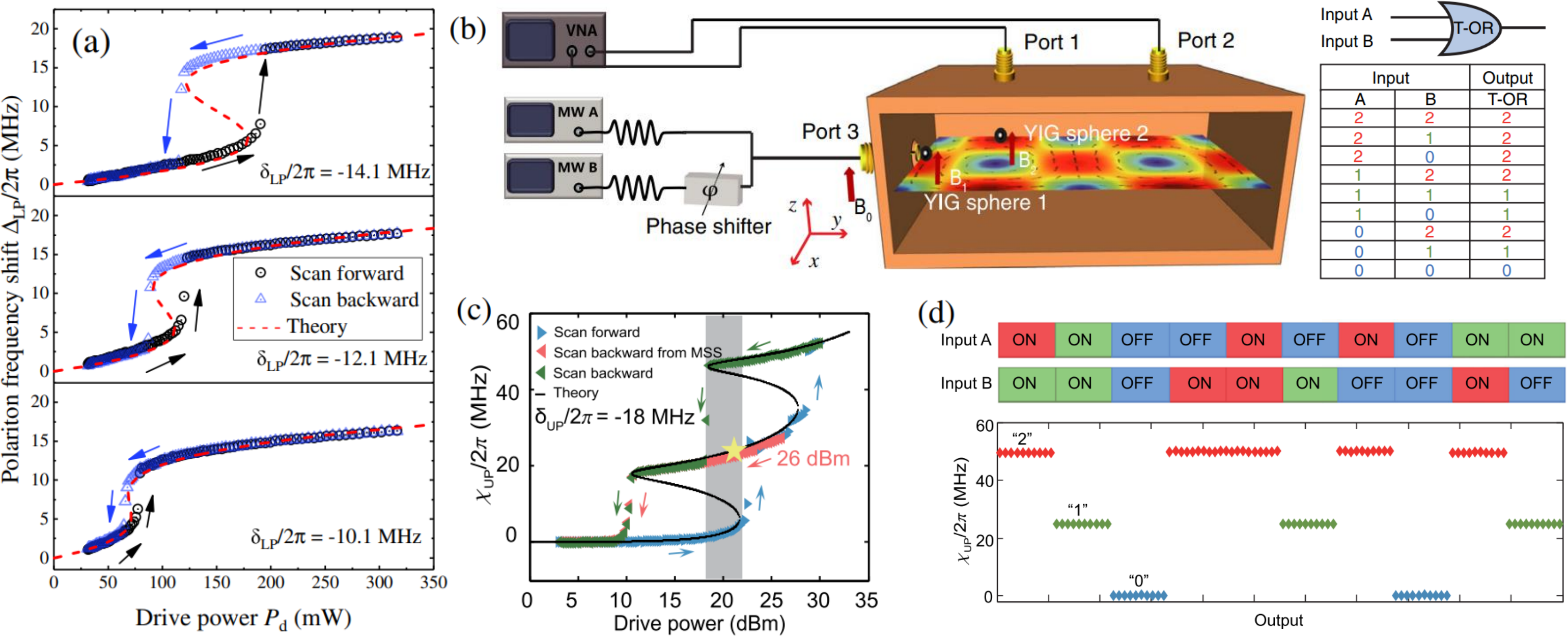}
	\caption{(a)~Bistability of the cavity magnon polaritons. From top panel to bottom panel, the bistable regions are adjusted by changing the drive detuning $\delta_{\rm{LP}}$. (b) Left: Schematic of the three-mode cavity magnonic system, which is employed to realize the multi-stability of the cavity magnon polaritons. The two YIG spheres are glued on the walls of a microwave cavity. The transmission of the system is measured by the vector network analyzer via ports 1 and 2. The drive field is loaded to port 3. Right: Symbol and truth table of the ternary OR (T-OR) gate. (c) Experimental and fitting results of the multi-stability. (d) Top panel: The sequences of the ternary logic OR gate A-B inputs. Bottom panel: The corresponding outputs of the logic gate. The well-separated output levels 0, 1, and 2 correspond to the zero, small, and large frequency shifts of the upper-branch CMP. (a) is reproduced with permission from Wang et al., Phys. Rev. Lett. 120, 057202 (2018). Copyright 2018 American Physical Society. (b-d) is reproduced with permission from Shen et al., Phys. Rev. Lett. 127, 183202 (2021). Copyright 2021 American Physical Society.}
	\label{fig:1}
\end{figure*}

Before observing bistability, a driving field must be applied to the magnon mode. In the experiment, a drive field of frequency $\omega_{\rm{d}}$ can be used to directly pump the YIG sphere. The corresponding interaction Hamiltonian is
\begin{equation}\label{YP-7}
	\begin{aligned}
		\hat{\mathcal{H}}_{\rm{d}}=\Omega_{\rm{s}}(\hat{S}^{+}+\hat{S}^{-})(e^{i \omega_{\rm{d}} t}+e^{-i \omega_{\rm{d}} t}),
	\end{aligned}
\end{equation}
where $\Omega_{\rm{s}}$ denotes the coupling strength between the drive field and the macrospin of the YIG sphere.

Moreover, placing the YIG sphere in the microwave cavity provides a more convenient way for monitoring. The cavity mode will coherently couple with the magnon mode in the YIG sphere. Then, we can read the information of the magnon mode through the cavity. The interaction Hamiltonian between the macrospin and cavity mode can be expressed as~\cite{Wang16, Soykal10}
\begin{equation}\label{YP-8}
		\hat{\mathcal{H}}_{\rm{int}}=g_{\rm{s}}(\hat{S}^{+}+\hat{S}^{-})(\hat{a}^{\dagger}+\hat{a}),
\end{equation}
where $\hat{a}^{\dag}(\hat{a})$ is the creation (annihilation) operator of the cavity photons and $g_{\rm{s}}$ is the coupling strength between the macrospin and the cavity mode. Further, expanding $\hat{S}^{+}=\sqrt{2 S-\hat{m}^{\dagger} \hat{m}}\hat{m}$ and $\hat{S}^{-}=\hat{m}^{\dagger} \sqrt{2 S-\hat{m}^{\dagger} \hat{m}}$ in Eqs.~(\ref{YP-7}) and (\ref{YP-8}) will give rise to higher-order terms of $\hat{m}^{\dag}\hat{m}$, which complicates the Hamiltonian of cavity-magnon coupling system. Fortunately, under the ordinary experimental conditions, such as when the drive power is 30 dBm, $\langle \hat{m}^{\dagger} \hat{m} \rangle \approx  10^{17}$~\cite{PhysRevLett.127.183202}. The excitation numbers are significantly smaller than the total spin $S$ of a 1 mm-diameter YIG sphere, $S=5.52\times 10^{18}$, calculated using $S=\rho V_{\rm{m}} s$, with $\rho=4.22\times 10^{27}~\rm{m^{-3}}$, $V_{\rm{m}}=\frac{4}{3}\pi R^3$, $R=1$~mm, and $s=\frac{5}{2}$~\cite{PhysRevLett.113.083603,PhysRevLett.113.156401}. Therefore, we get $\langle \hat{m}^{\dagger} \hat{m} \rangle/2S=0.89\%$. The proportion of the excited magnons relative to the total spin number is less than 1$\%$, hence higher-order terms can be neglected in the Holstein-Primakoff transformation. In this circumstance, we have $\hat{S}^{+}\approx \hat{m}\sqrt{2S}$ and $\hat{S}^{-}\approx \hat{m}^{\dag}\sqrt{2S}$. The total Hamiltonian of the system then becomes~\cite{PhysRevLett.120.057202,Zhang2019}
\begin{eqnarray}\label{YP-9}
	\hat{\mathcal{H}}&=&\omega_{\rm{c}}\hat{a}^{\dag}\hat{a}+\omega_{\rm{m}}\hat{m}^{\dag}\hat{m}+K\hat{m}^{\dag}\hat{m}\hat{m}^{\dag}\hat{m}+g_{\rm{m}}(\hat{a}^{\dag}\hat{m}+\hat{a}\hat{m}^{\dag})
	\nonumber\\
	&&+\Omega_{\rm{d}}(\hat{m}^{\dag}e^{-i\omega_{\rm{d}}t}+
	\hat{m}e^{i\omega_{\rm{d}}t}),	
\end{eqnarray}
where $g_{\rm{m}}$ is the coherent coupling strength between the magnon mode and the cavity mode, and $\Omega_{\rm{d}}$ is the drive-field strength.

We mentioned above that the magnon Kerr nonlinearity will cause the magnon frequency shift when a larger number of magnons are excited. Due to the coherent coupling between the cavity mode and the magnon mode, cavity magnon polariton (CMP) is formed. When the YIG sphere is driven, the frequency of the magnon polariton shifts accordingly. Using a quantum Langevin approach, a cubic equation for the cavity magnon polariton frequency shift $\Delta_{\rm{LP}}$ can be obtained~\cite{PhysRevLett.120.057202}
\begin{equation}\label{YP-10}
	\bigg[(\Delta_{\rm{LP}}+\delta_{\rm{LP}})^2+\bigg(\frac{\gamma_{\rm{LP}}}{2}\bigg)^{2}\bigg]\Delta_{\rm{LP}}-\kappa P_{\rm{d}}=0,
\end{equation}
where $\gamma_{\rm{LP}}$ is the damping rate of the {\it lower}-branch CMP and $\kappa$ is a coefficient characterizing the coupling strength between the drive field and the {\it lower}-branch CMPs. This cubic equation provides a steady-state solution for the frequency shift of the {\it lower}-branch CMPs as a function of both the drive-field frequency detuning $\delta_{\rm{LP}}$ and the drive power $P_{\rm d}$. Under appropriate conditions, it has three solutions, two of them stable and the additional one unstable. This corresponds to the bistability of the system. In Fig.~\ref{fig:1}(a), Wang et al. reported the experimental observation of polariton frequency shift bistability (blue dots). The theoretical results (dashed curves) obtained using Eq.~(\ref{YP-10}) fit the experimental results very well. The two stable solutions of $\Delta_{\rm{LP}}$ in Eq.~(\ref{YP-10}) correspond to {\it two} states of the system with {\it large} and {\it small} number of polaritons excited in the lower branch. Thus, in Fig.~\ref{fig:1}(a), each sharp switching of the frequency shift $\Delta_{\rm{LP}}$ is related to the transition between these two states. When the corresponding experimental parameters are adjusted, such as the drive detuning $\delta_{\rm{LP}}$ in Fig.~\ref{fig:1}(a), the size of the bistable region can be changed accordingly, showing the flexible adjustability of the nonlinear system. In fact, the system is a prototypical driven-dissipative system, and the bistable transition here corresponds to the driven-dissipative phase transition. Therefore, the nonlinear magnonic system provides an excellent experimental platform for studying driven-dissipative dynamics~\cite{PhysRevA.95.012128,PhysRevA.94.011401,PhysRevLett.116.235302}.

\begin{figure*}[ht!]
	\includegraphics[width=0.8\textwidth]{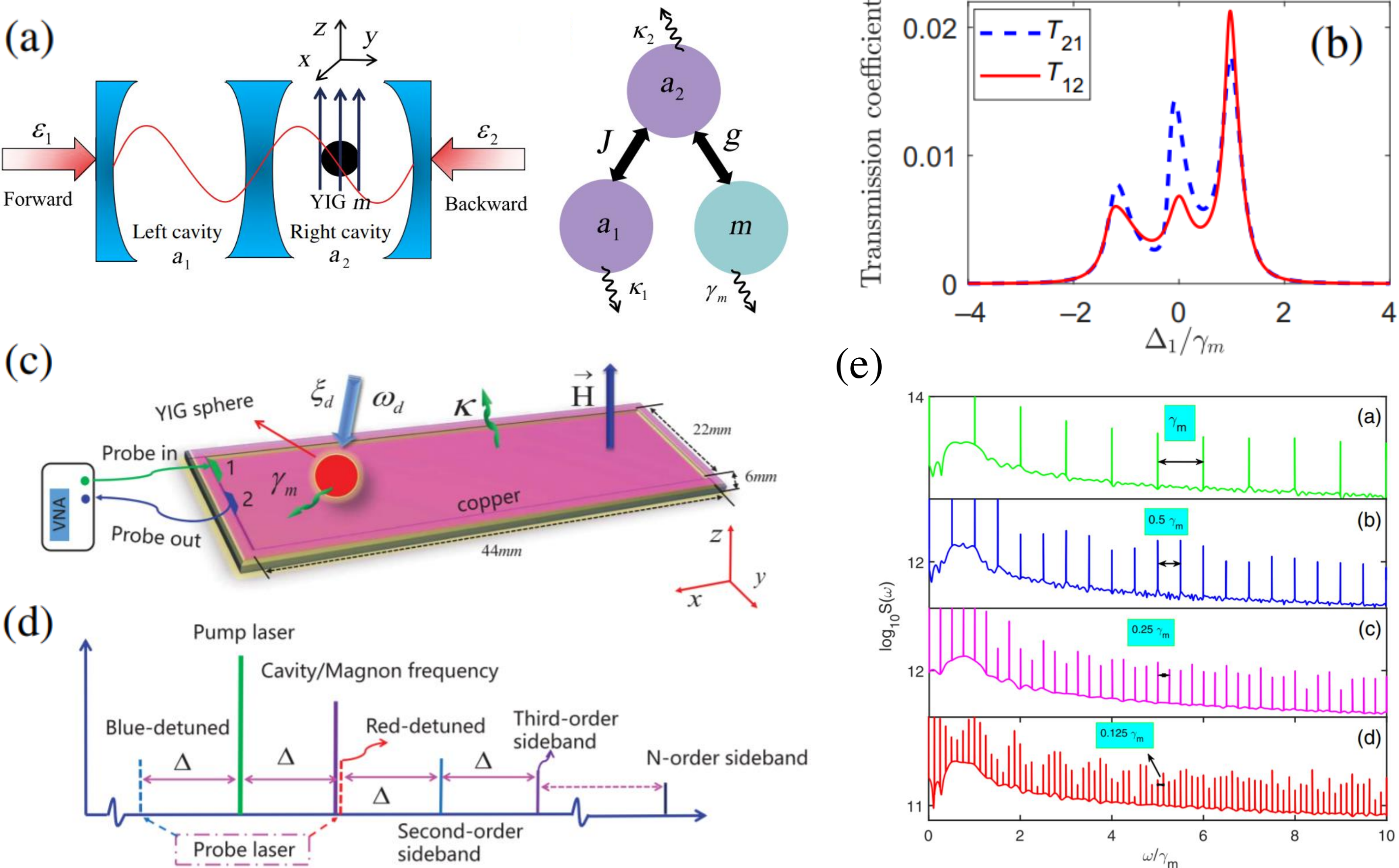}
	\caption{(a)~Schematic of the magnon Kerr effect-based non-reciprocal microwave transmission setup. The setup consists of two cavities and a YIG sphere, and the YIG sphere is placed in one of the cavities. (b) Theoretical results of the non-reciprocal transmission. (c) Schematic diagram of the cavity magnonic system for generating high-order sidebands. (d) Frequency spectrogram of the cavity magnonic system. The frequency of the pump laser is shown by the green line, while the purple line indicates the cavity (magnon) resonance frequency. The blue dashed (blue detuning) and red dotted lines (red detuning) represent the probe laser, and $\Delta$ is the beat frequency between the pump laser and the probe laser. (e) The magnon-induced high-order sideband generation spectra output with different beat frequencies $\Delta$. (a-b) is reproduced with permission from Kong et al., Phys. Rev. Appl. 12, 034001 (2019). Copyright 2019 American Physical Society. (c-e) is reproduced with permission from Liu et al., Opt. Lett. 44, 507 (2019). Copyright 2019 Optica Publishing Group.}
	\label{fig:2}
\end{figure*}

Bistability is a nonlinear effect with great application prospects. Two stable states can be defined as bits `0' and `1' carrying information, which can be applied to information storage and processing. In addition, bistable states can also be used in optical switches. Relevant demonstrations have been reported in a multi-stable cavity magnonic system~\cite{PhysRevLett.127.183202,PhysRevB.103.104411}. In the experiment~\cite{PhysRevLett.127.183202}, Shen et al. employed one cavity and two YIG spheres to achieve tristable states and demonstrates a ternary logic OR gate, as shown in Figs.~\ref{fig:1}(b-d). Experiments show that the cavity magnonic system possesses excellent switching characteristics and highly distinguishable logic states.

In addition to the self-Kerr effect of the magnon mode, there is another nonlinear effect when considering the phonon mode and the higher-order magnetostrictive effects in the YIG sphere. Due to the higher-order interaction terms, there will be a cross-Kerr effect between the magnon mode and the phonon mode~\cite{PhysRevLett.129.123601}. Shen et al. reported the simultaneous bistability of the frequency shift and linewidth of the phonon mode. This constitutes a novel method for manipulating phonon modes in magnetic materials.

Nonlinear effects as a resource can be utilized to implement a series of classical functions~\cite{rhoads2010nonlinear,duck2002nonlinear,doi:10.1063/1.5142397}. In the cavity magnonic system, relying on the bistability caused by the Kerr nonlinear effect, a theoretical strategy is provided for designing a non-reciprocal microwave transmission device with adjustable frequency and isolation~\cite{PhysRevApplied.12.034001}. The schematic is depicted in Fig.~\ref{fig:2}(a). The device consists of two microwave cavities and a YIG sphere, and the YIG sphere is placed in one of the microwave cavities. When microwave signals of the same power are incident from different cavities, the magnon excitation numbers will be substantially different, resulting in distinct magnon polariton frequency shifts and non-reciprocal microwave transmission.

Due to the presence of Kerr nonlinearity, in addition to the nonlinear frequency shift, a succession of nonlinear optical processes can be generated in magnetic material to generate new frequencies. Liu et al. proposed the generation of high-order sidebands in a cavity magnonic system with Kerr nonlinearity [Figs.~\ref{fig:2}(c)(d))]~\cite{Liu:18}. The authors also uncover that
the sideband spacing can be modified by adjusting the beat frequency between the pump laser and the probe laser [Fig.~\ref{fig:2}(e)], which is crucial for optical frequency metrology. The authors envision that these sidebands could be exploited for optical frequency metrology and optical communications. Recent theoretical research has determined that nonreciprocal higher-order sidebands can be produced in a two-cavity magnonic system by utilizing Kerr nonlinearity~\cite{PhysRevA.104.033708}. Besides, the nonlinearity of three-magnon process can be utilized to generate magnonic frequency comb, which shall be introduced in Sec. \ref{magnon_comb}.

Other theoretical works investigate the chaotic dynamics of the magnon mode induced by the intrinsic magnon Kerr nonlinearity~\cite{Liu:19,Wang_2019}. The authors find that the evolution of magnon mode experiences the transition from ordered to period-doubling bifurcation and finally enters chaos by adjusting the microwave driving power. The lifetime of transient chaos can be controlled by tuning the external magnetic field. These results pave the way towards exploring magnetic-field-controlled chaos and may find potential applications in secure communication.

Nonlinear effects are also a resource for implementing a variety of quantum effects~\cite{Chang2014,Du2021}. In nonlinear cavity magnonic system, some theoretical schemes have been proposed recently~\cite{PhysRevResearch.1.023021,PhysRevResearch.3.023126,PhysRevA.100.043831,PhysRevA.106.012419,doi:10.1063/5.0012072}. For example, it is reported that the quantum entanglement between magnon modes in two YIG spheres can be realized utilizing the magnon Kerr nonlinearity~\cite{PhysRevResearch.1.023021}. Single-photon sources are key aspects in quantum optics and quantum communications. Photons tend to reach detectors one by one. The photon blockade effect is a method for producing single-photon sources. Using the Kerr nonlinear effect, a corresponding photon blockade scheme is proposed~\cite{PhysRevA.100.043831}. In addition, the squeezed light is an important resource in quantum optics and quantum metrology. Theoretical schemes have proposed the generation of squeezed magnon modes utilizing the magnon Kerr effect in a cavity-magnon system~\cite{doi:10.1063/5.0012072,PhysRevB.105.245310}. More detailed introduction of the various quantum effects shall be given in Sec. \ref{qm_section}.

\subsection{Magnonic frequency comb} \label{magnon_comb}

Frequency comb is a spectrum consisting of a set of discrete and equally spaced spectral lines, which was originally proposed in optical systems \cite{UdemN2002}. Since its inception, optical frequency comb has dramatically improved the accuracy of frequency metrology and thus leads to the Nobel Prize in Physics in 2005. It has been widely used in many fields, such as the optical atomic clock \cite{LudlowRMP2015}, searching for dark matter \cite{BeloyN2021}, Earth-like planet hunting \cite{MetcalfO2019}, probing air pollution \cite{GiorgettaLPR2021}, following chemical reaction \cite{BjorkS2016}, and disease diagnosis \cite{ThorpeOE2008}. The success of optical frequency comb inspires scientists to search for frequency combs in other physical systems. Recently, an analogue of optical frequency comb in the acoustic system has been predicted theoretically \cite{CaoPRL2014} and realized in microscopic extensional mode resonators through a three-wave mixing process \cite{GanesanPRL2017}.
Magnons are akin to photons and phonons as bosonic quasiparticles and should have similar phenomena. However, there is no report about the magnonic frequency comb (MFC) for a long time. The main reason is that the nonlinearity playing a key role in the generation of frequency comb is very weak in a normal ferromagnet. Fortunately, it has been demonstrated that noncolinear magnetic textures can significantly enhance the nonlinear interaction in magnetic systems \cite{AristovPRB2016,ZhangPRB2018,KorberPRL2020}, which offers the possibility for the generation of magnonic frequency comb.

\begin{figure}[ht!]
  \centering
  \includegraphics[width=0.45\textwidth]{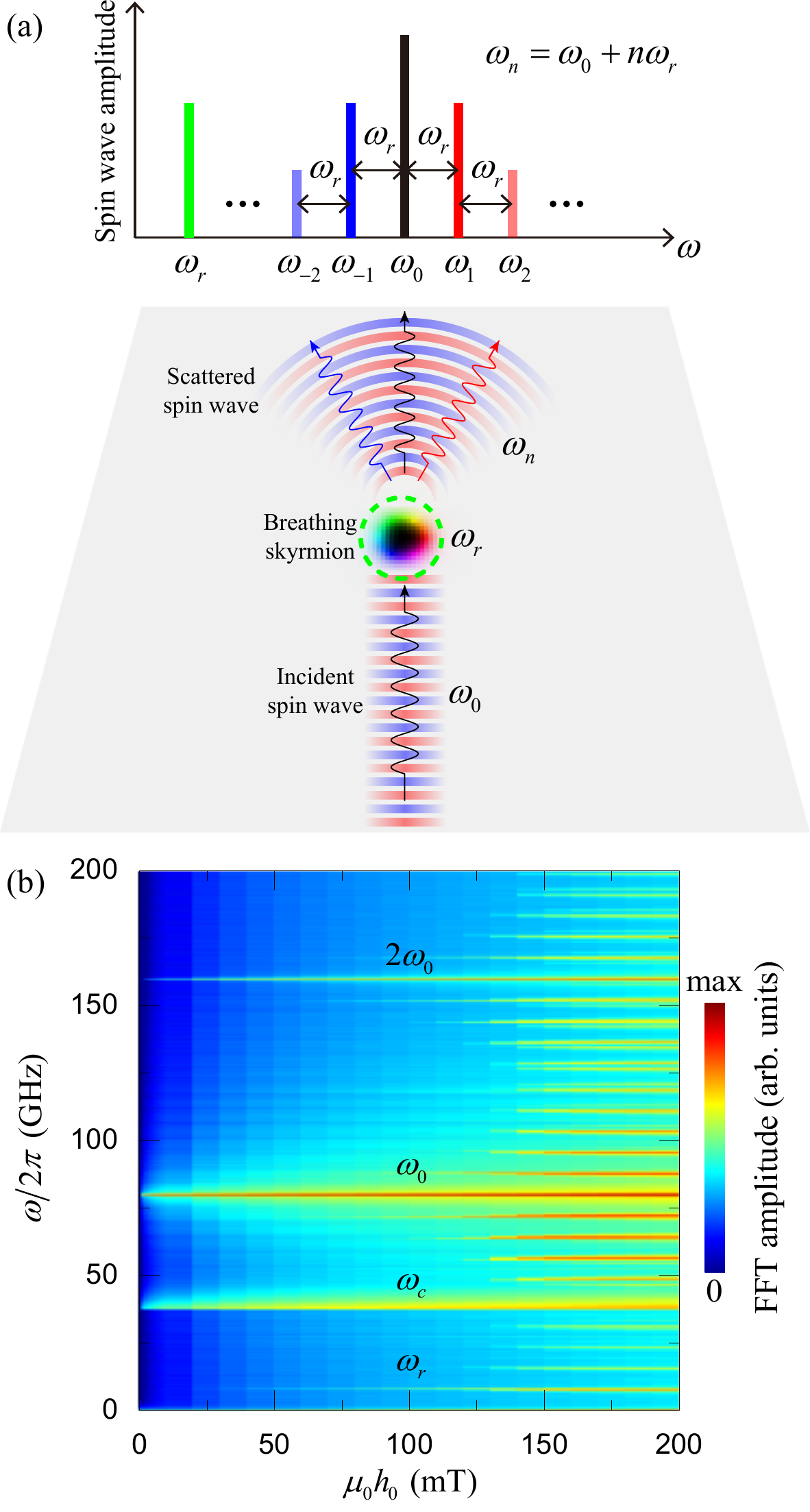}\\
  \caption{(a) Schematic illustration of nonlinear magnon-skyrmion scattering and the resulting MFC in a magnetic film. (b) The FFT amplitude of the system as a function of the driving field amplitude ($h_0$). The microwave field frequency is fixed at 80 GHz. Figures are reproduced with permission from Wang et al., Phys. Rev. Lett. 127, 037202 (2021). Copyright 2021 American Physical Society.}\label{fig2}
\end{figure}

In 2021, Wang et al. theoretically predicted the magnonic frequency comb effect by investigating the nonlinear magnon-skyrmion scattering in a chiral ferromagnetic film\cite{WangPRL2021}. The system Hamiltonian is written as
\begin{equation}
\hat{\mathcal{H}}=\int A(\nabla \mathbf{m})^2 - Km_z^2 +D[ m_z \nabla \cdot \mathbf{m} - (\mathbf{m} \cdot \nabla )m_z]d\mathbf{r},
\label{ham}
\end{equation}
where $\mathbf{m}$ is the normalized magnetization, $A$ is the exchange stiffness, $K$ is the easy-axis anisotropy coefficient, and $D$ is the strength of DMI. By Holstein-Primakoff transformation, the Hamiltonian can be rewritten as magnon bosonic operators ($\hat{a}_{k}, \hat{a}_{k}^{\dagger}$). The third-order term corresponds to three-magnon process and dominates in the MFC generation, as already introduced in Sec. \ref{mm_interaction}. Based on the quantized Hamiltonian, the dynamical equation of spin-wave modes in three-magnon process is derived as
\begin{subequations}
\begin{align}
&i\frac{d\hat{a}_k}{dt}=(\Delta_k-i\alpha_k\omega_k) \hat{a}_k + g_q \hat{a}_r \hat{a}_q + g_p \hat{a}_r^\dagger \hat{a}_p + h_0, \\
&i\frac{d\hat{a}_r}{dt}=(\omega_r -i\alpha_r\omega_r)\hat{a}_r + g_q \hat{a}_k \hat{a}_q^\dagger + g_p \hat{a}_k^\dagger \hat{a}_p,\\
&i\frac{d\hat{a}_p}{dt}=(\Delta_p -i\alpha_p\omega_p)\hat{a}_p + g_p \hat{a}_k \hat{a}_r, \\
&i\frac{d\hat{a}_q}{dt}=(\Delta_q -i\alpha_q\omega_q)\hat{a}_q + g_q \hat{a}_k \hat{a}_r^\dagger,
\end{align}\label{heisen}
\end{subequations}
where $\hat{a}_{k,r,p,q}$ are incident magnon, skyrmion breathing, sum-frequency, and difference-frequency modes with the frequencies $\omega_{k},\omega_{r},\omega_{p}=\omega_{k}+\omega_{r},\omega_{q}=\omega_{k}-\omega_{r}$, respectively. The incident spin waves are excited by a microwave field with the amplitude $h_0$ and frequency $\omega_0$. $g_p$ and $g_q$ are, respectively, the coupling strength of three-magnon confluence and splitting processes. $\Delta_{k,p,q}=\omega_{k,p,q}-\omega_0$ are the detuning terms and $\alpha_{k,r,p,q}$ are the damping rates of the corresponding spin-wave modes. The driving field threshold can be calculated from Eq. (\ref{heisen}), written as
\begin{equation}
h_{c}=\frac{\alpha ^2 \omega_0\sqrt{2\omega_0^3+\omega_0^2\omega_r-2\omega_0\omega_r^2-\omega_r^3}}{g\sqrt{4\omega_0 -2\omega_r}},
\end{equation}
where we have assumed the confluence and splitting processes have the same coupling strength $g$, and all spin-wave modes have the identical damping rate $\alpha$.

Above the threshold field amplitude ($h_c$), the internal breathing mode of skyrmion would be excited. Then, the breathing mode combines with the incoming spin wave to produce both sum-frequency and difference-frequency modes via three-magnon processes. The new-generated secondary signals can further hybrid with the breathing mode to generate more high-order frequency modes. Finally, a cascade of spin-wave mode excitations result in the generation of magnonic frequency comb, as illustrated in Fig. \ref{fig2}(a). Micromagnetic simulation results verify the above theoretical prediction [see Fig. \ref{fig2}(b)].

The mode spacing of the magnonic frequency comb is equal to the breathing frequency of skyrmion, which is inversely proportional to the square of the skyrmion size ($\omega_{r}\propto R_s^{-2}$) \cite{KravchukPRB2018}. Sun et al. numerically demonstrated that the tooth spacing of the magnonic frequency comb can be effectively modulated by a biaxial in-plane strain via magnetoelastic coupling \cite{SunAEM2022}. The physical principle of skyrmion-induced magnonic frequency comb can be extended to other magnetic textures. For example, Zhou et al. proposed a spin-wave frequency comb generated by interaction between propagating spin wave and oscillating domain wall \cite{ZhouJMMM2021}. Moreover, some other generation mechanisms of magnonic frequency comb were also proposed. Xiong presented a method for the generation of robust magnonic frequency comb in a YIG sphere via magnetostrictive effects \cite{XiongFR2022}. Hula et al. experimentally demonstrated the formation of spin-wave frequency combs based on the nonlinear four-magnon interaction in a 1D microstructured waveguide \cite{HulaAPL2022}. However, it is noted that the magnon spectrum induced by four-magnon processes is a continuous spectrum in two or higher dimensions, thus it is unable to generate spin-wave frequency combs \cite{WangPRL2021}.
Very recently, Rao et al. \cite{RaoPRL2023} reported the experimental observation of the mixing of the pump and probe and hence the magnon frequency comb in YIG spheres.

\begin{figure}[ht!]
  \centering
  \includegraphics[width=0.5\textwidth]{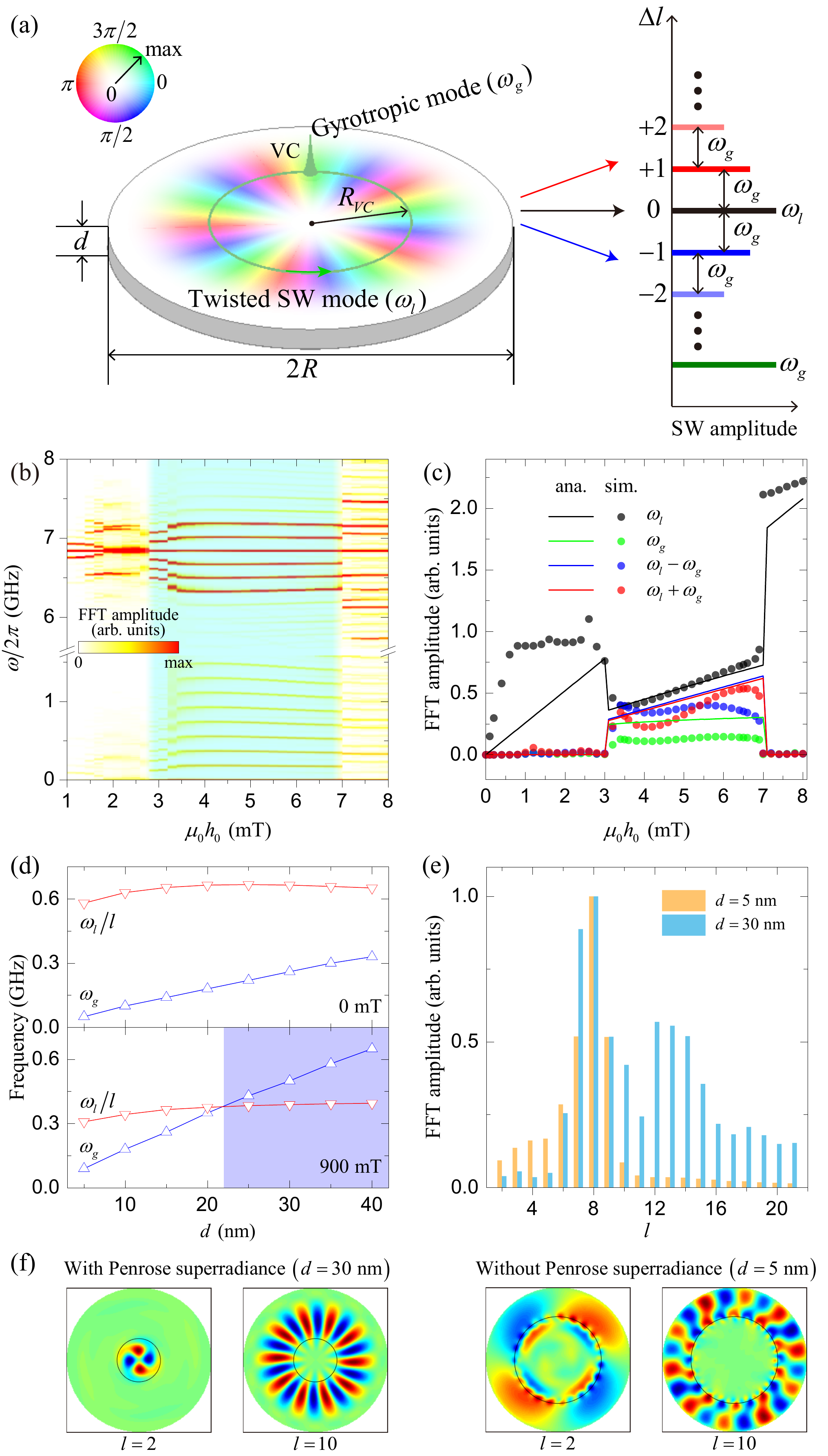}\\
  \caption{(a) Schematic illustration of nonlinear interactions between the VC and twisted spin waves and the resulting TMFC. (b) The FFT spectra of the vortex disk as a function of the driving field amplitude ($h_0$). (c) The amplitudes of four main peaks. Symbols are simulation results and curves are theoretical calculations. (d) The eigenfrequency of twisted magnon with $l=8$ and VC gyrotropic mode as a function of the disk thickness for two different bias fields. (e) Normalized amplitudes of the TMFC spectral lines under the bias field 900 mT. (f) Spatial distributions of magnons in the TMFC with and without Penrose superradiance. The black circles represent the VC gyrating orbit. Figures are reproduced with permission from Wang et al., Phys. Rev. Lett. 129, 107203 (2022). Copyright 2022 American Physical Society.}\label{fig3}
\end{figure}

Beside the spin angular momentum, magnon can also carry orbital angular momentum (OAM) in magnetic systems with rotational symmetry, such as the nanocylinder and nanodisk \cite{JiaNC2019,HuangJMMM2022}. Spin waves carrying OAM are dubbed as twisted magnons. The OAM carried by twisted magnons can take very large values and can be exploited for magnetic soliton manipulation \cite{JiangPRL2020} and multiplex magnonic communications \cite{JiaNC2019,JiaNCM2021}. To realize the OAM-based magnonic communications, similar to its optical counterpart \cite{WillnerAOP2015}, wide-band twisted magnons with adjustable multi-OAMs are desired. However, such twisted magnons are very difficult to be generated by the present exciting methods, such as applying an electric field through the Aharonov-Casher effect \cite{JiaNC2019,JiaNCM2021}, a spatially inhomogeneous microwave field \cite{JiangPRL2020,ChenAPL2020}, and a magnonic spiral phase plate \cite{JiaJO2019}. Generally, energy and momentum are conserved in three-magnon processes. Due to the unique rotational symmetry of twisted magnons, three-magnon processes of twisted spin waves should satisfy the conservation of energy and orbital angular momentum. Different spectral lines of twisted magnon frequency comb (TMFC) carry different OAMs, and would be a good magnonic source of twisted spin waves with multi-OAMs. Therefore, the question now becomes: How to generate a TMFC?

Previous studies on twisted magnons mainly focused on ferromagnetic systems with single domain, where the nonlinear three-magnon coupling is very weak. A noncolinear magnetic texture should be introduced to enhance the nonlinear coupling. Recently, three-magnon splitting has been observed in experiments, where a radial spin-wave mode splits to two azimuthal spin-wave modes with opposite OAMs \cite{SchultheissPRL2019}. The azimuthal spin-wave mode is actually a kind of twisted magnons. Thus, magnetic vortex is a good choice for generating TMFC.

In 2022, Wang et al. studied quantization effects of the nonlinear magnon-vortex interaction in ferromagnetic nanodisk and predicted an emerging TMFC generated by the three-magnon scattering of twisted magnons off the gyrating vortex core (VC) \cite{WangPRL2022}. There, the authors applied an in-plane rotating field to excite twisted spin waves. When the exciting field amplitude is above a critical value, the VC can be driven with a large-amplitude gyration by twisted magnons. Then, nonlinear interactions between the gyrating VC and twisted magnons generate the sum-frequency and difference-frequency modes. Further hybridizations produce higher-order spin-wave modes and finally lead to the formation of TMFC with the frequency spacing equal to the VC gyration frequency and the OAM difference by one between neighboring spectral lines, as illustrated in Fig. \ref{fig3}(a). The physical mechanism behind the TMFC generation is the same as that of the skyrmion-induced frequency comb. Numerical results agree well with the theoretical analysis and confirm the above physical pictures, as shown in Figs. \ref{fig3}(b) and \ref{fig3}(c).

Due to the frequency mismatch between TMFC modes and eigenmodes of magnetic vortex, the generated TMFC has limited comb lines and is not enough flat. To solve this issue, a mechanism called magnonic Penrose superradiance was recently proposed, which mimics the process that particles extract energy from a rotating black hole \cite{PenroseGRG2002}. The magnonic Penrose superradiance would occur when the phase velocity of twisted magnons is slower than the gyrating speed of the VC,
\begin{equation}\label{eq_Penrose}
  \frac{\omega_{l}}{l}<\omega_{g},
\end{equation}
where $\omega_{l}$ and $l$ are the frequency and OAM number of twisted magnons, and $\omega_{g}$ is the gyrotropic frequency of the VC. This condition can be derived theoretically from the dynamical susceptibility of the vortex disk in a rotating frame \cite{WangPRL2022}, written as
\begin{equation}\label{eq_susceptibility}
  \chi(\omega_{l})=\chi(0)\frac{\omega_{g}}{\omega_{g}-\omega^{\prime}-id_{\alpha}\omega^{\prime}},
\end{equation}
where $\chi(0)=2\gamma M_s/9\omega_g$ is the static susceptibility, $d_\alpha=-D/G$ is the effective damping parameter, and $\omega^{\prime}=\omega_l-l\omega_g$ due to the rotational Doppler frequency shift. When the damping term ($d_{\alpha}\omega^{\prime}$) becomes negative, an energy gain appears to amplify the twisted magnons. The condition of magnonic Penrose superradiance can be satisfied by applying a perpendicular bias field and increasing the disk thickness [see Fig. \ref{fig3}(d)]. For the case with Penrose superradiance, the higher-order modes of TMFC are significantly amplified, leading to a much more flat TMFC, shown in Fig. \ref{fig3} (e). The gyrating orbit of VC resembles an effective ergoregion, which traps the lower-order modes. In addition, the higher-order modes gain energy from the gyrating VC, therefore overcome the barrier to escape, as shown in Fig. \ref{fig3}(f). This finding opens a door to studying the fundamental physics of rotating black holes in magnonic platforms, which builds a bridge between microscopic magnonics and macroscopic cosmology.

\section{Quantum region} \label{qm_section}
With the rapid developments of both spintronics and quantum information science, the interdisciplinary field of quantum magnonics has stimulated widespread attention, which benefits from the unique advantages of the long lifetime of magnetic magnons and widely tunable coupling strength between magnons with various quantum platforms. It is worth noting that the definition of ``quantum magnonics" has slight differences in different literatures~\cite{Tabuchi2016,Lachance2019,Bunkov2020}. For example, Nakamura's group~\cite{Tabuchi2016,Lachance2019} refers the ``quantum magnonics" specifically to the branch concentrating on the interaction between magnetostatic modes and superconducting quantum circuits. Here we follow the broader definition put forward by Yuan et al.~\cite{YuanReview2022}, which points out that any study involving quantum states of magnons belongs to the category of quantum magnonics.

The reasons why we are especially interested in the field of quantum magnonics mainly consist of the following five points.
Firstly, both theoretical and experimental studies have confirmed that the magnons can be coupled to plentiful physical elements, including phonons~\cite{ZhangXufeng2016,LiJie2018PRL,DongChunhua2022PRL_MagOptomechanics}, microwave photons~\cite{Soykal2010,Hubel2013,Goryachev2014,ZhangXufeng2014,Tabuchi2014,Mergenthaler2017PRL_MagMicroPhoton,Bourhill2016PRB_MagMicroPhoton,Kostylev2016APL_MagMicroPhoton_SuperStrong}, optical photons~\cite{Haigh2015PRA_MO,Osada2016PRL_MO,ZhangXufeng2016PRL_MO,Silvia2016PRA_MO,Haigh2016PRL_MO,Osada2018PRL_MO,Parvini2019PRR_MO_antiferro}, as well as superconducting qubits~\cite{Tabuchi2015,Tabuchi2016,Lachance-Quirione2017NumberState,Lachance-Quirion2020SingleMagnon}. Therefore, quantum magnonic systems based on magnons offer great opportunities to serve as hybrid quantum systems possessing multitasking capabilities necessary and crucial for preferably implementing quantum information processing.
Secondly, the magnons can be coupled to the microwave cavity photons in the strong-coupling~\cite{Hubel2013,ZhangXufeng2014,Tabuchi2014,Mergenthaler2017PRL_MagMicroPhoton} even ultrastrong-coupling regime~\cite{Goryachev2014,Bourhill2016PRB_MagMicroPhoton,Kostylev2016APL_MagMicroPhoton_SuperStrong} owing to the distinguished advantage of the high spin density. This strong or ultrastrong light-matter interaction provide a promising experimental platform to deeply study novel phenomena that are in fairly contrast to optomechanics. For example, Yuan and Zheng et al. discovered a fancy entanglement enhancement phenomenon in the deep strong coupling regime~\cite{Yuan2020PRB,ZhengShasha2020SciChina}.
Thirdly, the frequency of the magnon mode is conveniently and widely tunable in the microwave range by adjusting the external magnetic field~\cite{Soykal2010,Hubel2013,ZhangXufeng2014,Tabuchi2014,BaiLihui2015SpinPumping,Chumak2015}, which is in distinct contrast to the case in optomechanical systems where the frequency of the mechanical mode is fixed. This flexible tunability of magnons offers great opportunities for experimentally observing quantum phenomena in hybrid magnonic systems.
Fourthly, magnons are ideal candidates for studies of macroscopic quantum effects, such as Schr\"{o}dinger cat states and quantum entanglement between massive objects. Macroscopic quantum systems can not only act as a fascinating platform to study fundamental quantum physics, such as large-scale quantum phenomena, quantum-classical transitions and quantum decoherence, but also hold great promise for applications in quantum memory, quantum networks, quantum computation, etc. Therefore, it has drawn widespread attention to investigate the characteristics of macroscopic quantum systems based on magnons in depth.
Last but not least, from the perspective of practical applications, hybrid quantum magnonic systems open novel avenues for the implementation of magnon gradient memory~\cite{ZhangXufeng2015NC_Memory}, random access memory~\cite{Kosub2017NC_Memory}, microwave-to-optical quantum transducers~\cite{Hisatomi2016PRB_MOTrans,ChaiChengzhe2022PhotonRes_MOTrans,DongChunhua2022PRL_MagOptomechanics}, and ultra-sensitive detection~\cite{Lachance2019}.

In this tutorial, the quantum magnonics first concentrates on the generation of non-classical state of a single magnon mode, such as single magnon state, magnon cat state;  in addition, magnonic in the quantum region simultaneously pays close attention to the quantum correlations between magnons with other typical quantum platforms including photons, phonons, and qubits. To distinguish with the recent review on quantum magnonics, we shall give more details and illustrations on how the nonlinear interactions in the hybrid magnonic platforms could generate magnon quantum states and the entanglement among magnons, photons and phonons.

\subsection{Quantum states of magnons}\label{magnon_qm_state}
\subsubsection{Single magnon state}\label{QuanMagnonics_SingleMagnonState}
A prior and crucial area of quantum magnonics is to deeply investigate the single magnon state. The concept of the single magnon state is an analogue of the single photon state widely studied in quantum optics.
Following the definition of single photon state in quantum optics~\cite{Lounis2005RPP_SinglePhoton,Milburn2015Review_SinglePhoton}, a single magnon state refers to a quantum state that one and only one magnon is included. This single magnon state is a pure quantum effect, which is the most critical step in the presentation of single magnon source. The quantum manipulation at the single magnon level is of great significance for the implementation of quantum information processing, quantum precise measurement, and quantum simulation, etc. In addition, single magnon states are fairly useful resources for investigating the fundamental quantum mechanism in macroscopic systems.

To characterize the generation of single magnon state, we need to investigate the statistical properties of magnons, which can be mathematically characterized by a second-order correlation function with zero-delay
\begin{equation}
g^{(2)}(0)=\frac{\left\langle \hat{m}^{\dagger} \hat{m}^{\dagger} \hat{m} \hat{m}\right\rangle}{\left\langle \hat{m}^{\dagger}\hat{m}\right\rangle^2},
\end{equation}
where $\hat{m}$ and $\hat{m}^{\dagger}$ represent the annihilation and creation operators of the magnon, respectively, without confusion with previous sections where we adopted $\hat{a}$ and $\hat{a}^\dagger$ for this purpose. The statistics of the magnon mode can be classified into three major types. (1) For the case of $g^{(2)}(0) > 1$, the magnon distribution is super-Poisson and the magnon bunching phenomenon emerges. (2) The condition $g^{(2)}(0) =1$ corresponds to the magnon Poisson distribution. (3) When $g^{(2)}(0) < 1$, the magnon distribution satisfies the sub-Poisson statistics, resulting in a nonclassical effect naming magnon antibunching. Particularly, the perfect single-magnon source is generated when $g^{(2)}(0) =0$.

Various proposals are presented for the preparation of a single magnon state originated from the nonlinear interaction between magnon and qubits, optical photons, etc. Here, we would like to enumerate some typical approaches classified by the contributing interaction responsible for the generation of single magnon state.

\begin{figure}[htbp!]
\centering
\includegraphics[width=0.98\columnwidth]{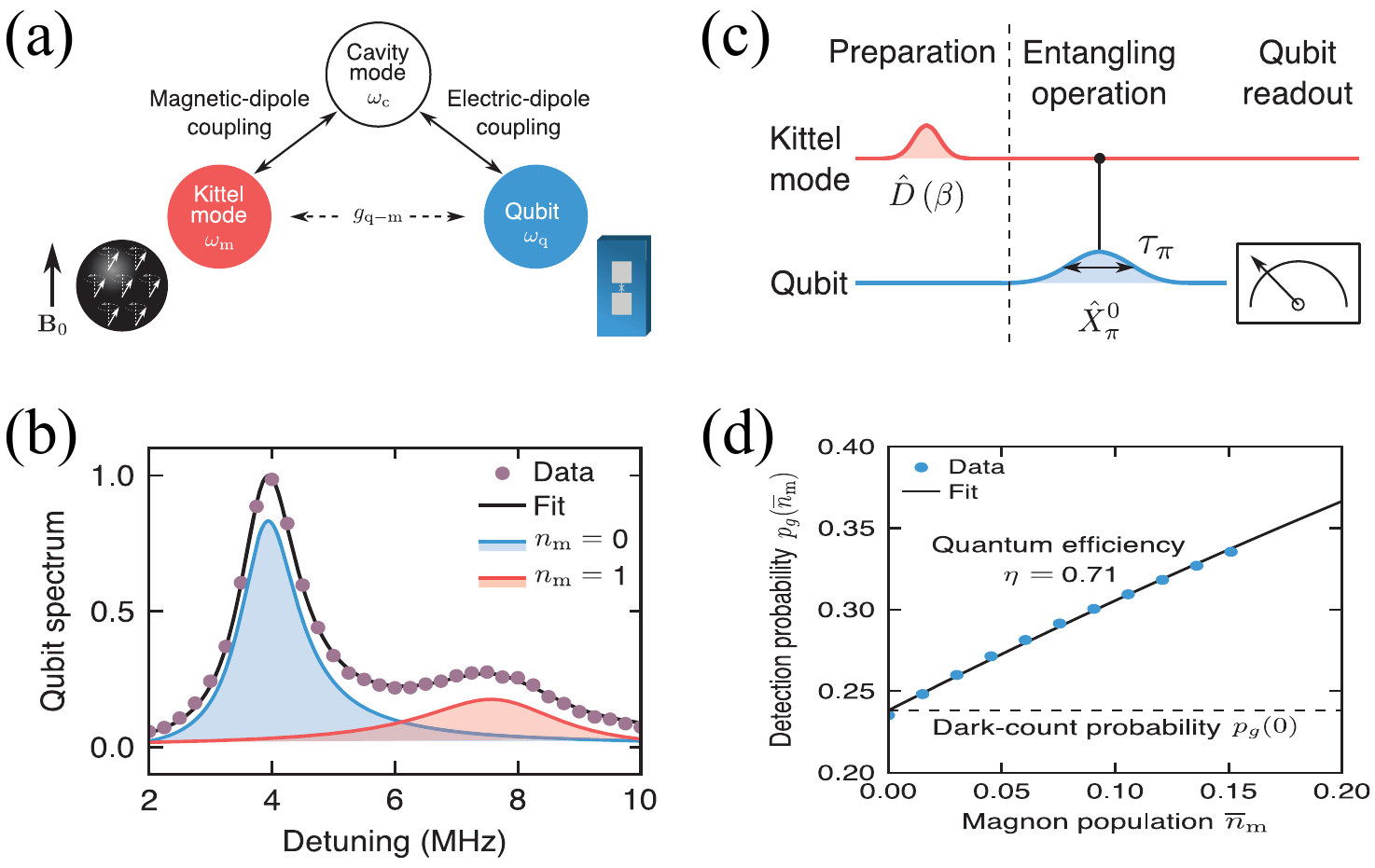}
\caption{(a) The schematic illustration of the hybrid magnon-qubit system mediated by the microwave photons~\protect\cite{Lachance-Quirion2020SingleMagnon}.
The frequency difference between the qubit and magnon is adjusted to be larger than the coherent magnon-qubit coupling strength to reach the non-resonant dispersive regime.
(b) The spectrum of the qubit. The blue curve together with the shaded area indicate the magnon vacuum state $|0\rangle$, while the red curve and shaded area denote the single magnon state $|1\rangle$.
(c) The protocol of detecting single magnon. The procedure mainly consists of three steps: the preparation of magnon coherent state, the generation of magnon-qubit entanglement through conditional excitation operation, the readout of the qubit.
(d) The detection probability $p_g\left(\bar{n}_{\mathrm{m}}\right)$ versus the magnon population $\bar{n}_{\mathrm{m}}$.
Reproduced with permission from Lachance-Quirion et al., Science 367, 425 (2020). Copyright 2020 The Author(s).
}
\label{Fig_QuanMagnonics_Lachance-Quirion2020Science_SingleMagnon}
\end{figure}

\textit{\textbf{Single-magnon state generation through nonlinear magnon-qubit interaction}}.
The most popular method of resolving single magnon seems to utilize the magnon-qubit interaction. First of all, it is necessary to clarify that the magnon mode can be coupled to the qubits through different kinds of interaction. For example, both theoretical and experimental studies show the magnons can be coherently coupled to the superconducting qubit described by a Jaynes-Cummings-type interaction $\hat{\mathcal{H}}_{\text {coh}}= g_{q m}\left(\hat{\sigma}_{+} \hat{m}+\hat{\sigma}_{-} \hat{m}^{\dagger}\right)$~\cite{Tabuchi2015,Tabuchi2016}. Besides, the dispersive coupling emerges when the magnon-qubit frequency difference is larger than the magnon-qubit coupling strength, where the dispersive Hamiltonian takes the form of $\hat{\mathcal{H}}_{\mathrm{disp}}=\chi_{mq} \hat{\mathrm{\sigma}}_z \hat{m}^{\dagger} \hat{m}$~\cite{Lachance-Quirione2017NumberState,Lachance-Quirion2020SingleMagnon}. Furthermore, the magnon can be directly coupled with transmon qubit through dipolar interaction
without the participation of the intermediate medium, such as the microwave photons. The direct magnon-qubit coupling comprises of both resonant magnon-qubit exchange interaction and nonlinear interaction, where $\hat{\mathcal{H}}_{\mathrm{dir}} = g_c\left(\hat{c}^{\dagger} \hat{m}+\hat{c} \hat{m}^{\dagger}\right)+g_{\mathrm{rp}} \hat{c}^{\dagger} \hat{c}\left(\hat{m}+\hat{m}^{\dagger}\right)$ as detailedly shown in Eq.~(\ref{QuanMagnonics_Kounalakis2022PRL_Hamiltonian_Component})~\cite{Kounalakis2022PRL_CatState}, where $\hat{c}$ ($\hat{c}^{\dagger}$) are the effective bosonic operators for the qubit system. These different types of magnon-qubit interaction are all possible of preparing a single magnon state, as detailly illustrated in the following.

We first focus on the single magnon state generated through nonlinear dispersive magnon-qubit coupling $\hat{\mathcal{H}}_{\mathrm{disp}}$.
As early as 2017, Lachance-Quirion and coauthors~\cite{Lachance-Quirione2017NumberState} exhibited the possibility of resolving magnon number states in a millimeter-sized ferromagnetic crystal coupled to a superconducting qubit mediated by the off-resonant dispersive coupling.
Furthermore, in the year of 2020, Lachance-Quirion et al.~\cite{Lachance-Quirion2020SingleMagnon} declared the experimental detection of a single magnon in a similar dispersive coupled magnon-qubit system, as indicated in Fig.~\ref{Fig_QuanMagnonics_Lachance-Quirion2020Science_SingleMagnon} (a). The single magnon state is achieved by firstly entangling the magnon with qubit and performing a single-shot measurement on the qubit. It is noticed that, in contrast to common coherent magnon-qubit interaction, the proposal in this work relies on the non-resonant dispersive coupling when the frequency detuning between the bare qubit and magnon mode is larger than the magnon-qubit coupling strength, i.e., $\omega_q-\omega_m\gg g_{mq}$. The magnon-qubit Hamiltonian in the dispersive regime takes the following form
\begin{equation}
\hat{\mathcal{H}}_{mq}^{\mathrm{disp}}=\chi_{mq} \hat{\mathrm{\sigma}}_z \hat{m}^{\dagger} \hat{m},
\end{equation}
where $ \hat{\mathrm{\sigma}}_z=|e\rangle\langle e|-|g\rangle\langle g|$ with $|g\rangle$ and $|e\rangle$ being the ground and excited state of the qubit, and $\chi_{mq}$ represents the magnon-qubit dispersive coupling strength. On the one hand, this dispersive coupling means that the effective frequency of the magnon is determined by the qubit state. On the other hand,  the magnon numbers would have an influence on the frequency of the qubit, which can thus be inferred through detecting the frequency shift of the qubit. Such dispersive coupling is confirmed by Fig.~~\ref{Fig_QuanMagnonics_Lachance-Quirion2020Science_SingleMagnon} (b).

The proposal of detecting single magnon is revealed in Fig.~\ref{Fig_QuanMagnonics_Lachance-Quirion2020Science_SingleMagnon} (c) mainly consisting of three steps. The first step is to prepare the magnon coherent state through a displacement operation $\hat{D}(\beta)=e^{\beta \hat{m}^{\dagger}-\beta^* \hat{m}}$, where such coherent  state can be expressed in terms of the magnon number states, i.e., $|\beta_m\rangle=\sum c_{n_{\mathrm{m}}}\left|n_{\mathrm{m}}\right\rangle$. Next, the conditional excitation operation $\hat{X}_\pi^0$ ($\pi$ pulse) is performed for the sake of entangling the magnon with qubit. The usage of this conditional excitation operation is reflected in conditionally exciting the qubit only with a magnon vacuum state, where the state is evolved to
\begin{equation}
\hat{X}_\pi^0|g \beta_m\rangle=c_0|e 0\rangle+\sum_{n_{\mathrm{m}}>0} c_{n_{\mathrm{m}}}\left|g n_{\mathrm{m}}\right\rangle.
\end{equation}
The last step is the qubit readout process through a high-power readout technique. The detection probability after three procedures reads
\begin{equation}
p_g\left(\bar{n}_{\mathrm{m}}\right)=\eta\left(1-e^{-\bar{n}_{\mathrm{m}}}\right)+p_g(0),
\end{equation}
where $\bar{n}_{\mathrm{m}}=|\beta|^2$, and $\eta$ denotes the quantum efficiency. The dark-count probability has been taken into account manifested by $p_g(0)$. Specifically, the detector clicks with a probability of $\eta+p_g(0)$ in the single-magnon case. Figure~\ref{Fig_QuanMagnonics_Lachance-Quirion2020Science_SingleMagnon} (d) displays the detection probability, in which the quantities of quantum efficiency and dark-count probability can be obtained as $\eta=0.71$ and $p_g(0)=0.24$. Then a single magnon state can be detected with a high probability of $\eta+p_g(0)=0.95$. This implies that the proposed single magnon detection method based on the entanglement and single-shot measurement utilizing the dispersive magnon-qubit coupling is effective. The single-magnon detection is an analogy of single-photon detector in quantum optics, which would be particularly promising for applications of quantum sensors.

With regard to the direct magnon-qubit coupling regime, the interaction Hamiltonian $\hat{\mathcal{H}}_{\mathrm{dir}} = g_c\left(\hat{c}^{\dagger} \hat{m}+\hat{c} \hat{m}^{\dagger}\right)+g_{\mathrm{rp}} \hat{c}^{\dagger} \hat{c}\left(\hat{m}+\hat{m}^{\dagger}\right)$ consists of a nonlinear term $\hat{\mathcal{H}}_{\mathrm{dir}}^{\mathrm{nl}} = g_{\mathrm{rp}} \hat{c}^{\dagger} \hat{c}\left(\hat{m}+\hat{m}^{\dagger}\right)$ which resembles the radiation pressure interaction in optomechanics. In the area of optomechanics, both theoretical~\cite{Galland2014PRL_PhononFock} and experimental~\cite{Riedinger2016Nature_PhononFock,Hong2017Science_PhononFock} studies have demonstrated the possibility of creating single photon state employing the nonlinear radiation pressure interaction under the regime of weak coupling and resolved sideband. In a similar way, the single magnon state can be analogously generated through the nonlinear radiation-pressure-like interaction $\hat{\mathcal{H}}_{\mathrm{dir}}^{\mathrm{nl}}$, however, the regime remains to be further confirmed as no one has worked on this before as far as we are concerned.

\begin{figure*}[htbp!]
\centering
\includegraphics[width=2\columnwidth]{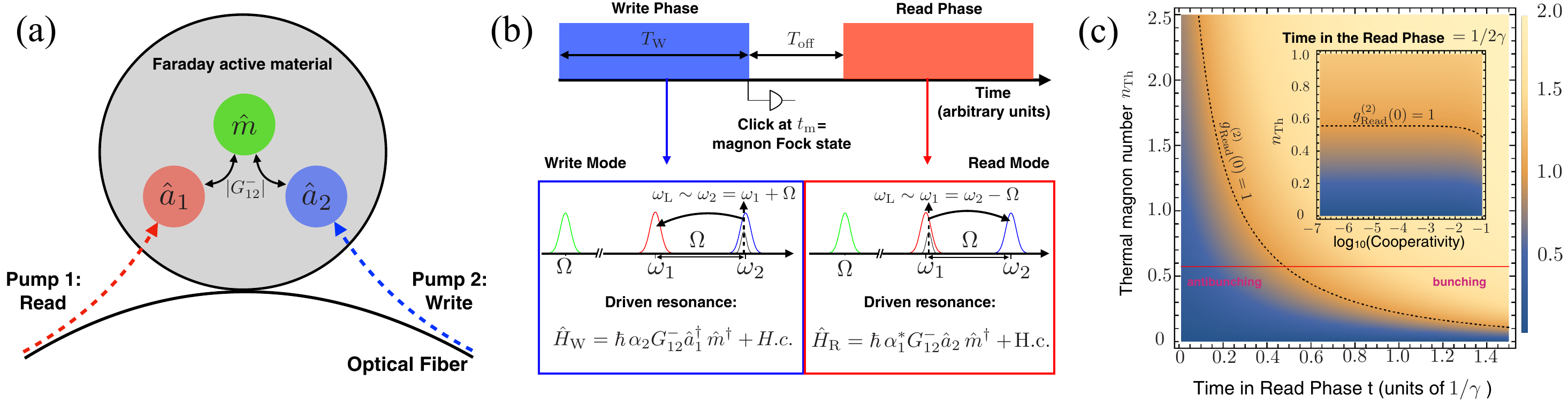}
\caption{(a) The schematic diagram of single magnon state generation through the nonlinear magneto-optical interaction~\protect\cite{Bittencourt2019PRA_MagnonFockState}. The magnon mode $\hat{m}$ are simultaneously coupled two optical modes $\hat{a}_1$ and $\hat{a}_2$. Each mode is individually driven to enhance the coupling strength.
(b) The simplified heralding protocol. The write pulse is firstly switched on to generate the two-mode squeezing interaction between $\hat{a}_1$ and $\hat{m}$ for a period of $T_W$, creating a single magnon state after a single photon measurement on $\hat{a}_1$. After a free evolution time $T_{\mathrm{off}}$, the read pulse is adopted by pumping cavity mode $\hat{a}_1$ to create beam-splitter-type interaction between mode $\hat{a}_2$ and $\hat{m}$ for purpose of reading out the single magnon state. (c) Second-order correlation function versus the duration time in reading process and thermal magnon number depending on the environmental temperature.
Reproduced with permission from Bittencourt et al., Phys. Rev. A 100, 013810 (2019). Copyright 2019 American Physical Society.
}
\label{Fig_QuanMagnonics_Silvia2019PRA_MagnonFockState}
\end{figure*}

Apart from the proposals based on the nonlinearity from the dispersive or direct nonlinear coupling, the well studied coherent magnon-qubit coupling without nonlinearity is the other helpful resource for generating a single magnon state. For example, Liu et al.~\cite{LiuZengXing2019PRB_MagnonBlockade}, Xie et al.~\cite{XieJiKun2020PRA_MagnonBlockade} and Xu et al.~\cite{XuYejun2021JOSAB_MagnonBlockade} theoretically demonstrated the possibility of generating a single magnon state employing the coherent magnon-qubit coupling in the form of Jaynes-Cummings coupling. These three schemes adopted a similar system comprising of a ferromagnetic YIG sphere and a superconducting qubit embodied in a microwave cavity, however, the underlying physical mechanisms to generate single magnon states are not the same. Specifically, Liu et al.~\cite{LiuZengXing2019PRB_MagnonBlockade} employed the conventional magnon blockade method, where the generation of a single magnon state originates from the anharmonicity mediated by the strong magnon-qubit coupling (i.e., the magnon-qubit coupling strength is much larger than the dissipation rates of magnon and qubit). The system is well described by the Hamiltonian $\hat{\mathcal{H}}_{\mathrm {eff}}=   \omega_q \hat{\sigma}_z/2+\omega_m \hat{m}^{\dagger} \hat{m}+g_{q m}\left(\hat{\sigma}_{+} \hat{m}+\hat{\sigma}_{-} \hat{m}^{\dagger}\right) +\Omega\left(\hat{\sigma}_{+} e^{-i \omega_d t}+\hat{\sigma}_{-} e^{i \omega_d t}\right)+\xi_p\left(\hat{m}^{\dagger} e^{-i \omega_p t}+\hat{m} e^{i \omega_p t}\right)$. In the strong coupling regime, the smallest second-order correlation function reaches $g^{2}(0)\approx 10^{-5}$, implying that a perfect single magnon source can be prepared~\cite{LiuZengXing2019PRB_MagnonBlockade}.
Furthermore, Xie et al.~\cite{XieJiKun2020PRA_MagnonBlockade} adopted the unconventional magnon blockade method, which utilizes the destructive quantum interference resulting from different pathways. The magnon-qubit coupling strength with smaller quantity is required to achieve magnon antibuching. Xu et al.~\cite{XuYejun2021JOSAB_MagnonBlockade} further showed that the magnon blockade effect can be implemented in a weak coupling regime when choosing appropriate parameters of the coupling strength and detuning.

\textit{\textbf{Single magnon state generation through nonlinear magneto-optical interaction}}.
In 2019, Bittencourt et al.~\cite{Bittencourt2019PRA_MagnonFockState} proposed a magnon-heralding scheme to implement a single magnon state by the measurement of an optical photon in a cavity optomagnonic system. The similar approach of preparing a single phonon state in quantum optomechanical systems has been confirmed feasible~\cite{Galland2014PRL_PhononFock,Riedinger2016Nature_PhononFock,Hong2017Science_PhononFock}.
In the protocol, two nondegenerate optical cavity modes $\hat{a}_1,~\hat{a}_2$ are simultaneously coupled to a magnon mode $\hat{m}$, playing the roles of the read mode and write mode, as illustrated in Fig.~\ref{Fig_QuanMagnonics_Silvia2019PRA_MagnonFockState}.
The interaction between these three modes takes the form of
\begin{equation}
\hat{\mathcal{H}}_{\mathrm{int}} \propto G_{12}^- \hat{a}_1^{\dagger} \hat{a}_2 \hat{m}^{\dagger}+\mathrm{h.c.},
\label{QuanMagnonics_Silvia2019PRA_Hamiltonian}
\end{equation}
with $G_{12}^-$ being the coupling strength, as introduced in Sec. \ref{magnon_photon_interaction}.
For the writing process, the optical mode $\hat{a}_2$ is required to be driven at frequency $\omega_L \approx \omega_2=\omega_1+\Omega$, where $\omega_1,~\omega_2,~\Omega$ are the frequency of modes $\hat{a}_1,~\hat{a}_2,~m$, respectively.
The frequency diagram is displayed in the left-hand panel of Fig.~\ref{Fig_QuanMagnonics_Silvia2019PRA_MagnonFockState}(b). The interaction Hamiltonian in Eq.~(\ref{QuanMagnonics_Silvia2019PRA_Hamiltonian}) is thus simplified as
\begin{equation}
\hat{\mathcal{H}}_{\mathrm{W}}\propto G_{\mathrm{W}} \hat{a}_1^{\dagger} \hat{m}^{\dagger}+\mathrm{h.c.},\label{QuanMagnonics_Silvia2019PRA_Hamiltonian_Write}
\end{equation}
 which is in the form of parametric-down-conversion-type coupling and enables the entanglement generation between optical mode $\hat{a}_1$ with $\hat{m}$. Starting from the initial ground state $|\psi_0\rangle=|0\rangle_{a_1}|0\rangle_{a_2}|0\rangle_m$, the state after a period $T$ becomes
 \begin{equation}
\left|\psi_{\mathrm{W}}(T)\right\rangle \simeq \frac{|0\rangle_{a_1}|0\rangle_{a_2}|0\rangle_m-\left(i G_{\mathrm{W}} T\right)|1\rangle_{a_1}|0\rangle_{a_2}|1\rangle_m}{\sqrt{1+\left|G_{\mathrm{W}}\right|^2 T^2}}.
\end{equation}
It can be apparently observed that the magnon state would evolve into a single magnon state $|1\rangle_m$ once the optical photons $\hat{a}_1$ is measured and yields the result of $|1\rangle_{a_1}$.

Subsequently, a reading pulse is introduced to read the state of the magnon. In terms of the reading process, the driving frequency of the optical mode $\hat{a}_1$ is chosen as $\omega_L \approx \omega_1=\omega_2-\Omega$, as manifested in the right-hand panel of Fig.~\ref{Fig_QuanMagnonics_Silvia2019PRA_MagnonFockState}(b). The interaction Hamiltonian in Eq.~(\ref{QuanMagnonics_Silvia2019PRA_Hamiltonian}) thus reads
\begin{equation}
\hat{\mathcal{H}}_{\mathrm{R}}\propto G_{\mathrm{R}} \hat{a}_2 \hat{m}^{\dagger}+\mathrm{h.c.},\label{QuanMagnonics_Silvia2019PRA_Hamiltonian_Read}
\end{equation}
which possesses the form of beam-splitter-type swap interaction and are thus able to read out the single magnon state.

Figure~\ref{Fig_QuanMagnonics_Silvia2019PRA_MagnonFockState}(c) shows the second-order correlation function of the read mode $g_{\text {Read }}^{(2)}(0)$ as a function of time duration of the reading process and the mean thermal magnon number. The red solid curve denotes the boundary where the heralded magnon number $n_m<1.1$. Therefore, the regime in the antibunching area below the horizontal red solid curve is capable of generating the single magnon state.

Notably, this magnon-heralding protocol to generate the single magnon state relies on the premise that the magnon mode is entangled with the photon mode resulting from the nonlinear magneto-optical interaction. The details of how entanglement is generated would be demonstrated in Sec.~\ref{QuanMagnonics_EntGeneration_MOptical}.

\begin{figure}[tbp!]
\centering
\includegraphics[width=0.98\columnwidth]{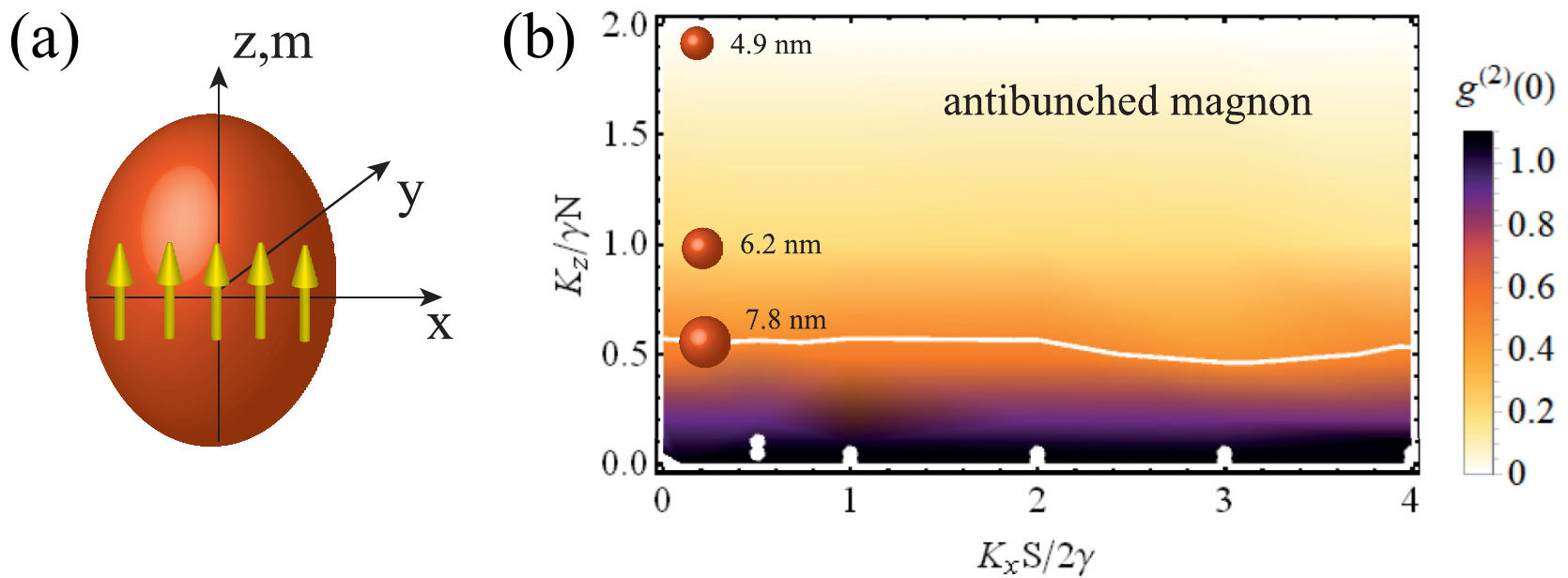}
\caption{(a) The schematic diagram of a biaxial nanomagnet~\protect\cite{YuanHY2020PRB_Antibunching}. $x$ and $z$ axises represent the hard- and easy-axis.
(b) The second-order correlation function versus easy- and hard-axis anisotropy coefficients. The white curve displays $g^{(2)}(0)=0.5$.
Reproduced with permission from Yuan et al., Phys. Rev. B 102, 100402(R) (2020). Copyright 2020 American Physical Society.
}
\label{Fig_QuanMagnonics_YuanHY2020PRB_MagnonFockState}
\end{figure}

\textit{\textbf{Single magnon state generation through nonlinear multi-magnon-magnon interaction}}.
Yuan et al.~\cite{YuanHY2020PRB_Antibunching} proposed an original method of generating single magnon state, which originated from the Kerr-type nonlinearity of multi-magnon-magnon interaction. The nonlinear system considers a biaxial nanomagnet, as shown in Fig.~\ref{Fig_QuanMagnonics_YuanHY2020PRB_MagnonFockState} (a), which is described by the effective Hamiltonian
\begin{eqnarray}
\hat{\mathcal{H}}&=&\omega_m \hat{m}^{\dagger} \hat{m}+w\left(\hat{m}^{\dagger} \hat{m}^{\dagger}+\hat{m} \hat{a}\right)+v\left(\hat{m}^{\dagger} \hat{m}\right)^2 \nonumber \\
&&+u\left(\hat{m}^{\dagger} \hat{m} \hat{m} \hat{m}+\hat{m}^{\dagger} \hat{m}^{\dagger} \hat{m}^{\dagger} \hat{m}\right)+\xi\left(\hat{m} e^{i \omega t}+\hat{m}^{\dagger} e^{-i \omega t}\right).
\label{QuanMagnonics_YuanHY2020PRB}
\end{eqnarray}
The parameters $\omega_a,~w,~v$, and $u$ are all related to the anisotropy coefficients of easy axis ($K_z$) and hard axis ($K_x$). The detailed derivation of this Hamiltonian can be found in Sec. \ref{mm_interaction}.

Figure~\ref{Fig_QuanMagnonics_YuanHY2020PRB_MagnonFockState} (b) plots the second-order correlation function of the magnon mode, in which the white curve corresponds to $g^{(2)}(0)=0.5$. It can be observed that the second-order correlation function approaches $g^{(2)}(0)\rightarrow 1$ along the horizontal line when the anisotropy is absent in the $z$ axis (i.e., $K_z=0$). In this case, the dominant Hamiltonian Eq.~(\ref{QuanMagnonics_YuanHY2020PRB}) collapses to $\hat{\mathcal{H}}_{K_z=0}=\omega_a \hat{m}^{\dagger} \hat{m}+w\left(\hat{m}^{\dagger} \hat{m}^{\dagger}+\hat{m} \hat{m}\right)+\xi\left(\hat{m} e^{i \omega t}+\hat{m}^{\dagger} e^{-i \omega t}\right)$, where two four-magnon processes are omitted for especially weak coupling strength. Thus we can conclude that the squeezing term $\left(\hat{m}^{\dagger} \hat{m}^{\dagger}+\hat{m} \hat{m}\right)$ seems not relevant to the generation of single magnon state.

On the other hand, it is easily found that, even when the anisotropy is missing in the $x$ axis (i.e., $K_x=0$), the antibunching effect of the magnon is apparently observed with large anisotropy in the $z$ axis. The Hamiltonian Eq.~(\ref{QuanMagnonics_YuanHY2020PRB}) in such case is reduced to $\hat{\mathcal{H}}_{K_x=0}=\omega_a \hat{m}^{\dagger} \hat{m}+v\left(\hat{m}^{\dagger} \hat{m}\right)^2+\xi\left(\hat{m} e^{i \omega t}+\hat{m}^{\dagger} e^{-i \omega t}\right)$, meaning that the multi-magnon-magnon term $\left(\hat{m}^{\dagger} \hat{m}\right)^2$ is the fundamental origination of the magnon nonclassical state. Furthermore, the antibunching phenomenon becomes more apparent (indicated by smaller $g^{(2)}(0)$) with larger anisotropic nonlinearity $v$. Particularly, the second-order correlation function approaches $g^{(2)}(0)\approx 0$ when the nonlinearity is as large as $K_z/\gamma N\approx 2$, which confirms the possibility of generating perfect single-magnon state.

To conclude, the nonlinear Kerr-type multi-magnon term with the form of $\left(\hat{m}^{\dagger} \hat{m}\right)^2$ is efficient in generating a single-magnon state, which has also been confirmed in quantum optics in terms of preparing a single-photon state~\cite{Paul1982RMP_PhotonAntibunching,Liew2010PRL_SinglePhoton,Rabl2011PRL_PhotonBlockade,Hoffman2011PRL_PhotonBlockade}. This approach offers great opportunities for the manipulation of magnons at a single quanta level.

\subsubsection{Cat state}\label{QuanMagnonics_CatState}
The concept of Schr\"{o}dinger cat state was first proposed by Schr\"{o}dinger~\cite{Schrodinger1935_CatState}, which represents a quantum superposition of macroscopically distinguishable states with high nonclassicality. The general form of cat state can be written as
\begin{equation}
| \mathrm{cat}\rangle =\mathcal{N} (| \alpha\rangle + e^{-i\phi}| -\alpha\rangle),
\end{equation}
where $| \alpha\rangle$ represents a coherent state with complex value $\alpha$, and $\mathcal{N}=1/\sqrt{2+2 \cos \phi \exp \left(-2|\alpha|^2\right)}$ denotes the normalization coefficient. Peculiarly, the odd and even cat state are obtained when the phase $\phi=0$ and $\phi=\pi$, respectively, which reads
\begin{eqnarray}
| \mathrm{cat}\rangle_{\mathrm{even}} =\mathcal{N} (| \alpha\rangle+| -\alpha\rangle), \nonumber\\
| \mathrm{cat}\rangle_{\mathrm{odd}} =\mathcal{N} (| \alpha\rangle-| -\alpha\rangle).
\end{eqnarray}

Schr\"{o}dinger cat state has stimulated much interest from the perspective of both
 fundamental and practical applications. On the one hand, cat states are valuable resource for testing the fundamental problems in quantum mechanics, for example, the quantum-to-classical transition~\cite{Wineland2013RMP_NobelLecture,Arndt2014NatPhys_FoundamentalTest}. On the other hand, cat states are of great importance for potential applications ranging from fault-tolerant quantum computation~\cite{Cochrane1999PRA_Cat_BosonicCode,Lund2008PRL_Cat_FaultTolerantComput,Neergaard-Nielsen2010PRL_Cat_QuanCompu,Ofek2016Nature_Cat_ErrorCorrelation}, quantum secret sharing~\cite{Karimipour2002PRA_Cat_QSS}, quantum repeater~\cite{Brask2010PRL_Cat_QuanRepeater,Sangouard2010JOSAB_Cat_QuanRepeater}, quantum teleportation~\cite{Phien2008PLA_Cat_Teleportation}, and quantum metrology~\cite{Joo2011PRL_Cat_Metrology}.

The exact cat state can be implemented through the Kerr effect in a nonlinear medium, however, it is extremely difficult because very strong Kerr nonlinearities are required~\cite{Yurke1986PRL_CatState,Glancy2008JOSAB_Cat_Kerr}. From then on, numerous representative approaches of generating approximate cat states are proposed, typically on squeezing followed by measurement and post-selection. For example, the schemes based on the backaction evasion measurement are feasible to generate cat state~\cite{SongShang1990PRA_Cat_BackactionEvading,Yurke1990PRA_Cat_BackactionEvading,Glancy2008JOSAB_Cat_Kerr}, which requires two stages of squeezing and may encounter high demands of strong squeezing and high efficiency photon counter. Later, a well-established photon subtraction scheme, a simplified version of backaction evasion measurement regime, is theoretically proven to be a straightforward and efficient approach for creating cat states~\cite{Dakna1997PRA_CatTheory_PhoSubstraction,Lund2004PRA_CatTheory_PhoSubstraction} and has been experimentally demonstrated~\cite{AlexeiOurjoumtsev2006Science_CatExperiment,Neergaard-Nielsen2006PRL_CatExperiment,Wakui2007OE_CatExperiment,Takahashi2008PRL_CatExperiment,Gerrits2010PRA_CatExp_Pluse}.
Note that only one stage of squeezing is demanded in the photon subtraction regime, which is in quite contrast to the regime based on the backaction evasion measurement~\cite{SongShang1990PRA_Cat_BackactionEvading,Yurke1990PRA_Cat_BackactionEvading,Glancy2008JOSAB_Cat_Kerr}, thus offering great convenience for experimental demonstration~\cite{AlexeiOurjoumtsev2006Science_CatExperiment,Neergaard-Nielsen2006PRL_CatExperiment,Wakui2007OE_CatExperiment,Takahashi2008PRL_CatExperiment,Gerrits2010PRA_CatExp_Pluse}.

However, a majority of above studies are theoretically designed or experimentally realized based on the optical systems, while the generation of cat states in massive hybrid magnonic systems are started to receive widespread attention ever since the flourishing development of quantum magnonics. Correspondingly, magnon Schr\"{o}dinger cat state represents a macroscopic quantum superposition of collective magnetic excitations of large number spins.
There are just few typical approaches of generating and manipulating magnon cat states in hybrid magnonic systems~\cite{Sharma2021PRB_CatState,SunFX2021PRL_CatState,Kounalakis2022PRL_CatState} , all of which are contributed from a certain nonlinearity. For example,  Sharma et al.~\cite{Sharma2021PRB_CatState} adopted the nonlinearity from the anisotropic magnon interaction in the form of the single-mode squeezing; Sun et al.~\cite{SunFX2021PRL_CatState} employed the intrinsic nonlinear interaction between magnons and optical photons naming magneto-optical interaction, which resembles closely to the form of radiation pressure force in optomechanics; Kounalakis et al.~\cite{Kounalakis2022PRL_CatState} utilized nonlinear magnon-qubit couplings mediated between magnons and transmon qubits, with similar form to magneto-optical interaction. In the following, we would like to thoroughly introduce these three methods taking advantage of different nonlinearities to generate magnon cat states.

\begin{figure}[tbp!]
\centering
\includegraphics[width=0.98\columnwidth]{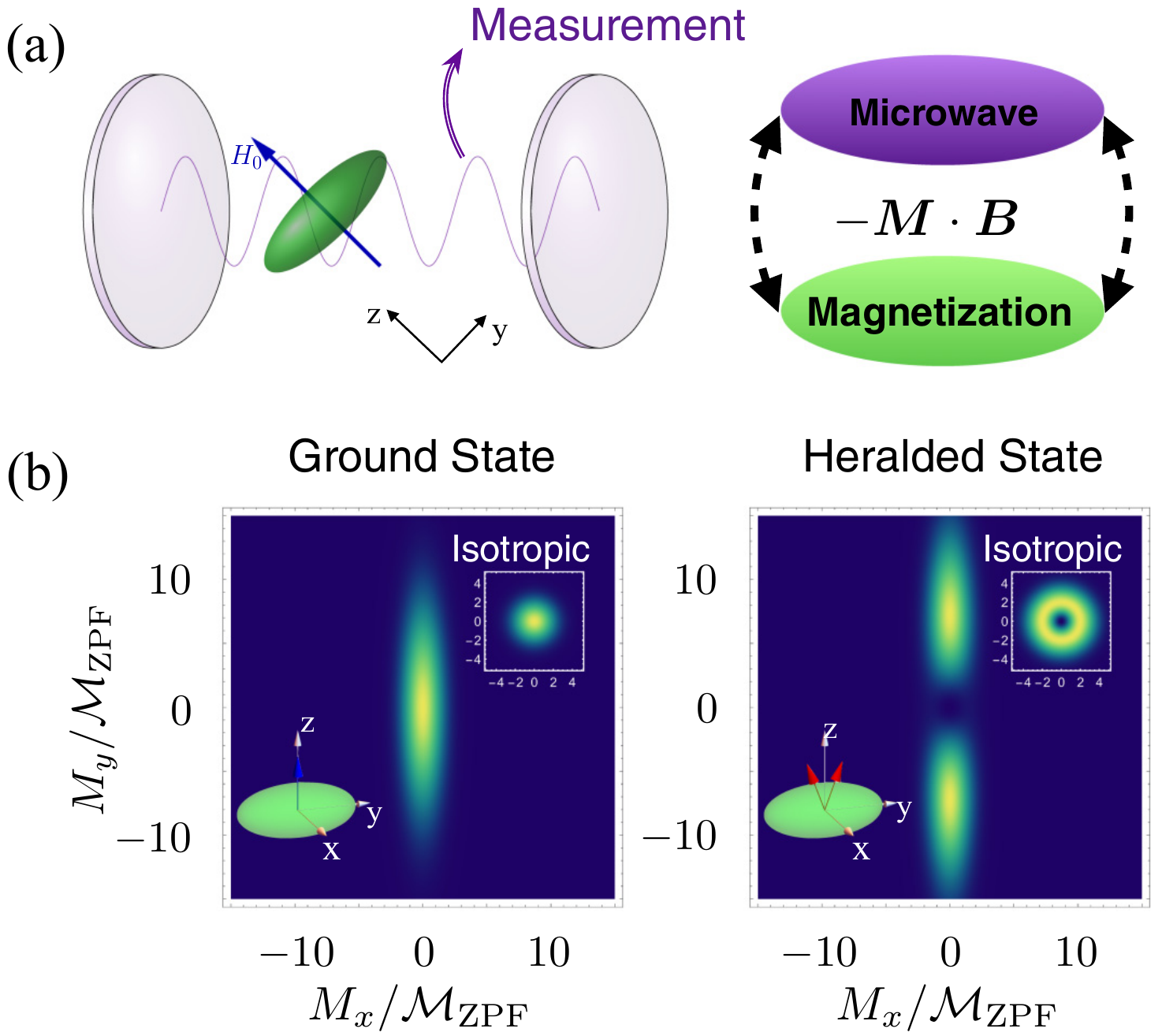}
\caption{(a) The schematic diagram of magnon cat state generation through the nonlinear single-mode squeezing of magnons~\protect\cite{Sharma2021PRB_CatState}. An anisotropic magnet with an ellipsoid shape is placed inside the microwave cavity, resulting the squeezing of the magnetization state. Performing a single photon or a parity measurement on the microwave photons would collapse the magnetic state to the cat state.
(b) The Husimi $Q$ functions of the projected magnon state for the squeezed vacuum state (left panel) and projected magnon cat state after performing a photon measurement on the mricrowave mode (right panel). The insets in (b) represent the isotropic case where cat state is absent even after a single-photon measurement.
Reproduced with permission from Sharma et al., Phys. Rev. B 103, L100403 (2021). Copyright 2021 American Physical Society.
}
\label{Fig_QuanMagnonics_Silvia2021PRB_CatState}
\end{figure}

\textit{\textbf{Magnon cat state generation through nonlinear magnon squeezing process}}.
In 2021, Sharma et al.~\cite{Sharma2021PRB_CatState} presented a theoretical proposal to generate a magnon cat state, expressed as a quantum superposition of two distinct magnon states, by applying an experimental feasible quantum magnonic system with an anisotropic-shape magnet inside the microwave cavity. This scheme follows the well established method in the area of quantum optics by adding or subtracting a photon to a squeezed optical vacuum~\cite{Huang2015PRL_Cat_QuanOptics,Dakna1997PRA_Cat_QuanOptics,Alexei2006Science_Cat_QuanOptics}. The procedure of preparing magnon cat states mainly consists of two steps. Firstly, a squeezed state of magnons is created taking advantage of the anisotropy of the ellipsoid shaped magnet. Secondly, coupling the magnon mode to a microwave cavity through Zeeman coupling and subsequently performing single photon or parity measurement on the microwave photons, projecting the magnetic state to a cat state, as displayed in Fig.~\ref{Fig_QuanMagnonics_Silvia2021PRB_CatState} (a). The effective Hamiltonian of such system reads
\begin{equation}
\hat{\mathcal{H}}=\omega_m \hat{m}^{\dagger} \hat{m}+\omega_a \hat{a}^{\dagger} \hat{a}+g\left(\hat{m} \hat{a}^{\dagger}+\hat{m}^{\dagger} \hat{a}\right)+g_m\left(\hat{m}^2+\hat{m}^{\dagger 2}\right),
\end{equation}
where $\hat{m}$ and $\hat{a}$ are operators annihilating a magnon and photon, respectively, and $g$ denotes the magnon-photon coupling strength. In the dispersive regime when $\omega_m,~g\ll\omega_a$, the
ground state of the system is approximately as
\begin{eqnarray}
|\Psi\rangle_0&& \approx\left(1+\sqrt{P} \frac{\hat{m}^{\dagger} \hat{a}^{\dagger}}{\cosh r}\right) \hat{S}(r)\left|0_m 0_a\right\rangle, \nonumber \\
&&=\hat{S}(r)\left|0_m 0_a\right\rangle+\frac{\sqrt{P}}{\cosh r} \hat{m}^{\dagger} \hat{S}(r)\left|0_m 1_a\right\rangle,
\label{QuanMagnonics_Sharma2021PRB_GroudState}
\end{eqnarray}
where $P$ represents the probability of finding a single photon state and $r$ denotes the effective magnetization squeezing parameter. The squeezing operator takes the form of $\hat{S}(r)=\mathrm{exp}[r(\hat{m}^2-\hat{m}^{\dagger 2})/2]$. By performing single photon or parity measurement on the microwave mode and yielding the results of $|1_a\rangle$, the magnon mode is projected to the state
\begin{eqnarray}
|\Psi\rangle_m=\frac{1}{\cosh r} \hat{m}^{\dagger} \hat{S}(r)\left|0_m\right\rangle.
\label{QuanMagnonics_Sharma2021PRB_MagnonState}
\end{eqnarray}

The underlying meaning of this state lies in that adding a magnon to the magnon squeezed vacuum state. It is well known that the squeezed vacuum state can be expressed in terms of the Fock state as $|\xi\rangle=\hat{S}(r)\left|0_m\right\rangle=\sum_m C_{2m}|2m\rangle$ with $C_{2 m}=\frac{1}{\sqrt{\cosh r}}(-1)^m\left(\frac{ \tanh r}{2}\right)^m \frac{\sqrt{(2 m) !}}{m !}$, meaning that only magnons with even number can be detected in the squeezed vacuum state. Therefore, the collapsed magnon state indicated by Eq.~(\ref{QuanMagnonics_Sharma2021PRB_MagnonState}) is a superposition of odd magnon numbers, which can be rewritten as
\begin{eqnarray}
|\Psi\rangle_m \propto |\Psi_+\rangle_m- |\Psi_-\rangle_m,
\label{QuanMagnonics_Sharma2021PRB_MagnonState_1}
\end{eqnarray}
where $\left|\Psi_{\pm}\right\rangle_m \propto \sum_{m=0}^{\infty} \frac{(\pm 1)^m \sqrt{m !}}{\Gamma(m / 2+1 / 2)}\left(\frac{-\tanh r}{2}\right)^{\frac{m-1}{2}}|m\rangle$ with $\Gamma$ being the gamma function. The above equation indicates that the projected magnon state is approximately the superposition of two semiclassical states, that is, the magnon odd cat state is generated through the single photon measurement on the microwave photons.

\begin{figure*}[tbp!]
\centering
\includegraphics[width=2\columnwidth]{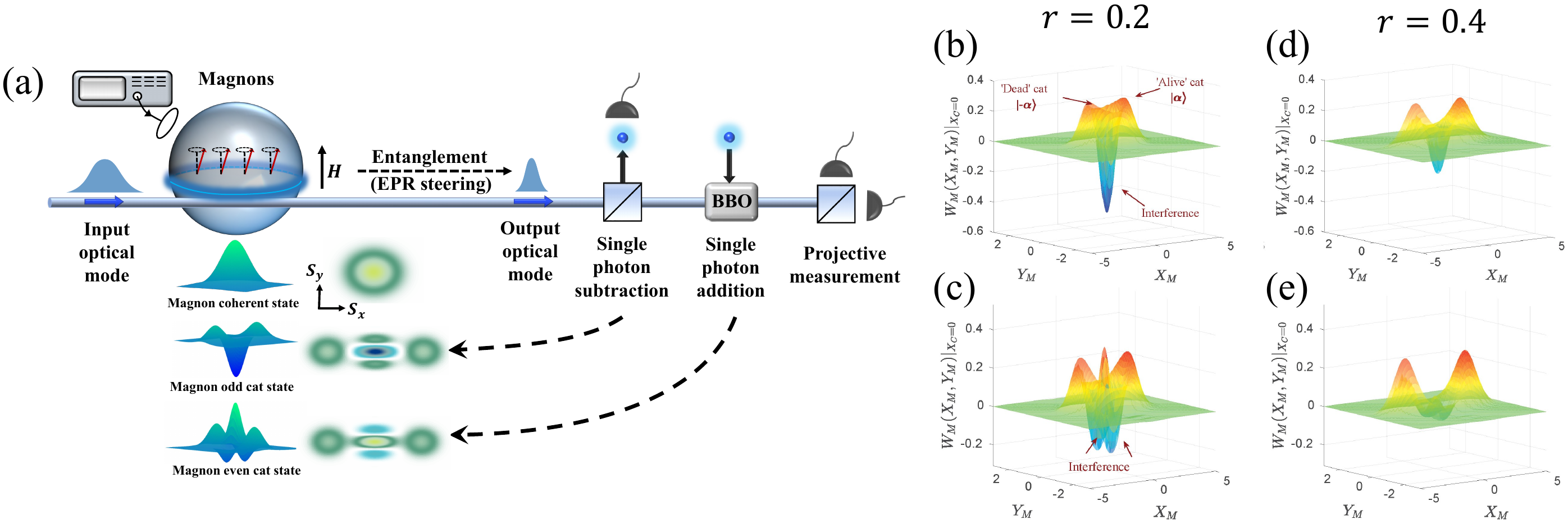}
\caption{(a) The schematic diagram of magnon cat state generation through the nonlinear magneto-optical process~\protect\cite{SunFX2021PRL_CatState}. Firstly, the nonlinear magneto-optical interaction is created to entangle the magnons with optical photons. Then, the single photon subtraction or addition operations are performed on the optical mode to remotely convert the coherent magnon state to a non-Gaussian state, which collapses to the even/odd cat state after the applying projective measurements. (b-e) Wigner functions of the projected magnon mode with outcome $X_C=0$ for squeezing parameters $r=0.2$ in (b-c) and $r=0.4$ in (d-e).(b) and (d) correspond to the case by subtracting single photon from the optical mode, obtaining an odd cat state; while (c) and (e) denote the performance of a single-photon subtraction and subsequent single-photon addition on the optical mode, arriving at even cat state. Here, $X_C$ denotes the amplitude quadrature of the optical field; while $X_M$ and $Y_M$ represent the amplitude and phase quadratures of the magnon mode, respectively.
Reproduced with permission from Sun et al., Phys. Rev. Lett. 127, 087203 (2021). Copyright 2021 American Physical Society.
}
\label{Fig_QuanMagnonics_SunFX2021PRL_CatState}
\end{figure*}

Figure~\ref{Fig_QuanMagnonics_Silvia2021PRB_CatState} (b) displays the Husimi $Q$ functions for the case of magnon squeezed vacuum state before the projection measurement shown in the left panel, as well as the projected magnon state $|\Psi\rangle_m$ after applying the single-photon or parity measurement shown in the right panel. Two distinct peaks appear in the right panel, indicating the generation of magnon cat state. The appearance of this magnon cat state generation can be all attributed to the anisotropic squeezing term of the magnons. Because when the magnetization is isotropic, the magnon cat state disappears as indicated in the insets.

\textit{\textbf{Magnon cat state generation through nonlinear magneto-optical interaction}}.
Notably, the approach of generating magnon cat states introduced above~\cite{Sharma2021PRB_CatState} concentrated on an ideal system where the decoherence effect is missing. However, it is a great challenge for experiments because of the inevitable coupling between the concerned system to the surrounding environment. Besides, this method mainly utilizes adding or subtracting an excitation to a squeezed vacuum state, which is merely capable of preparing cat states locally at the position where the local addition or subtraction operations are implemented. To overcome these two overriding problems, Sun and coauthors~\cite{SunFX2021PRL_CatState} proposed a novel method of remotely generating magnon cat states.

In the scheme, the pulsed nonlinear magneto-optical interaction is adopted to prepare strong quantum entanglement and EPR steering between magnon and photon pairs separately in remote distance, which is the underlying origination of the magnon cat states. Subsequently, even/odd cat states of the magnon mode is able to be prepared and manipulated by performing appropriate local non-Gaussian operations (single-photon operations and projective measurement) on the remote optical photons.

The detailed scheme is illustrated in Fig.~\ref{Fig_QuanMagnonics_SunFX2021PRL_CatState} (a). A pulsed light beam firstly propagates tangentially to the equator of the YIG sphere, resulting in the nonlinear magneto-optical interaction between the whispering gallery mode and the magnon mode stimulated by the external magnetic field. The pulsed nonlinear magneto-optical interaction is capable of preparing strong quantum entanglement and EPR steering between magnon and photon pairs separately in remote distance, which is the underlying origin of the magnon cat states. Then by performing appropriate single-photon operations and projective measurements on the remote optical mode, the magnon mode will collapse to even/odd cat states. The effective Hamiltonian of this system in the frame rotating at the driving frequency $\omega_p$ is given by
\begin{equation}
	\hat{\mathcal{H}}=\Delta \hat{c}^{\dagger}\hat{c}+ \omega_{m}\hat{m}^{\dagger}\hat{m}+g_{0}\hat{c}^{\dagger}\hat{c}(\hat{m}^{\dagger}+\hat{m}),
\label{QuanMagnonics_SunFX2021PRL}
\end{equation}
where $\hat{c}$ and $\hat{m}$ are respectively the annihilation operators for the optical mode and the Kittel magnon mode, $\Delta=\omega_{c}-\omega_{p}$ is the detuning of the cavity relative to the driving laser with $\omega_{c}$ being the frequency of the optical photons, $\omega_{m}$ represents the magnon's frequency while $g_0$ denotes the nonlinear optomagnonic coupling strength, as introduced in Sec. \ref{magnon_photon_interaction}.

When pumping the cavity mode with blue-detuned pulses, i.e., $\Delta=-\omega_{m}$, the magnon mode and the optical cavity mode can be strongly entangled, which has been demonstrated in optomechanics~\cite{Hofer2011PulseEnt,Palomaki710}. When the effective squeezing parameter is especially large, even strong EPR steering can be prepared. The details of how transient entanglement and EPR steering is generated and the characteristic of entanglement in such system would be shown later in Fig.~\ref{Fig_QuanMagnonics_SunFX2021PRL_CatState_Ent} in Sec.~\ref{QuanMagnonics_EntGeneration_MOptical} when introducing the approach of generating entanglement through the nonlinear magnero-optical interaction. Here we concentrate on the details of how magnon cat state is generated. It has been recently demonstrated that the remote generation of Wigner negativity on the steering subsystem is created by means of performing single-photon subtraction on the steered subsystem, in which EPR steering plays a significant role. On the other hand, cat states are typical non-Gaussian states possessing Wigner negativity, indicating that the EPR steering in the direction from the magnon mode to photon mode is possible to prepare cat states in magnon modes.

Figures~\ref{Fig_QuanMagnonics_SunFX2021PRL_CatState} (b) and (d) show the Wigner functions of the projected magnon mode with an outcome of $X_C=0$ after a single-photon subtraction performed on the output optical field. It can be clearly observed that two distinct peaks appear in the direction of $X_M$, besides, the amplitude of the Wigner function at the symmetry center ($X_M=Y_M=0$) is negative, revealing that the magnon collapses to an odd cat state. Moreover, when a photon subtraction and a subsequent single-photon addition are performed on the optical mode in sequence, an even cat state of the magnon mode with positive Wigner function at the symmetry center can be generated, as shown in Figs.~\ref{Fig_QuanMagnonics_SunFX2021PRL_CatState} (c) and (e). Nevertheless, compared Figs.~\ref{Fig_QuanMagnonics_SunFX2021PRL_CatState} (b-c) to (d-e), we find that the peaks of ``dead'' and ``alive'' states become more separated with larger squeezing parameter $r$ (that is, the longer optical pulse duration), which means that the generated magnon cat state possesses a larger cat size. However, the minimum value of Wigner function at the symmetry center approximates closer to zero with increasing squeezing parameter $r$, i.e., the interference fringe between the ``dead'' and ``alive'' states gradually becomes less apparent. This phenomena can be understood as follows. Stronger magnon-photon entanglement and EPR steering from the magnon to photon are prepared with larger squeezing parameter $r$, as shown in Fig.~\ref{Fig_QuanMagnonics_SunFX2021PRL_CatState_Ent} in Sec.~\ref{QuanMagnonics_EntGeneration_MOptical}, thus contributing to cat state generation with larger size. But larger squeezing parameter indicates stronger coupling to the surrounding environment, the decoherence effect is more apparent to decrease the interference pattern.

To conclude, the regime to remotely generate and manipulate magnon cat state by performing local non-Gaussian operations on the remote optical photons utilizing EPR steering strongly prepared by the pulsed nonlinear magneto-optical interaction is feasible and effective. Even or odd cat states are both capable of being prepared through performing approximate single-photon operations. In addition, the authors demonstrated that the generated magnon cat states are equipped with high fidelity, nonclassicality, and a large size when considering experimentally feasible parameters where the decoherence effect is present, which is useful for quantum technologies.

\begin{figure}[htbp!]
\centering
\includegraphics[width=0.98\columnwidth]{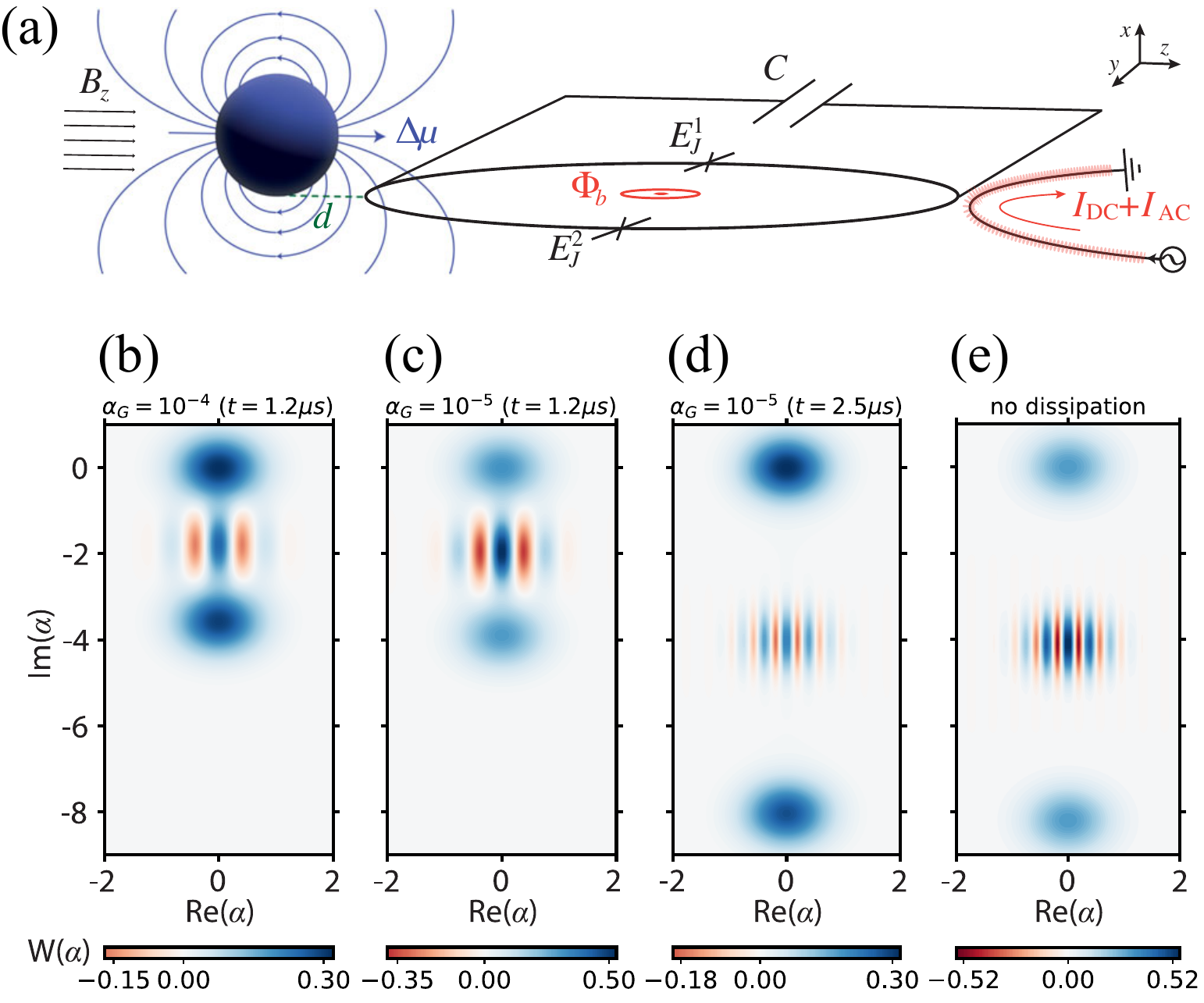}
\caption{(a) The schematic diagram of magnon cat state generation through the nonlinear magnon-qubit interaction~\protect\cite{Kounalakis2022PRL_CatState}.
A YIG sphere is placed close to a superconducting quantum interference device (SQUID). An in-place field $B_z$ is adopted to orient the uniform magnetization of the YIG particle. The magnetic fluctuations would modulate the inductive energy and frequency of the transmon qubit through inducing a flux in the SQUID loop, resulting in a direct magnon-qubit linear exchange interaction and radiation-pressure-like nonlinear interaction.
(b-e) The Wigner function of the generated magnon cat state with different Gilbert damping parameter in (b) $\alpha_G=10^{-4}$, (c,d) $\alpha_G=10^{-5}$ and (e) $\alpha_G=0$. (b) and (c) are calculated at time $t\simeq1.2\mu$s, while (d) and (e) are calculated at longer time $t\simeq2.5\mu$s.
Reproduced with permission from Kounalakis et al., Phys. Rev. Lett. 129, 037205 (2022). Copyright 2022 American Physical Society.
}
\label{Fig_QuanMagnonics_Kounalakis2022PRL_CatState}
\end{figure}

\textit{\textbf{Magnon cat state generation through nonlinear magneto-qubit interaction}}.
Apart from generating magnon cat state via the nonlinearity from multi-magnon-magnon interaction~\cite{Sharma2021PRB_CatState} and magneto-optical interaction~\cite{SunFX2021PRL_CatState}, the nonlinearity mediated by magnons and qubits is another approach of generating magnon cat states. Recently, Kounalakis et al.~\cite{Kounalakis2022PRL_CatState} put forward a theoretical scheme to directly couple the magnon to a transmon qubit in a planar superconducting circuit through magnetic flux. The system comprises of both the resonant magnon-qubit exchange coupling and nonlinear interactions, which is responsible for preparing large macroscopic quantum superpositions of magnetization with a high fidelity. The schematic diagram is shown in Fig.~\ref{Fig_QuanMagnonics_Kounalakis2022PRL_CatState} (a), and the total Hamiltonian of the magnon-qubit system is described by
\begin{equation}
\hat{\mathcal{H}}=\hat{\mathcal{H}}_0^M+\hat{\mathcal{H}}_0^T+\hat{\mathcal{H}}^{\mathrm{int}},
\label{QuanMagnonics_Kounalakis2022PRL_Hamiltonian}
\end{equation}
where $\hat{\mathcal{H}}_0^M$,~$\hat{\mathcal{H}}_0^T$ denote the free energy terms of the magnon mode and the transmon qubit, respectively; while $\hat{\mathcal{H}}^{\mathrm{int}}$ represent the magnon-qubit interaction. The explicit expressions of these three terms read
\begin{eqnarray}
\hat{\mathcal{H}}_0^T &=& \omega_c \hat{c}^{\dagger} \hat{c}-\left(E_C / 2\right) \hat{c}^{\dagger} \hat{c}^{\dagger} \hat{c} \hat{c}, \nonumber \\
\hat{\mathcal{H}}_0^M &=&  \omega_m \hat{m}^{\dagger} \hat{m}, \nonumber \\
\hat{\mathcal{H}}^{\mathrm{int}} &=& g_c\left(\hat{c}^{\dagger} \hat{m}+\hat{c} \hat{m}^{\dagger}\right)+g_{\mathrm{rp}} \hat{c}^{\dagger} \hat{c}\left(\hat{m}+\hat{m}^{\dagger}\right).
\label{QuanMagnonics_Kounalakis2022PRL_Hamiltonian_Component}
\end{eqnarray}
The annihilation operators $\hat{c}$ and $\hat{m}$ respectively indicate the transmon qubit and magnon modes. Note that the free Hamiltonian of the transmon qubit  is generally characterized by the superconducting phase difference and the number of tunneling Copper pairs, which can also be expressed with respect to the bosonic annihilation and creation operators in some certain cases when the zero-point fluctuations of the phase variable are especially small.
$\omega_c$ and $E_C$ denote the excitation energy and charging energy of the transmon qubit. $\omega_m$ is the frequency of the magnon mode, which is flexibly tuned via the magnetic field. The coupling between magnon and transmon qubit consists of two different types. On the one hand, the transmon qubit is coupled to the magnon through the coherent exchange interaction with coupling strength $g_c$, which resembles the indirect magnon-qubit coupling mediated by microwave photons~\cite{Tabuchi2015}. On the other hand, the transmon qubit and magnon are coupled through the nonlinear interaction in a similar form with optical photon-magnon coupling as well as the radiation pressure in optomechanics, where the coupling strength denotes $g_{\mathrm{rp}}$.

The generation protocol of magnon cat states is illustrated as follows. Firstly, the state of the transmon qubit is evolved into a superposition state $|+\rangle_T=\left(|0\rangle_T+|1\rangle_T\right) / \sqrt{2}$ after preforming a $R_{\hat{y},(\pi / 2)}$ pulse on the qubit. Subsequently, the state of the whole magnon-qubit coupled system reads  $\left(|0\rangle_T|0\rangle_m+e^{i \theta(t)}|1\rangle_T|\beta(t)\rangle_m\right) / \sqrt{2}$ according to the interaction Hamiltonian Eq.~(\ref{QuanMagnonics_Kounalakis2022PRL_Hamiltonian}) after applying the flux modulation. When the flux modulation is turned off at time $t=\tau$, the state finally turns to be highly entangled with a form of Bell-cat, describing by $\left\{|+\rangle_T\left[|0\rangle+e^{i \theta(\tau)}|\beta(\tau)\rangle\right]_m+|-\rangle_T\left[|0\rangle-e^{i \theta(\tau)}|\beta(\tau)\rangle]_m\right\}\right/2$. By further projective measurement, the magnon state collapses to the desired even/odd cat. Specifically, the magnon state evolved into an even (odd) cat state if a rotation $R_{\hat{y},(\pi / 2)}$ is applied and the projective measurement yields the result of $|0\rangle_T$ ($|1\rangle_T$), where the cat state is written as $|\Psi\rangle_{\mathrm{even}}=\left[|0\rangle+|\beta(\tau)\rangle\right]_m/\mathcal{N}$ ($|\Psi\rangle_{\mathrm{odd}}=\left[|0\rangle-|\beta(\tau)\rangle\right]_m/\mathcal{N}$) with $\mathcal{N}$ being the normalization constant. It is noteworthy that although the form of this macroscopic superposition state is slightly different to the familiar cat state, it is equivalent to $|\Psi\rangle=\left[|-\beta(\tau)/2\rangle\pm |\beta(\tau)/2\rangle\right]_m/\mathcal{N}$ followed by a coherent magnon displacement of amplitude $-\beta(\tau)/2$, which is in the standard form of even/odd cat states.

Figures~\ref{Fig_QuanMagnonics_Kounalakis2022PRL_CatState} (b-e) show the Wigner function of the magnon mode with different Gilbert damping parameters. It is clearly shown that two separate peaks is present in the direction of vertical axis, and the interference fringes with negative Wigner functions appear between the two peaks, indicating the occurrence of odd cat state. Notably, the fidelities under realistic experimental conditions are $F\simeq 80\%$ in Figs.~\ref{Fig_QuanMagnonics_Kounalakis2022PRL_CatState} (b,d) and  $F\simeq 95\%$ in (c).

\textit{\textbf{Brief summary}}.
To conclude, quantum magnonic systems are promising platforms to prepare and manipulate Schr\"{o}dinger cat state. Up to now, various schemes~\cite{Sharma2021PRB_CatState,SunFX2021PRL_CatState,Kounalakis2022PRL_CatState} are proposed to prepare macroscopic quantum superpositions of magnons, which is useful both from the fundamental perspective and practical applications in quantum information processing. The approaches of generating magnon cat states can be originated from the the nonlinearity of the anisotropic magnon interaction~\cite{Sharma2021PRB_CatState}, magneto-optical process~\cite{SunFX2021PRL_CatState}, and direct nonlinear transmon qubit-magnon interaction~\cite{Kounalakis2022PRL_CatState}.

\subsection{Decoherence time of magnons}

From the knowledge of qubit, it is known that the decoherence time is an important parameter for the quantum states. To maintain the superposition and entanglement properties of a quantum state, one has to perform quantum operations within the coherent time. The decoherence mechanism for qubit has been well known while it is yet to be studied for magnon quantum states. The most popular Landau-Lifshitz-Gilbert theory treats the relaxation of magnons as a phenomenological damping parameter so-called Gilbert damping \cite{Landau1935, Gilbert2004}. Here the decoherence time of magnons is $1/(\alpha \omega_m)$, with $\alpha$ the damping parameter and $\omega_m$ the magnon frequency. Such decoherence is accompanied by the magnon number relaxation, while the pure dephasing of magnon quantum states in the absence of magnon relaxation has drawn little attention until recently. Yuan et al. \cite{YuanDephasing2022} considered a Heisenberg magnet and showed that the four-magnon interaction can induce the pure dephasing of magnons, which becomes significant and can dominate the decoherence of magnon quantum states at several Kelvins.

To illustrate the essential physics of dephasing, we recall the effective Hamiltonian with four-magnon interaction derived in \eqref{four_magnon} that reads
\begin{equation} \label{exchange_ham}
\hat{\mathcal{H}}=\omega_r \hat{m}^\dagger \hat{m} +\sum_{\mathbf{k} \neq \mathbf{k}_0} \omega_\mathbf{k} \hat{m}_\mathbf{k}^\dagger \hat{m}_\mathbf{k} + \hat{m}^\dagger \hat{m} \sum_{\mathbf{k} \neq \mathbf{k}_0} g(\mathbf{k})\hat{m}_\mathbf{k}^\dagger \hat{m}_\mathbf{k},
\end{equation}
where we use label $\hat{m}_\mathbf{k}$ to label magnon operator to make it consistent throughout this section and we have defined $\hat{m} \equiv \hat{m}_{\mathbf{k}_0},\omega_r= \omega_{\mathbf{k}_0}, g(\mathbf{k}) = C(\mathbf{k}_0,\mathbf{k},\mathbf{q}=0)$ to make the notation simple.
According to Eq.~\eqref{exchange_ham}, the scattering of mode $\mathbf{k}_0$ with other magnons (third term in Eq.~\eqref{exchange_ham}) adds a random fluctuation $\zeta$ to the eigenfrequency $\omega_r$. After a sufficiently long time, even though the average frequency of magnon is still $\omega_r$, the phase fluctuations of the magnon states will vary with time as $\delta \varphi \propto \sqrt{t}$ according to the central limit theorem. When the phase uncertainty $\delta \varphi$ exceeds $2\pi$, the magnon mode has dephased. This is similar to the random walk of a Brownian particle \cite{FeynmanBook2015}.

To quantify the dephasing rate, one can trace out the information of all the other magnons following the standard Lindblad formalism and derive the master equation describing the evolution of mode $\mathbf{k}_0$ as \cite{YuanDephasing2022, YuanME2022}
\begin{equation}\label{me_dephase}
\frac{d \hat{\rho}}{dt}=i[\hat{\rho}, \hat{\mathcal{H}}_s] + \kappa_\mathrm{dp} \mathcal{L}_{\hat{n}\hat{n}}(\hat{\rho}),
\end{equation}
where $\hat{\rho}$ is the density matrix describing the mode $\mathbf{k}_0$, $\hat{\mathcal{H}}_s=\omega^{'}_r \hat{m}^\dagger \hat{m}$, $\mathcal{L}_{\hat{n}\hat{n}}(\rho) \equiv 2\hat{n} \hat{\rho} \hat{n}^\dagger - \hat{n}^\dagger \hat{n} \rho - \hat{\rho} \hat{n}^\dagger \hat{n} $ with $\hat{n}=\hat{m}^\dagger \hat{m}$. The parameter $\kappa_\mathrm{dp}$ is a coefficient that characterizes the strength of dephasing
\begin{equation}\label{exchange_integral}
\kappa_\mathrm{dp} = \int_0^\infty  |D(\omega)g(\omega)|^2 n_{\mathrm{th}}[n_{\mathrm{th}}+1]d\omega,
\end{equation}
where $D(\omega)$ is the density of states of magnons and $n_\mathrm{th}$ is the thermal occupation of magnons. By further including the relaxation channel, one should add the following terms to the master equation \cite{YuanME2022}
\begin{equation}
\Delta \mathcal{L}=\frac{\kappa}{2} (n_{\mathrm{th}}+1) \mathcal{L}_{\hat{m}\hat{m}} [\hat{\rho}]
+ \frac{\kappa}{2} n_{\mathrm{th}}\mathcal{L}_{\hat{m}^\dagger \hat{m}^\dagger} [\hat{\rho}],
\end{equation}
where $\kappa$ is the relaxation rate that may depend spin-orbit coupling, spin pumping and two-magnon scattering \cite{YuanReview2022}.

From the master equation, we can evaluate the evolution of $\langle \hat{m} \rangle $ as
\begin{equation}
\frac{d\langle \hat{m}(t) \rangle}{dt} = -i\left (\omega_a-i\frac{\kappa}{2} -i \gamma\right) \langle \hat{a}(t) \rangle.
\end{equation}
According to the quantum regression theorem \cite{CarmichaelBook2013}, the first-order coherence characterized by the function $\langle \hat{m}^\dagger (t)\hat{m}(t+\tau) \rangle$ satisfies the same dynamic equation as that of $\langle \hat{m}(t) \rangle$, i.e.,
\begin{equation}
\frac{d\langle \hat{m}^\dagger (t)\hat{m}(t+\tau) \rangle}{dt} = -i\left (\omega_a-i\frac{\kappa}{2} -i \gamma\right) \langle \hat{m}^\dagger (t)\hat{m}(t+\tau) \rangle.
\end{equation}
By solving this differential equation, we obtain
\begin{equation}
\langle \hat{m}^\dagger (t)\hat{m}(t+\tau) \rangle = \langle \hat{m}^\dagger (t)\hat{m}(t) \rangle \exp(-i\omega_0 -\gamma-\frac{\kappa}{2})\tau.
\end{equation}
Here the magnon density evolution can be solved explicitly from the master equation as
\begin{equation}
 \langle \hat{m}^\dagger (t)\hat{m}(t) \rangle =  \langle \hat{m}^\dagger (0)\hat{m}(0) \rangle e^{-\gamma t} + n_\mathrm{th}(1-e^{-\gamma t}).
\end{equation}
Then we can derive the first-order coherence at sufficiently longer time scale ($t\rightarrow \infty$) as
\begin{equation}
\langle \hat{m}^\dagger (t)\hat{m}(t+\tau) \rangle = n_\mathrm{th} \exp(-i\omega_0 + \gamma + \frac{\kappa}{2})\tau.
\end{equation}
The total decoherence time now includes both contributions from the relaxation and pure dephasing rate as,
\begin{equation}
\frac{1}{T_2}=\frac{1}{T_1} + \frac{1}{T_2^*},
\end{equation}
where $T_2^{-1} = \gamma + \kappa/2,~ T_1^{-1} = \kappa/2,~ T_2^{*-1} = \gamma$ are respectively the decoherence time, relaxation time and pure dephasing time. This relation resembles the spin decoherence in nuclear spin resonance and that of the spin qubit \cite{HornakBook1996}.

\subsection{Bipartite and multipartite entanglement among magnons, phonons and photons} \label{mpp_entanglement}
\subsubsection{The classification and quantification of entanglement}\label{QuanMagnonics_EntConcept}

\begin{figure}
\centering
\includegraphics[width=0.6\columnwidth]{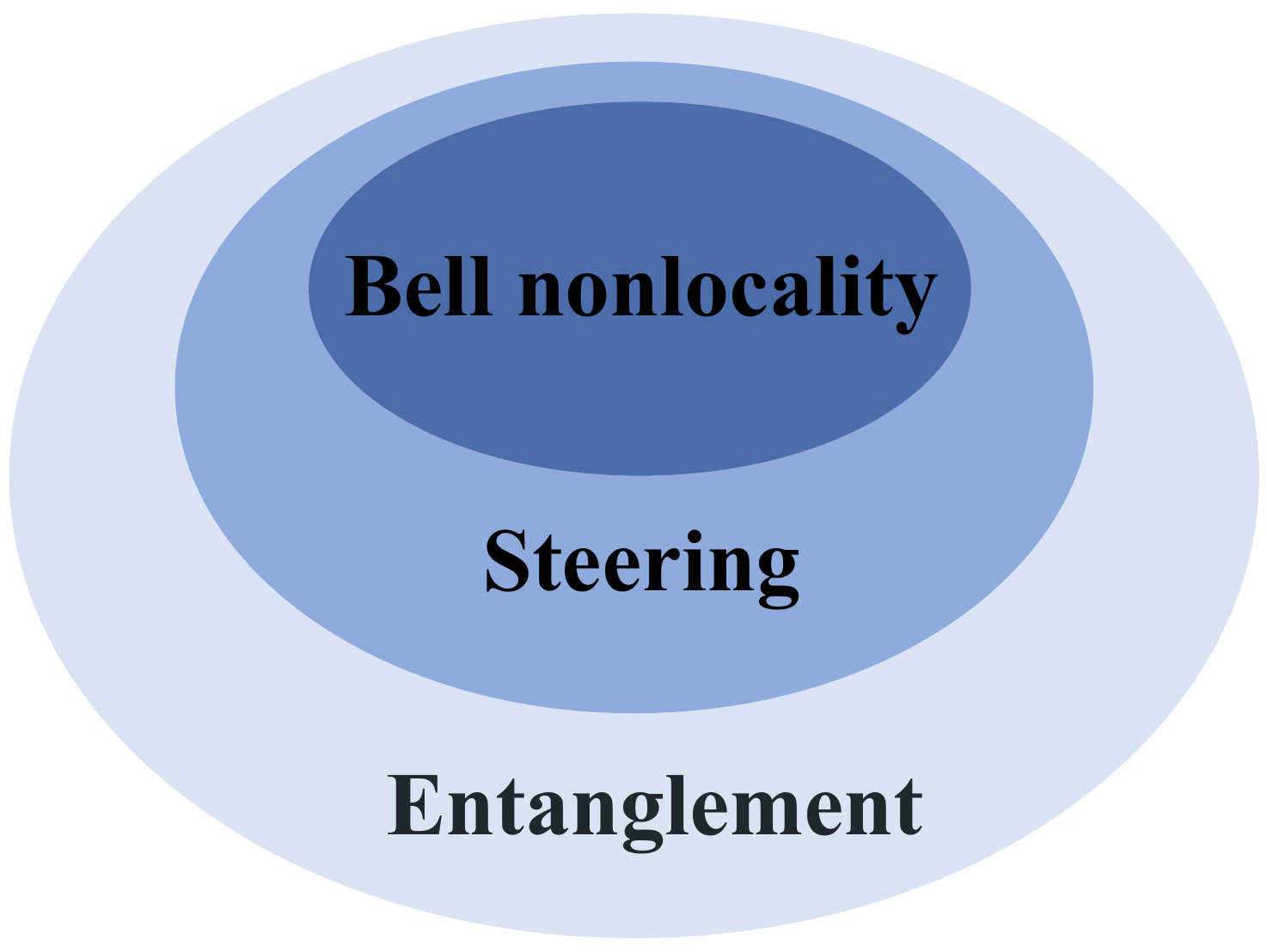}
\caption{Hierarchy of quantum entanglement. Quantum steering is a subset of quantum entanglement, providing extra accuracy for quantum information processing tasks. Besides, quantum steering is a superset of Bell nonlocality, which is friendly for experimental implementations.}
\label{EntHierarchy_QuanMagnonics}
\end{figure}

Quantum entanglement~\cite{Horodecki2009RMP_Ent}, which is the most intrinsic feature of quantum mechanics, possesses a variety of applications in quantum communication~\cite{Simon2017QuantumCommunication}, quantum simulation~\cite{Georgescu2014QuantumSimulation}, quantum computation~\cite{Nielsen2010QuantumComputation} and quantum sensing~\cite{Degen2017QuantumSensing,Pirandola2018QuantumSensing}. The concept of quantum entanglement was originally discovered by Schr\"{o}dinger~\cite{Schrodinger1935} in response to the famous EPR paradox predicted by Einstein, Podolsky, and Rosen~\cite{Einstein1935,Reid1989}, describing that the quantum state of the whole system cannot be factored as a product of each subsystem under any forms. It has been further rigorously defined by mathematical formulation from the perspective of quantum information tasks that quantum entanglement can be classified into three different classes, naming quantum nonseparability, quantum steering, and Bell nonlocality~\cite{Wiseman2007,Jones2007,Cavalcanti2009}. Figure~\ref{EntHierarchy_QuanMagnonics} manifests the hierarchy of three types of quantum correlations. It is apparently noted that quantum steering lies intermediately between the quantum entanglement and Bell nonlocality, more specifically, quantum steering is a strict subset of entanglement, while is a superset of Bell nonlocality~\cite{Uola2020RMP_SteeringReview}.

In contrast to entanglement, quantum steering~\cite{Reid2009,PhysRep2017} emphasizes the state of one particle can be remotely adjusted (steered) by performing local measurements on the other spatially separated particle when these two particles share a certain entanglement. This definition demonstrates a distinctive asymmetric feature of quantum steering with respect to entanglement and Bell nonlocality, that is, the degree of quantum steering from the direction of particle A to particle B is not the same as that in the opposite direction~\cite{Wagner2008}. Particularly, an extreme case of one-way steering appears when particle A can steer particle B, however, particle B is unable to steer particle A~\cite{Wiseman2007,Handchen2012}. The unique property of asymmetric steering has been theoretically confirmed~\cite{Olsen2008PRA,Midgley2010PRA,Olsen2013PRA,Schneeloch2013PRA,HeQY2013PRA_PulseOpto_Ent,HeQY2014PRA,Evans2014PRA,Bowles2014PRL,Skrzypczyk2014PRL,WangM2014PRA,Reidjosab2015,HeQY2015PRL,TanHT2015PRA,Marco2015PRA,Bowles2016PRA,Olsen2017PRL,Baker2018JOpt,ZhengShasha2019PRA,ZhengShasha2020SciChina}  and experimentally demonstrated with different configurations~\cite{Handchen2012,Armstrong2015,QinZZ2017PRA,SunKai2016PRL,Wollmann2016,XiaoYa2017PRL,Tischler2018PRL,Fadel2018Science}, among which some representative approaches to prepare directional steering are through employing asymmetric losses or noises on the subsystems~\cite{Wollmann2016,Tischler2018PRL,Handchen2012,Armstrong2015,QinZZ2017PRA}, as well as by an active manipulation method such as controlling the amplitude of signals injected into nondegenerate parametric oscillator~\cite{Olsen2017PRL} or adopting phase control mechanism in a closed-loop coupling structure~\cite{ZhengShasha2019PRA}.
The second distinguishing feature of quantum steering in contrast to quantum entanglement and Bell nonlocality can be demonstrated from the perspective of quantum information task. Quantum steering from the direction of Alice to Bob admits the verification of entanglement between Alice and Bob without assumptions of trust in the Alice's device. This so-called one-sided device independent feature promotes quantum steering as an essential resource and provides extra security for entanglement-based quantum information tasks in which the devices of some subsystems are not trusted~\cite{Opanchuk2014}, such as quantum teleportation~\cite{He2015PRL_Teleportation,Reid2013PRA,CMLi2016NPJ_QI}, quantum key distribution~\cite{Branciard2012QKD,Gehring2015QKD,Walk2016QKD}, quantum secret sharing~\cite{Armstrong2015QSS,Xiang2017QSS,Kogias2017QSS}, one-way quantum computing~\cite{Li2015OneWayComputation}, randomness generation~\cite{Skrzypczyk2018PRL_RandomnessGeneration,GuoYu2019PRL_RandomnessGeneration}, subchannel discrimination~\cite{Piani2015PRL}, and other related protocols.

A great deal of efforts have been devoted to generating and manipulating different kinds of quantum entanglement in a variety of quantum systems. In particular, the theoretical and experimental realizations of quantum entanglement among macroscopic size objects are more desirable. As indicated before, the hybrid quantum magnonic systems offer promising platforms for studying macroscopic quantum entanglement.
In order to confirm and further quantify the entanglement generation in the hybrid quantum magnonic systems, it is necessary to introduce some typical entanglement quantifiers in the following.
Generally, we can classify the hybrid quantum magnonic systems as two main categories, that is, discrete variable system and continuous variable system, respectively.
The so-called discrete (continuous) variable system means that the relevant observables are characterized by a discrete (continuous) eigenvalue spectrum. Specially, the quantum physical elements of cavity photons, phonons, as well as bosonized magnons can be considered as typical continuous variable systems; while the physical elements of single atom, qubit, etc., are classified into the discrete variable system.
In this tutorial, we mainly focus on the entanglement generation among magnons, cavity photons, and phonons, therefore, we emphatically introduce some typical entanglement quantifiers in the continuous variable systems.
Generally speaking, the orthogonal  amplitude and phase quadratures are usually employed to characterize such system. Among the continuous variable systems, the Gaussian system, referring to the system whose Wigner function obeys the Gaussian distribution, has received extensive attention due to its unique advantages of simplified calculation in theory and easy tunability in experiments. In a great amount of quantum magnonic systems without discrete physical elements (such as, qubits, atoms) as introduced in this tutorial, the input quantum noise of each mode possesses Gaussian characteristics and the dynamical equations of the system can be translated to linear, therefore the steady state of the system is guaranteed as a zero-mean Gaussian state. Note that the magneto-elastic, magneto-optical, multi-magnon-magnon interactions are all nonlinear as introduced in Sec.~\ref{preliminary}, however, the equations of motion only include linear terms while ignoring high-order terms through the standard linearization procedure when the driving laser intensity is fairly large.
For a continuous variable Gaussian system with $N$ subsystems, the system can be characterized by the vector $\hat{R}=(\hat{x}_{1},~\hat{p}_{1},~\hat{x}_{2},~\hat{p}_{2},\cdots,\hat{x}_{N},~\hat{p}_{N})^{\rm T}$, where $\hat{x}_{j}=(\hat{a}_{j}+\hat{a}_{j}^{\dagger})/\sqrt{2},~\hat{p}_{j}=(\hat{a}_{j}-\hat{a}_{j}^{\dagger})/\sqrt{2}i$ respectively represent the position and momentum quadratures of the $j$-th ($j=1,2,\cdots, N$) subsystems with $\hat{a}_{j}$ and $\hat{a}_{j}^{\dagger}$ denoting the corresponding annihilation and creation operators. Therefore, such Gaussian system can be characterized by the first-order moment
\begin{equation}
d_{j}=\langle \hat{R}_{j} \rangle_{\rho},
\end{equation}
as well as second-order moment (i.e., the covariance matrix)
\begin{equation}
V_{i, j}=\langle \hat{R}_{i}\hat{R}_{j}+\hat{R}_{j}\hat{R}_{i} \rangle_{\rho}/2-\langle \hat{R}_{i}\rangle_{\rho} \langle \hat{R}_{j} \rangle_{\rho}.
\label{QuanMagnonics_EntConcept_CM}
\end{equation}
Here $\langle \hat{O} \rangle_{\rho}=\mathrm{Tr}[\rho \hat{O}]$ denotes the expectation value of the operator $\hat{O}$ on the state $\rho$. For the Gaussian system, the first-order and second-order moments of the position and momentum quadratures are adequate to completely determine all properties inside the system~\cite{Holevo1975Gaussian, Holevo2011Gaussian}.

There exist various criteria for quantifying bipartite and multipartite quantum entanglement and EPR steering for Gaussian systems. The references introduced in this tutorial are specifically interested in quantum correlations among two parties or three parties, which mainly adopt the quantifier based on the covariance matrix of the system.

For the bipartite Gaussian system consisting of the $j$-th and $k$-th subsystems, it is enough to consider the $4\times 4$ reduced covariance matrix (CM)
\begin{equation}
V=\left(
   \begin{array}{cc}
   V_{j} & V_{j,k}\\
   V_{j,k}^{\mathrm{T}} & V_{k}
   \end{array}
   \right).
\end{equation}
Here, block matrices $V_{j}$ and $V_{k}$ with dimension $2\times 2$ are the corresponding covariance matrices of the reduced $j$-th and $k$-th subsystems (where we have abbreviated $V_{j,j}=V_{j}$ and $V_{k,k}=V_k$  ), and $2\times 2$ matrix $V_{j,k}$ represents the correlations among two subsystems, as defined in Eq.~(\ref{QuanMagnonics_EntConcept_CM}).

Generally, the logarithmic negativity~\cite{Vidal2002_LogNegativity,Adesso2004_LogNegativity} is frequently employed to measure the bipartite entanglement, which is given by
\begin{equation}\label{QuanMagnonics_LogNegativity}
E_N=\max{\{0,-\ln{2\eta^{-}}\}}.
\end{equation}
Here, $\eta^{-}\equiv \min \mathrm{eig}\left | i\Omega \widetilde{V}  \right|$ represents the minimum symplectic eigenvalue of the partial transposed covariance matrix $\widetilde{V}=PVP$, where $\Omega=\bigoplus_{i=1}^2 i\hat{\sigma}_y$ denotes the symplectic matrix with $\hat{\sigma}_y$ being the Pauli operator, and $P=\mathrm{diag}(1,-1,1,1)$ is used for realizing partial transposition.
Therefore, two subsystems of the $j$-th and $k$-th components are entangled if and only if $\eta^{-}<1/2$. The value of logarithmic negativity quantifies the degree of entanglement, that is, the higher the value of $E_N$ becomes, the stronger the bipartite entanglement appears.

To quantify EPR steering between bipartite subsystems, Kogias et al.~\cite{Kogias2015_BipartiteSteering} proposed a computable measure, which is valid for arbitrary bipartite Gaussian states under Gaussian measurements. The steering in two directions are given by
 \begin{eqnarray}\label{Eq_QuanMagnonics_BipartiteSteeringG}
&& \mathcal{G}^{j\rightarrow k}=\max{\{0,S(2V_{j})-S(2V)\}}, \nonumber \\
&& \mathcal{G}^{k\rightarrow j}=\max{\{0,S(2V_{k})-S(2V)\}},
 \end{eqnarray}
 where $S(\sigma)=\frac{1}{2}\ln{\det{\sigma}}$ corresponds to the R\'enyi-2 entropy. $\mathcal{G}^{j \rightarrow k}>0$ ($\mathcal{G}^{k \rightarrow j}>0$) indicates that bipartite Gaussian steering is present in the direction from the $j$-th ($k$-th) subsystem to $k$-th ($j$-th) subsystem, i.e., the $j$-th ($k$-th) subsystem has the ability to infer the amplitude or phase quadratures of the $k$-th ($j$-th) subsystem at a given accuracy. Besides, the larger value of $\mathcal{G}$ implies the stronger steering among two subsystems.

It is worth noting that the entanglement quantifier logarithmic negativity $E_N$~\cite{Vidal2002_LogNegativity,Adesso2004_LogNegativity} and steering quantifier $\mathcal{G}$~\cite{Kogias2015_BipartiteSteering} are both sufficient and necessary for two-mode Gaussian modes under Gaussian measurements, which is equivalent to the corresponding criteria of bipartite entanglement~\cite{DuanLuming2020_Ent_Inequality} and EPR steering~\cite{Reid1989,HeQY2013PRA_PulseOpto_Ent} identified by the violation of the inequality.

Apart from the bipartite quantum correlations, the entanglement and EPR steering criteria for tripartite systems have also been developed based on the covariance matrix. Generally, the residual contangle $E_\tau$ is adopted to signify the tripartite entanglement in the continuous variable system~\cite{Gerardo2006NJP_TripartiteEnt,Gerardo2007JPA_EntReview}, which is analogous to the tangle characterizing tripartite entanglement in discrete variable system. The tripartite entanglement among three subsystems labeled by $\{j,~k,~l\}$ is quantified by the minimum residual contangle, which reads~\cite{Gerardo2006NJP_TripartiteEnt,Gerardo2007JPA_EntReview}
\begin{equation}\label{Eq_QuanMagnonics_ResContangle}
E_\tau^{\min } \equiv \min \left[E_\tau^{j \mid k l}, E_\tau^{k \mid j l}, E_\tau^{l \mid j k}\right],
\end{equation}
where $E_\tau^{u \mid v w}=C_\tau^{u \mid v w}-C_\tau^{u \mid v}-C_\tau^{u \mid w}$  is the residual contangle with superscripts $\{u,~v,~w\}$ representing all the possible results when permuting three subsystems $\{j,~k,~l\}$. $C_\tau^{u \mid v w}=(E_N^{u \mid v w})^2,~C_\tau^{u \mid v}=(E_N^{u \mid v})^2,~C_\tau^{u \mid w}=(E_N^{u \mid w})^2$ represent the contangle of subsystems defined by a squared logarithmic negativity. The logarithmic negativity $E_N^{u \mid v}$ and $E_N^{u \mid w}$ can be computed according to Eq.~(\ref{QuanMagnonics_LogNegativity}), while the one-mode-versus-two-modes logarithmic negativity is determined by $E_N^{u \mid v w}=\max{\{0,-\ln{2(\eta^{-})^{u \mid v w}}\}}$ with $(\eta^{-})^{u \mid v w} \equiv \min \mathrm{eig}\left | i (\bigoplus_{i=1}^3 i\hat{\sigma}_y) P^{u \mid v w} V' P^{u \mid v w}  \right|$. $V'$ is the corresponding $6\times6$ reduced covariance matrix of the $\{j,~k,~l\}$-th subsystems, and $P^{j \mid k l} =\mathrm{diag}(1,-1,1,1,1,1)$, $P^{k \mid j l} =\mathrm{diag}(1,1,1,-1,1,1)$, $P^{l \mid j k} =\mathrm{diag}(1,1,1,1,1,-1)$.

With regard to tripartite steering, it has been confirmed that the residual Gaussian steering $\mathcal{G}_\tau^{\min } \equiv \min \left[\mathcal{G}_\tau^{j \mid k l}, \mathcal{G}_\tau^{k \mid j l}, \mathcal{G}_\tau^{l \mid j k}\right]$ with $\mathcal{G}_\tau^{j \mid k l}\equiv \mathcal{G}_\tau^{j \rightarrow k l}-\mathcal{G}_\tau^{j \rightarrow k}-\mathcal{G}_\tau^{j \rightarrow l}$ deriving from the Coffman-Kundu-Wootters (CKW) monogamy inequality is a quantitative indicator of genuine tripartite steering under Gaussian measurements, however, which is effective for pure three-mode Gaussian states~\cite{XiangYu2017PRA_SteeringMonogamy}. The realistic quantum magnonic system we consider is in a mixed state for the inevitable coupling between the hybrid system with the surrounding environment. In view of this situation, the genuine tripartite steering based on the violation of the inequality is a convenient and effective method~\cite{HeQY2013PRL_MultipartiteSteering}. The genuine tripartite steering among the $j,~k,~l$-th subsystems is present when
\begin{equation}\label{Eq_QuanMagnonics_TripartiteSteering}
S_{jkl}=S_{j|kl}+S_{k|jl}+S_{l|jk}<1/4,
\end{equation}
where $S_{u|vw} \equiv V_{\mathrm{inf}}(\hat{X}_u^{\theta_u}) V_{\mathrm{inf}}(\hat{P}_u^{\theta_u})(u \neq v \neq w \in \{j,~k,~l\})$ quantifies the steering from the $v,~w$-th collective modes to the $u$-th subsystem.
$\hat{X}_u^{\theta_u}=\hat{x}_u\cos{\theta_u}+\hat{p}_u\cos{\theta_u}$ and $\hat{P}_u^{\theta_u}=-\hat{x}_u\sin{\theta_u}+\hat{p}_u\sin{\theta_u}$ denote the generalized quadratures with $\hat{x}_u$ and $\hat{p}_u$ defined as the amplitude and phase quadratures of each mode. $V_{\mathrm{inf}}(\hat{X}_u^{\theta_u})$ and $V_{\mathrm{inf}}(\hat{P}_u^{\theta_u})$ are inferred variance, defined as $V_{\mathrm{inf}}(\hat{X}_u^{\theta_u})=V(\hat{X}_u^{\theta_u}+f_{x,v} \hat{X}_v^{\theta_v}+f_{x,w} \hat{X}_w^{\theta_w})$,~$V_{\mathrm{inf}}(\hat{P}_u^{\theta_u})=V(\hat{P}_u^{\theta_u}+f_{y,v} \hat{P}_v^{\theta_v}+f_{y,w} \hat{P}_w^{\theta_w})$.

\subsubsection{The significance and approaches of entanglement generation in quantum magnonic system}
Before we introduce the means of generating entanglement through nonlinear magnon process, we would like to briefly demonstrate the significance of investigating quantum entanglement in quantum magnonic system. As has been mentioned before, hybrid magnonic systems have emerged as a promising platform for studying macroscopic quantum phenomenon owing to the large size of the magnon. In a broad sense, generating nonclassical states especially the quantum correlations in these hybrid magnonic systems is fairly significant for both fundamental studies in terms of quantum-to-classical crossover~\cite{Zurek2003review_Decoherence}, macroscopic quantum effects~\cite{Florian2018RMP_MacroState}, wave-function collapse theory~\cite{Angelo2013RMP_WaveCollapse,ZhangJing2017PRA_WaveCollaspe} and near-term practical applications for plenty of quantum information tasks ranging from quantum cryptography, quantum teleportation, to quantum precise measurement~\cite{Horodecki2009RMP_Ent}. More specifically, it has been experimentally confirmed that the entanglement between a magnetostatic magnon mode and superconducting qubit is beneficial for the detection of a single magnon, which may open the avenues for quantum sensing of magnons in the area of quantum magnonics~\cite{Lachance-Quirion2020SingleMagnon}. Furthermore,
the entanglement inside the quantum magnonic system may serve as a complementary resource for deep investigation of the essential characteristics intrinsic to the magnetic material. For example, it is known that the magnetic resonance technique can provide useful information about the relaxation of the collective excitation in a ferromagnet, while cannot accurately determine the damping of each sublattice in a two-sublattice ferrimagnet or antiferromagnet. Surprisingly, Zheng et al.~\cite{ZhengShasha2020SciChina} showed that the asymmetric EPR steering among two magnons in a hybrid ferrimagnet-light system benefits the determination of the magnetic damping rates of the two sublattices, which further enables the deep understanding of the spin-pumping effect among the two sublattices.

Specifically, the generation and manipulation of the quantum correlations between two massive objects are extremely significant, and have stimulated increasing research interest.
There are abundant approaches to generate entanglement among magnons and other kinds of freedom, which can be primarily divided into two categories--entanglement generation with nonlinearity and without nonlinearity, respectively. As this tutorial mainly focuses on the topic of nonlinear magnonics, we will emphasize the entanglement generation by nonlinear magnon process, which is more extensively studied than the linear case. However, we would briefly introduce several representative schemes of generating entanglement without nonlinearity. Typically, Yuan et al.~\cite{Yuan2020PRB} proposed to generate the magnon-magnon entanglement in a two-sublattice antiferromagnetic system, and further enhance this entanglement through coupling the antiferromagnet to an additional light. The magnon-magnon entanglement is originated from the intrinsic exchange interaction between two spin sublattices with the form of parametric-down-conversion-type coupling, while none additional nonlinear interaction is required. The magnon-magnon entanglement in the antiferromagnetic system has also be considered by Kamra et al.~\cite{Kamra2019PRB_AntiferroEnt} Furthermore, Zheng et al.~\cite{ZhengShasha2020SciChina} deeply investigated the entanglement and EPR steering properties between two types of mangons in a hybrid ferrimagent-light system, finding that the intrinsic exchange interaction cannot prepare any steering between two magnon modes possessing the same damping rates of the two sublattices in the absence of light. While strong two-way asymmetric EPR steering is generated with identical dissipation with the assistance of the cavity field. It is noting that these two approaches~\cite{Yuan2020PRB,ZhengShasha2020SciChina} of generating magnon-magnon entanglement and quantum steering without nonlinearity is stronger than that of adopting nonlinearity~\cite{LiJie2019Magon-MagnonEnt,ZhangZhedong2019Magnon-MagnonEnt,Nair2020Magnon-MagnonEnt}, which would be introduced later in this subsection. Another important work of employing linearity to prepare entanglement is proposed by Yuan et al.~\cite{Yuan2020PRL_Bell}, where the authors studied the entanglement and Bell nonlocality properties between the magnons and cavity photons possessing dissipative coupling. This kind of linear dissipative coupling has been sufficiently experimentally confirmed~\cite{Harder2018PRL_LevelAttraction,Bhoi2019PRB_LevelAttraction,Boventer2020PRR_LevelAttraction}, fully characterized by the non-Hermitian Hamiltonian $\hat{\mathcal{H}}=\omega_r \hat{m}^{\dagger} \hat{m}+\omega_c \hat{c}^{\dagger} \hat{c}+g\left(\hat{m}^{\dagger}\hat{c}+e^{i \Phi} \hat{c}^{\dagger} \hat{m}\right)$ with $\hat{m}$ and $\hat{c}$ being the annihilation operators of the magnon and photon mode, respectively. It is shown that, for the level attraction case of $\Phi=\pi$, the magnon and photon could generate a high-fidelity steady Bell state being equipped with the maximum entanglement in the parity-time ($\mathcal{PT}$) broken phase ($|\Delta|=(\omega_r -\omega_c)/(2g)<1$), while the entanglement oscillates and no steady-state entanglement appears in the $\mathcal{PT}$-exact phase ($|\Delta|>1$). It seems somewhat confusing why this beam-splitter-type-like linear interaction could generate entanglement and Bell nonlocality in the steady state. This exceptional phenomenon can be explained by the non-Hermitian dynamics assuming the prerequisite of particle number conservation. Besides, Wang et. al~\cite{WangFei2022PRA_MagnonAtomEnt} proposed a scheme to create magnon-atom entanglement in a hybrid magnon-cavity-atom system, in which both a YIG sphere and a three-level $\Gamma$ type atomic ensemble are interacted with the microwave cavity through the Zeeman interaction. Kong and coauthors~\cite{KongDeyi2021PRB_MagnonAtomEnt} demonstrated the generation of magnon-magnon entanglement through the indirect coherent coupling between two magnons and a atom.

In addition to the approaches of entanglement generation with linear interaction, generating and manipulating different kinds of quantum correlations in quantum magnonic systems with the assistance of nonlinearity is an alternative way~\cite{LiJie2018PRL,Muhammad2022TriEntEnhancement_Feedback,TanHuatang2019PRR_Steering,LiJie2019Magon-MagnonEnt,ZhangZhedong2019Magnon-MagnonEnt,Nair2020Magnon-MagnonEnt,YuMei2020EntMagnon,YangZhibo2020OE_BiTripartite_KerrNonlinearity,SunFX2021PRL_CatState,Bittencourt2019PRA_MagnonFockState,ZhouYing2020SqueezingDrive_Bi-MultiEnt,JieLi2021QuanSciTec_TwoModeSqueezingDrive}. On the one hand, the nonlinearity can be generated by intrinsic nonlinear interactions, such as multi-magnon-magnon interaction process, magneto-elastic interaction, magneto-optical interaction, and magnet-qubit interaction, which are detailedly demonstrated in Sec.~\ref{preliminary}.
On the other hand, the nonlinearity intending for preparing different kinds of quantum correlations can be created externally and injected to the quantum magnonic system whose bare Hamiltonian is in the linear form, for example, injecting a single-mode or two-mode squeezing light generated externally by the parametric down-conversion process or Josephson parametric amplifier into the concerned quantum magnonic system. In the following, we would categorize the approaches of creating and manipulating different kinds of quantum correlations among quantum freedoms (including magnons, photons, phonons, etc.) from the perspective of different kinds of nonlinear interactions.

\begin{figure*}[tbp!]
\centering
\includegraphics[width=2\columnwidth]{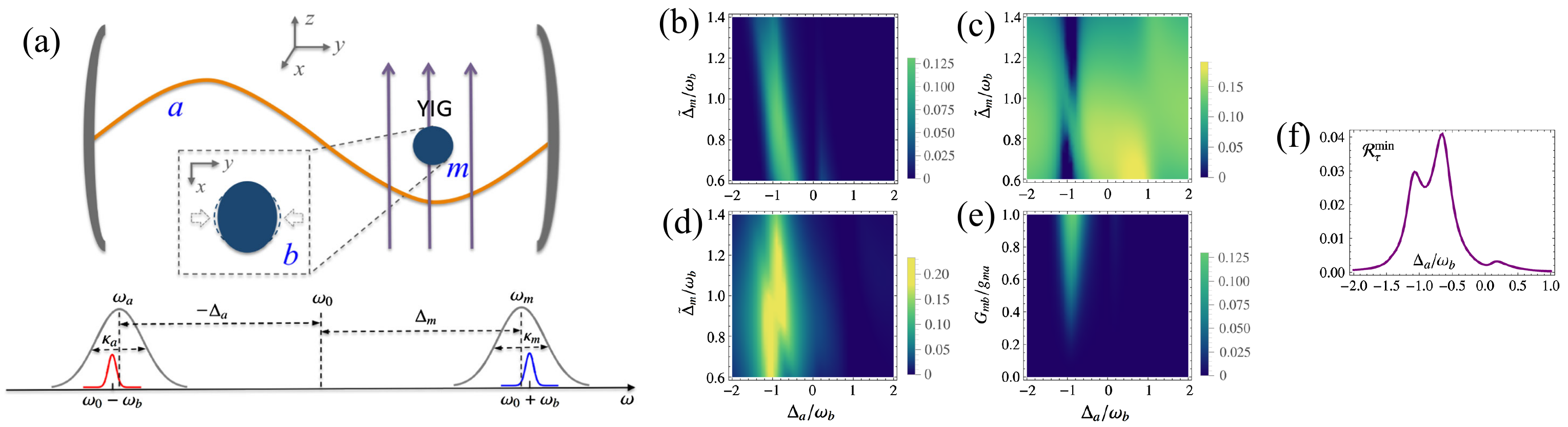}
\caption{(a) The top panel shows the schematic diagram of the hybrid cavity magnomechanical system~\protect\cite{LiJie2018PRL}. A YIG sphere is located at the position approaching the maximum magnetic field inside the cavity for the sake of generating large magnon-photon coupling. Besides, an additional uniform bias magnetic field is employed to flexibly tune the frequency of the magnon mode. The bottom panel shows the frequencies and linewidths of each mode. (b-d) The bipartite entanglement of magnon-photon $E_{am}$, magnon-phonon $E_{mb}$, photon-phonon $E_{ab}$ as a function of detuning $\Delta_a$ and $\Delta_m$, respectively. (e) The bipartite magnon-photon entanglement $E_{am}$ versus detuning $\Delta_a$ and the nonlinear magneto-elastic coupling strength $G_{mb}$. (f) The tripartite entanglement quantified by the minimum residual contangle versus $\Delta_a$. Reproduced with permission from Li et al., Phys. Rev. Lett. 121, 203601 (2018). Copyright 2018 American Physical Society.
}
\label{Fig_QuanMagnonics_JieLi2018PRL}
\end{figure*}

\subsubsection{The entanglement originated from the nonlinear magneto-elastic interaction}
\textit{\textbf{Generation of bipartite and tripartite entanglement}}. As far as we are concerned, the article put forward in 2018 by Li et al.~\cite{LiJie2018PRL} is the first work on generating genuine tripartite magnon-photon-phonon entanglement through the nonlinear magneto-elastic interaction by employing the cavity magnomechanical system, which represents a significant advance for the communities of both spintronics and quantum information science. Ever since from then on, macroscopic quantum entanglement in hybrid magnonic system has stimulated a wide range of research interests. Researchers make great efforts to prepare quantum entanglement between different parties, to enhance the amplitude of the quantum correlations, to extend the entanglement from bipartite to multipartite, etc. Here we would like to introduce and discuss the work \cite{LiJie2018PRL} thoroughly since the method and essential physics can be generalized to other hybrid magnonic systems.

Li et al.~\cite{LiJie2018PRL} studied an experimental feasible cavity magnomechanical system comprising of three different kinds of quantum physical elements, including the magnons, microwave photons and phonons, respectively, as displayed in Fig.~\ref{Fig_QuanMagnonics_JieLi2018PRL} (a). The Hamiltonian of the system reads
\begin{eqnarray}
\label{QuanMagnonics_JieLi2018PRL}
\hat{\mathcal{H}}&=&\omega_a \hat{a}^{\dagger} \hat{a}+\omega_m \hat{m}^{\dagger} \hat{m}+\frac{\omega_b}{2}\left(\hat{q}^2+\hat{p}^2\right)+g_{m b} \hat{m}^{\dagger} \hat{m} \hat{q} \nonumber \\
&&+g_{m a}\left(\hat{m}+\hat{m}^{\dagger}\right)\left(\hat{a}+\hat{a}^{\dagger}\right)+i \Omega\left(\hat{m}^{\dagger} e^{-i \omega_0 t}-\hat{m} e^{i \omega_0 t}\right). \nonumber \\
\end{eqnarray}
The first three terms in Eq.~(\ref{QuanMagnonics_JieLi2018PRL}) demonstrate the free energy of three different kinds of physical elements, in which $\hat{a}$ and $\hat{m}$ represent respectively the annihilation operators of the microwave photon mode and magnon mode, while $\hat{q}$ and $\hat{p}$ denote the dimensionless position and momentum operators for the phonon mode, respectively. The magnons couple to the microwave photons through Zeeman interaction with coupling strength $g_{m a}$, while interact with the phonons by means of the nonlinear magneto-elastic process with coupling strength $g_{m b}$, as introduced in Sec. \ref{magnon_phonon_interaction}. However, this single-magnon magnomechanical coupling strength $g_{m b}$ is especially small, which can be greatly enhanced through driving the magnon mode by a strong microwave source, as exhibited by the last term in Eq.~(\ref{QuanMagnonics_JieLi2018PRL}). Note that the coherent coupling between magnons and phonons are strongly suppressed in this case due to their frequency mismatch.

Translating the Hamiltonian Eq.~(\ref{QuanMagnonics_JieLi2018PRL}) to the time independent frame by means of frame rotating at the driving frequency $\omega_0$ and employing the Heisenberg equations of motion when taking the dissipation-fluctuation into account, the quantum Langevin equations can be naturally obtained. After the standard linearization procedure, the covariance matrix characterizing the correlations between the amplitude and phase quadratures can be determined by solving the Lyapunov equation. Therefore, according to the definitions of logarithmic negativity $E_N$ and minimum residual contangle $E_\tau$ determined in Eqs.~(\ref{QuanMagnonics_LogNegativity}) and~(\ref{Eq_QuanMagnonics_ResContangle}), respectively, the bipartite entanglement among arbitrarily two parties and tripartite magnon-photon-phonon entanglement can be well quantified. The numerical results of the dependence of the three pairs of bipartite entanglement on the detunings $\Delta_a$ and $\Delta_m$ are given in Figs.~\ref{Fig_QuanMagnonics_JieLi2018PRL} (b-d), which have adopted the experimentally feasible parameters~\cite{ZhangXufeng2016}. It is clearly demonstrated that all bipartite entanglement are generated when detunings satisfy $\Delta\simeq -\omega_b,~\Delta_m\simeq \omega_b$. In addition to the bipartite entanglement, the tripartite magnon-photon-phonon entanglement can be prepared, as indicated in Fig.~\ref{Fig_QuanMagnonics_JieLi2018PRL} (f).

\begin{figure*}[tbp!]
\centering
\includegraphics[width=1.90\columnwidth]{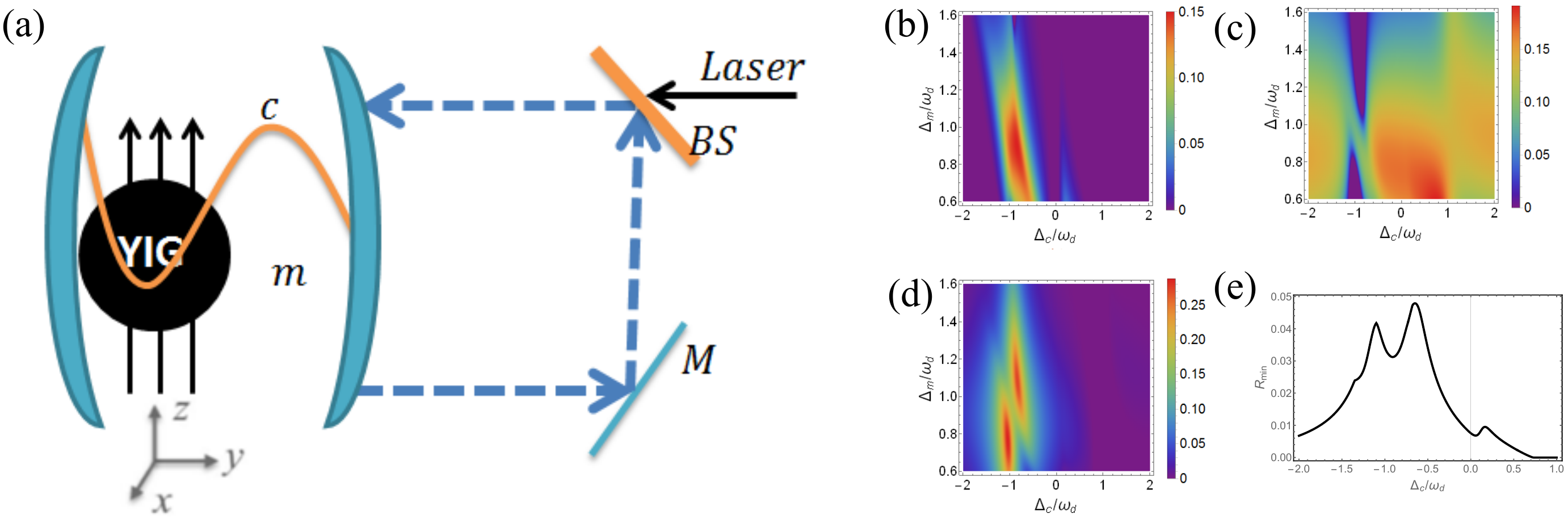}
\caption{(a) The schematic diagram of enhancing bipartite and tripartite entanglement in the hybrid cavity magnomechanical system~\protect\cite{Muhammad2022TriEntEnhancement_Feedback}. The schematic diagram is similar to Ref.~\protect\cite{LiJie2018PRL}, but with an additional feedback loop to reinject the photons into the cavity. (b-d) The bipartite magnon-photon, magnon-phonon, photon-phonon entanglement as a function of optical and magnonic detuning, respectively. (e) The tripartite entanglement quantified by the minimum residual contangle versus the optical detuning. Reproduced with permission from Amazioug et al., arXiv: 2211.17052 (2022).
}
\label{Fig_QuanMagnonics_Muhammad_Feedback}
\end{figure*}

Particularly worth mentioning is that one highlight of this work~\cite{LiJie2018PRL} relies on that the steady-state bipartite  entanglement among magnons and photons is generated, which is counterintuitive owing to the fact that the magnon mode are effectively coupled to the microwave photon mode through the beam-splitter-type interaction (i.e., with the form of $\hat{a}^\dagger \hat{m}+\hat{a}\hat{m}^\dagger$) after applying the rotating-wave approximation. Generally speaking, the individual beam-splitter-type coupling cannot generate continuous-variable entanglement and EPR steering in steady states without auxiliary operations on the desired system.
The essential physics of this counterintuitive phenomenon is well understood with the picture of entanglement transfer. Firstly, the bipartite magnon-phonon entanglement is initially created through the strong nonlinear magneto-elastic interaction, which is very similar to the phenomena in optomechanical systems~\cite{Vitali2007PRL_Optomechanics,Genes2008Ent_Optomechanics} as the magneto-elastic interaction possesses the similar form with the radiation pressure interaction. Subsequently, the bipartite magnon-phonon entanglement is partially transferred to the bipartite magnon-photon and photon-phonon entanglement. The emergence of the magnon-photon entanglement can be totally contributed to the nonlinear magneto-elastic interaction. The underlying physics can be confirmed by Fig.~\ref{Fig_QuanMagnonics_JieLi2018PRL} (e),
from which it is apparently manifested that the magnon-photon entanglement $E_{am}$ disappears when the nonlinear magneto-elastic interaction is absent.
This work has stimulated a wide range of research interests from both spintronics and quantum information science, and opens the door for studying macroscopic quantum phenomena in quantum magnonics.

\textit{\textbf{Enhancement of entanglement and EPR steering}}.
Ever since the entanglement preparation scheme in Ref.~\cite{LiJie2018PRL},  a variety of proposals are put forward to enhance different kinds of quantum correlations through the nonlinear magneto-elastic interaction.
Typically, Amazioug et al.~\cite{Muhammad2022TriEntEnhancement_Feedback} proposed a scheme to enhance the genuine tripartite magnon-photon-phonon entanglement in cavity magnomechanics by means of the coherent feedback control. This scheme is similar to that in the work by Li et al.~\cite{LiJie2018PRL},
where the main difference lies in the presence of an additional coherent feedback loop, as displayed in Fig.~\ref{Fig_QuanMagnonics_Muhammad_Feedback} (a). Figures~\ref{Fig_QuanMagnonics_Muhammad_Feedback} (b-d) and (e) show the three pairs of bipartite entanglement and tripartite magnon-photon-phonon entanglement, respectively. Similarly, the logarithmic negativity and minimum residual contangle are respectively adopted to quantify the bipartite and tripartite entanglement. Compared to Figs.~\ref{Fig_QuanMagnonics_JieLi2018PRL} (b-f) in Ref.~\cite{LiJie2018PRL}, it is apparent to find that the maximum value of both the bipartite and tripartite entanglement are greatly enhanced, which can be explained from the viewpoint that reinjecting the photons into the cavity through the coherent feedback loop could effectively increase the couplings between different bipartite modes. What is noteworthy is that the fundamental reason for the occurrence of entanglement still relies on the nonlinear magnon-elastic interaction between the magnon and mechanical modes, while the coherent feedback is just an effective method to improve the amplitudes of the quantum correlations.

\begin{figure}[tbp!]
\centering
\includegraphics[width=1\columnwidth]{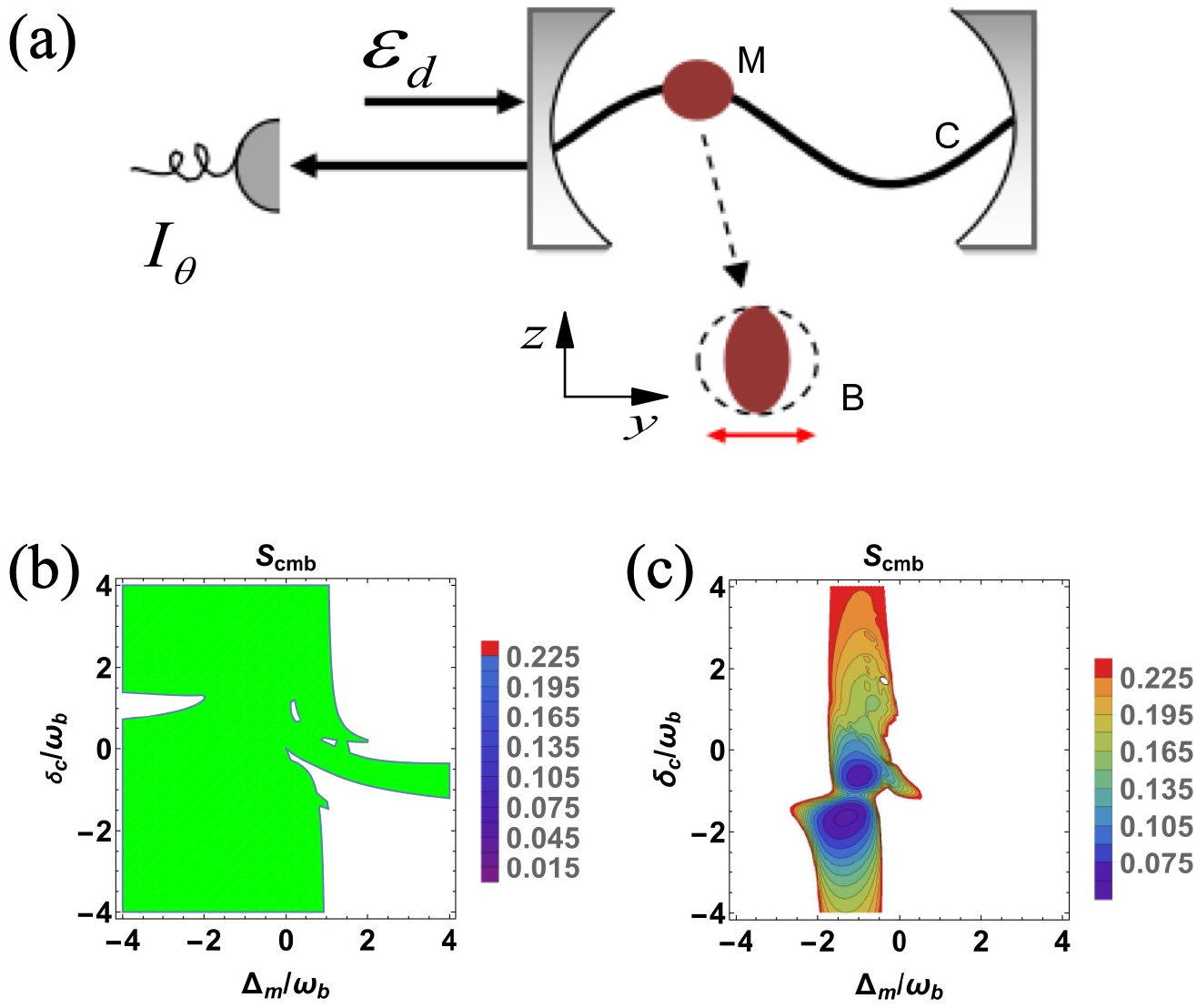}
\caption{(a) The schematic diagram of enhancing entanglement and preparing bipartite and tripartite steering in the hybrid cavity magnomechanical system~\protect\cite{TanHuatang2019PRR_Steering}. The schematic diagram is similar to that in Ref.~\protect\cite{LiJie2018PRL}, but with an additional time-continuous homodyne detection of the cavity output. (b-c) The genuine tripartite steering versus the optical and magnonic detunings. The green areas represent the system is unstable within this region, while the white areas signify the steering is absent in this area with steering quantifier $S_{cmb}>1/4$. Reproduced with permission from Tan et al., Phys. Rev. Res. 1, 033161 (2019). Copyright 2019 American Physical Society.
}
\label{Fig_QuanMagnonics_TanHuatang2019PRR}
\end{figure}

Besides, another alternative proposal to enhance bipartite and tripartite entanglement and further generate genuine tripartite steering (a strict subset of entanglement with stronger correlations, see details in Sec.~\ref{QuanMagnonics_EntConcept}) is put forward by Tan~\cite{TanHuatang2019PRR_Steering}. The cavity magnomechanical system consists of a YIG sphere inside the microwave cavity, in which the magnons are coupled to the photons and phonons via the Zeeman interaction and magnetostrictive interaction, respectively. It can be apparently seen that this system resembles closely to the regime presented in Ref.~\cite{LiJie2018PRL} shown in Fig.~\ref{Fig_QuanMagnonics_JieLi2018PRL} (a), however, with a remarkable modification in appending a time-continuous homodyne detection to the output field of the driven microwave cavity, as shown in Fig.~\ref{Fig_QuanMagnonics_TanHuatang2019PRR} (a).
Figures~\ref{Fig_QuanMagnonics_TanHuatang2019PRR} (b-c) display the tripartite steering without and with the continuous measurement, respectively, where the tripartite steering is quantified by Eq.~(\ref{Eq_QuanMagnonics_TripartiteSteering}) in Sec.~\ref{QuanMagnonics_EntConcept}. It is clearly observed from Fig.~\ref{Fig_QuanMagnonics_TanHuatang2019PRR} (b) that there exists no genuine tripartite steering in the steady-state scheme, although weak genuine tripartite magnon-photon-phonon entanglement is present as shown in Fig.~\ref{Fig_QuanMagnonics_JieLi2018PRL} (f) in literature~\cite{LiJie2018PRL}. In contrast, the bipartite steering among any two parties can be considerably enhanced (which is not shown in the figure) while the genuine tripartite steering can be further generated with the assistance of the time-continuous homodyne detection, manifested in Fig.~\ref{Fig_QuanMagnonics_TanHuatang2019PRR} (c). To deeply investigate the systematic Hamiltonian, we know that the entanglement and steering are totally originated from the nonlinear magnetostrictive interaction between magnons and phonons.

The authors considered that the output field of the driven microwave cavity is subject to a time-continuous homodyne detection. It is found that without the continuous measurement, only weak bipartite steering among the photons, magnons, and phonons can be obtained and the genuine tripartite steering is unachievable, although there exists weak genuine tripartite entanglement; when the continuous measurement is present, the bipartite steering is enhanced considerably and furthermore the genuine photon-magnon-phonon tripartite steering can be generated in the steady-state regime. The reason why the bipartite and tripartite steering can be enhanced is because the continuous measurement can greatly extend the stability area to the blue-detuned regime (the effective magnon detuning $\Delta_m<0$), as indicated in Figs.~\ref{Fig_QuanMagnonics_TanHuatang2019PRR} (b-c),
where larger magnon-phonon coupling strength is allowed to generate stronger bipartite and tripartite steering.
It is shown that the generated tripartite steering is robust against thermal fluctuations for realistic parameters. The scheme opens a promising route for exploring macroscopic quantum effects in such a macroscopic quantum interface of photons, magnons, and phonons.

\begin{figure*}[htbp!]
\centering
\includegraphics[width=1.96\columnwidth]{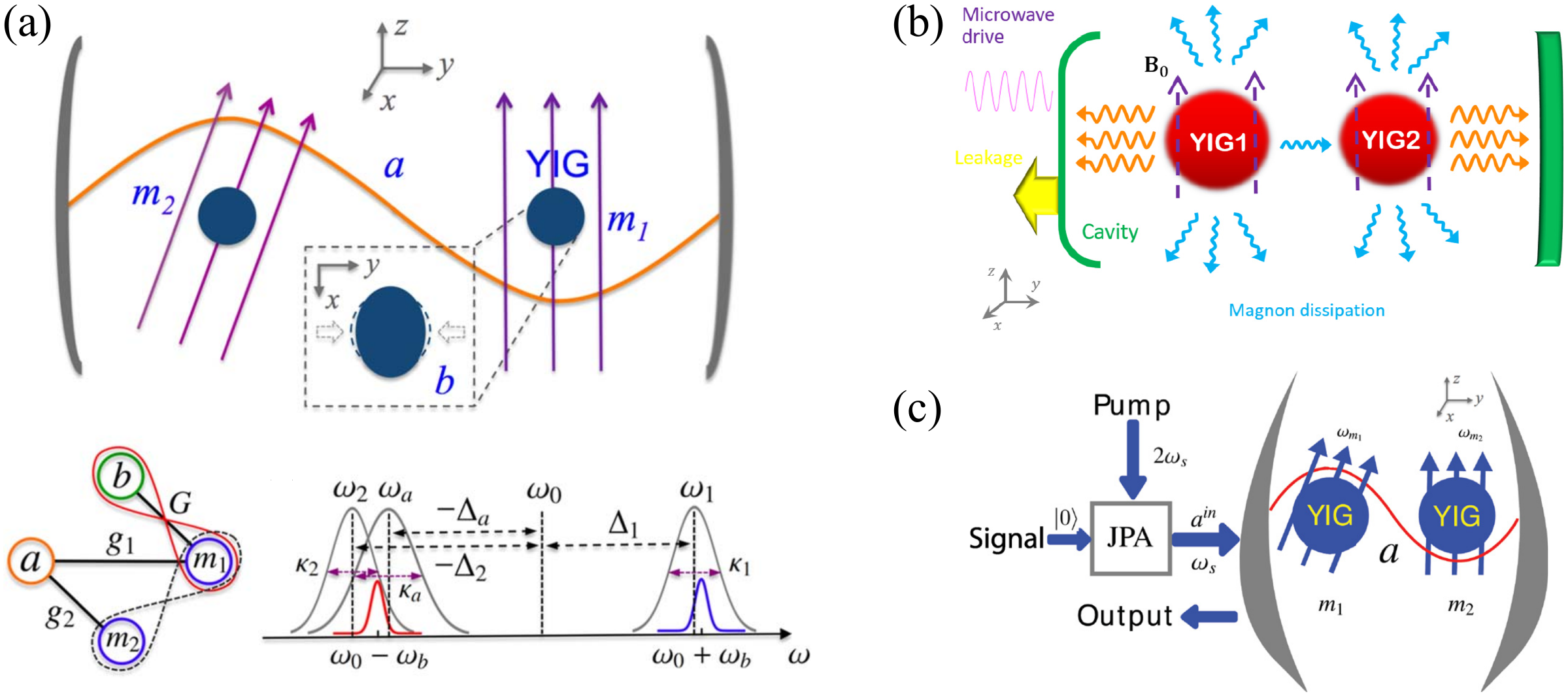}
\caption{Three typical approaches of generating magnon-magnon entanglement with a certain nonlinearity.
(a) The top panel shows schematic diagram of generating magnon-magnon entanglement based on the nonlinear magneto-elastic interaction~\protect\cite{LiJie2019Magon-MagnonEnt}. The bottom figures show the simplified interaction coupling in the system, and the frequencies and linewidths of each mode.
(b) Entanglement generation between two magnons through the magnon Kerr nonlinear interaction~\protect\cite{ZhangZhedong2019Magnon-MagnonEnt}.
(c) The scheme to generate magnon-magnon entanglement with the assistance of the nonlinearity from the externally injected light~\protect\cite{Nair2020Magnon-MagnonEnt}, while the bare interaction Hamiltonian of the quantum magnonic system is linear. [(a) is reproduced with permission from Li et al.,  New J. Phys. 21, 085001 (2019); Copyright 2019 The Author(s). (b) is reproduced with permission from Zhang et al., Phys. Rev. Res. 1, 023021 (2019); Copyright 2019 American Physical Society. (c) is reproduced with permission from Appl. Phys. Lett. 117, 084001 (2020),  Copyright 2020 The Author(s).]
}
\label{Fig_QuanMagnonics_MagnonMagnonEnt_Nonlinearity}
\end{figure*}

\textit{\textbf{Magnon-magnon entanglement}}. In addition to creating quantum entanglement and EPR steering between magnons with other kinds of freedom, such as photons or phonons, it is more attractive to generate and enhance quantum correlations between two macroscopic magnon modes from both fundamental significance (e.g., studying macroscopic quantum phenomena or the quantum-classical crossover) and practical motivation for applications in various quantum information processing tasks. When two YIG crystals are placed in a microwave cavity, the quantum entanglement between two magnon modes can be prepared by introducing a certain nonlinear interaction~\cite{LiJie2019Magon-MagnonEnt,ZhangZhedong2019Magnon-MagnonEnt,Nair2020Magnon-MagnonEnt}. For example, as listed in Fig.~\ref{Fig_QuanMagnonics_MagnonMagnonEnt_Nonlinearity},
Li et al.~\cite{LiJie2019Magon-MagnonEnt} introduced the nonlinear interaction provided by the magnetostrictive effect between the magnon and phonon modes; Zhang et al.~\cite{ZhangZhedong2019Magnon-MagnonEnt} adopted the Kerr nonlinear effect produced by the anisotropic field, which is equivalent to squeeze the intracavity magnon mode; Nair et al.~\cite{Nair2020Magnon-MagnonEnt} employed the externally injected squeezed light (externally injected squeezed light is also equivalent to introducing additional nonlinear terms to the system). Here in this part, we focus on the entanglement generation originated from the nonlinear magneto-elastic interaction. We would firstly go through the work of Ref. \cite{LiJie2019Magon-MagnonEnt} while the other two schemes~\cite{ZhangZhedong2019Magnon-MagnonEnt,Nair2020Magnon-MagnonEnt} would be introduced in the following sections ``The entanglement originated from the Kerr nonlinearity of magnons'' and ``The entanglement originated from the external nonlinear interaction'', respectively.

In the regime proposed by Li et al.~\cite{LiJie2019Magon-MagnonEnt}, the hybrid cavity magnomechanical system comprises of two magnon modes, a microwave cavity mode, and a mechanical mode, as manifested in Fig.~\ref{Fig_QuanMagnonics_MagnonMagnonEnt_Nonlinearity} (a). Two YIG spheres are still placed at the position near the maximum magnetic fields in the microwave cavity, and two magnon modes are excited through the uniform bias magnetic fields. However, the direction of the bias magnetic are specially tuned so that the mechanical mode $\hat{b}$ is excited merely in the surface of the right-hand YIG magnon mode $\hat{m}_1$, where these two modes ($\hat{m}_1$ and $\hat{b}$) are coupled through the nonlinear magneto-elastic interaction. Besides, two magnon modes ($\hat{m}_1,~\hat{m}_2$) respectively couple to the photon mode $\hat{a}$ via the Zeeman interaction. The interactions among the system are simplified as the left panel on the bottom of Fig.~\ref{Fig_QuanMagnonics_MagnonMagnonEnt_Nonlinearity} (a), resulting the Hamiltonian as
\begin{eqnarray}
\label{QuanMagnonics_JieLi2019MagMag}
\hat{\mathcal{H}}&= &\sum\limits_{j=1,2} \omega_j \hat{m}_j^{\dagger} \hat{m}_j+\frac{\omega_b}{2}\left(\hat{q}^2+\hat{p}^2\right)+ \omega_a \hat{a}^{\dagger} \hat{a}+G_0 \hat{m}_1^{\dagger} \hat{m}_1 \hat{q}  \nonumber\\
&& +\sum_{j=1,2} g_j\left(\hat{a}^{\dagger} \hat{m}_j+\hat{a} \hat{m}_j^{\dagger}\right)+i \Omega\left(\hat{m}_1^{\dagger} e^{-i \omega_0 t}-\hat{m}_1 e^{i \omega_0 t}\right), \nonumber\\
\end{eqnarray}
where the first three terms denote the free Hamiltonian of each mode; the fourth and fifth terms represent the nonlinear magneto-elastic interaction with coupling strength $G_0$ and magnon-photon  interaction with coupling strength $g_j~(j=1,2)$, respectively. The last term is introduced to enhance the effective coupling strength between the magnon mode and the mechanical mode.

\begin{figure}[htbp!]
\centering
\includegraphics[width=1\columnwidth]{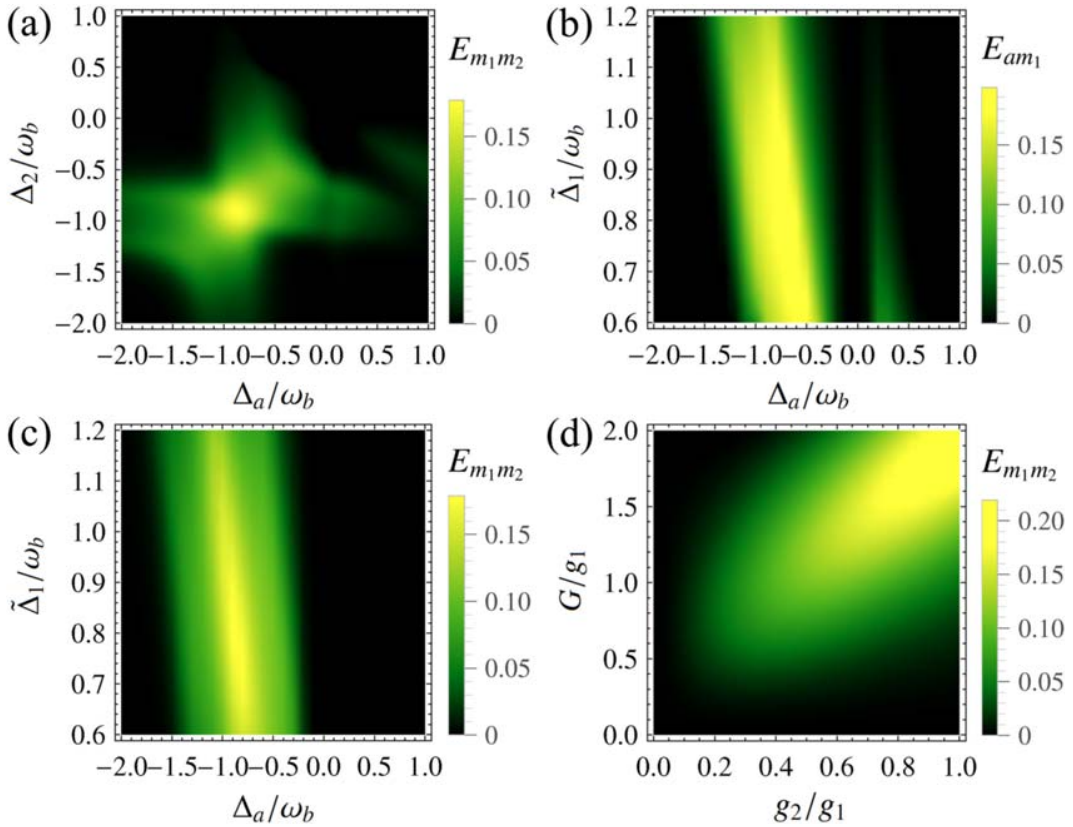}
\caption{Numerical results of magnon-magnon entanglement generated through the nonlinear magneto-elastic interaction~\protect\cite{LiJie2019Magon-MagnonEnt}. The schematic diagram is shown in Fig.~\ref{Fig_QuanMagnonics_MagnonMagnonEnt_Nonlinearity} (a).
The quantity of the magnon-magnon entanglement $E_{m_1 m_2}$ characterized by the logarithmic negativity versus (a) detunings $\Delta_a$ and $\Delta_2$; (c) detunings $\Delta_a$ and $\widetilde{\Delta}_1$; (d) the coupling strengths $g_2/g_1$ and $G/g_1$ with $g_1$ fixed. (b) The magnon-photon entanglement $E_{a m_1}$ as a function of detunings $\Delta_a$ and $\Delta_1$ when $g_2=0$. Reproduced with permission from Li et al.,  New J. Phys. 21, 085001 (2019). Copyright 2019 The Author(s).
}
\label{Fig_QuanMagnonics_JieLi2018NJP_MagMagEnt}
\end{figure}

Employing the experimental feasible parameters~\cite{ZhangXufeng2016}, Fig.~\ref{Fig_QuanMagnonics_JieLi2018NJP_MagMagEnt} manifests the steady-state magnon-magnon and the magnon-photon entanglement versus several key parameters of the system. It can be clearly seen from the Fig.~\ref{Fig_QuanMagnonics_JieLi2018NJP_MagMagEnt} (a) that the optimal magnon-magnon entanglement appears when the detunings satisfy $\widetilde{\Delta}_1\simeq\omega_b,~\Delta_2\simeq\Delta_a\simeq -\omega_b$.
Here, $\widetilde{\Delta}_1=\omega_1-\omega_0+G_0\langle \hat{q}\rangle$,~$\Delta_2=\omega_1-\omega_0$ and $\Delta_a=\omega_a-\omega_0$ denote the effective detunings of modes $\hat{m}_1$,~$\hat{m}_2$ and $\hat{a}$, respectively, where $\langle \hat{q} \rangle$ is the classical mean value of the position quadrature of the mechanical mode.
This phenomena is well understood as follows: Firstly, the nonlinear magneto-elastic interaction with strong coupling strength $G$ ($G=i\sqrt{2}G_0\langle \hat{m}_1\rangle$, where $\langle \hat{m}_1\rangle$ is the mean value of the magnon mode $\hat{m}_1$) is able to generate strong magnon-phonon entanglement $E_{m_1 b}$ when the detuning $\widetilde{\Delta}_1\simeq\omega_b$; this condition is required for the reason of significantly cooling the mechanical oscillator to be occupied with small thermal phonons so as to acquire maximal entanglement. Then, the magnon-phonon entanglement $E_{m_1 b}$ is partially transferred to the magnon-photon $E_{m_1 a}$ owing to the beam-splitter-type interaction between modes $\hat{m}_1$ and $\hat{a}$, which is prominent when detuning satisfies $\Delta_2\simeq -\omega_b$. Through the beam-splitter-type interaction between modes $\hat{m}_2$ and $\hat{a}$, the magnon-phonon entanglement $E_{m_1 a}$ is transferred to $E_{m_2 a}$ and finally to the magnon-magnon entanglement $E_{m_1 m_2}$. This entanglement transfer mechanism can be confirmed by Figs.~\ref{Fig_QuanMagnonics_JieLi2018NJP_MagMagEnt} (b-c), which show the magnon-photon entanglement $E_{m_1 a}$ when $g_2=0$ and the magnon-magnon entanglement $E_{m_1 m_2}$ when $g_2\neq0$, demonstrating that the magnon-photon entanglement $E_{m_1 a}$ is transferred to the magnon-magnon entanglement $E_{m_1 m_2}$. To summarize, the nonlinearity of the magneto-elastic interaction between the magnon and the mechanical modes of lattice vibration is responsible for  generating the magnon-magnon entanglement.

\textit{\textbf{Entanglement generation beyond magnons}}. Apart from the entanglement generation involving magnons, the magneto-elastic nonlinearity is also possible to prepare the entanglement merely involving photons, phonons and etc. For example, in 2020, Yu et al. ~\cite{YuMei2020EntMagnon} theoretically proposed a scheme to prepare quantum entanglement between two microwave fields in a cavity magnomechanical system utilizing the magnon mode as an intermediary. The system consists of a typical three-mode cavity magnomechanical system (comprised by a magnon mode, a mechanical mode, and a microwave cavity mode) as in Refs.~\cite{ZhangXufeng2016,LiJie2018PRL,Muhammad2022TriEntEnhancement_Feedback,TanHuatang2019PRR_Steering}, besides, an auxiliary microwave field is introduced to couple the magnon mode through the Zeeman interaction, where the Hamiltonian reads $ \hat{\mathcal{H}}= \sum_{j=1,2} \omega_j \hat{a}_j^{\dagger} \hat{a}_j+\omega_m \hat{m}^{\dagger} \hat{m}+\omega_b\left(\hat{q}^2+\hat{p}^2\right)/2+G_0 \hat{m}^{\dagger} \hat{m} \hat{q} +\sum_{j=1,2} g_j\left(\hat{a}_j^{\dagger} \hat{m}+\hat{a}_j \hat{m}^{\dagger}\right)+i \Omega\left(\hat{m}^{\dagger} e^{-i \omega_0 t}-\hat{m} e^{i \omega_0 t}\right)$. The annihilation operators $\hat{a}_1,~\hat{a}_2,~\hat{m}$ denote two microwave cavity modes and the magnon mode, respectively. And the dimensionless position and momentum operators $\hat{q},~\hat{p}$ represent the mechanical mode. The last term in the above Hamiltonian represents the driving of magnon to enhance the effective magnon-phonon coupling. By deeply analyzing the interaction type in this Hamiltonian, and taking the entanglement transfer mechanism into account, it can be theoretically confirmed that these two microwave cavity modes can be entangled, the details of which are omitted because the analysis is closely similar to that in generating magnon-magnon entanglement shown in Fig.~\ref{Fig_QuanMagnonics_MagnonMagnonEnt_Nonlinearity} (a) and Fig.~\ref{Fig_QuanMagnonics_JieLi2018NJP_MagMagEnt} when introducing the Ref~\cite{LiJie2019Magon-MagnonEnt}. Notably, the photon-photon entanglement in this system fundamentally results from the nonlinearity of magneto-elastic process in ferromagnetic crystals.

Let's briefly summarize this part. The magneto-elastic nonlinear interaction with the form of $\hat{\mathcal{H}}_{\mathrm{magneto-elastic}}=\hat{m}^{\dagger} \hat{m} \hat{q}$ possesses the ability to generate the steady-state entanglement and EPR steering between the magnon and the mechanical modes, which can be further partially transferred to the bipartite entanglement and quantum steering between X-Y (X and Y can be magnons, photons, or phonons) pairs~\cite{LiJie2018PRL,Muhammad2022TriEntEnhancement_Feedback,TanHuatang2019PRR_Steering,LiJie2019Magon-MagnonEnt,YuMei2020EntMagnon}, as well as genuine tripartite magnon-photon-phonon entanglement~\cite{LiJie2018PRL,Muhammad2022TriEntEnhancement_Feedback,TanHuatang2019PRR_Steering} by the essential mechanism of the entanglement transfer.

\subsubsection{The entanglement originated from the Kerr nonlinearity of magnons}
As indicated in Fig.~\ref{Fig_QuanMagnonics_MagnonMagnonEnt_Nonlinearity} (b), an optional approach to generate the magnon-magnon entanglement in quantum magnonics is to make use of the multi-magnon-magnon interaction, for example, the magnon Kerr interaction~\cite{ZhangZhedong2019Magnon-MagnonEnt}. This kind of magnon Kerr effect has been broadly studied both in experiments~\cite{WangYipu2016PRB_KerrExp,WangYipu2018PRL_KerrExpBistability} and in theoretical research (e.g., nonclassicality~\cite{JiangXi2022Photonics_KerrNonclassical}, nonreciprocity~\cite{KongCui2019PRAppl_KerrNonreciprocity}, and etc.)
The proposal put forward by Zhang et al.~\cite{ZhangZhedong2019Magnon-MagnonEnt} considers a hybrid cavity magnonic system consisting of two YIG spheres in a microwave cavity, in which the Kittel mode (i.e., spatially uniform mode) is excited by driving the cavity with a strong microwave field, as indicated in Fig.~\ref{Fig_QuanMagnonics_MagnonMagnonEnt_Nonlinearity} (b). The effective Hamiltonian is characterized by
\begin{eqnarray}
\hat{\mathcal{H}}&=&   \omega_c \hat{a}^{\dagger} \hat{a}+\sum_{j=1,2}\left[\omega_j \hat{m}_j^{\dagger} \hat{m}_j+g_j\left(\hat{m}_j^{\dagger} \hat{a}+\hat{m}_j \hat{a}^{\dagger}\right)\right. \nonumber \\
&& \left.+g_{k_j} \hat{m}_j^{\dagger} \hat{m}_j \hat{m}_j^{\dagger} \hat{m}_j\right]+i  \Omega\left(\hat{a}^{\dagger} e^{-i \omega_d t}-\hat{a} e^{i \omega_d t}\right),
\label{QuanMagnonics_ZhangZhedong2019_Ham}
\end{eqnarray}
where $\hat{a},~\hat{m}_1$, and $\hat{m}_2$ ($\hat{a}^{\dagger},~\hat{m}_1^{\dagger}$, and $\hat{m}_2^{\dagger}$) respectively represent the annihilation (creation) operators of the cavity mode, the left-hand and right-hand magnon modes. The multi-magnon process $g_{k_j}\hat{m}_j^{\dagger} \hat{m}_j \hat{m}_j^{\dagger} \hat{m}_j$ in the fourth term is the Kerr nonlinearity, resulting from the anisotropy of the magnetic medium, where $g_{k_j}$ ($j=1,2$) denotes the intensity of the Kerr effect. When the microcavity is strongly driven, the annihilation operators of each mode can be effectively expanded by the addition of the classical mean value and the quantum fluctuation, i.e., $\hat{O}=\langle \hat{O} \rangle+\delta \hat{O}$ ($\hat{O}=\hat{a},~\hat{m}_1,~\hat{m}_2$), which is the so-called linearization procedure. Inserting those linearized operators $\hat{O}$ into Eq. (\ref{QuanMagnonics_ZhangZhedong2019_Ham}) and neglecting the insusceptible high-order terms, the effective Hamiltonian is given by
\begin{eqnarray}
\hat{\mathcal{H}}_{\mathrm{eff}}&=&  \Delta_c  \hat{a}^{\dagger}  \hat{a}+\sum_{j=1,2}\left[\tilde{\delta}_j \hat{m}_j^{\dagger}  \hat{m}_j+\tilde{\Delta}_s  \hat{m}_j^{\dagger} \hat{m}_j^{\dagger}+\tilde{\Delta}_s^*  \hat{m}_j \hat{m}_j\right. \nonumber \\
&& \left.+g_j\left( \hat{m}_j^{\dagger}  \hat{a}+ \hat{m}_j  \hat{a}^{\dagger}\right)\right].
\end{eqnarray}
Here, the $\delta$ symbols in relevant quantum fluctuations have been omitted for simplicity. And the effective detunings denote $\Delta_c=\omega_c-\omega_d,~\tilde{\delta}_j =\omega_j-\omega_d+4g_{k_j}\left|\langle \hat{m}_j\rangle\right|^2,~\tilde{\Delta}_s=g_{k_j}\langle \hat{m}_j\rangle^2$. The four-magnon term is thus simplified as the quadratic items $\hat{m}_j^{\dagger} \hat{m}_j^{\dagger}$ and $\hat{m}_j \hat{m}_j$. This squeezing-like term is capable of squeezing the two magnon modes and further generating the entanglement between the magnons and photons. Furthermore, the beam-splitter-type interaction between magnons and photons results in the magnon-magnon entanglement with the picture of entanglement transfer.

\begin{figure}[tbp!]
\centering
\includegraphics[width=1\columnwidth]{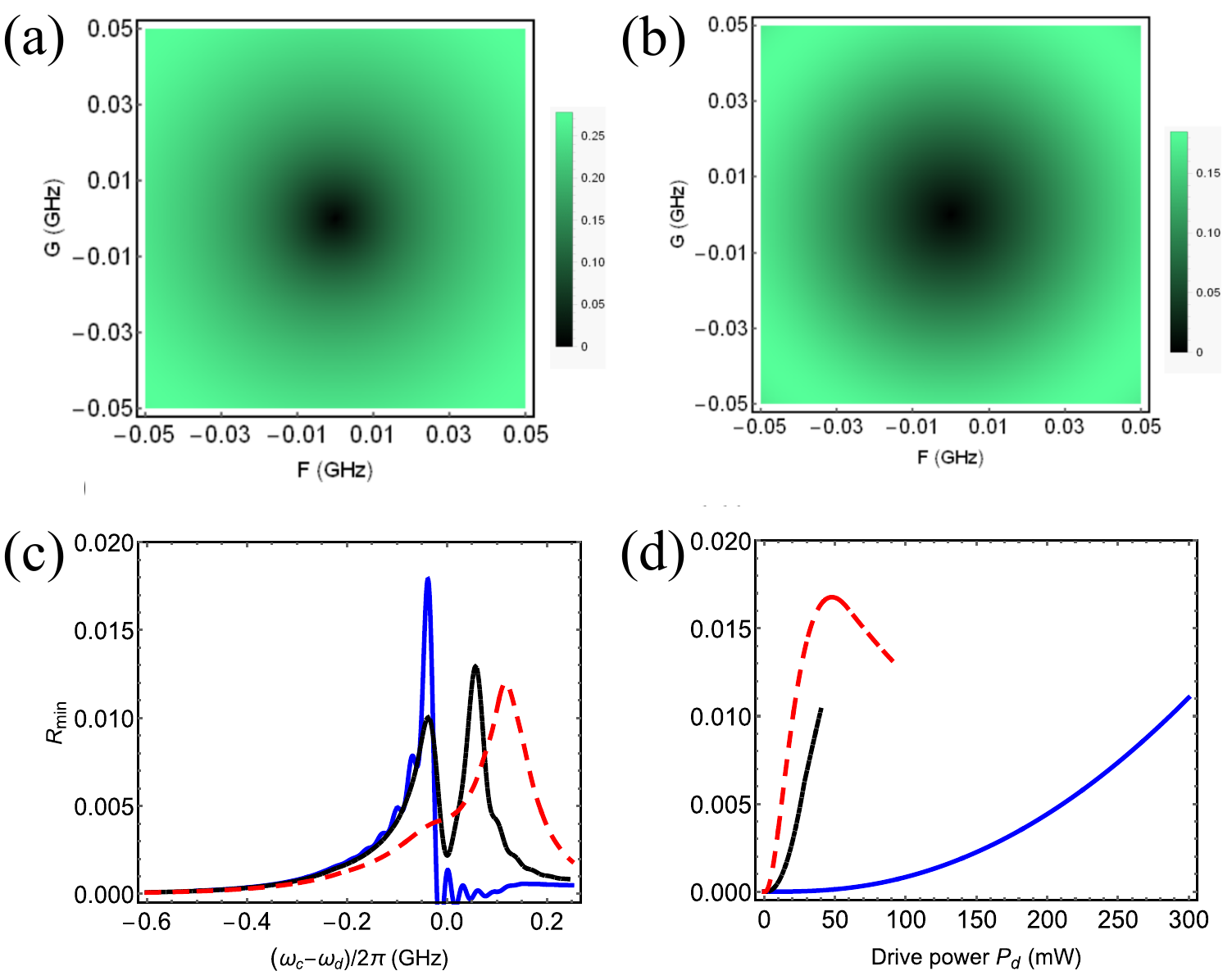}
\caption{Numerical results of magnon-magnon entanglement generated through the nonlinear multi-magnon-magnon process~\protect\cite{ZhangZhedong2019Magnon-MagnonEnt}. The schematic diagram corresponds to Fig.~\ref{Fig_QuanMagnonics_MagnonMagnonEnt_Nonlinearity} (b).
(a) The magnon-magnon entanglement $E_{m_1 m_2}$ and (b) the magnon-photon entanglement $E_{m_1 a}$ when $g_2=0$ versus the effective magnetocrystalline anisotropy quantities $F_j=2g_{k_j}\mathrm{Im}\langle \hat{m}_j\rangle^2$ and $G_j=2g_{k_j}\mathrm{Re}\langle \hat{m}_j\rangle^2$. Two magnon modes possessing the same parameters and are symmetrically coupled to the cavity photons, therefore $F_1=F_2,~G_1=G_2$.
(c-d) The tripartite magnon-magnon-photon entanglement quantified by the minimum residual contangle as a function of (c) the cavity detuning and (d) the driving power. The blue solid, red dashed and black dash-dotted curves represent different driving power in (c) and cavity detuning in (d).
Reproduced with permission from Zhang et al., Phys. Rev. Res. 1, 023021 (2019). Copyright 2019 American Physical Society.
}
\label{Fig_QuanMagnonics_ZhangZhedong2019PRR_MagMagEnt}
\end{figure}

Figures~\ref{Fig_QuanMagnonics_ZhangZhedong2019PRR_MagMagEnt} (a-b) show the magnon-magnon $E_{m_1 m_2}$ and magnon-photon entanglement $E_{m_1 a}$ characterized by the logarithmic negativity as a function of the effective magnetocrystalline anisotropy quantities $F$ and $G$, the definitions of which are explained in the caption. When the Kerr effect is absent (i.e., $g_{k_1}=g_{k_2}=0$, thus $F=G=0$), it can be apparently observed that neither the magnon-photon entanglement nor the magnon-magnon entanglement is present, owing to the fact that the individual beam-splitter-type interactions cannot generate any steady-state entanglement in the continuous variable systems. However, both the magnon-magnon $E_{m_1 m_2}$ and magnon-photon entanglement $E_{m_1 a}$ appear in the regime of $F=G\neq0$ when Kerr effect is taken into account, indicating that this Kerr nonlinear interaction is the fundamental origination of the entanglement.
In addition to the bipartite entanglement among two distinct magnon modes, the tripartite magnon-magnon-photon entanglement can be prepared and further manipulated by adjusting the cavity detuning and the microwave driving power, as manifested in Figs.~\ref{Fig_QuanMagnonics_ZhangZhedong2019PRR_MagMagEnt} (c-d). Similarly, the tripartite entanglement is also originated from the nonlinear Kerr effect of the magnons.

In addition, Yang et al.~\cite{YangZhibo2020OE_BiTripartite_KerrNonlinearity} proposed to generate both bipartite magnon-photon, photon-photon and tripartite entanglement employing the nonlinear Kerr effect resulting from the magnetocrystalline anisotropy. In the system, a YIG sphere is placed at the crossing point inside two intersecting cavity fields, described by the Hamiltonian
\begin{equation}
\begin{aligned}
\hat{\mathcal{H}}& =\omega_m \hat{m}^{\dagger} \hat{m}+\sum_{j=1,2}\left[\omega_j \hat{a}_j^{\dagger} \hat{a}_j+g_j\left(\hat{a}_j^{\dagger} \hat{m}+\hat{a}_j \hat{m}^{\dagger}\right)\right]\\
& ~~+g_k \hat{m}^{\dagger} \hat{m} \hat{m}^{\dagger} \hat{m}+i \Omega\left(\hat{m}^{\dagger} e^{-i \omega_d t}-\hat{m} e^{i \omega_d t}\right).
\label{QuanMagnonics_YangZhibo2020OE_Ham}
\end{aligned}
\end{equation}
The annihilation operators of two microwave photons and magnons are respectively written as $\hat{a}_1,~\hat{a}_2$, and $\hat{m}$. Notice that two microwave photon modes are coupled to the magnon mode through the beam-splitter-type interaction as indicated in Eq.~(\ref{QuanMagnonics_YangZhibo2020OE_Ham}), hence, it can be easily found that no entanglement can be prepared without the Kerr nonlinearity (i.e., $g_k=0$). The entanglement generation among different parties is all contributed from the Kerr nonlinearity. In the presence of the magnon Kerr nonlinearity (i.e., $g_k\neq0$), the magnon-photon and photon-photon bipartite entanglement can be generated. However, the maximal magnon-photon entanglement $E_{ma_1}\approx 0.065$, which is nearly four times smaller than that in Fig.~\ref{Fig_QuanMagnonics_ZhangZhedong2019PRR_MagMagEnt} (b) by Zhang et al~\cite{ZhangZhedong2019Magnon-MagnonEnt}. This may be contributed to the small Kerr nonlinearity. The entanglement can be further enhanced by increasing the strength of the Kerr nonlinearity.

However, compared to the approach of generating entanglement based on the nonlinear magneto-elastic interaction introduced in the last part, the entanglement generation originated from the multi-magnon process is not thoroughly investigated. An alternative direction of the research may lies at enhancing the bipartite entanglement to acquire stronger EPR steering, generating different types of multipartite quantum correlations  for applications in more complicated quantum information tasks.

\subsubsection{The entanglement originated from the nonlinear magneto-optical interaction}
\label{QuanMagnonics_EntGeneration_MOptical}
It has been confirmed from both theory and experiment that magnons can be coupled to the photons in the optical domain, which is in the form of $\hat{\mathcal{H}}_{mc}=\hat{c}^{\dagger}c(\hat{m}^\dagger + \hat{m})$ with $\hat{c}$ and $\hat{m}$ being the annihilation operators of the optical photons and magnons, respectively. The details of the derivation of the Hamiltonian how optical photons and magnons are coupled can refer to Sec.~\ref{magnon_photon_interaction}.
Note that such kind of interaction type resembles closely to the nonlinear magneto-elastic interaction between magnon-phonon pairs in quantum magnonics, as well as the radiation pressure interaction between photons-phonons pairs in optomechanics. Therefore, generating entanglement utilizing the nonlinear magneto-optical interaction is definitely feasible based on our preliminary knowledge. As a complement to the mostly studied approaches of generating quantum correlations in the steady states~\cite{LiJie2018PRL,Muhammad2022TriEntEnhancement_Feedback,TanHuatang2019PRR_Steering,LiJie2019Magon-MagnonEnt,YuMei2020EntMagnon,LiJie2018PRL,Muhammad2022TriEntEnhancement_Feedback,TanHuatang2019PRR_Steering}, we here introduce a novel method to create transient entanglement and EPR steering between magnon-photon pairs~\cite{SunFX2021PRL_CatState} with stronger quantum correlations than that in the steady states.

In Sec.~\ref{QuanMagnonics_EntGeneration_MOptical}, we have detailly introduce the theoretical proposal to remotely prepare magnon cat states through nonlinear magneto-optical interaction put forward by Sun and coauthors~\cite{SunFX2021PRL_CatState} as in Fig.~\ref{Fig_QuanMagnonics_SunFX2021PRL_CatState}, while the generation of quantum entanglement and EPR steering between magnons and photons are not yet shown. Here in this part, we would like to carefully analyze the specific procedure of entanglement generation as well as the effect of the nonlinear magneto-optical interaction on magnon-photon entanglement and EPR steering.

Since the method of calculating entanglement in the pulsed regime is quite different to that in the steady-state regime, here we detailedly introduce the way how to compute entanglement in the regime with pulsed magneto-optical interaction. Let's quickly recall the structure of the cavity magnonic system displayed in Fig.~\ref{Fig_QuanMagnonics_SunFX2021PRL_CatState}, which comprises of an optical photon mode and a magnon mode with these modes coupled through the nonlinear magneto-optical interaction. The effective Hamiltonian after frame rotating with respect to the driving frequency of the photons is described by Eq.~(\ref{QuanMagnonics_SunFX2021PRL}). For convenience, we again write down the Hamiltonian in this part, which reads $\hat{\mathcal{H}}=\Delta \hat{c}^{\dagger}\hat{c}+ \omega_{m}\hat{m}^{\dagger}\hat{m}+g_{0}\hat{c}^{\dagger}\hat{c}(\hat{m}^{\dagger}+\hat{m})$.

When pumping the cavity mode with a blue-detuning ($\Delta=-\omega_m$) and employing the standard linearization procedure as well as the rotating-wave approximation relative to $\omega_m$, the above Hamiltonian is simplified as
\begin{equation}
	\hat{\mathcal{H}}_{\mathrm{RWA}}=g (\hat{m}^{\dagger}\hat{c}^{\dagger}+\hat{m}\hat{c}),
\label{QuanMagnonics_SunFX2021PRL_RWA}
\end{equation}
where $g=g_0\langle \hat{c}\rangle$ is the effective magneto-optical coupling strength with  $\langle \hat{c}\rangle$ being the classical average value of the photon mode. Note that the annihilation operators $\hat{m}$ and $\hat{c}$ in Eq.~(\ref{QuanMagnonics_SunFX2021PRL_RWA}) represent the corresponding quantum fluctuations. The quantum Langevin equations are thus given by
\begin{align}
	\dot{\hat{m}}&=-\kappa_m \hat{m}-ig \hat{c}^{\dagger}-\sqrt{2\kappa_m}\hat{m}^{\mathrm{in}},\nonumber\\
	\dot{\hat{c}}&=-\kappa_c \hat{c}-ig \hat{m}^{\dagger}-\sqrt{2\kappa_c}\hat{c}^{\mathrm{in}}.
\end{align}
Here, $\kappa_m,~\kappa_c$ denote the dissipation rates of the magnon and photon modes; $\hat{m}^{\mathrm{in}}$ and $\hat{c}^{\mathrm{in}}$ represent the quantum noise of these two modes.

Furthermore, the adiabatic approximation of the cavity mode ($\dot{\hat{c}}=0$) is feasible when $\kappa_c\gg\kappa_m,~g$. Therefore, the optical mode can be analytically solved, i.e., $\hat{c}=-i\frac{g}{\kappa_c}\hat{m}^{\dagger}-\sqrt{\frac{2}{\kappa_c}}\hat{c}^{\mathrm{in}}$, and the magnon mode is rewritten as
\begin{equation}
	\dot{\hat{m}}=G\hat{m}+ig\sqrt{\frac{2}{\kappa_c}}\hat{c}^{\mathrm{in} \dagger}-\sqrt{2\kappa_m}\hat{m}^{\mathrm{in}},
\end{equation}
where $G=g^2/\kappa_c-\kappa_m$. The above process of deriving the equations of motion has no difference to that in the steady-state regime. However, since the optical mode is driven by a sequence of pulses, it is convenient to introduce normalized temporal operators for the input and output optical fields~\cite{Hofer2011PulseEnt,HeQY2013PRA_PulseOpto_Ent,Palomaki710}, which is in great contrast to regime in steady states. The normalized temporal operators are defined by~\cite{Hofer2011PulseEnt,HeQY2013PRA_PulseOpto_Ent,Palomaki710}
\begin{align}
	\hat{C}^{\mathrm{in}}&=\sqrt{\frac{2G}{1-e^{-2G\tau}}}\int_0^{\tau}e^{-Gt}\hat{c}^{\mathrm{in}}(t)dt,\nonumber\\
	\tilde{\hat{C}}^{\mathrm{in}}&=\sqrt{\frac{2G}{e^{2G\tau}-1}}\int_{0}^{\tau}e^{Gt}\hat{c}^{\mathrm{in}}(t)dt,\nonumber\\
	\hat{C}^{\mathrm{out}}&=\sqrt{\frac{2G}{e^{2G\tau}-1}}\int_{0}^{\tau}e^{Gt}\hat{c}^{\mathrm{out}}(t)dt,
	\label{temporal_mode_c}
\end{align}
in which $\hat{c}^{\mathrm{out}}(t)=\hat{c}^{\mathrm{in}}(t)+\sqrt{2\kappa_c}\hat{c}(t)$ denotes the input-output relation, and $\tau$ is the duration of the laser pulse. Similarly, the normalized operators for the magnon mode take the following form
\begin{align}
	\hat{M}_m&=\sqrt{\frac{2G}{1-e^{-2G\tau}}}\int_{0}^{\tau}e^{-Gt}\hat{m}^{\mathrm{in}}(t)dt,\nonumber\\
	\tilde{\hat{M}}_m&=\sqrt{\frac{2G}{e^{2G\tau}-1}}\int_{0}^{\tau}e^{Gt}\hat{m}^{\mathrm{in}}(t)dt,\nonumber\\
	\hat{M}^{\mathrm{in}}&=\hat{m}(0), \nonumber\\
	\hat{M}^{\mathrm{out}}&=\hat{m}(\tau).
	\label{temporal_mode_m}
\end{align}
At the end of the pulse, i.e., $t=\tau$, the output of the optical and magnon modes read,
\begin{align}
	\hat{C}^{\mathrm{out}}&=-i\sqrt{\frac{2g^2}{\kappa_c}}\sqrt{\frac{e^{2r}-1}{2G}}\hat{M}^{\mathrm{in}\dagger}-\frac{g^2}{G\kappa_c}(\hat{C}^{\mathrm{in}}e^r-\tilde{\hat{C}}^{\mathrm{in}})\nonumber\\
	&~~-\tilde{\hat{C}}^{\mathrm{in}}+i\frac{g}{G}\sqrt{\frac{\kappa_m}{\kappa_c}}(\hat{M}_m^\dagger e^r-\tilde{\hat{M}}_m^\dagger),\nonumber\\
	\hat{M}^{\mathrm{out}}&=e^{r}\hat{M}^{\mathrm{in}}+\sqrt{\frac{e^{2r}-1}{2G}}(i\sqrt{\frac{2g^2}{\kappa_c}}\hat{C}^{\mathrm{in}\dagger}-\sqrt{2\kappa_m}\hat{M}_m).
\label{QuanMagnonics_SunFX2021PRL_Output}
\end{align}
Here, we have defined the effective squeezing parameter $r=G\tau$, which is closely related to the quantity of the entanglement and the magnon cat size discussed in Fig.~\ref{Fig_QuanMagnonics_SunFX2021PRL_CatState} in Sec.~\ref{QuanMagnonics_CatState}.
The position and momentum quadratures of the photon and magnon output modes are defined as
$\hat{X}_{\hat{O}}=(\hat{O}+\hat{O}^\dagger)/\sqrt{2},~\hat{P}_{\hat{O}}=(\hat{O}-\hat{O}^\dagger)/\sqrt{2}i$ ($\hat{O}=\hat{c},~\hat{m}$), the solution of which can be directly derived according to Eq.~(\ref{QuanMagnonics_SunFX2021PRL_Output}). Therefore, the magnon-photon entanglement and EPR steering quantified by Eq.~(\ref{QuanMagnonics_LogNegativity}),~(\ref{Eq_QuanMagnonics_BipartiteSteeringG}) in Sec.~\ref{QuanMagnonics_EntConcept} as well as the Wigner function of magnon cat state can be obtained.

\begin{figure}[tbp!]
\centering
\includegraphics[width=0.75\columnwidth]{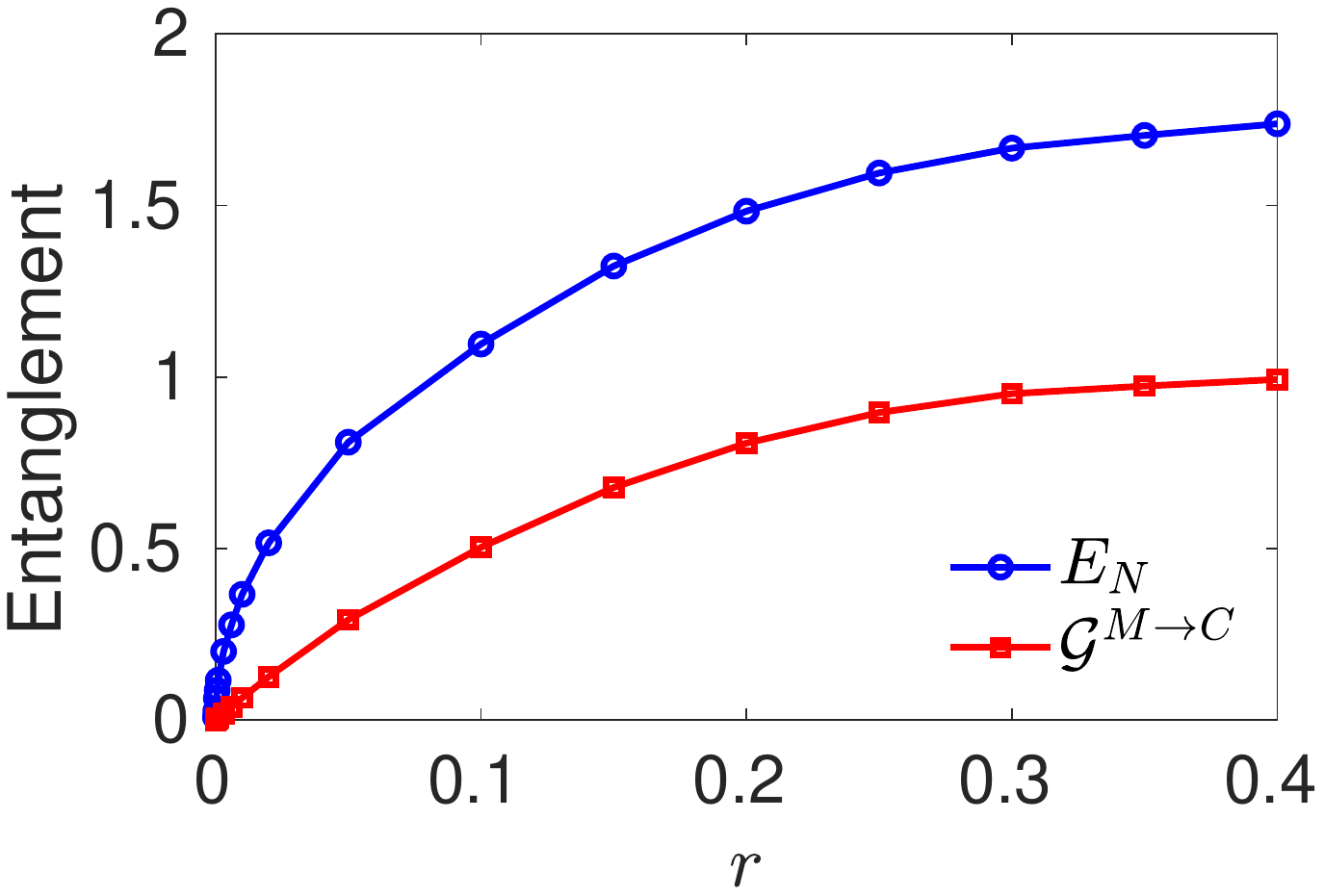}
\caption{Numerical results of magnon-photon entanglement and EPR steering generated through the nonlinear magneto-optical process~\protect\cite{SunFX2021PRL_CatState}. The schematic diagram corresponds to Fig.~\ref{Fig_QuanMagnonics_SunFX2021PRL_CatState} (a) in Sec.~\ref{QuanMagnonics_CatState}. $E_N$ denotes the bipartite magnon-photon entanglement characterized by logarithmic negativity in Eq.~(\ref{QuanMagnonics_LogNegativity}); $\mathcal{G}^{M\rightarrow C}$ represents the EPR steering from the magnon mode to photon mode quantified by Eq.~(\ref{Eq_QuanMagnonics_BipartiteSteeringG}). The horizontal axis denotes the squeezing parameter $r=G\tau$, representing the effective coupling strength of the two-mode squeezing generated by the nonlinear magneto-optical interaction.
Reproduced with permission from Sun et al., Phys. Rev. Lett. 127, 087203 (2021). Copyright 2021 American Physical Society.
}
\label{Fig_QuanMagnonics_SunFX2021PRL_CatState_Ent}
\end{figure}

Figure~\ref{Fig_QuanMagnonics_SunFX2021PRL_CatState_Ent} displays the numerical results of bipartite magnon-photon entanglement $E_N$ and steering from photons to magnons $\mathcal{G}^{M\rightarrow C}$ as a function of the squeezing parameter $r$. It is apparently observed that both the entanglement and EPR steering increase with increasing effective squeezing parameter $r$. Note that $E_N$ and $\mathcal{G}^{M\rightarrow C}$ nearly saturate at very large $r$ due to the accumulation of decoherence.
Most noteworthy, the maximum transient entanglement and EPR steering originated from the pulsed nonlinear magneto-optical interaction are fairly large approaching approximately to 1.75 and 1, respectively, which is nearly five times larger than the steady-state entanglement generated through intrinsic nonlinear coupling inside the quantum magnonic system displayed in Figs.~\ref{Fig_QuanMagnonics_JieLi2018PRL},~\ref{Fig_QuanMagnonics_Muhammad_Feedback},~\ref{Fig_QuanMagnonics_TanHuatang2019PRR},~\ref{Fig_QuanMagnonics_JieLi2018NJP_MagMagEnt},~\ref{Fig_QuanMagnonics_ZhangZhedong2019PRR_MagMagEnt} with the maximal entanglement being approximately 0.25. Therefore, the approach of employing the pulsed nonlinear magneto-optical interaction is an effective method to generate and significantly enhance the quantum correlations within different modes.

In addition to the work of generating transient optical photon-magnon entanglement in a magneto-optical system with only a optical mode and a magnon mode~\cite{SunFX2021PRL_CatState}, Bittencourt et al.~\cite{Bittencourt2019PRA_MagnonFockState} demonstrated that magnon-photon entanglement can also be created in a three-mode cavity optomagnonic consisting of two optical cavity modes and a magnon mode. The details of the systematic Hamiltonian can be referred to Eq.~(\ref{QuanMagnonics_Silvia2019PRA_Hamiltonian}) in Sec.~\ref{QuanMagnonics_SingleMagnonState} when introducing the generation of single magnon state through the nonlinear magneto-optical interaction. For the sake of reading, we repeat displaying the interaction Hamiltonian as
\begin{equation}
\hat{\mathcal{H}}_{\mathrm{int}} \propto G_{12}^- \hat{a}_1^{\dagger} \hat{a}_2 \hat{m}^{\dagger}+G_{12}^{-*} \hat{a}_1 \hat{a}_2^{\dagger} \hat{m}.
\label{QuanMagnonics_Silvia2019PRA_Hamiltonian_Ent}
\end{equation}
Pumping the optical cavity $\hat{a}_2$ on resonance while not pumping the optical cavity $\hat{a}_1$, the effective Hamiltonian is reduced to
\begin{equation}
\hat{\mathcal{H}}_{\mathrm{W}}\propto G_{\mathrm{W}} \hat{a}_1^{\dagger} \hat{m}^{\dagger}+\mathrm{h.c.},
\end{equation}
where the effective coupling strength $G_{\mathrm{W}}=G_{12}^- \langle a_2 \rangle$ with $\langle a_2 \rangle$ being the average value of mode $\hat{a}_2$.
 It is noted that this parametric-down-conversion type interaction is closely similar to Eq.~(\ref{QuanMagnonics_SunFX2021PRL_RWA}) mentioned above in work by Sun and coauthors~\cite{SunFX2021PRL_CatState}, which is responsible for transient optical photon-magnon entanglement generation there. In a similar fashion, the entanglement can be prepared between optical modes $\hat{a}_1$ and magnon mode $\hat{m}$ in
above three-mode cavity optomagnonic system, acting as a significant resource for the generation of heralding single magnon state~\cite{Bittencourt2019PRA_MagnonFockState}.
Apart from the magnon-photon entanglement, Xie et al.~\cite{XieHong2023OE_MO_Ent} demonstrated that entanglement between two optical modes can also be generated in a three-mode cavity optomagnonic system possessing the analogous Hamiltonian to Eq.~(\ref{QuanMagnonics_Silvia2019PRA_Hamiltonian_Ent}) as in Ref.~\onlinecite{Bittencourt2019PRA_MagnonFockState}. Different from the proposal of generating magnon-photon entanglement by pumping only one cavity mode $\hat{a}_2$,
in this regime, two cavity modes are both driven at resonance in order to generate photon-photon entanglement. The effective Hamiltonian after standard linearization procedure reads
\begin{equation}
\hat{\mathcal{H}}_{\mathrm{int}} \propto (G_{1} \hat{a}_1^{\dagger} \hat{m}^{\dagger}+G_{1}^* \hat{a}_1 \hat{m})+(G_{2} \hat{a}_2^{\dagger} \hat{m}+G_{2}^* \hat{a}_2 \hat{m}^{\dagger}),
\label{QuanMagnonics_XieHong2023OE_Hamiltonian}
\end{equation}
where the effective coupling strengths read $G_{1}=G_{12}^- \langle a_1 \rangle,~G_{2}=G_{12}^- \langle a_2 \rangle$ with $\langle a_1 \rangle$ and $\langle a_2 \rangle$ being the average value of modes $\hat{a}_1$ and $\hat{a}_2$, respectively. It can be found that the first term in Eq.~(\ref{QuanMagnonics_XieHong2023OE_Hamiltonian}) is in the form of parametric-down-conversion-type coupling, which is capable of entangling modes $\hat{a}_1$ and $\hat{m}$.
Then the entanglement can be transferred to two optical modes $\hat{a}_1$ and $\hat{a}_2$ through the beam-splitter-type swap interaction between $\hat{a}_2$ and $\hat{m}$.

\subsubsection{The entanglement originated from the external nonlinear interaction}
Apart from the nonlinearity generated in the quantum magnonic system through intrinsic nonlinear interaction (including magneto-elastic interaction, magneto-optical interaction, and magnet-qubit interaction, etc.), the nonlinearity resulting from the external nonlinear force acting on the quantum magnonic system is another significant means for preparing and manipulating different kinds of quantum correlations.
The supreme motivation of investigating this novel approach lies in that the entanglement can be more strongly prepared and easily manipulated, owing to the fact that the external injected nonlinearity is more tunable and stronger than the nonlinearity generated by the intrinsic nonlinear interaction listed above. Since the intrinsic nonlinear interactions in the quantum magnonic systems are relatively weak, the prepared steady-state entanglement and EPR steering among different parties are relatively small, as is displayed in Figs.~\ref{Fig_QuanMagnonics_JieLi2018PRL},~\ref{Fig_QuanMagnonics_Muhammad_Feedback},~\ref{Fig_QuanMagnonics_TanHuatang2019PRR},~\ref{Fig_QuanMagnonics_JieLi2018NJP_MagMagEnt},~\ref{Fig_QuanMagnonics_ZhangZhedong2019PRR_MagMagEnt} with the maximal entanglement being approximately 0.25~\cite{LiJie2018PRL,Muhammad2022TriEntEnhancement_Feedback,TanHuatang2019PRR_Steering,LiJie2019Magon-MagnonEnt,ZhangZhedong2019Magnon-MagnonEnt}. Although the strong transient magnon-photon entanglement (with maximal quantity near to 1.75) and EPR steering (with quantity near to 1)  is created by pulsed nonlinear magneto-optical interaction as shown in Fig.~\ref{Fig_QuanMagnonics_SunFX2021PRL_CatState_Ent}, the quantum correlations in steady states are more desirable to be generated and manipulated. Therefore, preparing strong quantum correlations beyond utilizing the intrinsic nonlinear interaction in the quantum magnonic system are necessarily required. The nonlinearity originated from the external nonlinear interaction is possibly a superior method.

\textit{\textbf{Entanglement generation through single-mode squeezing driving.}} For example, Nair et al.~\cite{Nair2020Magnon-MagnonEnt} recently proposed a theoretical scheme to entangle two magnons in two distant macroscopic ferromagnetic crystals utilizing externally injected squeezed light, which can be regarded as introducing additional nonlinear terms to the system. This kind of nonlinearity is totally responsible for the entanglement generation process, because the bare intrinsic quantum magnonic system is effectively characterized by a beam-splitter-type interaction between magnons and photons. The theoretical schematic diagram has been shown in Fig.~\ref{Fig_QuanMagnonics_MagnonMagnonEnt_Nonlinearity} (c) when classifying various approaches of generating magnon-magnon entanglement. The cavity-magnon system consists of two YIG spheres embodied near the maximum magnetic field inside the microwave cavity, in which the magnons are coupled to the microwave photons through the Zeeman interaction. Taking account of the rotating wave approximation, the Hamiltonian of such system is described by
\begin{eqnarray}
\hat{\mathcal{H} }&= & \omega_{m_1} \hat{m}_1^{\dagger} \hat{m}_1+\omega_{m_2} \hat{m}_2^{\dagger} \hat{m}_2+\omega_a \hat{a}^{\dagger} \hat{a} \nonumber \\
&& +g_{m_1 a}(\hat{m}_1^{\dagger} \hat{a}+\hat{m}_1 \hat{a}^{\dagger})+g_{m_2 a}(\hat{m}_2^{\dagger}\hat{a}+\hat{m}_2\hat{a}^{\dagger}),\nonumber \\
\end{eqnarray}
where the two magnon modes and a microwave photon mode are represented by the annihilation operators $\hat{m}_1,~\hat{m}_2,~\hat{a}$, respectively. It is apparently seen that there is no nonlinearity explicitly included in the systematic Hamiltonian, with only linear beam-splitter-type interactions between magnons and photons. The nonlinearity results from the squeezed vacuum field that drives the cavity externally from the left side of the cavity, which is generated through a degenerate parametric-down-conversion utilizing the inductance of Josephson junctions~\cite{ZhongL2013NJP_OneModeSqueezing_JPA,Fedorov2016PRL_OneModeSqueezing_JPA,Bienfait2017PRX_OneModeSqueezing_JPA,Malnou2018PRAppl_OneModeSqueezing_JPA}, as indicated in Fig.~\ref{Fig_QuanMagnonics_MagnonMagnonEnt_Nonlinearity} (c).
The effective influence of this externally injected nonlinearity can be simplified as that the microwave cavity mode is emerged in a squeezed environment, which could be exhibited in the equations of motion. In the frame rotating at  frequency $\omega_s$ (e.g., the frequency of the light injected into the cavity), the quantum Langevin equations considering the dissipation-fluctuation theory take the form of
\begin{eqnarray}
\dot{\hat{a}} & =&-\left(i \Delta_a+k_a\right) \hat{a}-i g_{m_1 a}\hat{m}_1-i g_{m_2 a} \hat{m}_2+\sqrt{2 k_a} \hat{a}^{\mathrm{in}},\nonumber \\
\dot{\hat{m}}_1 &=&-\left(i \Delta_{m_1}+k_{m_1}\right) \hat{m}_1-i g_{m_1 a} \hat{a}+\sqrt{2 k_{m_1}} \hat{m}_1^{\mathrm{in}}, \nonumber\\
\dot{\hat{m}}_2 & =&-\left(i \Delta_{m_2}+k_{m_2}\right) \hat{m}_2-i g_{m_2 a} \hat{a}+\sqrt{2 k_{m_2}} \hat{m}_2^{\mathrm{in}},
\end{eqnarray}
where detunings $\Delta_a=\omega_a-\omega_s,~\Delta_{m_i}=\omega_{m_i}-\omega_s$. Owing to the externally injected nonlinearity, the input quantum nosies of photons are characterized by
$\langle  \hat{a}^{{\rm in} \dagger}(t) \hat{a}^{\rm in}(t')\rangle=\delta(t-t')N_s$, $\langle \hat{a}^{ \rm in }(t) \hat{a}^{\rm in}(t')\rangle=\delta(t-t')M_s$, where the corresponding parameters are $N_s=\sinh^2 r$,~$M_s=e^{i\theta}\cosh r\sinh r$ with $r$ and $\theta$ being the external squeezing parameter and phase of the injected squeezed field. Besides, the correlations of the magnon input noises read $\langle \hat{m}^{{\rm in} \dagger}(t) \hat{m}^{\rm in}(t')\rangle=0$, $\langle \hat{m}^{{\rm in}}(t) \hat{m}^{\rm in \dagger}(t')\rangle=\delta(t-t')$. Note that the thermal photon and magnon number occupation are all nearly zero at low temperature $T=20~\mathrm{mK}$.

\begin{figure}[tbp!]
\centering
\includegraphics[width=0.98\columnwidth]{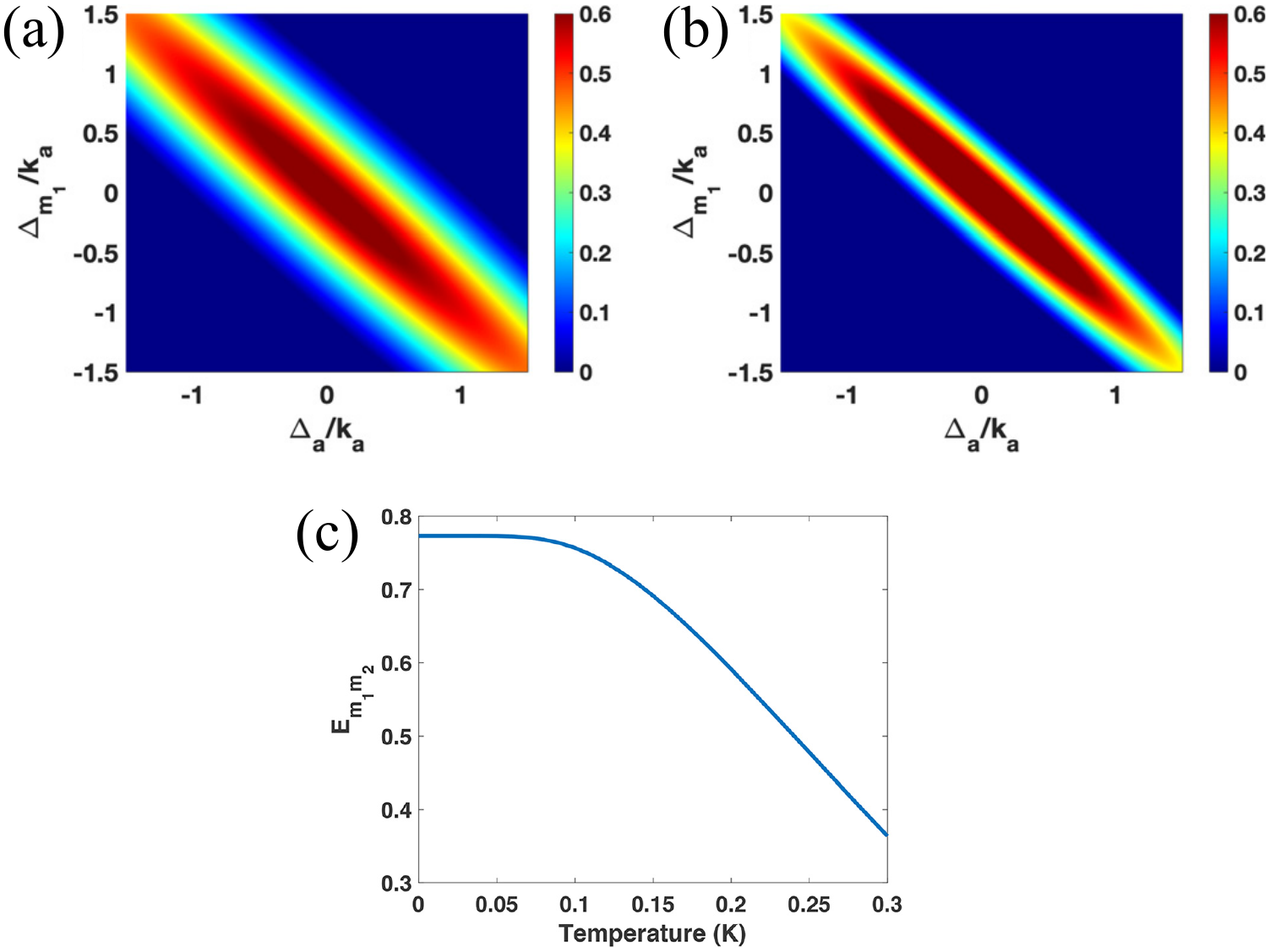}
\caption{Numerical results of bipartite magnon-magnon entanglement generated through the nonlinear  process~\protect\cite{Nair2020Magnon-MagnonEnt}. The schematic diagram corresponds to Fig.~\ref{Fig_QuanMagnonics_MagnonMagnonEnt_Nonlinearity} (c). (a-b) Magnon-magnon entanglement $E_{m_1 m_2}$ versus the detunings $\Delta_a$ and $\Delta_{m_1}=\Delta_{m_2}$ with squeezing parameter (a) $r=1$ and (b) $r=1.5$. (c) Bipartite entanglement as a function of environmental temperature. Reproduced with permission from Nair et al., Appl. Phys. Lett. 117, 084001 (2020). Copyright 2020 The Author(s). 
}
\label{Fig_QuanMagnonics_Nair2020APL_MagMagEnt}
\end{figure}

By carefully analyzing the above quantum Langevin equations, we know that the term $\langle \hat{a}^{ \rm in }(t) \hat{a}^{\rm in}(t')\rangle$ would firstly generate the squeezing of the microwave photons, and further create the magnon-magnon entanglement through a series of beam-splitter-type swap interactions mediated by magnons and photons. The bipartite entanglement, quantified by the logarithmic negativity, among two distinct magnon modes as a function of detunings ($\Delta_a,~\Delta_{m_1}$) is displayed in Fig.~\ref{Fig_QuanMagnonics_Nair2020APL_MagMagEnt}. We find that two magnon modes are strongly entangled with maximum logarithmic negativity at the position near resonance $\Delta_a=0,~\Delta_{m_1}=\Delta_{m_1}=0$. Besides, by comparing the entanglement in Fig.~\ref{Fig_QuanMagnonics_Nair2020APL_MagMagEnt} (a) $r=1$ and (b) $r=1.5$ with different squeezing parameters, it is observed that the larger the external injected nonlinearity is, the stronger the entanglement is generated. Most notably, the maximum magnon-magnon entanglement with quantity about 0.6 is nearly two times larger than that generated through intrinsic nonlinear interaction (e.g., magneto-elastic interaction, magneto-optical interaction, and multi-magnon-magnon interaction, etc.) inside the quantum magnonic system as shown in Figs.~\ref{Fig_QuanMagnonics_JieLi2018PRL}, ~\ref{Fig_QuanMagnonics_Muhammad_Feedback},~\ref{Fig_QuanMagnonics_TanHuatang2019PRR},~\ref{Fig_QuanMagnonics_JieLi2018NJP_MagMagEnt},~\ref{Fig_QuanMagnonics_ZhangZhedong2019PRR_MagMagEnt}. Besides, the strong entanglement is quite robust with the environmental temperature, which survives even at $T=0.3$ K, as manifested in Fig.~\ref{Fig_QuanMagnonics_Nair2020APL_MagMagEnt} (c).
The above two aspects show the advantages of utilizing the external nonlinear interaction to prepare entanglement.

In addition to the bipartite entanglement, different kinds of quantum correlations can also be generated and  manipulated through single-mode squeezing driving. For example,
Zhang et al.~\cite{ZhangWei2022OE_SqueezingDrive_Bi-EntSteering} recently confirmed that the one-way steering between photons and phonons can be prepared in a typical three-mode cavity magnomechanical system via externally injected squeezed light, in which the magnon mode simultaneously couples to the photon and phonon modes via the Zeeman interaction and magnetostrictive interaction, respectively.
Besides, Zhou et al.~\cite{ZhouYing2020SqueezingDrive_Bi-MultiEnt} presented a theoretical proposal to generate both bipartite and multipartite entanglement in a magnetic-cavity quantum electrodynamics system comprising of a ferromagnetic resonance (FMR) mode, a magnetostatic (MS) mode and a microwave cavity mode. The nonlinearity is introduced by placing a second-order nonlinear crystal inside the cavity to generate the nonlinear degenerate parametric down conversion with Hamiltonian $\hat{\mathcal{H}}_{\rm non}=\mu\hat{a}^2\hat{a}_p^{\dagger}+\mu^*\hat{a}^{\dagger 2}\hat{a}_p$, where $\hat{a},~\hat{a_p}$ are respectively the fundamental and second-order modes and $\mu$ denotes the single photon $\chi^{(2)}$ nonlinearity. In realistic parameters, the second-order can be treated classically, arriving the whole effective Hamiltonian of the system at $\hat{\mathcal{H}}= \Delta_{\mathrm{c}} \hat{a}^{\dagger} \hat{a}+\Delta_{m_1} \hat{m}_1^{\dagger} \hat{m}_1+\Delta_{m_2} \hat{m}_2^{\dagger} \hat{m}_2  +g_1(\hat{a}^{\dagger} \hat{m}_1+\hat{a} \hat{m}_1^{\dagger})+g_2(\hat{a}^{\dagger} \hat{m}_2+\hat{a} \hat{m}_2^{\dagger}) +\varepsilon(\hat{a}^2+\hat{a}^{\dagger 2})+\Omega(\hat{a}+\hat{a}^{\dagger})$. What is well known is that the term $\hat{a}^2+\hat{a}^{\dagger 2}$ is capable of generating squeezing among photons. Furthermore, similar to the underlying physical mechanism in the work of Nair et al.~\cite{Nair2020Magnon-MagnonEnt}, the squeezing of microwave photons can further prepare bipartite entanglement among any two parties and magnon-magnon-photon tripartite entanglement in the presence of the beam-splitter-type swap interaction. The numerical results show that the bipartite as well as tripartite entanglement increase gradually with nonlinear coefficient $\varepsilon$, and the entanglement is robust in a range of low temperature, the details of which are omitted here.

\begin{figure}[tbp!]
\centering
\includegraphics[width=0.75\columnwidth]{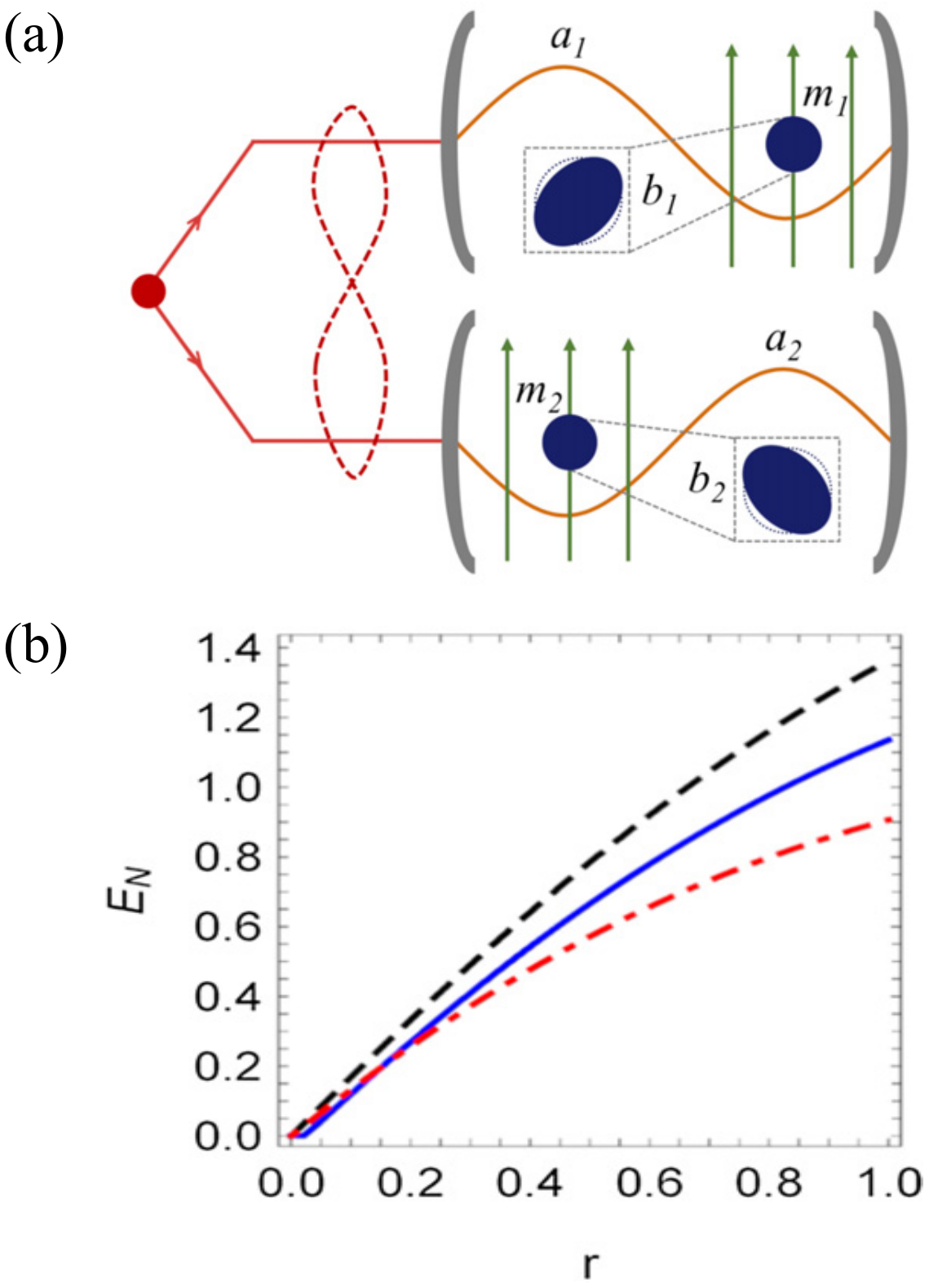}
\caption{(a) The schematic diagram of the dual-cavity magnomechanical system driven by a two-mode squeezed fields~\protect\cite{JieLi2021QuanSciTec_TwoModeSqueezingDrive}.
(b) Numerical results of bipartite photon-photon (black dashed), phonon-phonon (blue solid), and magnon-magnon (red dash-dotted) entanglement generated with the assistance of the nonlinearity from external two-mode squeezing driving.
 Reproduced with permission from Li et al., Quantum Sci. Technol. 6, 024005 (2021). Copyright 2021 The Author(s).
}
\label{Fig_QuanMagnonics_JieLi2021QuanSciTec_TwoModeSqueezingDrive}
\end{figure}

\textit{\textbf{Entanglement generation through two-mode squeezing driving.}}
Apart from preparing entanglement by means of a single-mode squeezing driving~\cite{Nair2020Magnon-MagnonEnt,ZhouYing2020SqueezingDrive_Bi-MultiEnt} shown above, Li et al. ~\cite{JieLi2021QuanSciTec_TwoModeSqueezingDrive} put forward a theoretical scheme to entangle the vibrational phonon modes of two massive ferromagnetic spheres in a dual-cavity magnomechanical system benefiting from a two-mode squeezing driving. This two-mode squeezing driving proposal is superior to the single-mode squeezing driving in terms of the degree of entanglement.
The schematic diagram is presented in Fig.~\ref{Fig_QuanMagnonics_JieLi2021QuanSciTec_TwoModeSqueezingDrive} (a).
The two-mode squeezing driving can be experimentally implemented by employing a Josephson parametric amplifier with Kerr nonlinearity~\cite{Eichler2011PRL_TwoModeSqueezing_JPA}, or a Josephson mixer based on the nonlinear three-wave mixing~\cite{Flurin2012PRL_TwoModeSqueezing_JPM}. The effective Hamiltonian in each individual cavity can be routinely derived as in Ref.~\cite{LiJie2018PRL}, thus the whole Hamiltonian in such dual-cavity magnomechanical system is easily given as
\begin{eqnarray}
\hat{\mathcal{H} }&=& \sum_{j=1,2}\Big\{\omega_{a_j} \hat{a}_j^{\dagger} \hat{a}_j+\omega_{m_j} \hat{m}_j^{\dagger} \hat{m}_j+\omega_{b_j} \hat{b}_j^{\dagger} \hat{b}_j \nonumber \\
&&+g_j\left(\hat{a}_j^{\dagger} \hat{m}_j+\hat{a}_j \hat{m}_j^{\dagger}\right)+G^{0}_j \hat{m}_j^{\dagger} \hat{m}_j\left(\hat{b}_j^{\dagger}+\hat{b}_j\right) \nonumber \\
&&+i \Omega_j\left(\hat{m}_j^{\dagger} \mathrm{e}^{-\mathrm{i} \omega^0_{ j} t}-\hat{m}_j \mathrm{e}^{\mathrm{i} \omega_j^{0 } t}\right)\Big\},
\label{QuanMagnonics_JieLi2021QST_Ham}
\end{eqnarray}
where $\hat{a}_j,~\hat{m}_j,~\hat{b}_j$ respectively represent the annihilation operators of the photon mode, magnon mode and mechanical mode in cavity $j$. The last three terms describe the magnon-photon swap interaction, magnon-phonon magnetostrictive interaction, and the driving of magnon in each cavity.

It is noted that, in this case, the entanglement between two magnon modes or two phonon modes is not attributed to the nonlinear magneto-elastic interaction. It is shown that this nonlinearity is responsible for generating bipartite and tripartite entanglement~\cite{LiJie2018PRL}, as depicted in Fig.~\ref{Fig_QuanMagnonics_JieLi2018PRL}. However, the entanglement is merely prepared between two or three parties inside an identical cavity, while two parties in two independent cavities cannot be entangled. The quantum correlations between two magnon modes, as well as two mechanical modes in two distinct cavities are created with the assistance of the two-mode squeezing driving, which establishes a tight correlation between two dependent cavity magnomechanical systems.

The treatment of quantifying entanglement is this two-mode squeezing driving scheme is similar to that in a single-mode squeezing driving scheme put forward by Nair et al.~\cite{Nair2020Magnon-MagnonEnt}, which has been detailly demonstrated just above. The only difference lies in that the input quantum noises of two cavity modes possess the form of two-mode squeezing, which are characterized by $\langle \hat{a}_j^{\mathrm{in}}(t) \hat{a}_j^{\mathrm{in}\dagger}(t^{\prime})\rangle =(N+1) \delta(t-t^{\prime})$,~$\langle \hat{a}_j^{\mathrm{in}\dagger}(t) \hat{a}_j^{\mathrm{in}}(t^{\prime})\rangle =N \delta(t-t^{\prime})$,~$\langle \hat{a}_j^{\mathrm{in}}(t) \hat{a}_k^{\mathrm{in}}(t^{\prime})\rangle =M e^{-i(\Delta_j t+\Delta_k t^{\prime})} \delta (t-t^{\prime})$,~$\langle \hat{a}_j^{\mathrm{in} \dagger}(t) \hat{a}_k^{\mathrm{in} \dagger}(t^{\prime})\rangle  =M^* e^{i(\Delta_j t+\Delta_k t^{\prime})} \delta (t-t^{\prime}) (j \neq k)$. Here, $N=\sinh^2 r$,~$M =\cosh r\sinh r$ with $r$ being the squeezing parameter of the two-mode squeezing driving. $\Delta_i (i=1,~2)$ represents the frequency detuning of photons, magnons, and phonons relative to the magnon drive frequency.

Figure~\ref{Fig_QuanMagnonics_JieLi2021QuanSciTec_TwoModeSqueezingDrive} (b) shows the bipartite entanglement among different parties generated in the presence of the nonlinearity from external two-mode squeezing driving. It can be apparently observed that, the quantum correlations among the driving fields is transferred through the embodied beam-splitter-type swap interaction to the quantum correlations between two magnon modes, two photon modes, as well as two mechanical modes in two distinct cavities.
Notably, the quantities of three bipartite entanglement are considerable, where photon-photon, magnon-magnon, and phonon-phonon entanglement are approximately $E_N=1.3,~1.1$, and $0.9$. Comparing the degree of magnon-magnon entanglement as shown in Fig.~\ref{Fig_QuanMagnonics_JieLi2021QuanSciTec_TwoModeSqueezingDrive} (b) with that in Fig.~\ref{Fig_QuanMagnonics_Nair2020APL_MagMagEnt}, we find that the maximum entanglement $E_N\approx 1.1$ resulting from the two-mode squeezing driving is greatly enhanced with respect to the case of the single-mode squeezing driving regime where $E_N\approx 0.6$. Furthermore, the degree of magnon-magnon entanglement in two-mode squeezing driving regime is nearly four times larger than that in Figs.~\ref{Fig_QuanMagnonics_JieLi2018NJP_MagMagEnt}, and~\ref{Fig_QuanMagnonics_ZhangZhedong2019PRR_MagMagEnt} resulting from the intrinsic nonlinearity from the magneto-elastic interaction, and multi-magnon-magnon interaction, respectively~\cite{LiJie2019Magon-MagnonEnt,ZhangZhedong2019Magnon-MagnonEnt} .


Yu et al.~\cite{YuMei2020JPB_MagMagEnt} considered a similar proposal to prepare magnon-magnon entanglement. The schematic diagram resembles closely to that in Fig.~\ref{Fig_QuanMagnonics_JieLi2021QuanSciTec_TwoModeSqueezingDrive}, consisting of a double-cavity magnomechanical system where two cavities are similarly driven by a two-mode squeezing field, however, with two mechanical modes not excited. That is, the Hamiltonian can be described by Eq.~(\ref{QuanMagnonics_JieLi2021QST_Ham}) with $G_j^0=0,~\Omega_j=0,~\omega_{b_j}=0$, which is also possible to create strong magnon-magnon entanglement.

To briefly conclude this part, theoretical studies show that the entanglement generation through external nonlinearity is able to overcome the difficulties of weak intrinsic nonlinearity within the system and the employment of the strong driving. The external nonlinearity can be achieved by injecting auxiliary single-mode or two-mode squeezing driving into the concerned quantum magnonic system. This method results in much stronger entanglement than that by means of intrinsic nonlinearity and is quite robust against the environmental temperature. Therefore, the nonlinearity injected externally into the quantum magnonic system is a desirable means of generating and manipulating entanglement.~\cite{Nair2020Magnon-MagnonEnt,ZhouYing2020SqueezingDrive_Bi-MultiEnt,JieLi2021QuanSciTec_TwoModeSqueezingDrive}

\subsection{Applications in quantum tasks}
\subsubsection{Quantum transducer} \label{QuanMagnonics_QuantumTransducer}

Quantum transducer for coherently converting microwave and optical photons with high efficiency is urgently demanded in advanced quantum information processing, especially towards the development of large-scale quantum networks~\cite{Schoelkopf2008Wiring,Xiang2013Hybrid}.
It is comprehensible that although high-fidelity and intrinsically scalable quantum processors have emerged in superconducting circuits~\cite{Clarke2008SuperconductingQubits}, the long-distance transmission of quantum states remains a great challenge at microwave frequency due to high thermal noises and large decay rates.
In contrast, optical communication channels allow for low transmission losses and negligible thermal noises~\cite{Obrien2009PhotonicQuantum,Lvovsky2009Optical}, however, the implementation of high-fidelity quantum gates is experiencing difficulties on account of the weak single-photon nonlinearities~\cite{Kok2007LinearOptical}.
In order to make use of the advantages of both sides so that quantum information can be processed by superconducting circuits and transmitted with optical photons, microwave to optical quantum conversion has been abundantly investigated and demonstrated in various platforms including optomechanical systems~\cite{Bochmann2013NatPhys_MOTrans_Optomechanics,Andrews2014NatPhys_MOTrans_Optomechanics,FanLinran2015NatCommun_MOTrans_Optomechanics,Lecocq2016PRL_MOTrans_Optomechanics,Balram2016NatPhoton_MOTrans_Optomechanics,FanLinRan2016NatPhoton_MOTrans_Optomechanics,ShaoLinbo2019Optica_MOTrans_Optomechanics,XuHan2020NatCommun_MOTrans_Optomechanics}, electro-optic systems~\cite{Tsang2010PRA_MOTrans_EO,Tsang2011PRA2_MOTrans_EO,Rueda2016Optica_MOTrans_EO,Javerzac-Galy2016PRA_MOTrans_EO,FanLinran2018SciAdv_MOTrans_EO,XuYuntao2021NatCommun_MOTrans_EO}, atomic systems~\cite{Hafezi2012PRA_MOTrans_Atom,Gard2017PRA_MOTrans_Atom,HanJingshan2018PRL_MOTrans_Atom}, etc. However, most systems lack tunability and thus limit their practical applications.
Nowadays, quantum magnonic systems have been noticed as promising platforms for achieving coherent conversion between microwave and optical photons~\cite{Hisatomi2016PRB_MOTrans,ZhuNa2020Optica_MOTrans}, benefitting from the unique advantages of the widely tunable frequency and coupling strength. This opens a novel avenue for future long-distance quantum communication, as well as quantum-noise-limited amplification of microwave signals~\cite{Bagci2014Nature_MicOptConvertor}.

\begin{figure}[tbp!]
\centering
\includegraphics[width=0.8\columnwidth]{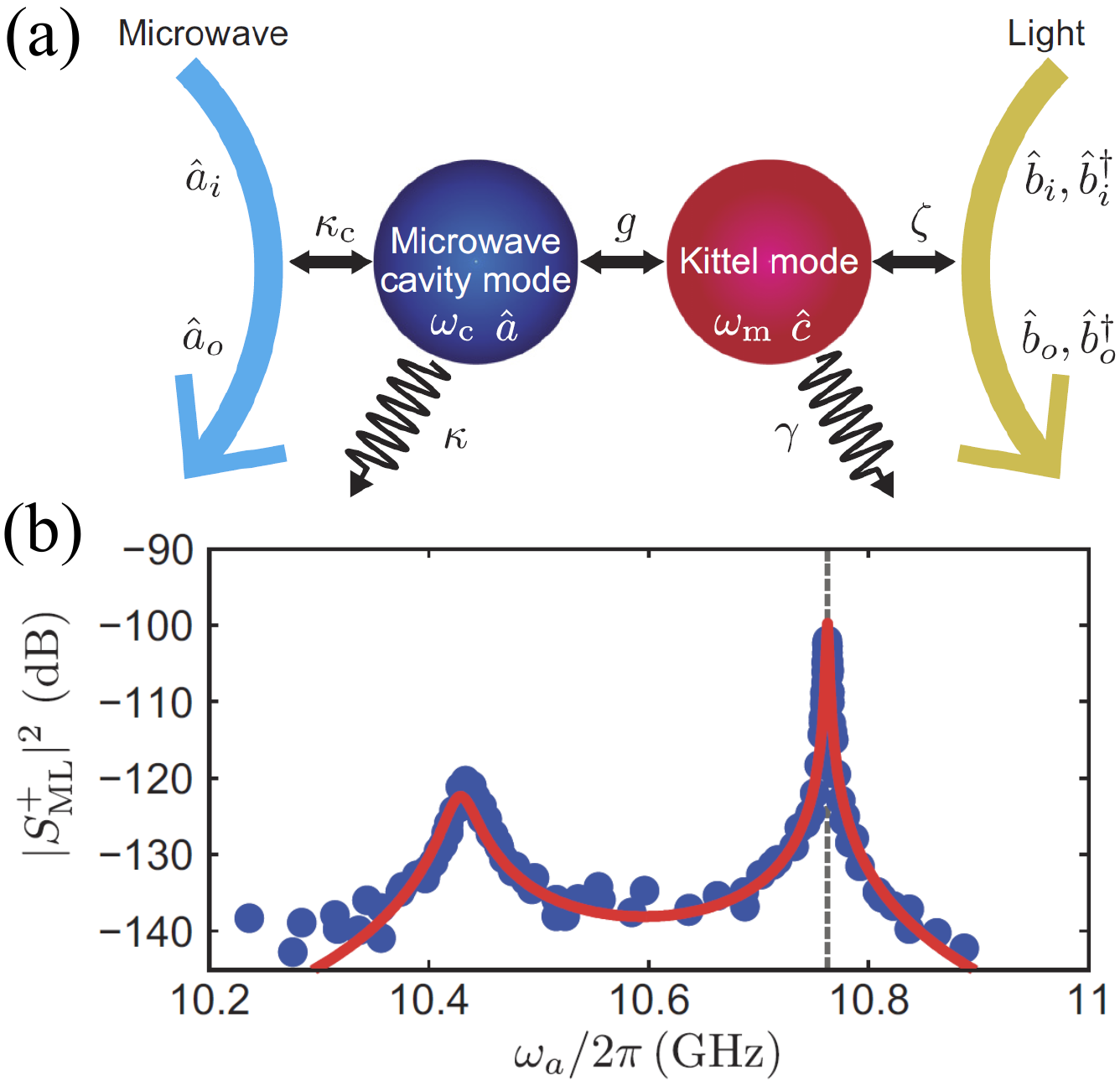}
\caption{(a) The schematic diagram of the microwave-light transducer~\protect\cite{Hisatomi2016PRB_MOTrans}. In the scheme, the microwave cavity mode $\hat{a}$ is strongly coupled to the Kittel mode $\hat{m}$. And the input itinerant microwave (optical) field is coupled to the microwave cavity mode $\hat{a}$ (Kittel mode $\hat{m}$). Here, we rewrite the annihilation operator of the Kittel mode as $\hat{m}$ for the consistent format in this tutorial.
(b) The conversion efficiency of optical photons to microwave photons defined by $\left|S_{\mathrm{ML}}^{+}\right|^2 \equiv\left|\left\langle\frac{\hat{a}_o \left(\omega_a\right)}{\hat{b}_i^{\dagger}(\Omega)}\right\rangle\right|^2$ versus the frequency difference $\omega_a$ between two phase-coherent
continuous-wave. Reproduced with permission from Hisatomi et al., Phys. Rev. B 93, 174427 (2016). Copyright 2016 American Physical Society.
}
\label{Fig_QuanMagnonics_Hisatomi2016PRB_MOTrans}
\end{figure}

Recently, Hisatomi et al.~\cite{Hisatomi2016PRB_MOTrans} experimentally achieved the bidirectional and coherent conversion of microwave and optical photons. The interaction diagram is shown in Fig.~\ref{Fig_QuanMagnonics_Hisatomi2016PRB_MOTrans} (a), with the effective Hamiltonian described by
\begin{eqnarray}
\hat{\mathcal{H}}&=&g\left(\hat{a} \hat{m}^{\dagger}+\hat{a}^{\dagger} \hat{m}\right)-i \sqrt{\kappa_c} \int_{-\infty}^{\infty} \frac{d \omega}{2 \pi}\left[\hat{a}^{\dagger} \hat{a}_i(\omega)-\hat{a}\hat{a}_i^{\dagger}(\omega) \right] \nonumber \\
&&-i  \sqrt{\zeta} \int_{-\infty}^{\infty} \frac{d \omega}{2 \pi}\left(\hat{m}+\hat{m}^{\dagger}\right)\left[\hat{b}_i(\omega) e^{i \Omega_0 t}-\hat{b}_i^{\dagger}(\omega) e^{-i \Omega_0 t}\right].\nonumber \\
\label{QuanMagnonics_Hisatomi2016PRB_Ham}
\end{eqnarray}
Here, the first term on the right hand side of Eq.~(\ref{QuanMagnonics_Hisatomi2016PRB_Ham}) represents the Zeeman interaction between the microwave cavity mode $\hat{a}$ and the uniform magnetostatic mode (i.e., the Kittel mode) $\hat{m}$ in YIG with $g$ being the relevant coupling strength. The second term denotes the coupling of an itinerant microwave field $\hat{a}_i$ and the microwave cavity mode $\hat{a}$ with coupling strength $\kappa_c$, while the third term describes the interaction of a traveling optical field $\hat{b}_i$ with the Kittel mode $\hat{m}$, where $\zeta$ is the coupling strength and $\Omega_0$ indicates the frequency of the optical driving field.
It can be easily known that such system is capable of achieving both microwave-to-light conversion and light-to-microwave conversion. On the one hand, the microwave photons of the input itinerant mode $\hat{a}_i$ can be converted to the optical photons of the output itinerant traveling mode $\hat{b}_o$. On the other hand, the optical photons of the input itinerant traveling mode $\hat{b}_i$ can also be converted to the microwave photons of the output itinerant microwave mode $\hat{a}_o$.
Figure~\ref{Fig_QuanMagnonics_Hisatomi2016PRB_MOTrans} (b) exhibits the conversion efficiency from light to microwave as a function of the frequency difference between two phase-coherent continuous-wave lasers (note that these two laser fields are designed for inducing the inverse Faraday effect). It can be apparently seen from Fig.~\ref{Fig_QuanMagnonics_Hisatomi2016PRB_MOTrans} (b)  that the maximum conversion efficiency is merely around $10^{-10}$, which is limited by the small magneto-optical interaction coupling strength $\zeta$. This conversion efficiency can be improved by employing an optical cavity to enhance the magnon-light interaction.

\begin{figure}[tbp!]
\centering
\includegraphics[width=0.8\columnwidth]{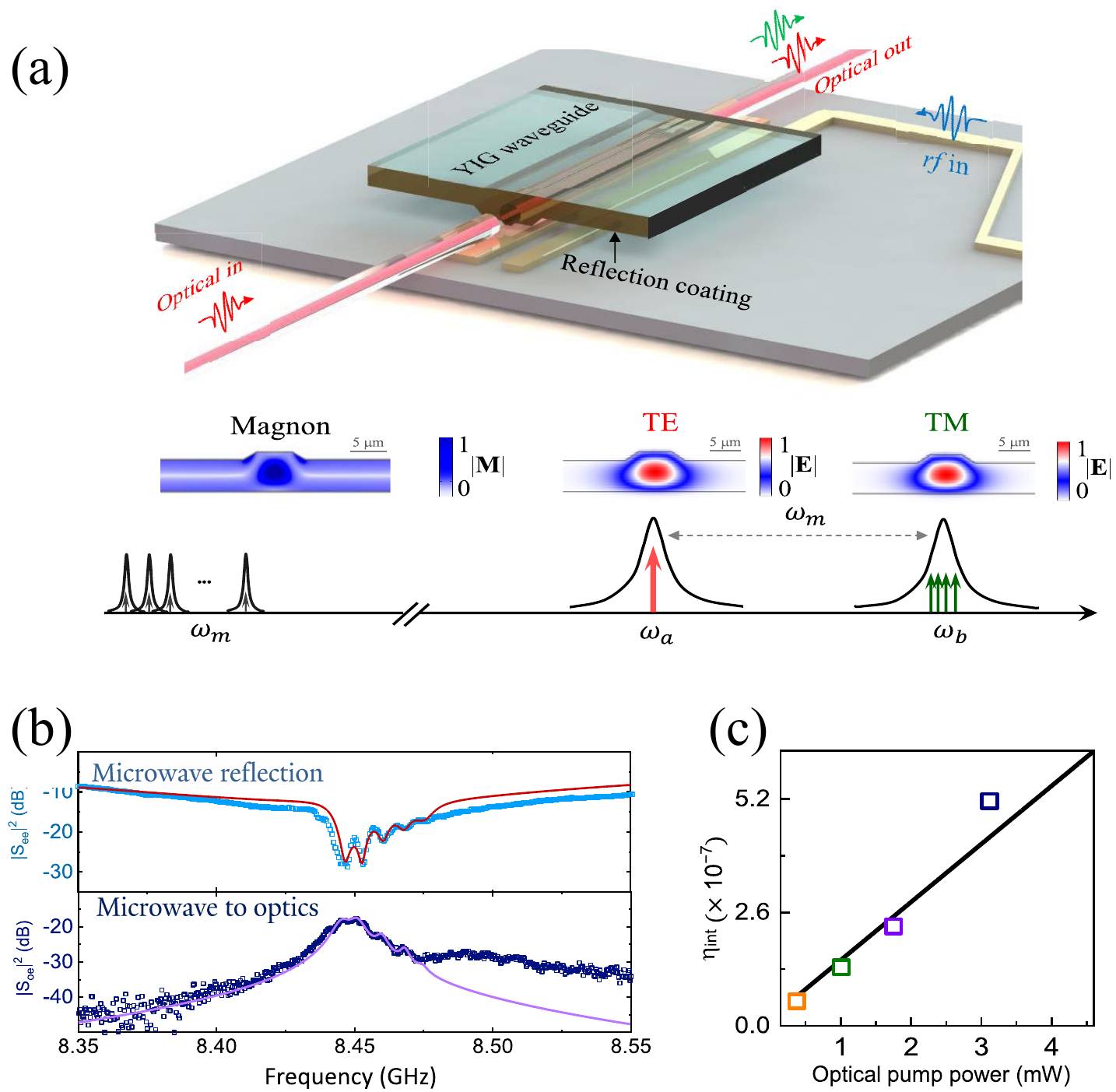}
\caption{(a) The top panel displays the schematic diagram of the microwave-light transducer~\protect\cite{ZhuNa2020Optica_MOTrans}. The optical light is coupled to the YIG waveguide.
The bottom panel shows the frequency diagram of triple-resonance process. The frequency difference of optical transverse-electric (TE) and transverse-magnetic (TM) modes are specially engineered to resonant with the magnon frequency. (b) The microwave reflection spectrum and the corresponding conversion spectrum from microwave photons to optical photons. (c) Conversion efficiency versus the optical pump power. The squared marks show the experimental data and solid curves show the fitting results. Reproduced with permission from Zhu et al., Optica 7, 1291 (2020). Copyright 2020 Optical
Society of America.
}
\label{Fig_QuanMagnonics_ZhuNa2020Optica_MOTrans}
\end{figure}

In contrast to the experimental implementation utilizing the YIG sphere~\cite{Hisatomi2016PRB_MOTrans}, Zhu et al.~\cite{ZhuNa2020Optica_MOTrans} adopted a single crystalline YIG thin film waveguide configuration, as shown in Fig.~\ref{Fig_QuanMagnonics_ZhuNa2020Optica_MOTrans}(a).
It is worth mentioning that the YIG waveguide simultaneously supports multiple magnon modes which are formed by forward volume magnetostatic standing waves (FVMSWs).
Thus, the system Hamiltonian reads
\begin{eqnarray}
\hat{\mathcal{H}}&=&\sum_i  g_{o, i}\left(\hat{a} \hat{b}^{\dagger}+\hat{a}^{\dagger} \hat{b}\right)\left(\hat{m}_i+\hat{m}_i^{\dagger}\right) \nonumber \\
&&+\sum_i  g_{\mathrm{e}, i}\left(\hat{c} \hat{m}_i^{\dagger}+\hat{c}^{\dagger} \hat{m}_i\right),
\end{eqnarray}
where $\hat{m}_i$ denotes the $i$th-order magnon mode, $\hat{a}$ and $\hat{b}$ represent the optical pump and signal modes, and $\hat{c}$ is the microwave mode. $g_{\mathrm{e}, i}$ ($g_{o, i}$) is the electro-magnonic (magneto-optical) coupling strength for the $i$th-order magnon mode.
As multiple magnon modes are coupled to the optical modes as well as the microwave mode, a broad frequency conversion bandwidth is expected, which is illustrated in Fig.~\ref{Fig_QuanMagnonics_ZhuNa2020Optica_MOTrans}(b). Here, the microwave-to-optical conversion spectrum $|S_{oe}|^2$ is obtained by comparing the photodetector measured optical beating signal and the microwave input power, while the microwave reflection spectrum $|S_{ee}|^2$ can be measured as well.
In the reflection spectrum $|S_{ee}|^2$, discrete magnon features can be observed where distinguishable dips appear due to the coupling with different magnon modes. Whereas, the microwave-to-optical conversion $|S_{oe}|^2$ spectrum shows a broadening effect instead of discrete peaks, as the optical cavity linewidths of signal and pump modes ($\sim1.6$ GHz) are much larger than the free spectral range ($\sim 7$ MHz) of the magnon modes.
Due to these factors, the microwave-to-optical conversion is actually a collective process assisted by multiple magnon modes, where a broad conversion bandwidth of $16.1$~MHz has been realized, which is around three orders of magnitude broader than conversion proposals in optomechanics~\cite{Andrews2014NatPhys_MOTrans_Optomechanics}.

In addition to the broad conversion bandwidth, a high conversion efficiency can be achieved. On the one hand, the single-photon magneto-optical interaction can be greatly enhanced because the magnon and optical photon modes are confined in a small mode volume.
Specifically, the highest single photon magneto-optical coupling strength $g_{o, i}/2\pi=17.2$~Hz has been reached, which is four orders of magnitude larger than the corresponding coupling strength employing YIG bulk crystals~\cite{Hisatomi2016PRB_MOTrans}.
Furthermore, the triple-resonance condition~\cite{Osada2016PRL_MO,Haigh2016PRL_MO,ZhangXufeng2016PRL_MO} is employed to enhance the total effective magneto-optical coupling, as displayed in Fig.~\ref{Fig_QuanMagnonics_ZhuNa2020Optica_MOTrans}(a).
Here, the optical mode $\hat{a}$ is strongly driven, which results in the effective Hamiltonian considering the rotating-wave approximation,
\begin{eqnarray}
\hat{\mathcal{H}}_{\text {int }}=\sum_i  \left[G_{o, i}\left(\hat{b}^{\dagger}\hat{m}_i+\hat{b}\hat{m}_i^{\dagger}\right)+  g_{\mathrm{e}, i}\left(\hat{c} \hat{m}_i^{\dagger}+\hat{c}^{\dagger} \hat{m}_i\right)\right], \nonumber \\
\end{eqnarray}
where $ G_{o, i}= g_{o, i} \sqrt{n_a}$ with $n_a=|\langle\hat{a} \rangle|^2$ being the intra-cavity photon number. It is apparently seen that the effective magneto-optical coupling strength can be greatly enhanced owing to the strong driving of mode $\hat{a}$ under the triple-resonance procedure.
As a consequence, the microwave-to-optical conversion efficiency can be greatly increased, as shown in Fig.~\ref{Fig_QuanMagnonics_ZhuNa2020Optica_MOTrans}(c). Here $\eta_{\mathrm{int}}=\eta/{(\xi_e \xi_o)}$ is the internal conversion efficiency, where $\eta\equiv\langle b^{\dagger}_{\mathrm{out}}b_{\mathrm{out}} \rangle/\langle c^{\dagger}_{\mathrm{in}}c_{\mathrm{in}} \rangle$ is the on-chip conversion efficiency, $\xi_e=\kappa_{e,e}/\kappa_e$ ($\xi_o=\kappa_{o,e}/\kappa_o$) is the electro-magnonic (magneto-optical) cooperativity with $\kappa_{e,e}$ and $\kappa_e$ ($\kappa_{o,e}$ and $\kappa_o$) being the external coupling and total dissipation rates for microwave (optical) signal modes.
It is found that the achieved microwave-to-optical conversion efficiency is nearly three orders of magnitude higher than that in Ref~\cite{Hisatomi2016PRB_MOTrans}.

\subsubsection{Quantum sensing}

Quantum states are intrinsically fragile under external perturbations, which can be leveraged in quantum sensing to detect weak signals~\cite{Degen2017QuantumSensing,Pirandola2018QuantumSensing}. In recent years, hybrid quantum systems based on magnonics have become a promising platform for quantum-enhanced sensing protocols. For example, sensing of magnons has been realized in cavity electromagnonics at millikelvin temperature, where the transverse relaxation of magnons can be observed with the average number of excited magnons much smaller than one~\cite{Tabuchi2014}. And the nonlinear magnon-qubit interaction also enables the quantum-enhanced sensing of magnons, where the magnon number states can be resolved with regards to the spectrum of the qubit~\cite{Lachance-Quirione2017NumberState,Lachance-Quirion2020SingleMagnon,Wolski2020QuantumSensing}, as shown in Fig.~\ref{Fig_QuanMagnonics_Lachance-Quirion2020Science_SingleMagnon}(b). The details have been presented in Sec.~\ref{QuanMagnonics_SingleMagnonState}.

\begin{figure}[tbp!]
\centering
\includegraphics[width=0.98\columnwidth]{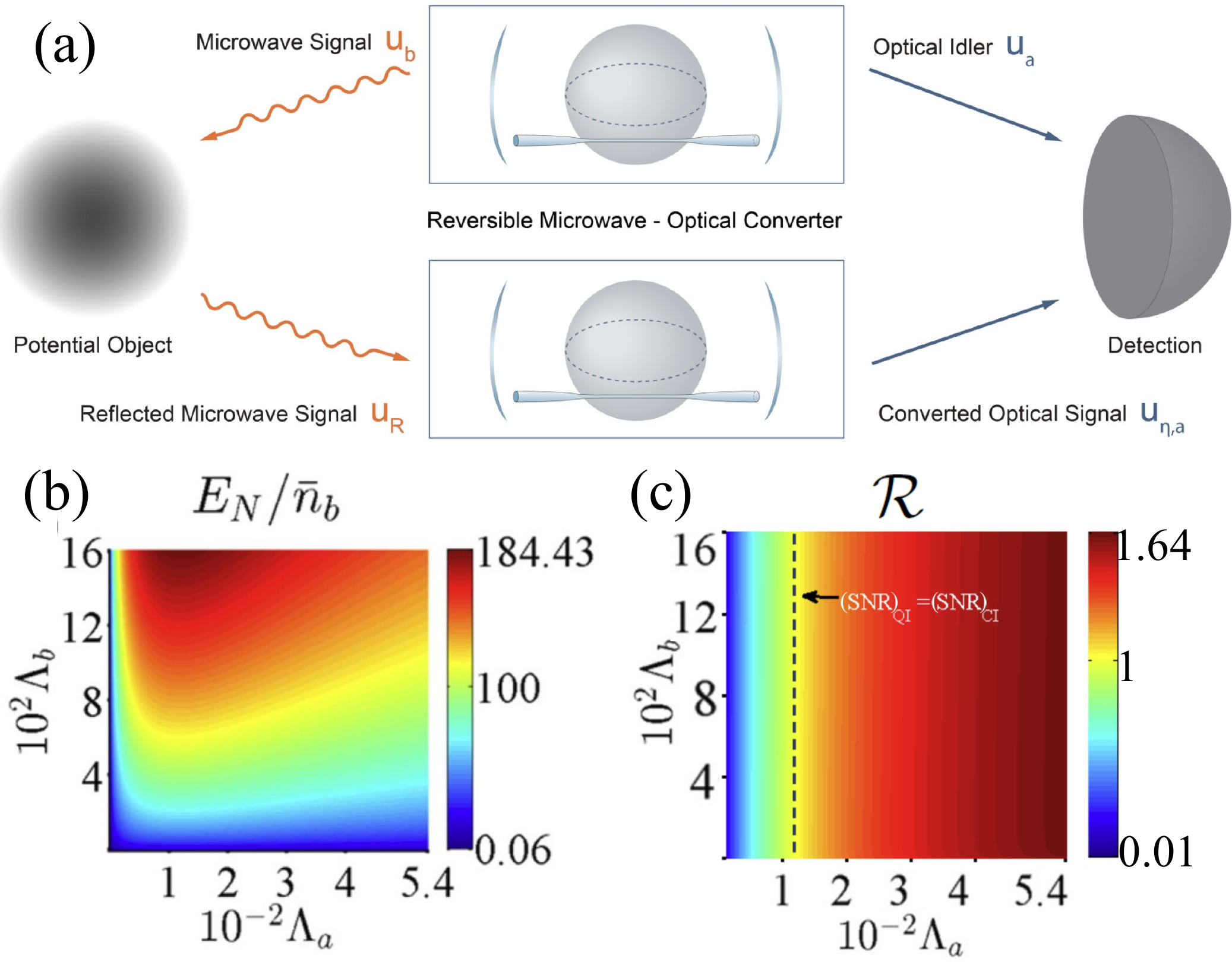}
\caption{(a) Schematic of the microwave-optical quantum illumination based on cavity magnonics~\protect\cite{Cai2021QuantumIllumination}. A YIG sphere is coupled to a microwave cavity and a optical nanofiber which serves as a magnonic reversible microwave-optical converter and generates entangled microwave-optical fields. The receiver collects the reflected microwave signal and upconverts it into the optical frequency. Then the converted optical signal together with the retained optical idler are fed into the detector for measuring the presence or absence of the target.
(b) The normalized logarithmic negativity $E_N/\bar{n}_b$ and (c) the ratio $\mathcal{R}=(\mathrm{SNR})_{\mathrm{QI}}/(\mathrm{SNR})_{\mathrm{CI}}$ versus optical cooperativity $\Lambda_a$ and microwave cooperativity $\Lambda_b$. Here $(\mathrm{SNR})_{\mathrm{QI}}$ and $(\mathrm{SNR})_{\mathrm{CI}}$ are the signal-to-noise ratios for quantum illumination and classical illumination, respectively.
Reproduced with permission from Cai et al., Phys. Rev. A 103, 052419 (2021). Copyright 2021 American Physical Society.
}
\label{Fig_QuanMagnonics_Cai2021QuantumIllumination}
\end{figure}

Another quantum sensing protocol known as quantum illumination is based on the entangled photon pairs to enhance the detection efficiency of low-reflectivity objects~\cite{Lloyd2008QuantumIllumination,Tan2008QuantumIllumination}. It has been shown that by using the entangled microwave-optical photon pairs, the advantages of high-performance photodetectors at optical frequency can be applied to enhance the detection sensitivity of target in the microwave regime~\cite{Barzanjeh2015QuantumIllumination}. Note that the cavity magnonics can be proposed as a transducer between microwave and optical waves as discussed in Sec.~\ref{QuanMagnonics_QuantumTransducer}, where microwave-optical entanglement can be established, the hybrid quantum system based on magnonic platform is promising for quantum illumination~\cite{Cai2021QuantumIllumination}. The simple sketch of the magnonic quantum illumination is shown in Fig.~\ref{Fig_QuanMagnonics_Cai2021QuantumIllumination}(a), where the reversible microwave-optical converter is proposed by inserting a YIG sphere into a microwave cavity and coupling it with an optical nanofiber at the same time. The nonlinear interaction Hamiltonian is then obtained as
\begin{equation}
\hat{\mathcal{H}}_{\mathrm{int}}= g_{m a}(\hat{a}_1 \hat{a}_2^{\dagger} \hat{m}^{\dagger}+\hat{a}_1^{\dagger} \hat{a}_2 \hat{m})+ g_{m b}(\hat{b}+\hat{b}^{\dagger})(\hat{m}+\hat{m}^{\dagger}).
\end{equation}
Here $\hat{a}_1$, $\hat{a}_2$, $\hat{b}$, and $\hat{m}$ are the annihilation operators of the optical TE mode, TM mode, microwave mode, and magnon mode, respectively. And $g_{ma}$ ($g_{mb}$) is the optomagnonic (electromagnonic) coupling rate.

By driving the TE mode with a coherent tone $\hat{a}_1\to\alpha$ and applying the rotating-wave approximation, the Hamiltonian can be linearized as $\hat{\mathcal{H}}^{\prime}_{\mathrm{int}}= G_{m a}\left(\hat{a}^{\dagger} \hat{m}^{\dagger}+\hat{a} \hat{m}\right)+ g_{m b}\left(\hat{b} \hat{m}^{\dagger}+\hat{b}^{\dagger} \hat{m}\right)$, where $\hat{a}\equiv\hat{a}_2$ and $G_{ma}\equiv g_{ma}\alpha$. As we have introduced in the previous sections, this type of Hamiltonian is capable of generating entangled microwave-optical photon pairs, which is illustrated with the normalized logarithmic negativity $E_N/\bar{n}_b$ in Fig.~\ref{Fig_QuanMagnonics_Cai2021QuantumIllumination}(b). Here $\bar{n}_b$ is the average photon number of the microwave mode, $\Lambda_a=G_{ma}^2/(\kappa_a \kappa_m)$ is the optical cooperativity and $\Lambda_b=g_{mb}^2/(\kappa_b \kappa_m)$ is the microwave cooperativity, where $\kappa_a$, $\kappa_b$, and $\kappa_m$ are the damping rates of the TM, microwave, and magnon modes, respectively.

In order to implement the quantum illumination, the microwave signal field is sent to probe the target and the entangled optical idler field is retained at the magnonic quantum source, as indicated in Fig.~\ref{Fig_QuanMagnonics_Cai2021QuantumIllumination}(a). Then the reflected microwave signal is collected by another reversible microwave-optical converter and upconverted into the optical domain. The converted signal will be fed into the detector together with the retained optical idler for measuring the presence or absence of the target. The quantum-enhanced sensitivity can be estimated by the signal-to-noise ratio of the system for quantum illumination, which is denoted as $(\mathrm{SNR})_{\mathrm{QI}}$. In Fig.~\ref{Fig_QuanMagnonics_Cai2021QuantumIllumination}(c), the advantage of the magnonic quantum illumination system over the classical coherent-state microwave radar is presented with $\mathcal{R}=(\mathrm{SNR})_{\mathrm{QI}}/(\mathrm{SNR})_{\mathrm{CI}}$, where $(\mathrm{SNR})_{\mathrm{CI}}$ is the signal-to-noise ratio of the latter system. It is shown that the quantum-enhanced sensitivity mainly depends on the optical cooperativity $\Lambda_a$ and exists with a slightly large $\Lambda_a$ where the intracavity entanglement can be transmitted to the propagating microwave and optical fields. For more details please refer to the Ref~\cite{Cai2021QuantumIllumination}.

\begin{figure}[tbp!]
\centering
\includegraphics[width=0.98\columnwidth]{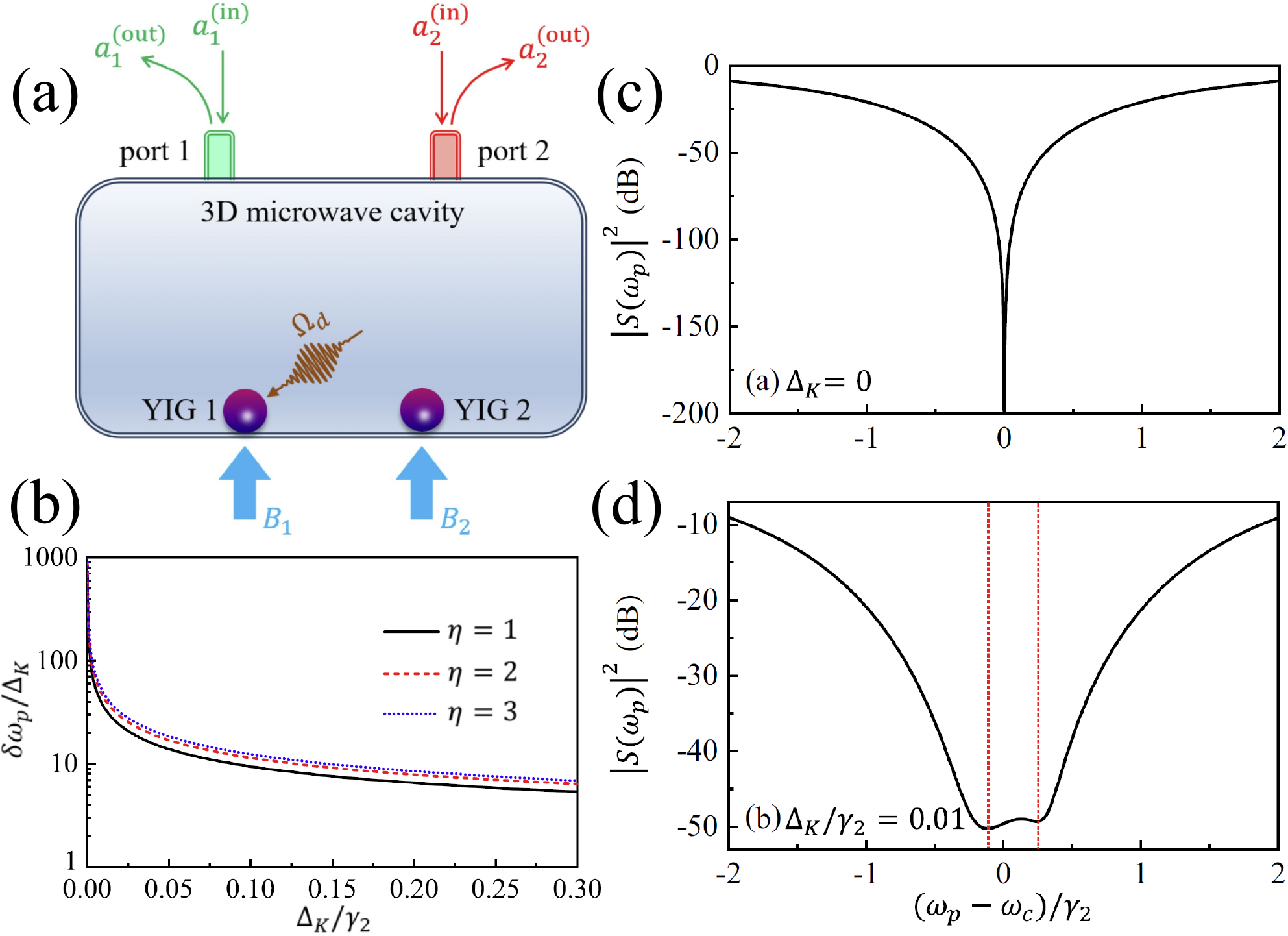}
\caption{(a) Schematic of the proposed setup for enhancing the detection
sensitivity of magnon Kerr nonlinearity~\protect\cite{Zhang2023PRB_detection}. Two YIG spheres are coupled to a microwave cavity, where YIG $1$ is driven by a microwave field with Rabi frequency $\Omega_d$. And two input fields $\hat{a}_1^{\text{(in)}}$ and $\hat{a}_2^{\text{(in)}}$ are fed into the cavity via ports $1$ and $2$, respectively, with output fields $\hat{a}_1^{\text{(out)}}$ and $\hat{a}_2^{\text{(out)}}$.
(b) The detection sensitivity enhancement factor $\delta\omega_p/\Delta_K$ changing with the magnon frequency shift $\Delta_K/\gamma_2$ for different $\eta$.
And the output spectrum $|S(\omega_p)|^2$ of the cavity near the third-order EP with (c) $\Delta_K=0$ and (d) $\Delta_K/\gamma_2=0.01$.
Reproduced with permission from Zhang et al., Phys. Rev. B 107, 064417 (2023). Copyright 2023 American Physical Society.
}
\label{Fig_QuanMagnonics_Zhang2023QuantumSensing}
\end{figure}

In addition to the above protocols, the so-called exceptional points (EPs), which correspond to the degenerate points of eigenvalues (and the related eigenvectors) in non-Hermitian systems, have been observed in cavity-magnon systems~\cite{Zhang2017Observation,Zhang2019HigherOrder}. It is well-known that EP has great potential for enhancing detection sensitivity, where the energy splitting follows a $\epsilon^{1/n}$ dependence around the $n$th-order EP subjected to a perturbation with strength $\epsilon$~\cite{Jan2014Enhancing,Liu2016Metrology,Chen2017Exceptional}. The enhanced sensitivity of linear external perturbations near EPs has been studied in cavity magnon polaritons~\cite{Cao2019EPsensitivity} and magnonic planar waveguides~\cite{Wang2021Enhanced}. Very recently, Zhang et.~al.~\cite{Zhang2023PRB_detection} proposed a scheme for quantum-enhanced sensing of magnon Kerr nonlinearity arround the third-order EP in cavity magnonics. In their proposal, two YIG spheres (denoted as YIG $1$ with magnon mode $\hat{m}_1$ and YIG $2$ with $\hat{m}_2$) are embedded in a microwave cavity as shown in Fig.~\ref{Fig_QuanMagnonics_Zhang2023QuantumSensing}(a). Hence, the Hamiltonian can be obtained as
\begin{eqnarray}
\hat{\mathcal{H}}&=& \sum_{j=1,2}\left[\omega_j \hat{m}_j^{\dagger} \hat{m}_j+K_j \hat{m}_j^{\dagger} \hat{m}_j \hat{m}_j^{\dagger} \hat{m}_j+g_j\left(\hat{a}^{\dagger} \hat{m}_j+\hat{a} \hat{m}_j^{\dagger}\right)\right] \nonumber \\
&& +\omega_c \hat{a}^{\dagger} \hat{a}+\Omega_d\left(\hat{m}_1^{\dagger} e^{-i \omega_{\mathrm{d}} t}+\hat{m}_1 e^{i \omega_{\mathrm{d}} t}\right).
\end{eqnarray}
Here $\hat{a}$ is the annihilation operator of the cavity mode at frequency $\omega_c$, $\omega_j$ is the frequency of the magnon mode in YIG $j$, $g_j$ is the coupling strength between the cavity mode $\hat{a}$ and the magnon mode $\hat{m}_j$, $\Omega_d$ is the strength of the driven field on YIG $1$ with frequency $\omega_d$, and $K_j$ is the strength of the Kerr nonlinearity. In this scheme, the magnon mode $\hat{m}_1$ is assumed to have the magnon Kerr nonlinearity ($K_1>0$) stemming from the magnetocrystalline anisotropy, while the auxiliary mode $\hat{m}_2$ is assumed to be linear with $K_2=0$. This is experimentally feasible as $K_j$ can be continuously tuned from negative to positive by controlling the angle between the crystallographic
axis of YIG $j$ and the bias magnetic field $B_j$~\cite{Stancil2009SpinWave}.

Supposed that the system is pumped by a strong driven field $\Omega_d$, where macroscopic number of magnons are excited in YIG $1$, $\langle \hat{m}_1^{\dagger} \hat{m}_1\rangle\gg 1$, the Kerr term in Hamiltonian can be simplified with the mean-field approximation, $\hat{\mathcal{H}}_K=K_1 \hat{m}_1^{\dagger} \hat{m}_1 \hat{m}_1^{\dagger} \hat{m}_1\approx \Delta_K \hat{m}_1^{\dagger} \hat{m}_1$ with $\Delta_K=2K_1 \langle \hat{m}_1^{\dagger} \hat{m}_1\rangle$. In the meanwhile, by considering the equations of motion of the cavity-magnon system, the expected values of the modes can be expressed as $\langle \hat{a}\rangle=\mathcal{A}e^{-i\omega_d t}$ and $\langle \hat{m}_j\rangle=\mathcal{M}_j e^{-i\omega_d t}$. Thus, the magnon frequency shift takes the form of $\Delta_K=2K_1|\mathcal{M}_1|^2$.

In order to measure the magnon frequency shift $\Delta_K$, two weak input fields $\hat{a}_1^{\text{(in)}}$ and $\hat{a}_2^{\text{(in)}}$ with same frequency $\omega_p$ are fed into the cavity. The changes of $\langle \hat{a}\rangle$ and $\langle \hat{m}_j\rangle$ can be treated as perturbations, which read $\langle \hat{a}\rangle=\mathcal{A}e^{-i\omega_d t}+Ae^{-i\omega_p t}$ and $\langle \hat{m}_j\rangle=\mathcal{M}_j e^{-i\omega_d t}+M_j e^{-i\omega_p t}$. In this circumstance, the system can be manipulated to achieve the coherent perfect absorption (CPA)~\cite{Zhang2019HigherOrder}, which is the phenomenon that two coherent waves are fed into a medium and there are no output waves from the medium, $\langle \hat{a}_1^{\text{(out)}}\rangle=\langle \hat{a}_2^{\text{(out)}}\rangle=0$. That is, the waves are completely absorbed by the medium, which results from the destructive interference between them and the medium dissipation~\cite{Chong2010CPA,Wan2011CPA}. Then by applying the input-output relation, the equation of motion can be rewritten for the perturbations, $(\dot{A},\dot{M}_1,\dot{M}_2)^T=-i \hat{\mathcal{H}}_{\mathrm{eff}}(A,M_1,M_2)^T$, where
\begin{equation}
\hat{\mathcal{H}}_{\mathrm{eff}}=\left(\begin{array}{ccc}
\delta_{\mathrm{cp}}+i \kappa_g & g_1 & g_2 \\
g_1 & \delta_{1 \mathrm{p}}+\Delta_K-i \gamma_1 & 0 \\
g_2 & 0 & \delta_{2 \mathrm{p}}-i \gamma_2
\end{array}\right).
\end{equation}
Here $\delta_{\mathrm{cp}}=\omega_c-\omega_p$, $\delta_{\mathrm{jp}}=\omega_j-\omega_p$, $\gamma_j$ is the decay rate of the magnon mode $\hat{m}_j$, and the effective gain of the cavity $\kappa_g=\kappa_1+\kappa_2-\kappa_{\mathrm{int}}$ where the decay rate $\kappa_j$ is induced by the port $j$ and $\kappa_{\mathrm{int}}$ is the intrinsic decay rate of the cavity mode. Then the eigenvalues of the non-Hermitian Hamiltonian $\hat{\mathcal{H}}_{\mathrm{eff}}$ can be obtained, which are denoted as $\Omega_0$ and $\Omega_{\pm}$ with $\mathrm{Re}[\Omega_{-}]\leq\mathrm{Re}[\Omega_{0}]\leq\mathrm{Re}[\Omega_{+}]$. It is directly checked that when $\Delta_K=0$ the three eigenvalues coalesce to the same point
\begin{equation}
\Omega_{\mathrm{EP} 3}=\delta_{\mathrm{cp}}-\frac{\sqrt{3}(\eta-1) \eta}{2 \eta^2+5 \eta+2} \gamma_2,
\end{equation}
which indicates it as a third-order EP. Here $\eta=\gamma_1/\gamma_2$. And in the case of $\xi=\Delta_K/\gamma_2\ll1$, the change of the eigenvalue near the third-order EP reads $\delta\Omega\propto \gamma_2\xi^{1/3}$. Hence, by introducing $\delta\omega_p=\mathrm{Re}[\Omega_+]-\mathrm{Re}[\Omega_-]$, the detection sensitivity enhancement factor $\delta\omega_p/\Delta_K$ can be displayed in Fig.~\ref{Fig_QuanMagnonics_Zhang2023QuantumSensing}(b). It shows that the detection sensitivity is largely enhanced in the regime where the magnon frequency shift $\Delta_K$ is much smaller than the linewidth $\gamma_2$ of the peaks in the transmission spectrum of the cavity (i.e., $\Delta_K\ll\gamma_2$), which is different from previous methods which are valid in the region $\Delta_K\geq\gamma_2 $~\cite{Haigh2015Dispersive,Yao2017Cooperative}.

It is notable that the quantity $\delta \omega_p$ is experimentally measurable. In the case of $\Delta_K=0$, the condition for implementing CPA at the third-order EP can be obtained as $\langle \hat{a}_2^{\text{(in)}}\rangle/\langle \hat{a}_1^{\text{(in)}}\rangle=\sqrt{\kappa_2/\kappa_1}$ and $\omega_p^{\mathrm{(CPA)}}=\Omega_{\mathrm{EP} 3}$. Therefore, by illustrating the output spectrum $|S(\omega_p)|^2$ with $\langle \hat{a}_j^{(\mathrm{out})}\rangle=S(\omega_p)\langle \hat{a}_j^{(\mathrm{in})}\rangle$ in Fig.~\ref{Fig_QuanMagnonics_Zhang2023QuantumSensing}(c), it can be found that there is only one CPA point with $|S(\omega_p)|=0$. Then in the presence of the magnon Kerr nonlinearity with $\Delta_K>0$, the CPA disappears and two dips emerge, which are highlighted by the two red dashed lines in Fig.~\ref{Fig_QuanMagnonics_Zhang2023QuantumSensing}(d). By analyzing the complex eigenvalues of the cavity-magnon system, the dips are localized at $\omega_p^{\mathrm{(dip1)}}=\mathrm{Re}[\Omega_-]$ (left dip) and $\omega_p^{\mathrm{(dip2)}}=\mathrm{Re}[\Omega_+]$ (right dip). Hence, $\delta\omega_p=\omega_p^{\mathrm{(dip2)}}-\omega_p^{\mathrm{(dip1)}}$ corresponds to the distance between the dips in the output spectrum, which makes this quantum sensing protocol experimentally feasible.

To briefly conclude this section, hybrid quantum systems based on magnonic platforms have been investigated for a variety of applications in quantum information processing, such as quantum transducer~\cite{Hisatomi2016PRB_MOTrans,ZhuNa2020Optica_MOTrans}, quantum sensing~\cite{Tabuchi2014,Lachance-Quirione2017NumberState,Lachance-Quirion2020SingleMagnon,Wolski2020QuantumSensing,Cai2021QuantumIllumination,Zhang2023PRB_detection}, etc.
Considering that magnons can simultaneously couple to the optical and microwave fields, the quantum magnonic systems are expected as promising platforms for coherent conversion between microwave and optical photons.
With the development of crystalline ferrimagnetic YIG thin film waveguide, a broad-bandwidth and high-efficiency microwave-to-optical conversion has been experimentally reported~\cite{ZhuNa2020Optica_MOTrans}.
And the ability of the quantum magnonic systems to generate microwave-optical entanglement will also facilitate the quantum illumination where the detection sensitivity of a low-reflectivity object can be greatly improved~\cite{Cai2021QuantumIllumination}.
Besides, with high-order EPs observed in cavity-magnon systems, an experimentally feasible scheme for quantum-enhanced sensing of magnon Kerr nonlinearity has been proposed~\cite{Zhang2023PRB_detection}, which serves as a significant complement in the regime where the magnon frequency shift is much smaller than the linewidth of the cavity mode.

\section{Conclusion and outlook}
In conclusion, we have introduced the interaction of magnons in magnetic ordered systems and the nonlinear magnonic process in hybrid quantum systems inducing photons, phonons and qubits. We show how these nonlinear interaction among magnons, photons, phonons and qubits can generate exotic magnon states in both the classical and quantum regime, together with their applications in spintronics and quantum information.

In the classical regime, one can keep exploring the exotic states of magnons by playing with the nonlinear scattering of magnons with other quasiparticles. Due to its great flexibility and rick nonlinearity, we expect many nonlinear phenomenon in optical and acoustic systems can be realized in magnonic systems. Besides, there are a few open questions based on the reported findings of magnon frequency comb and bistability.

\textit{\textbf{Experimental verification and application of frequency comb.}} While the magnon frequency comb through four-magnon interaction has been realized in experiments \cite{HulaAPL2022}, it remains open to verify the comb through three magnon process, as we discussed in Sec. \ref{magnon_comb}. One challenge may be the threshold of microwave power to launch the hybridization of driving microwave and the intrinsic magnon frequency comb, which may be overcome by choosing proper magnetic textures \cite{WangPRL2022}. Furthermore, it remains open to design spintronic devices with magnon frequency comb, for the spin-wave calibration, spin-wave excitation, accurate detection of topological defects and even finds its role in engineering phase-coherent magnon laser and magnon quantum computing. To be specific, (a) the spacing between the spectral lines of the MFC can range from MHz to GHz and can be potentially extended to THz in antiferromagnets. These uniformly distributed frequency modes can act as a ruler for spin-wave frequencies in a wide frequency range, and dramatically improve the accuracy of the frequency measurement in magnetism. (b) The MFC can be used to excite sustained coherent spin waves and to generate field-tunable caustic spin-wave beams with short-wavelength. The traditional method using optical combs requires magnetic materials with a strong magneto-optical coupling \cite{MuralidharPRL2021}. (c) The MFC with purely magnonic signals can exploit the ability of spin waves to provide precision information about the topological solitons and defects in the magnetic medium. For example, a magnon frequency comb excites and propagates to interact with the sample with magnetic solitons or defects. By analyzing the spectral response of the system, the information of the internal modes in the magnetic solitons or defects can be extracted.

\textit{\textbf{Bistability and multistability.}} Memory, switches, and logic devices wouldn't be possible without the precise discrimination, reading, and storage of 0 and 1 bits. The nonlinearity and bistability of the magnonic system offer new opportunities for the advancement of information technology. In the follow-up research of nonlinear magnonics, there are a number of areas that merit further investigation, including experimental research on the driven-dissipative system based on the cavity magnonic system, improvement of bistable switching speed, and the realization of larger-scale logic devices using nonlinear magnonic system.

\textit{\textbf{Nonlinear magnon transport.}} At the end of Sec. \ref{magnon_comb}, we show that a magnonic Penrose superradiance can be realized when the phase velocity of the twisted magnons exceeds that of the gyrating speed of vortex core, then the emission of spin waves can be enhanced. While an object extracts energy from a rotating black hole by emitting negative energy particle inside the ergosphere, it is not clear how to interpret the amplification by introducing negative-energy excitation of magnons. Harms et al. \cite{HarmsarXiv2022} showed that the negative excitation of magnetic systems correspond to antimagnons, i.e., antiparticle of magnons. This usually requires antiparallel magnetization to external field and external driving to sustain the stability of such systems. The manipulation of antimagnons can also be used to amplify spin waves. Further, how the magnon interacts with the magnetic textures and propagates in the nonlinear regime is another interesting direction. Jin et al. \cite{JinarXiv2023} made a step in this direction, showing that magnons may experience an additional Hall deflection when they pass through a skyrmion.

In the quantum regime, there have been many theoretical proposals to generate and manipulate different types of quantum correlations (including quantum entanglement, EPR steering, and Bell nonlocality) between magnons and other kinds of quantum systems. However, there are some open questions remains to be further solved.

\textit{\textbf{Experimental demonstration}}. The first one is the experimental realization and measurement of quantum entanglement in quantum magnonics. The experimental technical difficulties should be conquered for the purpose of flexibly preparing and manipulating different kinds of quantum correlations based on magnons, aiming at utilizing quantum entanglement as key quantum resources to promote the development of quantum information processing protocols.
In order to experimentally forecast entanglement and EPR steering between magnon mode with photons or phonons, it is necessary to measure the quadratures of these modes, which can be accomplished through balanced homodyne detection. Remarkably, the magnon mode needs to be coupled to an additional weak-microwave field with a dissipation rate significantly larger than that of the magnon modes through a beam-splitter-type interaction, therefore the magnon state can be read out via the balanced homodyne detection of the auxiliary-microwave-cavity mode so as to obtain the  corresponding covariance matrix.
On the other hand, the quadratures of the photon mode can be directly measured through the balanced homodyne detection process~\cite{LiJie2018PRL,Vitali2007PRL_Optomechanics,Palomaki710}.

\textit{\textbf{Practical applications}}. In addition to the problem of experimental realization, the unique and practicable applications of these magnoic entanglement in quantum information processing tasks are necessarily required to be proposed.
In general, magnonic entanglement not only is promising for studying fundamental macroscopic quantum physics, such as large-scale quantum phenomena, quantum-classical transitions and quantum decoherence, but also holds great promise for practical applications in quantum memory, quantum networks, quantum computation, etc.  However, the specific theoretical proposal together with experimental implementation for typical quantum information tasks taking advantage of magnonic entanglement need further investigation.

Despite the challenges mentioned above, the magnonic entanglement are still valuable to be deeply studied. Possible research directions in future can be listed in the following.

\textbf{\textit{Multipartite entanglement:} }
Multipartite entanglement owns the superior advantages of increasing communication capacity and noise resilience over quantum channels, thus is significant resource for quantum information processing. Most of the work concentrated on how to generate and further enhance the bipartite entanglement or quantum steering in hybrid magnonic systems~\cite{LiJie2018PRL,Muhammad2022TriEntEnhancement_Feedback,TanHuatang2019PRR_Steering,LiJie2019Magon-MagnonEnt,ZhangZhedong2019Magnon-MagnonEnt,Nair2020Magnon-MagnonEnt,YuMei2020EntMagnon,YangZhibo2020OE_BiTripartite_KerrNonlinearity,SunFX2021PRL_CatState,Bittencourt2019PRA_MagnonFockState,ZhouYing2020SqueezingDrive_Bi-MultiEnt,JieLi2021QuanSciTec_TwoModeSqueezingDrive}. Only few proposals pay attention to the generation of genuine tripartite magnon-photon-phonon entanglement~\cite{LiJie2018PRL,Muhammad2022TriEntEnhancement_Feedback,TanHuatang2019PRR_Steering,ZhangZhedong2019Magnon-MagnonEnt,YangZhibo2020OE_BiTripartite_KerrNonlinearity,ZhouYing2020SqueezingDrive_Bi-MultiEnt}, however, the degree of entanglement is especially weak which is not large enough for practical applications in quantum information processing.
The approach of generating useful and tunable multipartite entanglement remains to be thoroughly investigated. 

\textbf{\textit{Novel quantum phenomena in (ultra-,~deep-,~super-) strong coupling regimes.}}
 One interesting and necessary area is to deeply investigate the novel quantum phenomena in the regime of strong, ultra-strong, deep-strong, even super-strong coupling. Researchers from the area of optomechanics frequently ask the question that the quantum entanglement characteristics have already been thoroughly studied in optomechanical systems since the last century, why you still need to study quantum entanglement in the quantum magnonics. To better answer this question, we could first compare the respective features of each system. It is true that quantum magnonic systems have many features similar to quantum optomechanical systems. However, quantum magnonic systems possess superior characteristics of long lifetime and easy tunability; besides, one unique feature lies in that the systems are easier to generate
 (ultra-,~deep-,~super-) strong coupling strength, which makes it possible to study the novel quantum phenomena that are difficult to be observed in the quantum optomechanics.
For example, Yuan and Zheng et al.~\cite{ZhengShasha2020SciChina,Yuan2020PRB} have showed that the abnormal entanglement enhancement occurs in the deep strong coupling regime in a two-sublattice antiferromagnet/ferrimagnet-light system. More novel physical mechanisms of manipulating quantum entanglement remain to be discovered.

\textbf{\textit{The relationship between entanglement and intrinsic magnetic parameters.}}
From the above discussion, we know that the entanglement among magnons, photons and phonons can be realized at a proper parameter regimes of the system.
On the other hand, the quantification of entanglement can be further employed to infer the intrinsic magnetic parameters that is difficult to be experimentally measured via traditional measurement methods. For example, Zheng et al.~\cite{ZhengShasha2020SciChina} showed that the observation of asymmetric EPR steering among two magnons in a hybrid ferrimagnet-light system can serve as a useful alternative method for determining the magnetic damping rates of the two sublattices. In a long term, we may build a relationship between entanglement and intrinsic magnetic parameters.

\section*{Acknowledgements} This work was funded by the National Key Research Development
Program under Contract No. 2022YFA1402802 and the European Union’s Horizon 2020 research
and innovation programme under Marie Skłodowska-Curie Grant Agreement SPINCAT No. 101018193. S. S. Zheng acknowledges the financial support from National Natural Science Foundation of China (NSFC) (Grants No. 12204441). F. X. Sun acknowledges NSFC (Grants No. 12147148), and the China Postdoctoral Science Foundation (Grant No. 2020M680186). Q. Y. He acknowledges NSFC (Grants
No. 11975026, 12125402), the Key R\&D Program of Guangdong Province (Grant No. 2018B030329001) and the Innovation Program for Quantum Science and Technology (No. 2021ZD0301500). Y.-P. Wang is supported by the NSFC (Grants No. 12174329 and 92265202) and the Fundamental Research Funds for the Central Universities (No. 2021FZZX001-02).
\vspace{6pt}

\bibliography{magnonics_ref}

\begin{thebibliography}{336}%
\makeatletter
\providecommand \@ifxundefined [1]{%
 \@ifx{#1\undefined}
}%
\providecommand \@ifnum [1]{%
 \ifnum #1\expandafter \@firstoftwo
 \else \expandafter \@secondoftwo
 \fi
}%
\providecommand \@ifx [1]{%
 \ifx #1\expandafter \@firstoftwo
 \else \expandafter \@secondoftwo
 \fi
}%
\providecommand \natexlab [1]{#1}%
\providecommand \enquote  [1]{``#1''}%
\providecommand \bibnamefont  [1]{#1}%
\providecommand \bibfnamefont [1]{#1}%
\providecommand \citenamefont [1]{#1}%
\providecommand \href@noop [0]{\@secondoftwo}%
\providecommand \href [0]{\begingroup \@sanitize@url \@href}%
\providecommand \@href[1]{\@@startlink{#1}\@@href}%
\providecommand \@@href[1]{\endgroup#1\@@endlink}%
\providecommand \@sanitize@url [0]{\catcode `\\12\catcode `\$12\catcode
  `\&12\catcode `\#12\catcode `\^12\catcode `\_12\catcode `\%12\relax}%
\providecommand \@@startlink[1]{}%
\providecommand \@@endlink[0]{}%
\providecommand \url  [0]{\begingroup\@sanitize@url \@url }%
\providecommand \@url [1]{\endgroup\@href {#1}{\urlprefix }}%
\providecommand \urlprefix  [0]{URL }%
\providecommand \Eprint [0]{\href }%
\providecommand \doibase [0]{http://dx.doi.org/}%
\providecommand \selectlanguage [0]{\@gobble}%
\providecommand \bibinfo  [0]{\@secondoftwo}%
\providecommand \bibfield  [0]{\@secondoftwo}%
\providecommand \translation [1]{[#1]}%
\providecommand \BibitemOpen [0]{}%
\providecommand \bibitemStop [0]{}%
\providecommand \bibitemNoStop [0]{.\EOS\space}%
\providecommand \EOS [0]{\spacefactor3000\relax}%
\providecommand \BibitemShut  [1]{\csname bibitem#1\endcsname}%
\let\auto@bib@innerbib\@empty
\bibitem [{\citenamefont {\ifmmode \check{Z}\else
  \v{Z}\fi{}uti\ifmmode~\acute{c}\else \'{c}\fi{}}, \citenamefont {Fabian},\
  and\ \citenamefont {Das~Sarma}(2004)}]{DasmaRMP2004}%
  \BibitemOpen
  \bibfield  {author} {\bibinfo {author} {\bibfnamefont {I.}~\bibnamefont
  {\ifmmode \check{Z}\else \v{Z}\fi{}uti\ifmmode~\acute{c}\else \'{c}\fi{}}},
  \bibinfo {author} {\bibfnamefont {J.}~\bibnamefont {Fabian}}, \ and\ \bibinfo
  {author} {\bibfnamefont {S.}~\bibnamefont {Das~Sarma}},\ }\bibfield  {title}
  {\enquote {\bibinfo {title} {Spintronics: Fundamentals and applications},}\
  }\href {\doibase 10.1103/RevModPhys.76.323} {\bibfield  {journal} {\bibinfo
  {journal} {Rev. Mod. Phys.}\ }\textbf {\bibinfo {volume} {76}},\ \bibinfo
  {pages} {323--410} (\bibinfo {year} {2004})}\BibitemShut {NoStop}%
\bibitem [{\citenamefont {Fert}(2008)}]{FertRMP2008}%
  \BibitemOpen
  \bibfield  {author} {\bibinfo {author} {\bibfnamefont {A.}~\bibnamefont
  {Fert}},\ }\bibfield  {title} {\enquote {\bibinfo {title} {Nobel lecture:
  Origin, development, and future of spintronics},}\ }\href {\doibase
  10.1103/RevModPhys.80.1517} {\bibfield  {journal} {\bibinfo  {journal} {Rev.
  Mod. Phys.}\ }\textbf {\bibinfo {volume} {80}},\ \bibinfo {pages}
  {1517--1530} (\bibinfo {year} {2008})}\BibitemShut {NoStop}%
\bibitem [{\citenamefont {Hirohata}\ \emph {et~al.}(2020)\citenamefont
  {Hirohata}, \citenamefont {Yamada}, \citenamefont {Nakatani}, \citenamefont
  {Prejbeanu}, \citenamefont {Diény}, \citenamefont {Pirro},\ and\
  \citenamefont {Hillebrands}}]{HirohataReview2020}%
  \BibitemOpen
  \bibfield  {author} {\bibinfo {author} {\bibfnamefont {A.}~\bibnamefont
  {Hirohata}}, \bibinfo {author} {\bibfnamefont {K.}~\bibnamefont {Yamada}},
  \bibinfo {author} {\bibfnamefont {Y.}~\bibnamefont {Nakatani}}, \bibinfo
  {author} {\bibfnamefont {I.-L.}\ \bibnamefont {Prejbeanu}}, \bibinfo {author}
  {\bibfnamefont {B.}~\bibnamefont {Diény}}, \bibinfo {author} {\bibfnamefont
  {P.}~\bibnamefont {Pirro}}, \ and\ \bibinfo {author} {\bibfnamefont
  {B.}~\bibnamefont {Hillebrands}},\ }\bibfield  {title} {\enquote {\bibinfo
  {title} {Review on spintronics: Principles and device applications},}\ }\href
  {\doibase https://doi.org/10.1016/j.jmmm.2020.166711} {\bibfield  {journal}
  {\bibinfo  {journal} {J. Magn. Magn. Mater.}\ }\textbf {\bibinfo {volume}
  {509}},\ \bibinfo {pages} {166711} (\bibinfo {year} {2020})}\BibitemShut
  {NoStop}%
\bibitem [{\citenamefont {Chumak}\ \emph
  {et~al.}(2015{\natexlab{a}})\citenamefont {Chumak}, \citenamefont
  {Vasyuchka}, \citenamefont {Serga},\ and\ \citenamefont
  {Hillebrands}}]{ChumakNP2015}%
  \BibitemOpen
  \bibfield  {author} {\bibinfo {author} {\bibfnamefont {A.~V.}\ \bibnamefont
  {Chumak}}, \bibinfo {author} {\bibfnamefont {V.~I.}\ \bibnamefont
  {Vasyuchka}}, \bibinfo {author} {\bibfnamefont {A.~A.}\ \bibnamefont
  {Serga}}, \ and\ \bibinfo {author} {\bibfnamefont {B.}~\bibnamefont
  {Hillebrands}},\ }\bibfield  {title} {\enquote {\bibinfo {title} {Magnon
  spintronics},}\ }\href {\doibase 10.1038/nphys3347} {\bibfield  {journal}
  {\bibinfo  {journal} {Nat. Phys.}\ }\textbf {\bibinfo {volume} {11}},\
  \bibinfo {pages} {453--461} (\bibinfo {year}
  {2015}{\natexlab{a}})}\BibitemShut {NoStop}%
\bibitem [{\citenamefont {Yuan}\ \emph
  {et~al.}(2022{\natexlab{a}})\citenamefont {Yuan}, \citenamefont {Cao},
  \citenamefont {Kamra}, \citenamefont {Duine},\ and\ \citenamefont
  {Yan}}]{YuanReview2022}%
  \BibitemOpen
  \bibfield  {author} {\bibinfo {author} {\bibfnamefont {H.}~\bibnamefont
  {Yuan}}, \bibinfo {author} {\bibfnamefont {Y.}~\bibnamefont {Cao}}, \bibinfo
  {author} {\bibfnamefont {A.}~\bibnamefont {Kamra}}, \bibinfo {author}
  {\bibfnamefont {R.~A.}\ \bibnamefont {Duine}}, \ and\ \bibinfo {author}
  {\bibfnamefont {P.}~\bibnamefont {Yan}},\ }\bibfield  {title} {\enquote
  {\bibinfo {title} {Quantum magnonics: When magnon spintronics meets quantum
  information science},}\ }\href {\doibase
  https://doi.org/10.1016/j.physrep.2022.03.002} {\bibfield  {journal}
  {\bibinfo  {journal} {Phys. Rep.}\ }\textbf {\bibinfo {volume} {965}},\
  \bibinfo {pages} {1--74} (\bibinfo {year} {2022}{\natexlab{a}})}\BibitemShut
  {NoStop}%
\bibitem [{\citenamefont {{Zare Rameshti}}\ \emph {et~al.}(2022)\citenamefont
  {{Zare Rameshti}}, \citenamefont {{Viola Kusminskiy}}, \citenamefont {Haigh},
  \citenamefont {Usami}, \citenamefont {Lachance-Quirion}, \citenamefont
  {Nakamura}, \citenamefont {Hu}, \citenamefont {Tang}, \citenamefont {Bauer},\
  and\ \citenamefont {Blanter}}]{BabakReview2022}%
  \BibitemOpen
  \bibfield  {author} {\bibinfo {author} {\bibfnamefont {B.}~\bibnamefont
  {{Zare Rameshti}}}, \bibinfo {author} {\bibfnamefont {S.}~\bibnamefont
  {{Viola Kusminskiy}}}, \bibinfo {author} {\bibfnamefont {J.~A.}\ \bibnamefont
  {Haigh}}, \bibinfo {author} {\bibfnamefont {K.}~\bibnamefont {Usami}},
  \bibinfo {author} {\bibfnamefont {D.}~\bibnamefont {Lachance-Quirion}},
  \bibinfo {author} {\bibfnamefont {Y.}~\bibnamefont {Nakamura}}, \bibinfo
  {author} {\bibfnamefont {C.-M.}\ \bibnamefont {Hu}}, \bibinfo {author}
  {\bibfnamefont {H.~X.}\ \bibnamefont {Tang}}, \bibinfo {author}
  {\bibfnamefont {G.~E.}\ \bibnamefont {Bauer}}, \ and\ \bibinfo {author}
  {\bibfnamefont {Y.~M.}\ \bibnamefont {Blanter}},\ }\bibfield  {title}
  {\enquote {\bibinfo {title} {Cavity magnonics},}\ }\href {\doibase
  https://doi.org/10.1016/j.physrep.2022.06.001} {\bibfield  {journal}
  {\bibinfo  {journal} {Phys. Rep.}\ }\textbf {\bibinfo {volume} {979}},\
  \bibinfo {pages} {1--61} (\bibinfo {year} {2022})}\BibitemShut {NoStop}%
\bibitem [{\citenamefont {Wang}\ \emph
  {et~al.}(2021{\natexlab{a}})\citenamefont {Wang}, \citenamefont {Yuan},
  \citenamefont {Cao}, \citenamefont {Li}, \citenamefont {Duine},\ and\
  \citenamefont {Yan}}]{WangPRL2021}%
  \BibitemOpen
  \bibfield  {author} {\bibinfo {author} {\bibfnamefont {Z.}~\bibnamefont
  {Wang}}, \bibinfo {author} {\bibfnamefont {H.~Y.}\ \bibnamefont {Yuan}},
  \bibinfo {author} {\bibfnamefont {Y.}~\bibnamefont {Cao}}, \bibinfo {author}
  {\bibfnamefont {Z.-X.}\ \bibnamefont {Li}}, \bibinfo {author} {\bibfnamefont
  {R.~A.}\ \bibnamefont {Duine}}, \ and\ \bibinfo {author} {\bibfnamefont
  {P.}~\bibnamefont {Yan}},\ }\bibfield  {title} {\enquote {\bibinfo {title}
  {Magnonic frequency comb through nonlinear magnon-skyrmion scattering},}\
  }\href {\doibase 10.1103/PhysRevLett.127.037202} {\bibfield  {journal}
  {\bibinfo  {journal} {Phys. Rev. Lett.}\ }\textbf {\bibinfo {volume} {127}},\
  \bibinfo {pages} {037202} (\bibinfo {year} {2021}{\natexlab{a}})}\BibitemShut
  {NoStop}%
\bibitem [{\citenamefont {Sheng}\ \emph {et~al.}(2023)\citenamefont {Sheng},
  \citenamefont {Elyasi}, \citenamefont {Chen}, \citenamefont {He},
  \citenamefont {Wang}, \citenamefont {Wang}, \citenamefont {Feng},
  \citenamefont {Zhang}, \citenamefont {Medlej}, \citenamefont {Liu},
  \citenamefont {Jiang}, \citenamefont {Han}, \citenamefont {Yu}, \citenamefont
  {Ansermet}, \citenamefont {Bauer},\ and\ \citenamefont {Yu}}]{ShengPRL2023}%
  \BibitemOpen
  \bibfield  {author} {\bibinfo {author} {\bibfnamefont {L.}~\bibnamefont
  {Sheng}}, \bibinfo {author} {\bibfnamefont {M.}~\bibnamefont {Elyasi}},
  \bibinfo {author} {\bibfnamefont {J.}~\bibnamefont {Chen}}, \bibinfo {author}
  {\bibfnamefont {W.}~\bibnamefont {He}}, \bibinfo {author} {\bibfnamefont
  {Y.}~\bibnamefont {Wang}}, \bibinfo {author} {\bibfnamefont {H.}~\bibnamefont
  {Wang}}, \bibinfo {author} {\bibfnamefont {H.}~\bibnamefont {Feng}}, \bibinfo
  {author} {\bibfnamefont {Y.}~\bibnamefont {Zhang}}, \bibinfo {author}
  {\bibfnamefont {I.}~\bibnamefont {Medlej}}, \bibinfo {author} {\bibfnamefont
  {S.}~\bibnamefont {Liu}}, \bibinfo {author} {\bibfnamefont {W.}~\bibnamefont
  {Jiang}}, \bibinfo {author} {\bibfnamefont {X.}~\bibnamefont {Han}}, \bibinfo
  {author} {\bibfnamefont {D.}~\bibnamefont {Yu}}, \bibinfo {author}
  {\bibfnamefont {J.-P.}\ \bibnamefont {Ansermet}}, \bibinfo {author}
  {\bibfnamefont {G.~E.~W.}\ \bibnamefont {Bauer}}, \ and\ \bibinfo {author}
  {\bibfnamefont {H.}~\bibnamefont {Yu}},\ }\bibfield  {title} {\enquote
  {\bibinfo {title} {Nonlocal detection of interlayer three-magnon coupling},}\
  }\href {\doibase 10.1103/PhysRevLett.130.046701} {\bibfield  {journal}
  {\bibinfo  {journal} {Phys. Rev. Lett.}\ }\textbf {\bibinfo {volume} {130}},\
  \bibinfo {pages} {046701} (\bibinfo {year} {2023})}\BibitemShut {NoStop}%
\bibitem [{\citenamefont {Holstein}\ and\ \citenamefont
  {Primakoff}(1940)}]{HP_PR_1940}%
  \BibitemOpen
  \bibfield  {author} {\bibinfo {author} {\bibfnamefont {T.}~\bibnamefont
  {Holstein}}\ and\ \bibinfo {author} {\bibfnamefont {H.}~\bibnamefont
  {Primakoff}},\ }\bibfield  {title} {\enquote {\bibinfo {title} {Field
  dependence of the intrinsic domain magnetization of a ferromagnet},}\ }\href
  {\doibase 10.1103/PhysRev.58.1098} {\bibfield  {journal} {\bibinfo  {journal}
  {Phys. Rev.}\ }\textbf {\bibinfo {volume} {58}},\ \bibinfo {pages}
  {1098--1113} (\bibinfo {year} {1940})}\BibitemShut {NoStop}%
\bibitem [{\citenamefont {Kittel}(1949)}]{KittelRMP1949}%
  \BibitemOpen
  \bibfield  {author} {\bibinfo {author} {\bibfnamefont {C.}~\bibnamefont
  {Kittel}},\ }\bibfield  {title} {\enquote {\bibinfo {title} {Physical theory
  of ferromagnetic domains},}\ }\href {\doibase 10.1103/RevModPhys.21.541}
  {\bibfield  {journal} {\bibinfo  {journal} {Rev. Mod. Phys.}\ }\textbf
  {\bibinfo {volume} {21}},\ \bibinfo {pages} {541--583} (\bibinfo {year}
  {1949})}\BibitemShut {NoStop}%
\bibitem [{\citenamefont {Dreher}\ \emph {et~al.}(2012)\citenamefont {Dreher},
  \citenamefont {Weiler}, \citenamefont {Pernpeintner}, \citenamefont {Huebl},
  \citenamefont {Gross}, \citenamefont {Brandt},\ and\ \citenamefont
  {Goennenwein}}]{DreherPRB2012}%
  \BibitemOpen
  \bibfield  {author} {\bibinfo {author} {\bibfnamefont {L.}~\bibnamefont
  {Dreher}}, \bibinfo {author} {\bibfnamefont {M.}~\bibnamefont {Weiler}},
  \bibinfo {author} {\bibfnamefont {M.}~\bibnamefont {Pernpeintner}}, \bibinfo
  {author} {\bibfnamefont {H.}~\bibnamefont {Huebl}}, \bibinfo {author}
  {\bibfnamefont {R.}~\bibnamefont {Gross}}, \bibinfo {author} {\bibfnamefont
  {M.~S.}\ \bibnamefont {Brandt}}, \ and\ \bibinfo {author} {\bibfnamefont
  {S.~T.~B.}\ \bibnamefont {Goennenwein}},\ }\bibfield  {title} {\enquote
  {\bibinfo {title} {Surface acoustic wave driven ferromagnetic resonance in
  nickel thin films: Theory and experiment},}\ }\href {\doibase
  10.1103/PhysRevB.86.134415} {\bibfield  {journal} {\bibinfo  {journal} {Phys.
  Rev. B}\ }\textbf {\bibinfo {volume} {86}},\ \bibinfo {pages} {134415}
  (\bibinfo {year} {2012})}\BibitemShut {NoStop}%
\bibitem [{\citenamefont {R\"uckriegel}\ \emph {et~al.}(2014)\citenamefont
  {R\"uckriegel}, \citenamefont {Kopietz}, \citenamefont {Bozhko},
  \citenamefont {Serga},\ and\ \citenamefont
  {Hillebrands}}]{RuckriegelPRB2014}%
  \BibitemOpen
  \bibfield  {author} {\bibinfo {author} {\bibfnamefont {A.}~\bibnamefont
  {R\"uckriegel}}, \bibinfo {author} {\bibfnamefont {P.}~\bibnamefont
  {Kopietz}}, \bibinfo {author} {\bibfnamefont {D.~A.}\ \bibnamefont {Bozhko}},
  \bibinfo {author} {\bibfnamefont {A.~A.}\ \bibnamefont {Serga}}, \ and\
  \bibinfo {author} {\bibfnamefont {B.}~\bibnamefont {Hillebrands}},\
  }\bibfield  {title} {\enquote {\bibinfo {title} {Magnetoelastic modes and
  lifetime of magnons in thin yttrium iron garnet films},}\ }\href {\doibase
  10.1103/PhysRevB.89.184413} {\bibfield  {journal} {\bibinfo  {journal} {Phys.
  Rev. B}\ }\textbf {\bibinfo {volume} {89}},\ \bibinfo {pages} {184413}
  (\bibinfo {year} {2014})}\BibitemShut {NoStop}%
\bibitem [{\citenamefont {Kamra}\ and\ \citenamefont
  {Bauer}(2014)}]{KamraSSC2014}%
  \BibitemOpen
  \bibfield  {author} {\bibinfo {author} {\bibfnamefont {A.}~\bibnamefont
  {Kamra}}\ and\ \bibinfo {author} {\bibfnamefont {G.~E.}\ \bibnamefont
  {Bauer}},\ }\bibfield  {title} {\enquote {\bibinfo {title} {Actuation,
  propagation, and detection of transverse magnetoelastic waves in
  ferromagnets},}\ }\href {\doibase https://doi.org/10.1016/j.ssc.2013.10.007}
  {\bibfield  {journal} {\bibinfo  {journal} {Solid State Commun.}\ }\textbf
  {\bibinfo {volume} {198}},\ \bibinfo {pages} {35--39} (\bibinfo {year}
  {2014})},\ \bibinfo {note} {sI: Spin Mechanics}\BibitemShut {NoStop}%
\bibitem [{\citenamefont {Aspelmeyer}, \citenamefont {Kippenberg},\ and\
  \citenamefont {Marquardt}(2014)}]{AspelmeyerRMP2014}%
  \BibitemOpen
  \bibfield  {author} {\bibinfo {author} {\bibfnamefont {M.}~\bibnamefont
  {Aspelmeyer}}, \bibinfo {author} {\bibfnamefont {T.~J.}\ \bibnamefont
  {Kippenberg}}, \ and\ \bibinfo {author} {\bibfnamefont {F.}~\bibnamefont
  {Marquardt}},\ }\bibfield  {title} {\enquote {\bibinfo {title} {Cavity
  optomechanics},}\ }\href {\doibase 10.1103/RevModPhys.86.1391} {\bibfield
  {journal} {\bibinfo  {journal} {Rev. Mod. Phys.}\ }\textbf {\bibinfo {volume}
  {86}},\ \bibinfo {pages} {1391--1452} (\bibinfo {year} {2014})}\BibitemShut
  {NoStop}%
\bibitem [{\citenamefont {McMichael}\ and\ \citenamefont
  {Kunz}(2002)}]{MichaelJAP2002}%
  \BibitemOpen
  \bibfield  {author} {\bibinfo {author} {\bibfnamefont {R.~D.}\ \bibnamefont
  {McMichael}}\ and\ \bibinfo {author} {\bibfnamefont {A.}~\bibnamefont
  {Kunz}},\ }\bibfield  {title} {\enquote {\bibinfo {title} {Calculation of
  damping rates in thin inhomogeneous ferromagnetic films due to coupling to
  lattice vibrations},}\ }\href {\doibase 10.1063/1.1450831} {\bibfield
  {journal} {\bibinfo  {journal} {J. App. Phys.}\ }\textbf {\bibinfo {volume}
  {91}},\ \bibinfo {pages} {8650--8652} (\bibinfo {year} {2002})},\ \Eprint
  {http://arxiv.org/abs/https://aip.scitation.org/doi/pdf/10.1063/1.1450831}
  {https://aip.scitation.org/doi/pdf/10.1063/1.1450831} \BibitemShut {NoStop}%
\bibitem [{\citenamefont {Safonov}\ and\ \citenamefont
  {Bertram}(2003)}]{Safonov2003}%
  \BibitemOpen
  \bibfield  {author} {\bibinfo {author} {\bibfnamefont {V.~L.}\ \bibnamefont
  {Safonov}}\ and\ \bibinfo {author} {\bibfnamefont {H.~N.}\ \bibnamefont
  {Bertram}},\ }\bibfield  {title} {\enquote {\bibinfo {title} {Linear
  stochastic magnetization dynamics and microscopic relaxation mechanisms},}\
  }\href {\doibase 10.1063/1.1581349} {\bibfield  {journal} {\bibinfo
  {journal} {J. App. Phys.}\ }\textbf {\bibinfo {volume} {94}},\ \bibinfo
  {pages} {529--538} (\bibinfo {year} {2003})},\ \Eprint
  {http://arxiv.org/abs/https://doi.org/10.1063/1.1581349}
  {https://doi.org/10.1063/1.1581349} \BibitemShut {NoStop}%
\bibitem [{\citenamefont {Zhang}\ \emph
  {et~al.}(2016{\natexlab{a}})\citenamefont {Zhang}, \citenamefont {Zou},
  \citenamefont {Jiang},\ and\ \citenamefont {Tang}}]{ZhangSA2016}%
  \BibitemOpen
  \bibfield  {author} {\bibinfo {author} {\bibfnamefont {X.}~\bibnamefont
  {Zhang}}, \bibinfo {author} {\bibfnamefont {C.-L.}\ \bibnamefont {Zou}},
  \bibinfo {author} {\bibfnamefont {L.}~\bibnamefont {Jiang}}, \ and\ \bibinfo
  {author} {\bibfnamefont {H.~X.}\ \bibnamefont {Tang}},\ }\bibfield  {title}
  {\enquote {\bibinfo {title} {Cavity magnomechanics},}\ }\href {\doibase
  10.1126/sciadv.1501286} {\bibfield  {journal} {\bibinfo  {journal} {Sci.
  Adv.}\ }\textbf {\bibinfo {volume} {2}},\ \bibinfo {pages} {e1501286}
  (\bibinfo {year} {2016}{\natexlab{a}})},\ \Eprint
  {http://arxiv.org/abs/https://www.science.org/doi/pdf/10.1126/sciadv.1501286}
  {https://www.science.org/doi/pdf/10.1126/sciadv.1501286} \BibitemShut
  {NoStop}%
\bibitem [{\citenamefont {Harder}\ \emph
  {et~al.}(2018{\natexlab{a}})\citenamefont {Harder}, \citenamefont {Yang},
  \citenamefont {Yao}, \citenamefont {Yu}, \citenamefont {Rao}, \citenamefont
  {Gui}, \citenamefont {Stamps},\ and\ \citenamefont {Hu}}]{HarderPRL2018}%
  \BibitemOpen
  \bibfield  {author} {\bibinfo {author} {\bibfnamefont {M.}~\bibnamefont
  {Harder}}, \bibinfo {author} {\bibfnamefont {Y.}~\bibnamefont {Yang}},
  \bibinfo {author} {\bibfnamefont {B.~M.}\ \bibnamefont {Yao}}, \bibinfo
  {author} {\bibfnamefont {C.~H.}\ \bibnamefont {Yu}}, \bibinfo {author}
  {\bibfnamefont {J.~W.}\ \bibnamefont {Rao}}, \bibinfo {author} {\bibfnamefont
  {Y.~S.}\ \bibnamefont {Gui}}, \bibinfo {author} {\bibfnamefont {R.~L.}\
  \bibnamefont {Stamps}}, \ and\ \bibinfo {author} {\bibfnamefont {C.-M.}\
  \bibnamefont {Hu}},\ }\bibfield  {title} {\enquote {\bibinfo {title} {Level
  attraction due to dissipative magnon-photon coupling},}\ }\href {\doibase
  10.1103/PhysRevLett.121.137203} {\bibfield  {journal} {\bibinfo  {journal}
  {Phys. Rev. Lett.}\ }\textbf {\bibinfo {volume} {121}},\ \bibinfo {pages}
  {137203} (\bibinfo {year} {2018}{\natexlab{a}})}\BibitemShut {NoStop}%
\bibitem [{\citenamefont {Yu}\ \emph {et~al.}(2019)\citenamefont {Yu},
  \citenamefont {Wang}, \citenamefont {Yuan},\ and\ \citenamefont
  {Xiao}}]{YuPRL2019}%
  \BibitemOpen
  \bibfield  {author} {\bibinfo {author} {\bibfnamefont {W.}~\bibnamefont
  {Yu}}, \bibinfo {author} {\bibfnamefont {J.}~\bibnamefont {Wang}}, \bibinfo
  {author} {\bibfnamefont {H.~Y.}\ \bibnamefont {Yuan}}, \ and\ \bibinfo
  {author} {\bibfnamefont {J.}~\bibnamefont {Xiao}},\ }\bibfield  {title}
  {\enquote {\bibinfo {title} {Prediction of attractive level crossing via a
  dissipative mode},}\ }\href {\doibase 10.1103/PhysRevLett.123.227201}
  {\bibfield  {journal} {\bibinfo  {journal} {Phys. Rev. Lett.}\ }\textbf
  {\bibinfo {volume} {123}},\ \bibinfo {pages} {227201} (\bibinfo {year}
  {2019})}\BibitemShut {NoStop}%
\bibitem [{\citenamefont {Borovik-Romanov}\ and\ \citenamefont
  {Kreines}(1982)}]{BorovikPR1982}%
  \BibitemOpen
  \bibfield  {author} {\bibinfo {author} {\bibfnamefont {A.}~\bibnamefont
  {Borovik-Romanov}}\ and\ \bibinfo {author} {\bibfnamefont {N.}~\bibnamefont
  {Kreines}},\ }\bibfield  {title} {\enquote {\bibinfo {title}
  {Brillouin-mandelstam scattering from thermal and excited magnons},}\ }\href
  {\doibase https://doi.org/10.1016/0370-1573(82)90118-1} {\bibfield  {journal}
  {\bibinfo  {journal} {Phys. Rep.}\ }\textbf {\bibinfo {volume} {81}},\
  \bibinfo {pages} {351--408} (\bibinfo {year} {1982})}\BibitemShut {NoStop}%
\bibitem [{\citenamefont {Sebastian}\ \emph {et~al.}(2015)\citenamefont
  {Sebastian}, \citenamefont {Schultheiss}, \citenamefont {Obry}, \citenamefont
  {Hillebrands},\ and\ \citenamefont {Schultheiss}}]{SebastianFP2015}%
  \BibitemOpen
  \bibfield  {author} {\bibinfo {author} {\bibfnamefont {T.}~\bibnamefont
  {Sebastian}}, \bibinfo {author} {\bibfnamefont {K.}~\bibnamefont
  {Schultheiss}}, \bibinfo {author} {\bibfnamefont {B.}~\bibnamefont {Obry}},
  \bibinfo {author} {\bibfnamefont {B.}~\bibnamefont {Hillebrands}}, \ and\
  \bibinfo {author} {\bibfnamefont {H.}~\bibnamefont {Schultheiss}},\
  }\bibfield  {title} {\enquote {\bibinfo {title} {Micro-focused brillouin
  light scattering: imaging spin waves at the nanoscale},}\ }\href {\doibase
  10.3389/fphy.2015.00035} {\bibfield  {journal} {\bibinfo  {journal} {Front.
  Phys.}\ }\textbf {\bibinfo {volume} {3}} (\bibinfo {year} {2015}),\
  10.3389/fphy.2015.00035}\BibitemShut {NoStop}%
\bibitem [{\citenamefont {Fleury}\ and\ \citenamefont
  {Loudon}(1968)}]{FleuryPR1968}%
  \BibitemOpen
  \bibfield  {author} {\bibinfo {author} {\bibfnamefont {P.~A.}\ \bibnamefont
  {Fleury}}\ and\ \bibinfo {author} {\bibfnamefont {R.}~\bibnamefont
  {Loudon}},\ }\bibfield  {title} {\enquote {\bibinfo {title} {Scattering of
  light by one- and two-magnon excitations},}\ }\href {\doibase
  10.1103/PhysRev.166.514} {\bibfield  {journal} {\bibinfo  {journal} {Phys.
  Rev.}\ }\textbf {\bibinfo {volume} {166}},\ \bibinfo {pages} {514--530}
  (\bibinfo {year} {1968})}\BibitemShut {NoStop}%
\bibitem [{\citenamefont {Rezende}\ and\ \citenamefont
  {de~Aguiar}(1990)}]{RezendePIEEE1990}%
  \BibitemOpen
  \bibfield  {author} {\bibinfo {author} {\bibfnamefont {S.~M.}\ \bibnamefont
  {Rezende}}\ and\ \bibinfo {author} {\bibfnamefont {F.~M.}\ \bibnamefont
  {de~Aguiar}},\ }\bibfield  {title} {\enquote {\bibinfo {title} {Spin-wave
  instabilities, auto-oscillations, and chaos in yttrium-iron-garnet},}\ }\href
  {\doibase 10.1109/5.56906} {\bibfield  {journal} {\bibinfo  {journal}
  {Proceedings of the IEEE}\ }\textbf {\bibinfo {volume} {78}},\ \bibinfo
  {pages} {893--908} (\bibinfo {year} {1990})}\BibitemShut {NoStop}%
\bibitem [{\citenamefont {Sandweg}\ \emph {et~al.}(2011)\citenamefont
  {Sandweg}, \citenamefont {Kajiwara}, \citenamefont {Chumak}, \citenamefont
  {Serga}, \citenamefont {Vasyuchka}, \citenamefont {Jungfleisch},
  \citenamefont {Saitoh},\ and\ \citenamefont {Hillebrands}}]{SandwegPRL2011}%
  \BibitemOpen
  \bibfield  {author} {\bibinfo {author} {\bibfnamefont {C.~W.}\ \bibnamefont
  {Sandweg}}, \bibinfo {author} {\bibfnamefont {Y.}~\bibnamefont {Kajiwara}},
  \bibinfo {author} {\bibfnamefont {A.~V.}\ \bibnamefont {Chumak}}, \bibinfo
  {author} {\bibfnamefont {A.~A.}\ \bibnamefont {Serga}}, \bibinfo {author}
  {\bibfnamefont {V.~I.}\ \bibnamefont {Vasyuchka}}, \bibinfo {author}
  {\bibfnamefont {M.~B.}\ \bibnamefont {Jungfleisch}}, \bibinfo {author}
  {\bibfnamefont {E.}~\bibnamefont {Saitoh}}, \ and\ \bibinfo {author}
  {\bibfnamefont {B.}~\bibnamefont {Hillebrands}},\ }\bibfield  {title}
  {\enquote {\bibinfo {title} {Spin pumping by parametrically excited exchange
  magnons},}\ }\href {\doibase 10.1103/PhysRevLett.106.216601} {\bibfield
  {journal} {\bibinfo  {journal} {Phys. Rev. Lett.}\ }\textbf {\bibinfo
  {volume} {106}},\ \bibinfo {pages} {216601} (\bibinfo {year}
  {2011})}\BibitemShut {NoStop}%
\bibitem [{\citenamefont {Kurebayashi}\ \emph {et~al.}(2011)\citenamefont
  {Kurebayashi}, \citenamefont {Dzyapko}, \citenamefont {Demidov},
  \citenamefont {Fang}, \citenamefont {Ferguson},\ and\ \citenamefont
  {Demokritov}}]{KurebayashiAPL2011}%
  \BibitemOpen
  \bibfield  {author} {\bibinfo {author} {\bibfnamefont {H.}~\bibnamefont
  {Kurebayashi}}, \bibinfo {author} {\bibfnamefont {O.}~\bibnamefont
  {Dzyapko}}, \bibinfo {author} {\bibfnamefont {V.~E.}\ \bibnamefont
  {Demidov}}, \bibinfo {author} {\bibfnamefont {D.}~\bibnamefont {Fang}},
  \bibinfo {author} {\bibfnamefont {A.~J.}\ \bibnamefont {Ferguson}}, \ and\
  \bibinfo {author} {\bibfnamefont {S.~O.}\ \bibnamefont {Demokritov}},\
  }\bibfield  {title} {\enquote {\bibinfo {title} {Spin pumping by
  parametrically excited short-wavelength spin waves},}\ }\href {\doibase
  10.1063/1.3652911} {\bibfield  {journal} {\bibinfo  {journal} {Appl. Phys.
  Lett.}\ }\textbf {\bibinfo {volume} {99}},\ \bibinfo {pages} {162502}
  (\bibinfo {year} {2011})}\BibitemShut {NoStop}%
\bibitem [{\citenamefont {Bloembergen}\ and\ \citenamefont
  {Damon}(1952)}]{BloembergenPRB1952}%
  \BibitemOpen
  \bibfield  {author} {\bibinfo {author} {\bibfnamefont {N.}~\bibnamefont
  {Bloembergen}}\ and\ \bibinfo {author} {\bibfnamefont {R.~W.}\ \bibnamefont
  {Damon}},\ }\bibfield  {title} {\enquote {\bibinfo {title} {Relaxation
  effects in ferromagnetic resonance},}\ }\href {\doibase
  10.1103/PhysRev.85.699} {\bibfield  {journal} {\bibinfo  {journal} {Phys.
  Rev.}\ }\textbf {\bibinfo {volume} {85}},\ \bibinfo {pages} {699--699}
  (\bibinfo {year} {1952})}\BibitemShut {NoStop}%
\bibitem [{\citenamefont {Bloembergen}\ and\ \citenamefont
  {Wang}(1954)}]{BloembergenPR1954}%
  \BibitemOpen
  \bibfield  {author} {\bibinfo {author} {\bibfnamefont {N.}~\bibnamefont
  {Bloembergen}}\ and\ \bibinfo {author} {\bibfnamefont {S.}~\bibnamefont
  {Wang}},\ }\bibfield  {title} {\enquote {\bibinfo {title} {Relaxation effects
  in para- and ferromagnetic resonance},}\ }\href {\doibase
  10.1103/PhysRev.93.72} {\bibfield  {journal} {\bibinfo  {journal} {Phys.
  Rev.}\ }\textbf {\bibinfo {volume} {93}},\ \bibinfo {pages} {72--83}
  (\bibinfo {year} {1954})}\BibitemShut {NoStop}%
\bibitem [{\citenamefont {Anderson}\ and\ \citenamefont
  {Suhl}(1955{\natexlab{a}})}]{AndersonPR1955}%
  \BibitemOpen
  \bibfield  {author} {\bibinfo {author} {\bibfnamefont {P.~W.}\ \bibnamefont
  {Anderson}}\ and\ \bibinfo {author} {\bibfnamefont {H.}~\bibnamefont
  {Suhl}},\ }\bibfield  {title} {\enquote {\bibinfo {title} {Instability in the
  motion of ferromagnets at high microwave power levels},}\ }\href {\doibase
  10.1103/PhysRev.100.1788} {\bibfield  {journal} {\bibinfo  {journal} {Phys.
  Rev.}\ }\textbf {\bibinfo {volume} {100}},\ \bibinfo {pages} {1788--1789}
  (\bibinfo {year} {1955}{\natexlab{a}})}\BibitemShut {NoStop}%
\bibitem [{\citenamefont {Schl{\"o}mann}, \citenamefont {Green},\ and\
  \citenamefont {Milano}(1960)}]{SchlomannJAP1960}%
  \BibitemOpen
  \bibfield  {author} {\bibinfo {author} {\bibfnamefont {E.}~\bibnamefont
  {Schl{\"o}mann}}, \bibinfo {author} {\bibfnamefont {J.~J.}\ \bibnamefont
  {Green}}, \ and\ \bibinfo {author} {\bibfnamefont {U.}~\bibnamefont
  {Milano}},\ }\bibfield  {title} {\enquote {\bibinfo {title} {Recent
  developments in ferromagnetic resonance at high power levels},}\ }\href
  {\doibase 10.1063/1.1984759} {\bibfield  {journal} {\bibinfo  {journal} {J.
  Appl. Phys.}\ }\textbf {\bibinfo {volume} {31}},\ \bibinfo {pages}
  {S386--S395} (\bibinfo {year} {1960})}\BibitemShut {NoStop}%
\bibitem [{\citenamefont {Brächer}, \citenamefont {Pirro},\ and\ \citenamefont
  {Hillebrands}(2017)}]{BracherPR2017}%
  \BibitemOpen
  \bibfield  {author} {\bibinfo {author} {\bibfnamefont {T.}~\bibnamefont
  {Brächer}}, \bibinfo {author} {\bibfnamefont {P.}~\bibnamefont {Pirro}}, \
  and\ \bibinfo {author} {\bibfnamefont {B.}~\bibnamefont {Hillebrands}},\
  }\bibfield  {title} {\enquote {\bibinfo {title} {Parallel pumping for magnon
  spintronics: Amplification and manipulation of magnon spin currents on the
  micron-scale},}\ }\href {\doibase
  https://doi.org/10.1016/j.physrep.2017.07.003} {\bibfield  {journal}
  {\bibinfo  {journal} {Phys. Rep.}\ }\textbf {\bibinfo {volume} {699}},\
  \bibinfo {pages} {1--34} (\bibinfo {year} {2017})},\ \bibinfo {note}
  {parallel pumping for magnon spintronics: Amplification and manipulation of
  magnon spin currents on the micron-scale}\BibitemShut {NoStop}%
\bibitem [{\citenamefont {Okano}\ and\ \citenamefont
  {Nozaki}(2019)}]{OkanoPRB2019}%
  \BibitemOpen
  \bibfield  {author} {\bibinfo {author} {\bibfnamefont {G.}~\bibnamefont
  {Okano}}\ and\ \bibinfo {author} {\bibfnamefont {Y.}~\bibnamefont {Nozaki}},\
  }\bibfield  {title} {\enquote {\bibinfo {title} {Spin waves parametrically
  excited via three-magnon scattering in narrow nife strips},}\ }\href
  {\doibase 10.1103/PhysRevB.100.104424} {\bibfield  {journal} {\bibinfo
  {journal} {Phys. Rev. B}\ }\textbf {\bibinfo {volume} {100}},\ \bibinfo
  {pages} {104424} (\bibinfo {year} {2019})}\BibitemShut {NoStop}%
\bibitem [{\citenamefont {Suhl}(1957)}]{SuhlJPCS1957}%
  \BibitemOpen
  \bibfield  {author} {\bibinfo {author} {\bibfnamefont {H.}~\bibnamefont
  {Suhl}},\ }\bibfield  {title} {\enquote {\bibinfo {title} {The theory of
  ferromagnetic resonance at high signal powers},}\ }\href
  {https://www.sciencedirect.com/science/article/pii/0022369757900100}
  {\bibfield  {journal} {\bibinfo  {journal} {J. Phys. Chem. Solids}\ }\textbf
  {\bibinfo {volume} {1}},\ \bibinfo {pages} {209--227} (\bibinfo {year}
  {1957})}\BibitemShut {NoStop}%
\bibitem [{\citenamefont {Verba}\ \emph {et~al.}(2014)\citenamefont {Verba},
  \citenamefont {Tiberkevich}, \citenamefont {Krivorotov},\ and\ \citenamefont
  {Slavin}}]{VerbaPRA2014}%
  \BibitemOpen
  \bibfield  {author} {\bibinfo {author} {\bibfnamefont {R.}~\bibnamefont
  {Verba}}, \bibinfo {author} {\bibfnamefont {V.}~\bibnamefont {Tiberkevich}},
  \bibinfo {author} {\bibfnamefont {I.}~\bibnamefont {Krivorotov}}, \ and\
  \bibinfo {author} {\bibfnamefont {A.}~\bibnamefont {Slavin}},\ }\bibfield
  {title} {\enquote {\bibinfo {title} {Parametric excitation of spin waves by
  voltage-controlled magnetic anisotropy},}\ }\href {\doibase
  10.1103/PhysRevApplied.1.044006} {\bibfield  {journal} {\bibinfo  {journal}
  {Phys. Rev. Appl.}\ }\textbf {\bibinfo {volume} {1}},\ \bibinfo {pages}
  {044006} (\bibinfo {year} {2014})}\BibitemShut {NoStop}%
\bibitem [{\citenamefont {Chen}\ \emph
  {et~al.}(2017{\natexlab{a}})\citenamefont {Chen}, \citenamefont {Lee},
  \citenamefont {Verba}, \citenamefont {Katine}, \citenamefont {Barsukov},
  \citenamefont {Tiberkevich}, \citenamefont {Xiao}, \citenamefont {Slavin},\
  and\ \citenamefont {Krivorotov}}]{ChenNL2017}%
  \BibitemOpen
  \bibfield  {author} {\bibinfo {author} {\bibfnamefont {Y.-J.}\ \bibnamefont
  {Chen}}, \bibinfo {author} {\bibfnamefont {H.~K.}\ \bibnamefont {Lee}},
  \bibinfo {author} {\bibfnamefont {R.}~\bibnamefont {Verba}}, \bibinfo
  {author} {\bibfnamefont {J.~A.}\ \bibnamefont {Katine}}, \bibinfo {author}
  {\bibfnamefont {I.}~\bibnamefont {Barsukov}}, \bibinfo {author}
  {\bibfnamefont {V.}~\bibnamefont {Tiberkevich}}, \bibinfo {author}
  {\bibfnamefont {J.~Q.}\ \bibnamefont {Xiao}}, \bibinfo {author}
  {\bibfnamefont {A.~N.}\ \bibnamefont {Slavin}}, \ and\ \bibinfo {author}
  {\bibfnamefont {I.~N.}\ \bibnamefont {Krivorotov}},\ }\bibfield  {title}
  {\enquote {\bibinfo {title} {Parametric resonance of magnetization excited by
  electric field},}\ }\href {\doibase 10.1021/acs.nanolett.6b04725} {\bibfield
  {journal} {\bibinfo  {journal} {Nano Lett.}\ }\textbf {\bibinfo {volume}
  {17}},\ \bibinfo {pages} {572--577} (\bibinfo {year}
  {2017}{\natexlab{a}})}\BibitemShut {NoStop}%
\bibitem [{\citenamefont {Alekseev}\ \emph {et~al.}(2020)\citenamefont
  {Alekseev}, \citenamefont {Dizhur}, \citenamefont {Polzikova}, \citenamefont
  {Luzanov}, \citenamefont {Raevskiy}, \citenamefont {Orlov}, \citenamefont
  {Kotov},\ and\ \citenamefont {Nikitov}}]{AlekseevAPL2020}%
  \BibitemOpen
  \bibfield  {author} {\bibinfo {author} {\bibfnamefont {S.~G.}\ \bibnamefont
  {Alekseev}}, \bibinfo {author} {\bibfnamefont {S.~E.}\ \bibnamefont
  {Dizhur}}, \bibinfo {author} {\bibfnamefont {N.~I.}\ \bibnamefont
  {Polzikova}}, \bibinfo {author} {\bibfnamefont {V.~A.}\ \bibnamefont
  {Luzanov}}, \bibinfo {author} {\bibfnamefont {A.~O.}\ \bibnamefont
  {Raevskiy}}, \bibinfo {author} {\bibfnamefont {A.~P.}\ \bibnamefont {Orlov}},
  \bibinfo {author} {\bibfnamefont {V.~A.}\ \bibnamefont {Kotov}}, \ and\
  \bibinfo {author} {\bibfnamefont {S.~A.}\ \bibnamefont {Nikitov}},\
  }\bibfield  {title} {\enquote {\bibinfo {title} {Magnons parametric pumping
  in bulk acoustic waves resonator},}\ }\href {\doibase 10.1063/5.0022267}
  {\bibfield  {journal} {\bibinfo  {journal} {Appl. Phys. Lett.}\ }\textbf
  {\bibinfo {volume} {117}},\ \bibinfo {pages} {072408} (\bibinfo {year}
  {2020})}\BibitemShut {NoStop}%
\bibitem [{\citenamefont {Gurevich}\ and\ \citenamefont
  {Melkov}(1996)}]{Gurevich}%
  \BibitemOpen
  \bibfield  {author} {\bibinfo {author} {\bibfnamefont {A.~G.}\ \bibnamefont
  {Gurevich}}\ and\ \bibinfo {author} {\bibfnamefont {G.~A.}\ \bibnamefont
  {Melkov}},\ }\href@noop {} {\emph {\bibinfo {title} {Magnetization
  oscillations and waves}}}\ (\bibinfo  {publisher} {CRC press, Boca Raton},\
  \bibinfo {year} {1996})\BibitemShut {NoStop}%
\bibitem [{\citenamefont {Spaldin}(2010)}]{spaldin2010magnetic}%
  \BibitemOpen
  \bibfield  {author} {\bibinfo {author} {\bibfnamefont {N.~A.}\ \bibnamefont
  {Spaldin}},\ }\href@noop {} {\emph {\bibinfo {title} {Magnetic materials:
  fundamentals and applications}}}\ (\bibinfo  {publisher} {Cambridge
  university press},\ \bibinfo {year} {2010})\BibitemShut {NoStop}%
\bibitem [{\citenamefont {Kong}, \citenamefont {Xiong},\ and\ \citenamefont
  {Wu}(2019{\natexlab{a}})}]{PhysRevApplied.12.034001}%
  \BibitemOpen
  \bibfield  {author} {\bibinfo {author} {\bibfnamefont {C.}~\bibnamefont
  {Kong}}, \bibinfo {author} {\bibfnamefont {H.}~\bibnamefont {Xiong}}, \ and\
  \bibinfo {author} {\bibfnamefont {Y.}~\bibnamefont {Wu}},\ }\bibfield
  {title} {\enquote {\bibinfo {title} {Magnon-induced nonreciprocity based on
  the magnon kerr effect},}\ }\href {\doibase 10.1103/PhysRevApplied.12.034001}
  {\bibfield  {journal} {\bibinfo  {journal} {Phys. Rev. Appl.}\ }\textbf
  {\bibinfo {volume} {12}},\ \bibinfo {pages} {034001} (\bibinfo {year}
  {2019}{\natexlab{a}})}\BibitemShut {NoStop}%
\bibitem [{\citenamefont {Liu}\ \emph {et~al.}(2018)\citenamefont {Liu},
  \citenamefont {Wang}, \citenamefont {Xiong},\ and\ \citenamefont
  {Wu}}]{Liu:18}%
  \BibitemOpen
  \bibfield  {author} {\bibinfo {author} {\bibfnamefont {Z.-X.}\ \bibnamefont
  {Liu}}, \bibinfo {author} {\bibfnamefont {B.}~\bibnamefont {Wang}}, \bibinfo
  {author} {\bibfnamefont {H.}~\bibnamefont {Xiong}}, \ and\ \bibinfo {author}
  {\bibfnamefont {Y.}~\bibnamefont {Wu}},\ }\bibfield  {title} {\enquote
  {\bibinfo {title} {Magnon-induced high-order sideband generation},}\ }\href
  {\doibase 10.1364/OL.43.003698} {\bibfield  {journal} {\bibinfo  {journal}
  {Opt. Lett.}\ }\textbf {\bibinfo {volume} {43}},\ \bibinfo {pages}
  {3698--3701} (\bibinfo {year} {2018})}\BibitemShut {NoStop}%
\bibitem [{\citenamefont {Zhao}\ \emph {et~al.}(2022)\citenamefont {Zhao},
  \citenamefont {Yang}, \citenamefont {Peng}, \citenamefont {Yang},
  \citenamefont {Li},\ and\ \citenamefont {Zhou}}]{PhysRevApplied.18.044074}%
  \BibitemOpen
  \bibfield  {author} {\bibinfo {author} {\bibfnamefont {C.}~\bibnamefont
  {Zhao}}, \bibinfo {author} {\bibfnamefont {Z.}~\bibnamefont {Yang}}, \bibinfo
  {author} {\bibfnamefont {R.}~\bibnamefont {Peng}}, \bibinfo {author}
  {\bibfnamefont {J.}~\bibnamefont {Yang}}, \bibinfo {author} {\bibfnamefont
  {C.}~\bibnamefont {Li}}, \ and\ \bibinfo {author} {\bibfnamefont
  {L.}~\bibnamefont {Zhou}},\ }\bibfield  {title} {\enquote {\bibinfo {title}
  {Dissipative-coupling-induced transparency and high-order sidebands with kerr
  nonlinearity in a cavity-magnonics system},}\ }\href {\doibase
  10.1103/PhysRevApplied.18.044074} {\bibfield  {journal} {\bibinfo  {journal}
  {Phys. Rev. Appl.}\ }\textbf {\bibinfo {volume} {18}},\ \bibinfo {pages}
  {044074} (\bibinfo {year} {2022})}\BibitemShut {NoStop}%
\bibitem [{\citenamefont {Wang}\ \emph
  {et~al.}(2021{\natexlab{b}})\citenamefont {Wang}, \citenamefont {Kong},
  \citenamefont {Sun}, \citenamefont {Zhang}, \citenamefont {Wu},\ and\
  \citenamefont {Zheng}}]{PhysRevA.104.033708}%
  \BibitemOpen
  \bibfield  {author} {\bibinfo {author} {\bibfnamefont {M.}~\bibnamefont
  {Wang}}, \bibinfo {author} {\bibfnamefont {C.}~\bibnamefont {Kong}}, \bibinfo
  {author} {\bibfnamefont {Z.-Y.}\ \bibnamefont {Sun}}, \bibinfo {author}
  {\bibfnamefont {D.}~\bibnamefont {Zhang}}, \bibinfo {author} {\bibfnamefont
  {Y.-Y.}\ \bibnamefont {Wu}}, \ and\ \bibinfo {author} {\bibfnamefont {L.-L.}\
  \bibnamefont {Zheng}},\ }\bibfield  {title} {\enquote {\bibinfo {title}
  {Nonreciprocal high-order sidebands induced by magnon kerr nonlinearity},}\
  }\href {\doibase 10.1103/PhysRevA.104.033708} {\bibfield  {journal} {\bibinfo
   {journal} {Phys. Rev. A}\ }\textbf {\bibinfo {volume} {104}},\ \bibinfo
  {pages} {033708} (\bibinfo {year} {2021}{\natexlab{b}})}\BibitemShut
  {NoStop}%
\bibitem [{\citenamefont {Liu}\ \emph {et~al.}(2019)\citenamefont {Liu},
  \citenamefont {You}, \citenamefont {Wang}, \citenamefont {Xiong},\ and\
  \citenamefont {Wu}}]{Liu:19}%
  \BibitemOpen
  \bibfield  {author} {\bibinfo {author} {\bibfnamefont {Z.-X.}\ \bibnamefont
  {Liu}}, \bibinfo {author} {\bibfnamefont {C.}~\bibnamefont {You}}, \bibinfo
  {author} {\bibfnamefont {B.}~\bibnamefont {Wang}}, \bibinfo {author}
  {\bibfnamefont {H.}~\bibnamefont {Xiong}}, \ and\ \bibinfo {author}
  {\bibfnamefont {Y.}~\bibnamefont {Wu}},\ }\bibfield  {title} {\enquote
  {\bibinfo {title} {Phase-mediated magnon chaos-order transition in cavity
  optomagnonics},}\ }\href {\doibase 10.1364/OL.44.000507} {\bibfield
  {journal} {\bibinfo  {journal} {Opt. Lett.}\ }\textbf {\bibinfo {volume}
  {44}},\ \bibinfo {pages} {507--510} (\bibinfo {year} {2019})}\BibitemShut
  {NoStop}%
\bibitem [{\citenamefont {Wang}\ \emph {et~al.}(2019)\citenamefont {Wang},
  \citenamefont {Kong}, \citenamefont {Liu}, \citenamefont {Xiong},\ and\
  \citenamefont {Wu}}]{Wang_2019}%
  \BibitemOpen
  \bibfield  {author} {\bibinfo {author} {\bibfnamefont {B.}~\bibnamefont
  {Wang}}, \bibinfo {author} {\bibfnamefont {C.}~\bibnamefont {Kong}}, \bibinfo
  {author} {\bibfnamefont {Z.-X.}\ \bibnamefont {Liu}}, \bibinfo {author}
  {\bibfnamefont {H.}~\bibnamefont {Xiong}}, \ and\ \bibinfo {author}
  {\bibfnamefont {Y.}~\bibnamefont {Wu}},\ }\bibfield  {title} {\enquote
  {\bibinfo {title} {Magnetic-field-controlled magnon chaos in an active
  cavity-magnon system},}\ }\href {\doibase 10.1088/1612-202X/ab09e5}
  {\bibfield  {journal} {\bibinfo  {journal} {Laser Phys. Lett.}\ }\textbf
  {\bibinfo {volume} {16}},\ \bibinfo {pages} {045208} (\bibinfo {year}
  {2019})}\BibitemShut {NoStop}%
\bibitem [{\citenamefont {Slonczewski}(1958)}]{PhysRev.110.1341}%
  \BibitemOpen
  \bibfield  {author} {\bibinfo {author} {\bibfnamefont {J.~C.}\ \bibnamefont
  {Slonczewski}},\ }\bibfield  {title} {\enquote {\bibinfo {title} {Origin of
  magnetic anisotropy in cobalt-substituted magnetite},}\ }\href {\doibase
  10.1103/PhysRev.110.1341} {\bibfield  {journal} {\bibinfo  {journal} {Phys.
  Rev.}\ }\textbf {\bibinfo {volume} {110}},\ \bibinfo {pages} {1341--1348}
  (\bibinfo {year} {1958})}\BibitemShut {NoStop}%
\bibitem [{\citenamefont {Slonczewski}(1961)}]{doi:10.1063/1.2000425}%
  \BibitemOpen
  \bibfield  {author} {\bibinfo {author} {\bibfnamefont {J.~C.}\ \bibnamefont
  {Slonczewski}},\ }\bibfield  {title} {\enquote {\bibinfo {title} {Anisotropy
  and magnetostriction in magnetic oxides},}\ }\href {\doibase
  10.1063/1.2000425} {\bibfield  {journal} {\bibinfo  {journal} {J. Appl.
  Phys.}\ }\textbf {\bibinfo {volume} {32}},\ \bibinfo {pages} {S253--S263}
  (\bibinfo {year} {1961})}\BibitemShut {NoStop}%
\bibitem [{\citenamefont {Blundell}(2001)}]{Blundell01}%
  \BibitemOpen
  \bibfield  {author} {\bibinfo {author} {\bibfnamefont {S.}~\bibnamefont
  {Blundell}},\ }\href@noop {} {\emph {\bibinfo {title} {Magnetism in condensed
  matter}}}\ (\bibinfo  {publisher} {Oxford University Press, Oxford},\
  \bibinfo {year} {2001})\BibitemShut {NoStop}%
\bibitem [{\citenamefont {Wang}\ \emph
  {et~al.}(2016{\natexlab{a}})\citenamefont {Wang}, \citenamefont {Zhang},
  \citenamefont {Zhang}, \citenamefont {Luo}, \citenamefont {Xiong},
  \citenamefont {Wang}, \citenamefont {Li}, \citenamefont {Hu},\ and\
  \citenamefont {You}}]{Wang16}%
  \BibitemOpen
  \bibfield  {author} {\bibinfo {author} {\bibfnamefont {Y.-P.}\ \bibnamefont
  {Wang}}, \bibinfo {author} {\bibfnamefont {G.-Q.}\ \bibnamefont {Zhang}},
  \bibinfo {author} {\bibfnamefont {D.}~\bibnamefont {Zhang}}, \bibinfo
  {author} {\bibfnamefont {X.-Q.}\ \bibnamefont {Luo}}, \bibinfo {author}
  {\bibfnamefont {W.}~\bibnamefont {Xiong}}, \bibinfo {author} {\bibfnamefont
  {S.-P.}\ \bibnamefont {Wang}}, \bibinfo {author} {\bibfnamefont {T.-F.}\
  \bibnamefont {Li}}, \bibinfo {author} {\bibfnamefont {C.-M.}\ \bibnamefont
  {Hu}}, \ and\ \bibinfo {author} {\bibfnamefont {J.~Q.}\ \bibnamefont {You}},\
  }\bibfield  {title} {\enquote {\bibinfo {title} {Magnon kerr effect in a
  strongly coupled cavity-magnon system},}\ }\href {\doibase
  10.1103/PhysRevB.94.224410} {\bibfield  {journal} {\bibinfo  {journal} {Phys.
  Rev. B}\ }\textbf {\bibinfo {volume} {94}},\ \bibinfo {pages} {224410}
  (\bibinfo {year} {2016}{\natexlab{a}})}\BibitemShut {NoStop}%
\bibitem [{\citenamefont {Prabhakar}\ and\ \citenamefont
  {Stancil}(2009)}]{stancil2009spin}%
  \BibitemOpen
  \bibfield  {author} {\bibinfo {author} {\bibfnamefont {A.}~\bibnamefont
  {Prabhakar}}\ and\ \bibinfo {author} {\bibfnamefont {D.~D.}\ \bibnamefont
  {Stancil}},\ }\href@noop {} {\emph {\bibinfo {title} {Spin waves: Theory and
  applications}}}\ (\bibinfo  {publisher} {Springer, New York},\ \bibinfo
  {year} {2009})\BibitemShut {NoStop}%
\bibitem [{\citenamefont {Anderson}\ and\ \citenamefont
  {Suhl}(1955{\natexlab{b}})}]{PhysRev.100.1788}%
  \BibitemOpen
  \bibfield  {author} {\bibinfo {author} {\bibfnamefont {P.~W.}\ \bibnamefont
  {Anderson}}\ and\ \bibinfo {author} {\bibfnamefont {H.}~\bibnamefont
  {Suhl}},\ }\bibfield  {title} {\enquote {\bibinfo {title} {Instability in the
  motion of ferromagnets at high microwave power levels},}\ }\href {\doibase
  10.1103/PhysRev.100.1788} {\bibfield  {journal} {\bibinfo  {journal} {Phys.
  Rev.}\ }\textbf {\bibinfo {volume} {100}},\ \bibinfo {pages} {1788--1789}
  (\bibinfo {year} {1955}{\natexlab{b}})}\BibitemShut {NoStop}%
\bibitem [{\citenamefont {Soykal}\ and\ \citenamefont
  {Flatt\'e}(2010{\natexlab{a}})}]{Soykal10}%
  \BibitemOpen
  \bibfield  {author} {\bibinfo {author} {\bibfnamefont {O.~O.}\ \bibnamefont
  {Soykal}}\ and\ \bibinfo {author} {\bibfnamefont {M.~E.}\ \bibnamefont
  {Flatt\'e}},\ }\bibfield  {title} {\enquote {\bibinfo {title} {Strong field
  interactions between a nanomagnet and a photonic cavity},}\ }\href {\doibase
  10.1103/PhysRevLett.104.077202} {\bibfield  {journal} {\bibinfo  {journal}
  {Phys. Rev. Lett.}\ }\textbf {\bibinfo {volume} {104}},\ \bibinfo {pages}
  {077202} (\bibinfo {year} {2010}{\natexlab{a}})}\BibitemShut {NoStop}%
\bibitem [{\citenamefont {Gu}\ \emph {et~al.}(2017)\citenamefont {Gu},
  \citenamefont {Kockum}, \citenamefont {Miranowicz}, \citenamefont {xi~Liu},\
  and\ \citenamefont {Nori}}]{GU20171}%
  \BibitemOpen
  \bibfield  {author} {\bibinfo {author} {\bibfnamefont {X.}~\bibnamefont
  {Gu}}, \bibinfo {author} {\bibfnamefont {A.~F.}\ \bibnamefont {Kockum}},
  \bibinfo {author} {\bibfnamefont {A.}~\bibnamefont {Miranowicz}}, \bibinfo
  {author} {\bibfnamefont {Y.}~\bibnamefont {xi~Liu}}, \ and\ \bibinfo {author}
  {\bibfnamefont {F.}~\bibnamefont {Nori}},\ }\bibfield  {title} {\enquote
  {\bibinfo {title} {Microwave photonics with superconducting quantum
  circuits},}\ }\href {\doibase https://doi.org/10.1016/j.physrep.2017.10.002}
  {\bibfield  {journal} {\bibinfo  {journal} {Phys. Rep.}\ }\textbf {\bibinfo
  {volume} {718-719}},\ \bibinfo {pages} {1--102} (\bibinfo {year} {2017})},\
  \bibinfo {note} {microwave photonics with superconducting quantum
  circuits}\BibitemShut {NoStop}%
\bibitem [{\citenamefont {Krantz}\ \emph {et~al.}(2019)\citenamefont {Krantz},
  \citenamefont {Kjaergaard}, \citenamefont {Yan}, \citenamefont {Orlando},
  \citenamefont {Gustavsson},\ and\ \citenamefont
  {Oliver}}]{doi:10.1063/1.5089550}%
  \BibitemOpen
  \bibfield  {author} {\bibinfo {author} {\bibfnamefont {P.}~\bibnamefont
  {Krantz}}, \bibinfo {author} {\bibfnamefont {M.}~\bibnamefont {Kjaergaard}},
  \bibinfo {author} {\bibfnamefont {F.}~\bibnamefont {Yan}}, \bibinfo {author}
  {\bibfnamefont {T.~P.}\ \bibnamefont {Orlando}}, \bibinfo {author}
  {\bibfnamefont {S.}~\bibnamefont {Gustavsson}}, \ and\ \bibinfo {author}
  {\bibfnamefont {W.~D.}\ \bibnamefont {Oliver}},\ }\bibfield  {title}
  {\enquote {\bibinfo {title} {A quantum engineer's guide to superconducting
  qubits},}\ }\href {\doibase 10.1063/1.5089550} {\bibfield  {journal}
  {\bibinfo  {journal} {Appl. Phys. Rev.}\ }\textbf {\bibinfo {volume} {6}},\
  \bibinfo {pages} {021318} (\bibinfo {year} {2019})}\BibitemShut {NoStop}%
\bibitem [{\citenamefont {Tabuchi}\ \emph
  {et~al.}(2015{\natexlab{a}})\citenamefont {Tabuchi}, \citenamefont {Ishino},
  \citenamefont {Noguchi}, \citenamefont {Ishikawa}, \citenamefont {Yamazaki},
  \citenamefont {Usami},\ and\ \citenamefont
  {Nakamura}}]{doi:10.1126/science.aaa3693}%
  \BibitemOpen
  \bibfield  {author} {\bibinfo {author} {\bibfnamefont {Y.}~\bibnamefont
  {Tabuchi}}, \bibinfo {author} {\bibfnamefont {S.}~\bibnamefont {Ishino}},
  \bibinfo {author} {\bibfnamefont {A.}~\bibnamefont {Noguchi}}, \bibinfo
  {author} {\bibfnamefont {T.}~\bibnamefont {Ishikawa}}, \bibinfo {author}
  {\bibfnamefont {R.}~\bibnamefont {Yamazaki}}, \bibinfo {author}
  {\bibfnamefont {K.}~\bibnamefont {Usami}}, \ and\ \bibinfo {author}
  {\bibfnamefont {Y.}~\bibnamefont {Nakamura}},\ }\bibfield  {title} {\enquote
  {\bibinfo {title} {Coherent coupling between a ferromagnetic magnon and a
  superconducting qubit},}\ }\href {\doibase 10.1126/science.aaa3693}
  {\bibfield  {journal} {\bibinfo  {journal} {Science}\ }\textbf {\bibinfo
  {volume} {349}},\ \bibinfo {pages} {405--408} (\bibinfo {year}
  {2015}{\natexlab{a}})}\BibitemShut {NoStop}%
\bibitem [{\citenamefont {Lachance-Quirion}\ \emph
  {et~al.}(2017{\natexlab{a}})\citenamefont {Lachance-Quirion}, \citenamefont
  {Tabuchi}, \citenamefont {Ishino}, \citenamefont {Noguchi}, \citenamefont
  {Ishikawa}, \citenamefont {Yamazaki},\ and\ \citenamefont
  {Nakamura}}]{doi:10.1126/sciadv.1603150}%
  \BibitemOpen
  \bibfield  {author} {\bibinfo {author} {\bibfnamefont {D.}~\bibnamefont
  {Lachance-Quirion}}, \bibinfo {author} {\bibfnamefont {Y.}~\bibnamefont
  {Tabuchi}}, \bibinfo {author} {\bibfnamefont {S.}~\bibnamefont {Ishino}},
  \bibinfo {author} {\bibfnamefont {A.}~\bibnamefont {Noguchi}}, \bibinfo
  {author} {\bibfnamefont {T.}~\bibnamefont {Ishikawa}}, \bibinfo {author}
  {\bibfnamefont {R.}~\bibnamefont {Yamazaki}}, \ and\ \bibinfo {author}
  {\bibfnamefont {Y.}~\bibnamefont {Nakamura}},\ }\bibfield  {title} {\enquote
  {\bibinfo {title} {Resolving quanta of collective spin excitations in a
  millimeter-sized ferromagnet},}\ }\href {\doibase 10.1126/sciadv.1603150}
  {\bibfield  {journal} {\bibinfo  {journal} {Sci. Adv.}\ }\textbf {\bibinfo
  {volume} {3}},\ \bibinfo {pages} {e1603150} (\bibinfo {year}
  {2017}{\natexlab{a}})}\BibitemShut {NoStop}%
\bibitem [{\citenamefont {Lachance-Quirion}\ \emph
  {et~al.}(2020{\natexlab{a}})\citenamefont {Lachance-Quirion}, \citenamefont
  {Wolski}, \citenamefont {Tabuchi}, \citenamefont {Kono}, \citenamefont
  {Usami},\ and\ \citenamefont {Nakamura}}]{doi:10.1126/science.aaz9236}%
  \BibitemOpen
  \bibfield  {author} {\bibinfo {author} {\bibfnamefont {D.}~\bibnamefont
  {Lachance-Quirion}}, \bibinfo {author} {\bibfnamefont {S.~P.}\ \bibnamefont
  {Wolski}}, \bibinfo {author} {\bibfnamefont {Y.}~\bibnamefont {Tabuchi}},
  \bibinfo {author} {\bibfnamefont {S.}~\bibnamefont {Kono}}, \bibinfo {author}
  {\bibfnamefont {K.}~\bibnamefont {Usami}}, \ and\ \bibinfo {author}
  {\bibfnamefont {Y.}~\bibnamefont {Nakamura}},\ }\bibfield  {title} {\enquote
  {\bibinfo {title} {Entanglement-based single-shot detection of a single
  magnon with a superconducting qubit},}\ }\href {\doibase
  10.1126/science.aaz9236} {\bibfield  {journal} {\bibinfo  {journal}
  {Science}\ }\textbf {\bibinfo {volume} {367}},\ \bibinfo {pages} {425--428}
  (\bibinfo {year} {2020}{\natexlab{a}})}\BibitemShut {NoStop}%
\bibitem [{\citenamefont {Xu}\ \emph {et~al.}(2022)\citenamefont {Xu},
  \citenamefont {Gu}, \citenamefont {Li}, \citenamefont {Weng}, \citenamefont
  {Wang}, \citenamefont {Li}, \citenamefont {Wang}, \citenamefont {Zhu},\ and\
  \citenamefont {You}}]{xu2022quantum}%
  \BibitemOpen
  \bibfield  {author} {\bibinfo {author} {\bibfnamefont {D.}~\bibnamefont
  {Xu}}, \bibinfo {author} {\bibfnamefont {X.-K.}\ \bibnamefont {Gu}}, \bibinfo
  {author} {\bibfnamefont {H.-K.}\ \bibnamefont {Li}}, \bibinfo {author}
  {\bibfnamefont {Y.-C.}\ \bibnamefont {Weng}}, \bibinfo {author}
  {\bibfnamefont {Y.-P.}\ \bibnamefont {Wang}}, \bibinfo {author}
  {\bibfnamefont {J.}~\bibnamefont {Li}}, \bibinfo {author} {\bibfnamefont
  {H.}~\bibnamefont {Wang}}, \bibinfo {author} {\bibfnamefont {S.-Y.}\
  \bibnamefont {Zhu}}, \ and\ \bibinfo {author} {\bibfnamefont
  {J.}~\bibnamefont {You}},\ }\bibfield  {title} {\enquote {\bibinfo {title}
  {Quantum control of a single magnon in a macroscopic spin system},}\
  }\href@noop {} {\bibfield  {journal} {\bibinfo  {journal} {arXiv preprint
  arXiv:2211.06644}\ } (\bibinfo {year} {2022})}\BibitemShut {NoStop}%
\bibitem [{\citenamefont {Shen}\ \emph {et~al.}(2021)\citenamefont {Shen},
  \citenamefont {Wang}, \citenamefont {Li}, \citenamefont {Zhu}, \citenamefont
  {Agarwal},\ and\ \citenamefont {You}}]{PhysRevLett.127.183202}%
  \BibitemOpen
  \bibfield  {author} {\bibinfo {author} {\bibfnamefont {R.-C.}\ \bibnamefont
  {Shen}}, \bibinfo {author} {\bibfnamefont {Y.-P.}\ \bibnamefont {Wang}},
  \bibinfo {author} {\bibfnamefont {J.}~\bibnamefont {Li}}, \bibinfo {author}
  {\bibfnamefont {S.-Y.}\ \bibnamefont {Zhu}}, \bibinfo {author} {\bibfnamefont
  {G.~S.}\ \bibnamefont {Agarwal}}, \ and\ \bibinfo {author} {\bibfnamefont
  {J.~Q.}\ \bibnamefont {You}},\ }\bibfield  {title} {\enquote {\bibinfo
  {title} {Long-time memory and ternary logic gate using a multistable cavity
  magnonic system},}\ }\href {\doibase 10.1103/PhysRevLett.127.183202}
  {\bibfield  {journal} {\bibinfo  {journal} {Phys. Rev. Lett.}\ }\textbf
  {\bibinfo {volume} {127}},\ \bibinfo {pages} {183202} (\bibinfo {year}
  {2021})}\BibitemShut {NoStop}%
\bibitem [{\citenamefont {Tabuchi}\ \emph
  {et~al.}(2014{\natexlab{a}})\citenamefont {Tabuchi}, \citenamefont {Ishino},
  \citenamefont {Ishikawa}, \citenamefont {Yamazaki}, \citenamefont {Usami},\
  and\ \citenamefont {Nakamura}}]{PhysRevLett.113.083603}%
  \BibitemOpen
  \bibfield  {author} {\bibinfo {author} {\bibfnamefont {Y.}~\bibnamefont
  {Tabuchi}}, \bibinfo {author} {\bibfnamefont {S.}~\bibnamefont {Ishino}},
  \bibinfo {author} {\bibfnamefont {T.}~\bibnamefont {Ishikawa}}, \bibinfo
  {author} {\bibfnamefont {R.}~\bibnamefont {Yamazaki}}, \bibinfo {author}
  {\bibfnamefont {K.}~\bibnamefont {Usami}}, \ and\ \bibinfo {author}
  {\bibfnamefont {Y.}~\bibnamefont {Nakamura}},\ }\bibfield  {title} {\enquote
  {\bibinfo {title} {Hybridizing ferromagnetic magnons and microwave photons in
  the quantum limit},}\ }\href {\doibase 10.1103/PhysRevLett.113.083603}
  {\bibfield  {journal} {\bibinfo  {journal} {Phys. Rev. Lett.}\ }\textbf
  {\bibinfo {volume} {113}},\ \bibinfo {pages} {083603} (\bibinfo {year}
  {2014}{\natexlab{a}})}\BibitemShut {NoStop}%
\bibitem [{\citenamefont {Zhang}\ \emph
  {et~al.}(2014{\natexlab{a}})\citenamefont {Zhang}, \citenamefont {Zou},
  \citenamefont {Jiang},\ and\ \citenamefont {Tang}}]{PhysRevLett.113.156401}%
  \BibitemOpen
  \bibfield  {author} {\bibinfo {author} {\bibfnamefont {X.}~\bibnamefont
  {Zhang}}, \bibinfo {author} {\bibfnamefont {C.-L.}\ \bibnamefont {Zou}},
  \bibinfo {author} {\bibfnamefont {L.}~\bibnamefont {Jiang}}, \ and\ \bibinfo
  {author} {\bibfnamefont {H.~X.}\ \bibnamefont {Tang}},\ }\bibfield  {title}
  {\enquote {\bibinfo {title} {Strongly coupled magnons and cavity microwave
  photons},}\ }\href {\doibase 10.1103/PhysRevLett.113.156401} {\bibfield
  {journal} {\bibinfo  {journal} {Phys. Rev. Lett.}\ }\textbf {\bibinfo
  {volume} {113}},\ \bibinfo {pages} {156401} (\bibinfo {year}
  {2014}{\natexlab{a}})}\BibitemShut {NoStop}%
\bibitem [{\citenamefont {Wang}\ \emph
  {et~al.}(2018{\natexlab{a}})\citenamefont {Wang}, \citenamefont {Zhang},
  \citenamefont {Zhang}, \citenamefont {Li}, \citenamefont {Hu},\ and\
  \citenamefont {You}}]{PhysRevLett.120.057202}%
  \BibitemOpen
  \bibfield  {author} {\bibinfo {author} {\bibfnamefont {Y.-P.}\ \bibnamefont
  {Wang}}, \bibinfo {author} {\bibfnamefont {G.-Q.}\ \bibnamefont {Zhang}},
  \bibinfo {author} {\bibfnamefont {D.}~\bibnamefont {Zhang}}, \bibinfo
  {author} {\bibfnamefont {T.-F.}\ \bibnamefont {Li}}, \bibinfo {author}
  {\bibfnamefont {C.-M.}\ \bibnamefont {Hu}}, \ and\ \bibinfo {author}
  {\bibfnamefont {J.~Q.}\ \bibnamefont {You}},\ }\bibfield  {title} {\enquote
  {\bibinfo {title} {Bistability of cavity magnon polaritons},}\ }\href
  {\doibase 10.1103/PhysRevLett.120.057202} {\bibfield  {journal} {\bibinfo
  {journal} {Phys. Rev. Lett.}\ }\textbf {\bibinfo {volume} {120}},\ \bibinfo
  {pages} {057202} (\bibinfo {year} {2018}{\natexlab{a}})}\BibitemShut
  {NoStop}%
\bibitem [{\citenamefont {Zhang}, \citenamefont {Wang},\ and\ \citenamefont
  {You}(2019)}]{Zhang2019}%
  \BibitemOpen
  \bibfield  {author} {\bibinfo {author} {\bibfnamefont {G.}~\bibnamefont
  {Zhang}}, \bibinfo {author} {\bibfnamefont {Y.}~\bibnamefont {Wang}}, \ and\
  \bibinfo {author} {\bibfnamefont {J.}~\bibnamefont {You}},\ }\bibfield
  {title} {\enquote {\bibinfo {title} {Theory of the magnon kerr effect in
  cavity magnonics},}\ }\href {\doibase 10.1007/s11433-018-9344-8} {\bibfield
  {journal} {\bibinfo  {journal} {Sci. China Phys. Mech.}\ }\textbf {\bibinfo
  {volume} {62}},\ \bibinfo {pages} {987511} (\bibinfo {year}
  {2019})}\BibitemShut {NoStop}%
\bibitem [{\citenamefont {Casteels}, \citenamefont {Fazio},\ and\ \citenamefont
  {Ciuti}(2017)}]{PhysRevA.95.012128}%
  \BibitemOpen
  \bibfield  {author} {\bibinfo {author} {\bibfnamefont {W.}~\bibnamefont
  {Casteels}}, \bibinfo {author} {\bibfnamefont {R.}~\bibnamefont {Fazio}}, \
  and\ \bibinfo {author} {\bibfnamefont {C.}~\bibnamefont {Ciuti}},\ }\bibfield
   {title} {\enquote {\bibinfo {title} {Critical dynamical properties of a
  first-order dissipative phase transition},}\ }\href {\doibase
  10.1103/PhysRevA.95.012128} {\bibfield  {journal} {\bibinfo  {journal} {Phys.
  Rev. A}\ }\textbf {\bibinfo {volume} {95}},\ \bibinfo {pages} {012128}
  (\bibinfo {year} {2017})}\BibitemShut {NoStop}%
\bibitem [{\citenamefont {\ifmmode \check{S}\else
  \v{S}\fi{}ibali\ifmmode~\acute{c}\else \'{c}\fi{}}\ \emph
  {et~al.}(2016)\citenamefont {\ifmmode \check{S}\else
  \v{S}\fi{}ibali\ifmmode~\acute{c}\else \'{c}\fi{}}, \citenamefont {Wade},
  \citenamefont {Adams}, \citenamefont {Weatherill},\ and\ \citenamefont
  {Pohl}}]{PhysRevA.94.011401}%
  \BibitemOpen
  \bibfield  {author} {\bibinfo {author} {\bibfnamefont {N.}~\bibnamefont
  {\ifmmode \check{S}\else \v{S}\fi{}ibali\ifmmode~\acute{c}\else \'{c}\fi{}}},
  \bibinfo {author} {\bibfnamefont {C.~G.}\ \bibnamefont {Wade}}, \bibinfo
  {author} {\bibfnamefont {C.~S.}\ \bibnamefont {Adams}}, \bibinfo {author}
  {\bibfnamefont {K.~J.}\ \bibnamefont {Weatherill}}, \ and\ \bibinfo {author}
  {\bibfnamefont {T.}~\bibnamefont {Pohl}},\ }\bibfield  {title} {\enquote
  {\bibinfo {title} {Driven-dissipative many-body systems with mixed power-law
  interactions: Bistabilities and temperature-driven nonequilibrium phase
  transitions},}\ }\href {\doibase 10.1103/PhysRevA.94.011401} {\bibfield
  {journal} {\bibinfo  {journal} {Phys. Rev. A}\ }\textbf {\bibinfo {volume}
  {94}},\ \bibinfo {pages} {011401} (\bibinfo {year} {2016})}\BibitemShut
  {NoStop}%
\bibitem [{\citenamefont {Labouvie}\ \emph {et~al.}(2016)\citenamefont
  {Labouvie}, \citenamefont {Santra}, \citenamefont {Heun},\ and\ \citenamefont
  {Ott}}]{PhysRevLett.116.235302}%
  \BibitemOpen
  \bibfield  {author} {\bibinfo {author} {\bibfnamefont {R.}~\bibnamefont
  {Labouvie}}, \bibinfo {author} {\bibfnamefont {B.}~\bibnamefont {Santra}},
  \bibinfo {author} {\bibfnamefont {S.}~\bibnamefont {Heun}}, \ and\ \bibinfo
  {author} {\bibfnamefont {H.}~\bibnamefont {Ott}},\ }\bibfield  {title}
  {\enquote {\bibinfo {title} {Bistability in a driven-dissipative
  superfluid},}\ }\href {\doibase 10.1103/PhysRevLett.116.235302} {\bibfield
  {journal} {\bibinfo  {journal} {Phys. Rev. Lett.}\ }\textbf {\bibinfo
  {volume} {116}},\ \bibinfo {pages} {235302} (\bibinfo {year}
  {2016})}\BibitemShut {NoStop}%
\bibitem [{\citenamefont {Bi}\ \emph {et~al.}(2021)\citenamefont {Bi},
  \citenamefont {Yan}, \citenamefont {Zhang},\ and\ \citenamefont
  {Xiao}}]{PhysRevB.103.104411}%
  \BibitemOpen
  \bibfield  {author} {\bibinfo {author} {\bibfnamefont {M.~X.}\ \bibnamefont
  {Bi}}, \bibinfo {author} {\bibfnamefont {X.~H.}\ \bibnamefont {Yan}},
  \bibinfo {author} {\bibfnamefont {Y.}~\bibnamefont {Zhang}}, \ and\ \bibinfo
  {author} {\bibfnamefont {Y.}~\bibnamefont {Xiao}},\ }\bibfield  {title}
  {\enquote {\bibinfo {title} {Tristability of cavity magnon polaritons},}\
  }\href {\doibase 10.1103/PhysRevB.103.104411} {\bibfield  {journal} {\bibinfo
   {journal} {Phys. Rev. B}\ }\textbf {\bibinfo {volume} {103}},\ \bibinfo
  {pages} {104411} (\bibinfo {year} {2021})}\BibitemShut {NoStop}%
\bibitem [{\citenamefont {Shen}\ \emph
  {et~al.}(2022{\natexlab{a}})\citenamefont {Shen}, \citenamefont {Li},
  \citenamefont {Fan}, \citenamefont {Wang},\ and\ \citenamefont
  {You}}]{PhysRevLett.129.123601}%
  \BibitemOpen
  \bibfield  {author} {\bibinfo {author} {\bibfnamefont {R.-C.}\ \bibnamefont
  {Shen}}, \bibinfo {author} {\bibfnamefont {J.}~\bibnamefont {Li}}, \bibinfo
  {author} {\bibfnamefont {Z.-Y.}\ \bibnamefont {Fan}}, \bibinfo {author}
  {\bibfnamefont {Y.-P.}\ \bibnamefont {Wang}}, \ and\ \bibinfo {author}
  {\bibfnamefont {J.~Q.}\ \bibnamefont {You}},\ }\bibfield  {title} {\enquote
  {\bibinfo {title} {Mechanical bistability in kerr-modified cavity
  magnomechanics},}\ }\href {\doibase 10.1103/PhysRevLett.129.123601}
  {\bibfield  {journal} {\bibinfo  {journal} {Phys. Rev. Lett.}\ }\textbf
  {\bibinfo {volume} {129}},\ \bibinfo {pages} {123601} (\bibinfo {year}
  {2022}{\natexlab{a}})}\BibitemShut {NoStop}%
\bibitem [{\citenamefont {Rhoads}, \citenamefont {Shaw},\ and\ \citenamefont
  {Turner}(2010)}]{rhoads2010nonlinear}%
  \BibitemOpen
  \bibfield  {author} {\bibinfo {author} {\bibfnamefont {J.~F.}\ \bibnamefont
  {Rhoads}}, \bibinfo {author} {\bibfnamefont {S.~W.}\ \bibnamefont {Shaw}}, \
  and\ \bibinfo {author} {\bibfnamefont {K.~L.}\ \bibnamefont {Turner}},\
  }\bibfield  {title} {\enquote {\bibinfo {title} {Nonlinear dynamics and its
  applications in micro-and nanoresonators},}\ }\href@noop {} {\bibfield
  {journal} {\bibinfo  {journal} {Journal of dynamic systems, measurement, and
  control}\ }\textbf {\bibinfo {volume} {132}} (\bibinfo {year}
  {2010})}\BibitemShut {NoStop}%
\bibitem [{\citenamefont {Duck}(2002)}]{duck2002nonlinear}%
  \BibitemOpen
  \bibfield  {author} {\bibinfo {author} {\bibfnamefont {F.~A.}\ \bibnamefont
  {Duck}},\ }\bibfield  {title} {\enquote {\bibinfo {title} {Nonlinear
  acoustics in diagnostic ultrasound},}\ }\href@noop {} {\bibfield  {journal}
  {\bibinfo  {journal} {Ultrasound in medicine \& biology}\ }\textbf {\bibinfo
  {volume} {28}},\ \bibinfo {pages} {1--18} (\bibinfo {year}
  {2002})}\BibitemShut {NoStop}%
\bibitem [{\citenamefont {Smirnova}\ \emph {et~al.}(2020)\citenamefont
  {Smirnova}, \citenamefont {Leykam}, \citenamefont {Chong},\ and\
  \citenamefont {Kivshar}}]{doi:10.1063/1.5142397}%
  \BibitemOpen
  \bibfield  {author} {\bibinfo {author} {\bibfnamefont {D.}~\bibnamefont
  {Smirnova}}, \bibinfo {author} {\bibfnamefont {D.}~\bibnamefont {Leykam}},
  \bibinfo {author} {\bibfnamefont {Y.}~\bibnamefont {Chong}}, \ and\ \bibinfo
  {author} {\bibfnamefont {Y.}~\bibnamefont {Kivshar}},\ }\bibfield  {title}
  {\enquote {\bibinfo {title} {Nonlinear topological photonics},}\ }\href
  {\doibase 10.1063/1.5142397} {\bibfield  {journal} {\bibinfo  {journal}
  {Applied Physics Reviews}\ }\textbf {\bibinfo {volume} {7}},\ \bibinfo
  {pages} {021306} (\bibinfo {year} {2020})}\BibitemShut {NoStop}%
\bibitem [{\citenamefont {Chang}, \citenamefont {Vuleti{\'{c}}},\ and\
  \citenamefont {Lukin}(2014)}]{Chang2014}%
  \BibitemOpen
  \bibfield  {author} {\bibinfo {author} {\bibfnamefont {D.~E.}\ \bibnamefont
  {Chang}}, \bibinfo {author} {\bibfnamefont {V.}~\bibnamefont
  {Vuleti{\'{c}}}}, \ and\ \bibinfo {author} {\bibfnamefont {M.~D.}\
  \bibnamefont {Lukin}},\ }\bibfield  {title} {\enquote {\bibinfo {title}
  {Quantum nonlinear optics --- photon by photon},}\ }\href {\doibase
  10.1038/nphoton.2014.192} {\bibfield  {journal} {\bibinfo  {journal} {Nat.
  Photon.}\ }\textbf {\bibinfo {volume} {8}},\ \bibinfo {pages} {685--694}
  (\bibinfo {year} {2014})}\BibitemShut {NoStop}%
\bibitem [{\citenamefont {Du}, \citenamefont {Lu},\ and\ \citenamefont
  {Xie}(2021)}]{Du2021}%
  \BibitemOpen
  \bibfield  {author} {\bibinfo {author} {\bibfnamefont {Z.~Z.}\ \bibnamefont
  {Du}}, \bibinfo {author} {\bibfnamefont {H.-Z.}\ \bibnamefont {Lu}}, \ and\
  \bibinfo {author} {\bibfnamefont {X.~C.}\ \bibnamefont {Xie}},\ }\bibfield
  {title} {\enquote {\bibinfo {title} {Nonlinear hall effects},}\ }\href
  {\doibase 10.1038/s42254-021-00359-6} {\bibfield  {journal} {\bibinfo
  {journal} {Nature Reviews Physics}\ }\textbf {\bibinfo {volume} {3}},\
  \bibinfo {pages} {744--752} (\bibinfo {year} {2021})}\BibitemShut {NoStop}%
\bibitem [{\citenamefont {Zhang}, \citenamefont {Scully},\ and\ \citenamefont
  {Agarwal}(2019{\natexlab{a}})}]{PhysRevResearch.1.023021}%
  \BibitemOpen
  \bibfield  {author} {\bibinfo {author} {\bibfnamefont {Z.}~\bibnamefont
  {Zhang}}, \bibinfo {author} {\bibfnamefont {M.~O.}\ \bibnamefont {Scully}}, \
  and\ \bibinfo {author} {\bibfnamefont {G.~S.}\ \bibnamefont {Agarwal}},\
  }\bibfield  {title} {\enquote {\bibinfo {title} {Quantum entanglement between
  two magnon modes via kerr nonlinearity driven far from equilibrium},}\ }\href
  {\doibase 10.1103/PhysRevResearch.1.023021} {\bibfield  {journal} {\bibinfo
  {journal} {Phys. Rev. Res.}\ }\textbf {\bibinfo {volume} {1}},\ \bibinfo
  {pages} {023021} (\bibinfo {year} {2019}{\natexlab{a}})}\BibitemShut
  {NoStop}%
\bibitem [{\citenamefont {Yang}\ \emph {et~al.}(2021)\citenamefont {Yang},
  \citenamefont {Jin}, \citenamefont {Jin}, \citenamefont {Liu}, \citenamefont
  {Liu},\ and\ \citenamefont {Yang}}]{PhysRevResearch.3.023126}%
  \BibitemOpen
  \bibfield  {author} {\bibinfo {author} {\bibfnamefont {Z.-B.}\ \bibnamefont
  {Yang}}, \bibinfo {author} {\bibfnamefont {H.}~\bibnamefont {Jin}}, \bibinfo
  {author} {\bibfnamefont {J.-W.}\ \bibnamefont {Jin}}, \bibinfo {author}
  {\bibfnamefont {J.-Y.}\ \bibnamefont {Liu}}, \bibinfo {author} {\bibfnamefont
  {H.-Y.}\ \bibnamefont {Liu}}, \ and\ \bibinfo {author} {\bibfnamefont
  {R.-C.}\ \bibnamefont {Yang}},\ }\bibfield  {title} {\enquote {\bibinfo
  {title} {Bistability of squeezing and entanglement in cavity magnonics},}\
  }\href {\doibase 10.1103/PhysRevResearch.3.023126} {\bibfield  {journal}
  {\bibinfo  {journal} {Phys. Rev. Res.}\ }\textbf {\bibinfo {volume} {3}},\
  \bibinfo {pages} {023126} (\bibinfo {year} {2021})}\BibitemShut {NoStop}%
\bibitem [{\citenamefont {Gao}\ \emph {et~al.}(2019)\citenamefont {Gao},
  \citenamefont {Liu}, \citenamefont {Wang}, \citenamefont {Cao},\ and\
  \citenamefont {Wang}}]{PhysRevA.100.043831}%
  \BibitemOpen
  \bibfield  {author} {\bibinfo {author} {\bibfnamefont {Y.-P.}\ \bibnamefont
  {Gao}}, \bibinfo {author} {\bibfnamefont {X.-F.}\ \bibnamefont {Liu}},
  \bibinfo {author} {\bibfnamefont {T.-J.}\ \bibnamefont {Wang}}, \bibinfo
  {author} {\bibfnamefont {C.}~\bibnamefont {Cao}}, \ and\ \bibinfo {author}
  {\bibfnamefont {C.}~\bibnamefont {Wang}},\ }\bibfield  {title} {\enquote
  {\bibinfo {title} {Photon excitation and photon-blockade effects in
  optomagnonic microcavities},}\ }\href {\doibase 10.1103/PhysRevA.100.043831}
  {\bibfield  {journal} {\bibinfo  {journal} {Phys. Rev. A}\ }\textbf {\bibinfo
  {volume} {100}},\ \bibinfo {pages} {043831} (\bibinfo {year}
  {2019})}\BibitemShut {NoStop}%
\bibitem [{\citenamefont {Yang}\ \emph {et~al.}(2022)\citenamefont {Yang},
  \citenamefont {Wu}, \citenamefont {Li}, \citenamefont {Wang},\ and\
  \citenamefont {You}}]{PhysRevA.106.012419}%
  \BibitemOpen
  \bibfield  {author} {\bibinfo {author} {\bibfnamefont {Z.-B.}\ \bibnamefont
  {Yang}}, \bibinfo {author} {\bibfnamefont {W.-J.}\ \bibnamefont {Wu}},
  \bibinfo {author} {\bibfnamefont {J.}~\bibnamefont {Li}}, \bibinfo {author}
  {\bibfnamefont {Y.-P.}\ \bibnamefont {Wang}}, \ and\ \bibinfo {author}
  {\bibfnamefont {J.~Q.}\ \bibnamefont {You}},\ }\bibfield  {title} {\enquote
  {\bibinfo {title} {Steady-entangled-state generation via the cross-kerr
  effect in a ferrimagnetic crystal},}\ }\href {\doibase
  10.1103/PhysRevA.106.012419} {\bibfield  {journal} {\bibinfo  {journal}
  {Phys. Rev. A}\ }\textbf {\bibinfo {volume} {106}},\ \bibinfo {pages}
  {012419} (\bibinfo {year} {2022})}\BibitemShut {NoStop}%
\bibitem [{\citenamefont {Haghshenasfard}\ and\ \citenamefont
  {Cottam}(2020)}]{doi:10.1063/5.0012072}%
  \BibitemOpen
  \bibfield  {author} {\bibinfo {author} {\bibfnamefont {Z.}~\bibnamefont
  {Haghshenasfard}}\ and\ \bibinfo {author} {\bibfnamefont {M.~G.}\
  \bibnamefont {Cottam}},\ }\bibfield  {title} {\enquote {\bibinfo {title}
  {Sub-poissonian statistics and squeezing of magnons due to the kerr effect in
  a hybrid coupled cavity–magnon system},}\ }\href {\doibase
  10.1063/5.0012072} {\bibfield  {journal} {\bibinfo  {journal} {Journal of
  Applied Physics}\ }\textbf {\bibinfo {volume} {128}},\ \bibinfo {pages}
  {033901} (\bibinfo {year} {2020})}\BibitemShut {NoStop}%
\bibitem [{\citenamefont {Xiong}\ \emph {et~al.}(2022)\citenamefont {Xiong},
  \citenamefont {Tian}, \citenamefont {Zhang},\ and\ \citenamefont
  {You}}]{PhysRevB.105.245310}%
  \BibitemOpen
  \bibfield  {author} {\bibinfo {author} {\bibfnamefont {W.}~\bibnamefont
  {Xiong}}, \bibinfo {author} {\bibfnamefont {M.}~\bibnamefont {Tian}},
  \bibinfo {author} {\bibfnamefont {G.-Q.}\ \bibnamefont {Zhang}}, \ and\
  \bibinfo {author} {\bibfnamefont {J.~Q.}\ \bibnamefont {You}},\ }\bibfield
  {title} {\enquote {\bibinfo {title} {Strong long-range spin-spin coupling via
  a kerr magnon interface},}\ }\href {\doibase 10.1103/PhysRevB.105.245310}
  {\bibfield  {journal} {\bibinfo  {journal} {Phys. Rev. B}\ }\textbf {\bibinfo
  {volume} {105}},\ \bibinfo {pages} {245310} (\bibinfo {year}
  {2022})}\BibitemShut {NoStop}%
\bibitem [{\citenamefont {Udem}, \citenamefont {Holzwarth},\ and\ \citenamefont
  {H{\"a}nsch}(2002)}]{UdemN2002}%
  \BibitemOpen
  \bibfield  {author} {\bibinfo {author} {\bibfnamefont {T.}~\bibnamefont
  {Udem}}, \bibinfo {author} {\bibfnamefont {R.}~\bibnamefont {Holzwarth}}, \
  and\ \bibinfo {author} {\bibfnamefont {T.~W.}\ \bibnamefont {H{\"a}nsch}},\
  }\bibfield  {title} {\enquote {\bibinfo {title} {Optical frequency
  metrology},}\ }\href {\doibase 10.1038/416233a} {\bibfield  {journal}
  {\bibinfo  {journal} {Nature}\ }\textbf {\bibinfo {volume} {416}},\ \bibinfo
  {pages} {233--237} (\bibinfo {year} {2002})}\BibitemShut {NoStop}%
\bibitem [{\citenamefont {Ludlow}\ \emph {et~al.}(2015)\citenamefont {Ludlow},
  \citenamefont {Boyd}, \citenamefont {Ye}, \citenamefont {Peik},\ and\
  \citenamefont {Schmidt}}]{LudlowRMP2015}%
  \BibitemOpen
  \bibfield  {author} {\bibinfo {author} {\bibfnamefont {A.~D.}\ \bibnamefont
  {Ludlow}}, \bibinfo {author} {\bibfnamefont {M.~M.}\ \bibnamefont {Boyd}},
  \bibinfo {author} {\bibfnamefont {J.}~\bibnamefont {Ye}}, \bibinfo {author}
  {\bibfnamefont {E.}~\bibnamefont {Peik}}, \ and\ \bibinfo {author}
  {\bibfnamefont {P.~O.}\ \bibnamefont {Schmidt}},\ }\bibfield  {title}
  {\enquote {\bibinfo {title} {Optical atomic clocks},}\ }\href {\doibase
  10.1103/RevModPhys.87.637} {\bibfield  {journal} {\bibinfo  {journal} {Rev.
  Mod. Phys.}\ }\textbf {\bibinfo {volume} {87}},\ \bibinfo {pages} {637--701}
  (\bibinfo {year} {2015})}\BibitemShut {NoStop}%
\bibitem [{\citenamefont {Beloy}\ \emph {et~al.}(2021)\citenamefont {Beloy},
  \citenamefont {Bodine}, \citenamefont {Bothwell}, \citenamefont {Brewer},
  \citenamefont {Bromley}, \citenamefont {Chen}, \citenamefont {Desch{\^e}nes},
  \citenamefont {Diddams}, \citenamefont {Fasano}, \citenamefont {Fortier},
  \citenamefont {Hassan}, \citenamefont {Hume}, \citenamefont {Kedar},
  \citenamefont {Kennedy}, \citenamefont {Khader}, \citenamefont {Koepke},
  \citenamefont {Leibrandt}, \citenamefont {Leopardi}, \citenamefont {Ludlow},
  \citenamefont {McGrew}, \citenamefont {Milner}, \citenamefont {Newbury},
  \citenamefont {Nicolodi}, \citenamefont {Oelker}, \citenamefont {Parker},
  \citenamefont {Robinson}, \citenamefont {Romisch}, \citenamefont
  {Sch{\"a}ffer}, \citenamefont {Sherman}, \citenamefont {Sinclair},
  \citenamefont {Sonderhouse}, \citenamefont {Swann}, \citenamefont {Yao},
  \citenamefont {Ye}, \citenamefont {Zhang},\ and\ \citenamefont
  {Collaboration}}]{BeloyN2021}%
  \BibitemOpen
  \bibfield  {author} {\bibinfo {author} {\bibfnamefont {K.}~\bibnamefont
  {Beloy}}, \bibinfo {author} {\bibfnamefont {M.~I.}\ \bibnamefont {Bodine}},
  \bibinfo {author} {\bibfnamefont {T.}~\bibnamefont {Bothwell}}, \bibinfo
  {author} {\bibfnamefont {S.~M.}\ \bibnamefont {Brewer}}, \bibinfo {author}
  {\bibfnamefont {S.~L.}\ \bibnamefont {Bromley}}, \bibinfo {author}
  {\bibfnamefont {J.-S.}\ \bibnamefont {Chen}}, \bibinfo {author}
  {\bibfnamefont {J.-D.}\ \bibnamefont {Desch{\^e}nes}}, \bibinfo {author}
  {\bibfnamefont {S.~A.}\ \bibnamefont {Diddams}}, \bibinfo {author}
  {\bibfnamefont {R.~J.}\ \bibnamefont {Fasano}}, \bibinfo {author}
  {\bibfnamefont {T.~M.}\ \bibnamefont {Fortier}}, \bibinfo {author}
  {\bibfnamefont {Y.~S.}\ \bibnamefont {Hassan}}, \bibinfo {author}
  {\bibfnamefont {D.~B.}\ \bibnamefont {Hume}}, \bibinfo {author}
  {\bibfnamefont {D.}~\bibnamefont {Kedar}}, \bibinfo {author} {\bibfnamefont
  {C.~J.}\ \bibnamefont {Kennedy}}, \bibinfo {author} {\bibfnamefont
  {I.}~\bibnamefont {Khader}}, \bibinfo {author} {\bibfnamefont
  {A.}~\bibnamefont {Koepke}}, \bibinfo {author} {\bibfnamefont {D.~R.}\
  \bibnamefont {Leibrandt}}, \bibinfo {author} {\bibfnamefont {H.}~\bibnamefont
  {Leopardi}}, \bibinfo {author} {\bibfnamefont {A.~D.}\ \bibnamefont
  {Ludlow}}, \bibinfo {author} {\bibfnamefont {W.~F.}\ \bibnamefont {McGrew}},
  \bibinfo {author} {\bibfnamefont {W.~R.}\ \bibnamefont {Milner}}, \bibinfo
  {author} {\bibfnamefont {N.~R.}\ \bibnamefont {Newbury}}, \bibinfo {author}
  {\bibfnamefont {D.}~\bibnamefont {Nicolodi}}, \bibinfo {author}
  {\bibfnamefont {E.}~\bibnamefont {Oelker}}, \bibinfo {author} {\bibfnamefont
  {T.~E.}\ \bibnamefont {Parker}}, \bibinfo {author} {\bibfnamefont {J.~M.}\
  \bibnamefont {Robinson}}, \bibinfo {author} {\bibfnamefont {S.}~\bibnamefont
  {Romisch}}, \bibinfo {author} {\bibfnamefont {S.~A.}\ \bibnamefont
  {Sch{\"a}ffer}}, \bibinfo {author} {\bibfnamefont {J.~A.}\ \bibnamefont
  {Sherman}}, \bibinfo {author} {\bibfnamefont {L.~C.}\ \bibnamefont
  {Sinclair}}, \bibinfo {author} {\bibfnamefont {L.}~\bibnamefont
  {Sonderhouse}}, \bibinfo {author} {\bibfnamefont {W.~C.}\ \bibnamefont
  {Swann}}, \bibinfo {author} {\bibfnamefont {J.}~\bibnamefont {Yao}}, \bibinfo
  {author} {\bibfnamefont {J.}~\bibnamefont {Ye}}, \bibinfo {author}
  {\bibfnamefont {X.}~\bibnamefont {Zhang}}, \ and\ \bibinfo {author}
  {\bibfnamefont {B.~A. C. O. N.~B.}\ \bibnamefont {Collaboration}},\
  }\bibfield  {title} {\enquote {\bibinfo {title} {Frequency ratio measurements
  at 18-digit accuracy using an optical clock network},}\ }\href {\doibase
  10.1038/s41586-021-03253-4} {\bibfield  {journal} {\bibinfo  {journal}
  {Nature}\ }\textbf {\bibinfo {volume} {591}},\ \bibinfo {pages} {564--569}
  (\bibinfo {year} {2021})}\BibitemShut {NoStop}%
\bibitem [{\citenamefont {Metcalf}\ \emph {et~al.}(2019)\citenamefont
  {Metcalf}, \citenamefont {Anderson}, \citenamefont {Bender}, \citenamefont
  {Blakeslee}, \citenamefont {Brand}, \citenamefont {Carlson}, \citenamefont
  {Cochran}, \citenamefont {Diddams}, \citenamefont {Endl}, \citenamefont
  {Fredrick}, \citenamefont {Halverson}, \citenamefont {Hickstein},
  \citenamefont {Hearty}, \citenamefont {Jennings}, \citenamefont {Kanodia},
  \citenamefont {Kaplan}, \citenamefont {Levi}, \citenamefont {Lubar},
  \citenamefont {Mahadevan}, \citenamefont {Monson}, \citenamefont {Ninan},
  \citenamefont {Nitroy}, \citenamefont {Osterman}, \citenamefont {Papp},
  \citenamefont {Quinlan}, \citenamefont {Ramsey}, \citenamefont {Robertson},
  \citenamefont {Roy}, \citenamefont {Schwab}, \citenamefont {Sigurdsson},
  \citenamefont {Srinivasan}, \citenamefont {Stefansson}, \citenamefont
  {Sterner}, \citenamefont {Terrien}, \citenamefont {Wolszczan}, \citenamefont
  {Wright},\ and\ \citenamefont {Ycas}}]{MetcalfO2019}%
  \BibitemOpen
  \bibfield  {author} {\bibinfo {author} {\bibfnamefont {A.~J.}\ \bibnamefont
  {Metcalf}}, \bibinfo {author} {\bibfnamefont {T.}~\bibnamefont {Anderson}},
  \bibinfo {author} {\bibfnamefont {C.~F.}\ \bibnamefont {Bender}}, \bibinfo
  {author} {\bibfnamefont {S.}~\bibnamefont {Blakeslee}}, \bibinfo {author}
  {\bibfnamefont {W.}~\bibnamefont {Brand}}, \bibinfo {author} {\bibfnamefont
  {D.~R.}\ \bibnamefont {Carlson}}, \bibinfo {author} {\bibfnamefont {W.~D.}\
  \bibnamefont {Cochran}}, \bibinfo {author} {\bibfnamefont {S.~A.}\
  \bibnamefont {Diddams}}, \bibinfo {author} {\bibfnamefont {M.}~\bibnamefont
  {Endl}}, \bibinfo {author} {\bibfnamefont {C.}~\bibnamefont {Fredrick}},
  \bibinfo {author} {\bibfnamefont {S.}~\bibnamefont {Halverson}}, \bibinfo
  {author} {\bibfnamefont {D.~D.}\ \bibnamefont {Hickstein}}, \bibinfo {author}
  {\bibfnamefont {F.}~\bibnamefont {Hearty}}, \bibinfo {author} {\bibfnamefont
  {J.}~\bibnamefont {Jennings}}, \bibinfo {author} {\bibfnamefont
  {S.}~\bibnamefont {Kanodia}}, \bibinfo {author} {\bibfnamefont {K.~F.}\
  \bibnamefont {Kaplan}}, \bibinfo {author} {\bibfnamefont {E.}~\bibnamefont
  {Levi}}, \bibinfo {author} {\bibfnamefont {E.}~\bibnamefont {Lubar}},
  \bibinfo {author} {\bibfnamefont {S.}~\bibnamefont {Mahadevan}}, \bibinfo
  {author} {\bibfnamefont {A.}~\bibnamefont {Monson}}, \bibinfo {author}
  {\bibfnamefont {J.~P.}\ \bibnamefont {Ninan}}, \bibinfo {author}
  {\bibfnamefont {C.}~\bibnamefont {Nitroy}}, \bibinfo {author} {\bibfnamefont
  {S.}~\bibnamefont {Osterman}}, \bibinfo {author} {\bibfnamefont {S.~B.}\
  \bibnamefont {Papp}}, \bibinfo {author} {\bibfnamefont {F.}~\bibnamefont
  {Quinlan}}, \bibinfo {author} {\bibfnamefont {L.}~\bibnamefont {Ramsey}},
  \bibinfo {author} {\bibfnamefont {P.}~\bibnamefont {Robertson}}, \bibinfo
  {author} {\bibfnamefont {A.}~\bibnamefont {Roy}}, \bibinfo {author}
  {\bibfnamefont {C.}~\bibnamefont {Schwab}}, \bibinfo {author} {\bibfnamefont
  {S.}~\bibnamefont {Sigurdsson}}, \bibinfo {author} {\bibfnamefont
  {K.}~\bibnamefont {Srinivasan}}, \bibinfo {author} {\bibfnamefont
  {G.}~\bibnamefont {Stefansson}}, \bibinfo {author} {\bibfnamefont {D.~A.}\
  \bibnamefont {Sterner}}, \bibinfo {author} {\bibfnamefont {R.}~\bibnamefont
  {Terrien}}, \bibinfo {author} {\bibfnamefont {A.}~\bibnamefont {Wolszczan}},
  \bibinfo {author} {\bibfnamefont {J.~T.}\ \bibnamefont {Wright}}, \ and\
  \bibinfo {author} {\bibfnamefont {G.}~\bibnamefont {Ycas}},\ }\bibfield
  {title} {\enquote {\bibinfo {title} {Stellar spectroscopy in the
  near-infrared with a laser frequency comb},}\ }\href {\doibase
  10.1364/OPTICA.6.000233} {\bibfield  {journal} {\bibinfo  {journal} {Optica}\
  }\textbf {\bibinfo {volume} {6}},\ \bibinfo {pages} {233--239} (\bibinfo
  {year} {2019})}\BibitemShut {NoStop}%
\bibitem [{\citenamefont {Giorgetta}\ \emph {et~al.}(2021)\citenamefont
  {Giorgetta}, \citenamefont {Peischl}, \citenamefont {Herman}, \citenamefont
  {Ycas}, \citenamefont {Coddington}, \citenamefont {Newbury},\ and\
  \citenamefont {Cossel}}]{GiorgettaLPR2021}%
  \BibitemOpen
  \bibfield  {author} {\bibinfo {author} {\bibfnamefont {F.~R.}\ \bibnamefont
  {Giorgetta}}, \bibinfo {author} {\bibfnamefont {J.}~\bibnamefont {Peischl}},
  \bibinfo {author} {\bibfnamefont {D.~I.}\ \bibnamefont {Herman}}, \bibinfo
  {author} {\bibfnamefont {G.}~\bibnamefont {Ycas}}, \bibinfo {author}
  {\bibfnamefont {I.}~\bibnamefont {Coddington}}, \bibinfo {author}
  {\bibfnamefont {N.~R.}\ \bibnamefont {Newbury}}, \ and\ \bibinfo {author}
  {\bibfnamefont {K.~C.}\ \bibnamefont {Cossel}},\ }\bibfield  {title}
  {\enquote {\bibinfo {title} {Open-path dual-comb spectroscopy for
  multispecies trace gas detection in the 4.5-5 µm spectral region},}\ }\href
  {\doibase https://doi.org/10.1002/lpor.202000583} {\bibfield  {journal}
  {\bibinfo  {journal} {Laser Photon. Rev.}\ }\textbf {\bibinfo {volume}
  {15}},\ \bibinfo {pages} {2000583} (\bibinfo {year} {2021})}\BibitemShut
  {NoStop}%
\bibitem [{\citenamefont {Bjork}\ \emph {et~al.}(2016)\citenamefont {Bjork},
  \citenamefont {Bui}, \citenamefont {Heckl}, \citenamefont {Changala},
  \citenamefont {Spaun}, \citenamefont {Heu}, \citenamefont {Follman},
  \citenamefont {Deutsch}, \citenamefont {Cole}, \citenamefont {Aspelmeyer},
  \citenamefont {Okumura},\ and\ \citenamefont {Ye}}]{BjorkS2016}%
  \BibitemOpen
  \bibfield  {author} {\bibinfo {author} {\bibfnamefont {B.~J.}\ \bibnamefont
  {Bjork}}, \bibinfo {author} {\bibfnamefont {T.~Q.}\ \bibnamefont {Bui}},
  \bibinfo {author} {\bibfnamefont {O.~H.}\ \bibnamefont {Heckl}}, \bibinfo
  {author} {\bibfnamefont {P.~B.}\ \bibnamefont {Changala}}, \bibinfo {author}
  {\bibfnamefont {B.}~\bibnamefont {Spaun}}, \bibinfo {author} {\bibfnamefont
  {P.}~\bibnamefont {Heu}}, \bibinfo {author} {\bibfnamefont {D.}~\bibnamefont
  {Follman}}, \bibinfo {author} {\bibfnamefont {C.}~\bibnamefont {Deutsch}},
  \bibinfo {author} {\bibfnamefont {G.~D.}\ \bibnamefont {Cole}}, \bibinfo
  {author} {\bibfnamefont {M.}~\bibnamefont {Aspelmeyer}}, \bibinfo {author}
  {\bibfnamefont {M.}~\bibnamefont {Okumura}}, \ and\ \bibinfo {author}
  {\bibfnamefont {J.}~\bibnamefont {Ye}},\ }\bibfield  {title} {\enquote
  {\bibinfo {title} {Direct frequency comb measurement of $ \mathrm{OD} +
  \mathrm{CO} \rightarrow \mathrm{DOCO}$ kinetics},}\ }\href {\doibase
  10.1126/science.aag1862} {\bibfield  {journal} {\bibinfo  {journal}
  {Science}\ }\textbf {\bibinfo {volume} {354}},\ \bibinfo {pages} {444--448}
  (\bibinfo {year} {2016})}\BibitemShut {NoStop}%
\bibitem [{\citenamefont {Thorpe}\ \emph {et~al.}(2008)\citenamefont {Thorpe},
  \citenamefont {Balslev-Clausen}, \citenamefont {Kirchner},\ and\
  \citenamefont {Ye}}]{ThorpeOE2008}%
  \BibitemOpen
  \bibfield  {author} {\bibinfo {author} {\bibfnamefont {M.~J.}\ \bibnamefont
  {Thorpe}}, \bibinfo {author} {\bibfnamefont {D.}~\bibnamefont
  {Balslev-Clausen}}, \bibinfo {author} {\bibfnamefont {M.~S.}\ \bibnamefont
  {Kirchner}}, \ and\ \bibinfo {author} {\bibfnamefont {J.}~\bibnamefont
  {Ye}},\ }\bibfield  {title} {\enquote {\bibinfo {title} {Cavity-enhanced
  optical frequency comb spectroscopy: application to human breath analysis},}\
  }\href {\doibase 10.1364/OE.16.002387} {\bibfield  {journal} {\bibinfo
  {journal} {Opt. Express}\ }\textbf {\bibinfo {volume} {16}},\ \bibinfo
  {pages} {2387--2397} (\bibinfo {year} {2008})}\BibitemShut {NoStop}%
\bibitem [{\citenamefont {Cao}\ \emph {et~al.}(2014)\citenamefont {Cao},
  \citenamefont {Qi}, \citenamefont {Peng}, \citenamefont {Wang},\ and\
  \citenamefont {Schmelcher}}]{CaoPRL2014}%
  \BibitemOpen
  \bibfield  {author} {\bibinfo {author} {\bibfnamefont {L.~S.}\ \bibnamefont
  {Cao}}, \bibinfo {author} {\bibfnamefont {D.~X.}\ \bibnamefont {Qi}},
  \bibinfo {author} {\bibfnamefont {R.~W.}\ \bibnamefont {Peng}}, \bibinfo
  {author} {\bibfnamefont {M.}~\bibnamefont {Wang}}, \ and\ \bibinfo {author}
  {\bibfnamefont {P.}~\bibnamefont {Schmelcher}},\ }\bibfield  {title}
  {\enquote {\bibinfo {title} {Phononic frequency combs through nonlinear
  resonances},}\ }\href {\doibase 10.1103/PhysRevLett.112.075505} {\bibfield
  {journal} {\bibinfo  {journal} {Phys. Rev. Lett.}\ }\textbf {\bibinfo
  {volume} {112}},\ \bibinfo {pages} {075505} (\bibinfo {year}
  {2014})}\BibitemShut {NoStop}%
\bibitem [{\citenamefont {Ganesan}, \citenamefont {Do},\ and\ \citenamefont
  {Seshia}(2017)}]{GanesanPRL2017}%
  \BibitemOpen
  \bibfield  {author} {\bibinfo {author} {\bibfnamefont {A.}~\bibnamefont
  {Ganesan}}, \bibinfo {author} {\bibfnamefont {C.}~\bibnamefont {Do}}, \ and\
  \bibinfo {author} {\bibfnamefont {A.}~\bibnamefont {Seshia}},\ }\bibfield
  {title} {\enquote {\bibinfo {title} {Phononic frequency comb via intrinsic
  three-wave mixing},}\ }\href {\doibase 10.1103/PhysRevLett.118.033903}
  {\bibfield  {journal} {\bibinfo  {journal} {Phys. Rev. Lett.}\ }\textbf
  {\bibinfo {volume} {118}},\ \bibinfo {pages} {033903} (\bibinfo {year}
  {2017})}\BibitemShut {NoStop}%
\bibitem [{\citenamefont {Aristov}\ and\ \citenamefont
  {Matveeva}(2016)}]{AristovPRB2016}%
  \BibitemOpen
  \bibfield  {author} {\bibinfo {author} {\bibfnamefont {D.~N.}\ \bibnamefont
  {Aristov}}\ and\ \bibinfo {author} {\bibfnamefont {P.~G.}\ \bibnamefont
  {Matveeva}},\ }\bibfield  {title} {\enquote {\bibinfo {title} {Stability of a
  skyrmion and interaction of magnons},}\ }\href {\doibase
  10.1103/PhysRevB.94.214425} {\bibfield  {journal} {\bibinfo  {journal} {Phys.
  Rev. B}\ }\textbf {\bibinfo {volume} {94}},\ \bibinfo {pages} {214425}
  (\bibinfo {year} {2016})}\BibitemShut {NoStop}%
\bibitem [{\citenamefont {Zhang}\ \emph {et~al.}(2018)\citenamefont {Zhang},
  \citenamefont {Wang}, \citenamefont {Cao}, \citenamefont {Yan},\ and\
  \citenamefont {Wang}}]{ZhangPRB2018}%
  \BibitemOpen
  \bibfield  {author} {\bibinfo {author} {\bibfnamefont {B.}~\bibnamefont
  {Zhang}}, \bibinfo {author} {\bibfnamefont {Z.}~\bibnamefont {Wang}},
  \bibinfo {author} {\bibfnamefont {Y.}~\bibnamefont {Cao}}, \bibinfo {author}
  {\bibfnamefont {P.}~\bibnamefont {Yan}}, \ and\ \bibinfo {author}
  {\bibfnamefont {X.~R.}\ \bibnamefont {Wang}},\ }\bibfield  {title} {\enquote
  {\bibinfo {title} {Eavesdropping on spin waves inside the domain-wall
  nanochannel via three-magnon processes},}\ }\href {\doibase
  10.1103/PhysRevB.97.094421} {\bibfield  {journal} {\bibinfo  {journal} {Phys.
  Rev. B}\ }\textbf {\bibinfo {volume} {97}},\ \bibinfo {pages} {094421}
  (\bibinfo {year} {2018})}\BibitemShut {NoStop}%
\bibitem [{\citenamefont {K\"orber}\ \emph {et~al.}(2020)\citenamefont
  {K\"orber}, \citenamefont {Schultheiss}, \citenamefont {Hula}, \citenamefont
  {Verba}, \citenamefont {Fassbender}, \citenamefont {K\'akay},\ and\
  \citenamefont {Schultheiss}}]{KorberPRL2020}%
  \BibitemOpen
  \bibfield  {author} {\bibinfo {author} {\bibfnamefont {L.}~\bibnamefont
  {K\"orber}}, \bibinfo {author} {\bibfnamefont {K.}~\bibnamefont
  {Schultheiss}}, \bibinfo {author} {\bibfnamefont {T.}~\bibnamefont {Hula}},
  \bibinfo {author} {\bibfnamefont {R.}~\bibnamefont {Verba}}, \bibinfo
  {author} {\bibfnamefont {J.}~\bibnamefont {Fassbender}}, \bibinfo {author}
  {\bibfnamefont {A.}~\bibnamefont {K\'akay}}, \ and\ \bibinfo {author}
  {\bibfnamefont {H.}~\bibnamefont {Schultheiss}},\ }\bibfield  {title}
  {\enquote {\bibinfo {title} {Nonlocal stimulation of three-magnon splitting
  in a magnetic vortex},}\ }\href {\doibase 10.1103/PhysRevLett.125.207203}
  {\bibfield  {journal} {\bibinfo  {journal} {Phys. Rev. Lett.}\ }\textbf
  {\bibinfo {volume} {125}},\ \bibinfo {pages} {207203} (\bibinfo {year}
  {2020})}\BibitemShut {NoStop}%
\bibitem [{\citenamefont {Kravchuk}\ \emph {et~al.}(2018)\citenamefont
  {Kravchuk}, \citenamefont {Sheka}, \citenamefont {R\"o\ss{}ler},
  \citenamefont {van~den Brink},\ and\ \citenamefont
  {Gaididei}}]{KravchukPRB2018}%
  \BibitemOpen
  \bibfield  {author} {\bibinfo {author} {\bibfnamefont {V.~P.}\ \bibnamefont
  {Kravchuk}}, \bibinfo {author} {\bibfnamefont {D.~D.}\ \bibnamefont {Sheka}},
  \bibinfo {author} {\bibfnamefont {U.~K.}\ \bibnamefont {R\"o\ss{}ler}},
  \bibinfo {author} {\bibfnamefont {J.}~\bibnamefont {van~den Brink}}, \ and\
  \bibinfo {author} {\bibfnamefont {Y.}~\bibnamefont {Gaididei}},\ }\bibfield
  {title} {\enquote {\bibinfo {title} {Spin eigenmodes of magnetic skyrmions
  and the problem of the effective skyrmion mass},}\ }\href {\doibase
  10.1103/PhysRevB.97.064403} {\bibfield  {journal} {\bibinfo  {journal} {Phys.
  Rev. B}\ }\textbf {\bibinfo {volume} {97}},\ \bibinfo {pages} {064403}
  (\bibinfo {year} {2018})}\BibitemShut {NoStop}%
\bibitem [{\citenamefont {Sun}, \citenamefont {Shi},\ and\ \citenamefont
  {Wang}(2022)}]{SunAEM2022}%
  \BibitemOpen
  \bibfield  {author} {\bibinfo {author} {\bibfnamefont {J.}~\bibnamefont
  {Sun}}, \bibinfo {author} {\bibfnamefont {S.}~\bibnamefont {Shi}}, \ and\
  \bibinfo {author} {\bibfnamefont {J.}~\bibnamefont {Wang}},\ }\bibfield
  {title} {\enquote {\bibinfo {title} {Strain modulation of magnonic frequency
  comb by magnon–skyrmion interaction in ferromagnetic materials},}\ }\href
  {\doibase https://doi.org/10.1002/adem.202101245} {\bibfield  {journal}
  {\bibinfo  {journal} {Adv. Eng. Mater.}\ }\textbf {\bibinfo {volume} {24}},\
  \bibinfo {pages} {2101245} (\bibinfo {year} {2022})}\BibitemShut {NoStop}%
\bibitem [{\citenamefont {Zhou}\ \emph {et~al.}(2021)\citenamefont {Zhou},
  \citenamefont {Wang}, \citenamefont {Nie}, \citenamefont {Xia},\ and\
  \citenamefont {Guo}}]{ZhouJMMM2021}%
  \BibitemOpen
  \bibfield  {author} {\bibinfo {author} {\bibfnamefont {Z.-W.}\ \bibnamefont
  {Zhou}}, \bibinfo {author} {\bibfnamefont {X.-G.}\ \bibnamefont {Wang}},
  \bibinfo {author} {\bibfnamefont {Y.-Z.}\ \bibnamefont {Nie}}, \bibinfo
  {author} {\bibfnamefont {Q.-L.}\ \bibnamefont {Xia}}, \ and\ \bibinfo
  {author} {\bibfnamefont {G.-H.}\ \bibnamefont {Guo}},\ }\bibfield  {title}
  {\enquote {\bibinfo {title} {Spin wave frequency comb generated through
  interaction between propagating spin wave and oscillating domain wall},}\
  }\href {\doibase https://doi.org/10.1016/j.jmmm.2021.168046} {\bibfield
  {journal} {\bibinfo  {journal} {J. Magn. Magn. Mater.}\ }\textbf {\bibinfo
  {volume} {534}},\ \bibinfo {pages} {168046} (\bibinfo {year}
  {2021})}\BibitemShut {NoStop}%
\bibitem [{\citenamefont {Xiong}(2023)}]{XiongFR2022}%
  \BibitemOpen
  \bibfield  {author} {\bibinfo {author} {\bibfnamefont {H.}~\bibnamefont
  {Xiong}},\ }\bibfield  {title} {\enquote {\bibinfo {title} {Magnonic
  frequency combs based on the resonantly enhanced magnetostrictive effect},}\
  }\href {\doibase https://doi.org/10.1016/j.fmre.2022.08.017} {\bibfield
  {journal} {\bibinfo  {journal} {Fundamental Research}\ }\textbf {\bibinfo
  {volume} {3}},\ \bibinfo {pages} {8--14} (\bibinfo {year}
  {2023})}\BibitemShut {NoStop}%
\bibitem [{\citenamefont {Hula}\ \emph {et~al.}(2022)\citenamefont {Hula},
  \citenamefont {Schultheiss}, \citenamefont {Gonçalves}, \citenamefont
  {Körber}, \citenamefont {Bejarano}, \citenamefont {Copus}, \citenamefont
  {Flacke}, \citenamefont {Liensberger}, \citenamefont {Buzdakov},
  \citenamefont {Kákay}, \citenamefont {Weiler}, \citenamefont {Camley},
  \citenamefont {Fassbender},\ and\ \citenamefont {Schultheiss}}]{HulaAPL2022}%
  \BibitemOpen
  \bibfield  {author} {\bibinfo {author} {\bibfnamefont {T.}~\bibnamefont
  {Hula}}, \bibinfo {author} {\bibfnamefont {K.}~\bibnamefont {Schultheiss}},
  \bibinfo {author} {\bibfnamefont {F.~J.~T.}\ \bibnamefont {Gonçalves}},
  \bibinfo {author} {\bibfnamefont {L.}~\bibnamefont {Körber}}, \bibinfo
  {author} {\bibfnamefont {M.}~\bibnamefont {Bejarano}}, \bibinfo {author}
  {\bibfnamefont {M.}~\bibnamefont {Copus}}, \bibinfo {author} {\bibfnamefont
  {L.}~\bibnamefont {Flacke}}, \bibinfo {author} {\bibfnamefont
  {L.}~\bibnamefont {Liensberger}}, \bibinfo {author} {\bibfnamefont
  {A.}~\bibnamefont {Buzdakov}}, \bibinfo {author} {\bibfnamefont
  {A.}~\bibnamefont {Kákay}}, \bibinfo {author} {\bibfnamefont
  {M.}~\bibnamefont {Weiler}}, \bibinfo {author} {\bibfnamefont
  {R.}~\bibnamefont {Camley}}, \bibinfo {author} {\bibfnamefont
  {J.}~\bibnamefont {Fassbender}}, \ and\ \bibinfo {author} {\bibfnamefont
  {H.}~\bibnamefont {Schultheiss}},\ }\bibfield  {title} {\enquote {\bibinfo
  {title} {Spin-wave frequency combs},}\ }\href {\doibase 10.1063/5.0090033}
  {\bibfield  {journal} {\bibinfo  {journal} {Appl. Phys. Lett.}\ }\textbf
  {\bibinfo {volume} {121}},\ \bibinfo {pages} {112404} (\bibinfo {year}
  {2022})}\BibitemShut {NoStop}%
\bibitem [{\citenamefont {Rao}\ \emph {et~al.}(2023)\citenamefont {Rao},
  \citenamefont {Yao}, \citenamefont {Wang}, \citenamefont {Zhang},
  \citenamefont {Yu},\ and\ \citenamefont {Lu}}]{RaoPRL2023}%
  \BibitemOpen
  \bibfield  {author} {\bibinfo {author} {\bibfnamefont {J.~W.}\ \bibnamefont
  {Rao}}, \bibinfo {author} {\bibfnamefont {B.}~\bibnamefont {Yao}}, \bibinfo
  {author} {\bibfnamefont {C.~Y.}\ \bibnamefont {Wang}}, \bibinfo {author}
  {\bibfnamefont {C.}~\bibnamefont {Zhang}}, \bibinfo {author} {\bibfnamefont
  {T.}~\bibnamefont {Yu}}, \ and\ \bibinfo {author} {\bibfnamefont
  {W.}~\bibnamefont {Lu}},\ }\bibfield  {title} {\enquote {\bibinfo {title}
  {Unveiling a pump-induced magnon mode via its strong interaction with walker
  modes},}\ }\href {\doibase 10.1103/PhysRevLett.130.046705} {\bibfield
  {journal} {\bibinfo  {journal} {Phys. Rev. Lett.}\ }\textbf {\bibinfo
  {volume} {130}},\ \bibinfo {pages} {046705} (\bibinfo {year}
  {2023})}\BibitemShut {NoStop}%
\bibitem [{\citenamefont {Jia}\ \emph {et~al.}(2019{\natexlab{a}})\citenamefont
  {Jia}, \citenamefont {Ma}, \citenamefont {Sch{\"a}ffer},\ and\ \citenamefont
  {Berakdar}}]{JiaNC2019}%
  \BibitemOpen
  \bibfield  {author} {\bibinfo {author} {\bibfnamefont {C.}~\bibnamefont
  {Jia}}, \bibinfo {author} {\bibfnamefont {D.}~\bibnamefont {Ma}}, \bibinfo
  {author} {\bibfnamefont {A.~F.}\ \bibnamefont {Sch{\"a}ffer}}, \ and\
  \bibinfo {author} {\bibfnamefont {J.}~\bibnamefont {Berakdar}},\ }\bibfield
  {title} {\enquote {\bibinfo {title} {Twisted magnon beams carrying orbital
  angular momentum},}\ }\href {\doibase 10.1038/s41467-019-10008-3} {\bibfield
  {journal} {\bibinfo  {journal} {Nat. Commun.}\ }\textbf {\bibinfo {volume}
  {10}},\ \bibinfo {pages} {2077} (\bibinfo {year}
  {2019}{\natexlab{a}})}\BibitemShut {NoStop}%
\bibitem [{\citenamefont {Huang}\ and\ \citenamefont
  {Wang}(2022)}]{HuangJMMM2022}%
  \BibitemOpen
  \bibfield  {author} {\bibinfo {author} {\bibfnamefont {P.}~\bibnamefont
  {Huang}}\ and\ \bibinfo {author} {\bibfnamefont {R.}~\bibnamefont {Wang}},\
  }\bibfield  {title} {\enquote {\bibinfo {title} {Excitation modes of twisted
  spin-waves in thick ferromagnetic nanodisks},}\ }\href {\doibase
  https://doi.org/10.1016/j.jmmm.2022.169762} {\bibfield  {journal} {\bibinfo
  {journal} {J. Magn. Magn. Mater.}\ }\textbf {\bibinfo {volume} {562}},\
  \bibinfo {pages} {169762} (\bibinfo {year} {2022})}\BibitemShut {NoStop}%
\bibitem [{\citenamefont {Jiang}\ \emph {et~al.}(2020)\citenamefont {Jiang},
  \citenamefont {Yuan}, \citenamefont {Li}, \citenamefont {Wang}, \citenamefont
  {Zhang}, \citenamefont {Cao},\ and\ \citenamefont {Yan}}]{JiangPRL2020}%
  \BibitemOpen
  \bibfield  {author} {\bibinfo {author} {\bibfnamefont {Y.}~\bibnamefont
  {Jiang}}, \bibinfo {author} {\bibfnamefont {H.~Y.}\ \bibnamefont {Yuan}},
  \bibinfo {author} {\bibfnamefont {Z.-X.}\ \bibnamefont {Li}}, \bibinfo
  {author} {\bibfnamefont {Z.}~\bibnamefont {Wang}}, \bibinfo {author}
  {\bibfnamefont {H.~W.}\ \bibnamefont {Zhang}}, \bibinfo {author}
  {\bibfnamefont {Y.}~\bibnamefont {Cao}}, \ and\ \bibinfo {author}
  {\bibfnamefont {P.}~\bibnamefont {Yan}},\ }\bibfield  {title} {\enquote
  {\bibinfo {title} {Twisted magnon as a magnetic tweezer},}\ }\href {\doibase
  10.1103/PhysRevLett.124.217204} {\bibfield  {journal} {\bibinfo  {journal}
  {Phys. Rev. Lett.}\ }\textbf {\bibinfo {volume} {124}},\ \bibinfo {pages}
  {217204} (\bibinfo {year} {2020})}\BibitemShut {NoStop}%
\bibitem [{\citenamefont {Jia}\ \emph {et~al.}(2021)\citenamefont {Jia},
  \citenamefont {Chen}, \citenamefont {Sch{\"a}ffer},\ and\ \citenamefont
  {Berakdar}}]{JiaNCM2021}%
  \BibitemOpen
  \bibfield  {author} {\bibinfo {author} {\bibfnamefont {C.}~\bibnamefont
  {Jia}}, \bibinfo {author} {\bibfnamefont {M.}~\bibnamefont {Chen}}, \bibinfo
  {author} {\bibfnamefont {A.~F.}\ \bibnamefont {Sch{\"a}ffer}}, \ and\
  \bibinfo {author} {\bibfnamefont {J.}~\bibnamefont {Berakdar}},\ }\bibfield
  {title} {\enquote {\bibinfo {title} {Chiral logic computing with twisted
  antiferromagnetic magnon modes},}\ }\href {\doibase
  10.1038/s41524-021-00570-0} {\bibfield  {journal} {\bibinfo  {journal} {npj
  Comput. Mater.}\ }\textbf {\bibinfo {volume} {7}},\ \bibinfo {pages} {101}
  (\bibinfo {year} {2021})}\BibitemShut {NoStop}%
\bibitem [{\citenamefont {Willner}\ \emph {et~al.}(2015)\citenamefont
  {Willner}, \citenamefont {Huang}, \citenamefont {Yan}, \citenamefont {Ren},
  \citenamefont {Ahmed}, \citenamefont {Xie}, \citenamefont {Bao},
  \citenamefont {Li}, \citenamefont {Cao}, \citenamefont {Zhao}, \citenamefont
  {Wang}, \citenamefont {Lavery}, \citenamefont {Tur}, \citenamefont
  {Ramachandran}, \citenamefont {Molisch}, \citenamefont {Ashrafi},\ and\
  \citenamefont {Ashrafi}}]{WillnerAOP2015}%
  \BibitemOpen
  \bibfield  {author} {\bibinfo {author} {\bibfnamefont {A.~E.}\ \bibnamefont
  {Willner}}, \bibinfo {author} {\bibfnamefont {H.}~\bibnamefont {Huang}},
  \bibinfo {author} {\bibfnamefont {Y.}~\bibnamefont {Yan}}, \bibinfo {author}
  {\bibfnamefont {Y.}~\bibnamefont {Ren}}, \bibinfo {author} {\bibfnamefont
  {N.}~\bibnamefont {Ahmed}}, \bibinfo {author} {\bibfnamefont
  {G.}~\bibnamefont {Xie}}, \bibinfo {author} {\bibfnamefont {C.}~\bibnamefont
  {Bao}}, \bibinfo {author} {\bibfnamefont {L.}~\bibnamefont {Li}}, \bibinfo
  {author} {\bibfnamefont {Y.}~\bibnamefont {Cao}}, \bibinfo {author}
  {\bibfnamefont {Z.}~\bibnamefont {Zhao}}, \bibinfo {author} {\bibfnamefont
  {J.}~\bibnamefont {Wang}}, \bibinfo {author} {\bibfnamefont {M.~P.~J.}\
  \bibnamefont {Lavery}}, \bibinfo {author} {\bibfnamefont {M.}~\bibnamefont
  {Tur}}, \bibinfo {author} {\bibfnamefont {S.}~\bibnamefont {Ramachandran}},
  \bibinfo {author} {\bibfnamefont {A.~F.}\ \bibnamefont {Molisch}}, \bibinfo
  {author} {\bibfnamefont {N.}~\bibnamefont {Ashrafi}}, \ and\ \bibinfo
  {author} {\bibfnamefont {S.}~\bibnamefont {Ashrafi}},\ }\bibfield  {title}
  {\enquote {\bibinfo {title} {Optical communications using orbital angular
  momentum beams},}\ }\href {\doibase 10.1364/AOP.7.000066} {\bibfield
  {journal} {\bibinfo  {journal} {Adv. Opt. Photon.}\ }\textbf {\bibinfo
  {volume} {7}},\ \bibinfo {pages} {66--106} (\bibinfo {year}
  {2015})}\BibitemShut {NoStop}%
\bibitem [{\citenamefont {Chen}\ \emph {et~al.}(2020)\citenamefont {Chen},
  \citenamefont {Schäffer}, \citenamefont {Berakdar},\ and\ \citenamefont
  {Jia}}]{ChenAPL2020}%
  \BibitemOpen
  \bibfield  {author} {\bibinfo {author} {\bibfnamefont {M.}~\bibnamefont
  {Chen}}, \bibinfo {author} {\bibfnamefont {A.~F.}\ \bibnamefont {Schäffer}},
  \bibinfo {author} {\bibfnamefont {J.}~\bibnamefont {Berakdar}}, \ and\
  \bibinfo {author} {\bibfnamefont {C.}~\bibnamefont {Jia}},\ }\bibfield
  {title} {\enquote {\bibinfo {title} {Generation, electric detection, and
  orbital-angular momentum tunneling of twisted magnons},}\ }\href {\doibase
  10.1063/5.0005764} {\bibfield  {journal} {\bibinfo  {journal} {Appl. Phys.
  Lett.}\ }\textbf {\bibinfo {volume} {116}},\ \bibinfo {pages} {172403}
  (\bibinfo {year} {2020})}\BibitemShut {NoStop}%
\bibitem [{\citenamefont {Jia}\ \emph {et~al.}(2019{\natexlab{b}})\citenamefont
  {Jia}, \citenamefont {Ma}, \citenamefont {Schäffer},\ and\ \citenamefont
  {Berakdar}}]{JiaJO2019}%
  \BibitemOpen
  \bibfield  {author} {\bibinfo {author} {\bibfnamefont {C.}~\bibnamefont
  {Jia}}, \bibinfo {author} {\bibfnamefont {D.}~\bibnamefont {Ma}}, \bibinfo
  {author} {\bibfnamefont {A.~F.}\ \bibnamefont {Schäffer}}, \ and\ \bibinfo
  {author} {\bibfnamefont {J.}~\bibnamefont {Berakdar}},\ }\bibfield  {title}
  {\enquote {\bibinfo {title} {Twisting and tweezing the spin wave: on
  vortices, skyrmions, helical waves, and the magnonic spiral phase plate},}\
  }\href {\doibase 10.1088/2040-8986/ab4f8e} {\bibfield  {journal} {\bibinfo
  {journal} {J. Opt.}\ }\textbf {\bibinfo {volume} {21}},\ \bibinfo {pages}
  {124001} (\bibinfo {year} {2019}{\natexlab{b}})}\BibitemShut {NoStop}%
\bibitem [{\citenamefont {Schultheiss}\ \emph {et~al.}(2019)\citenamefont
  {Schultheiss}, \citenamefont {Verba}, \citenamefont {Wehrmann}, \citenamefont
  {Wagner}, \citenamefont {K\"orber}, \citenamefont {Hula}, \citenamefont
  {Hache}, \citenamefont {K\'akay}, \citenamefont {Awad}, \citenamefont
  {Tiberkevich}, \citenamefont {Slavin}, \citenamefont {Fassbender},\ and\
  \citenamefont {Schultheiss}}]{SchultheissPRL2019}%
  \BibitemOpen
  \bibfield  {author} {\bibinfo {author} {\bibfnamefont {K.}~\bibnamefont
  {Schultheiss}}, \bibinfo {author} {\bibfnamefont {R.}~\bibnamefont {Verba}},
  \bibinfo {author} {\bibfnamefont {F.}~\bibnamefont {Wehrmann}}, \bibinfo
  {author} {\bibfnamefont {K.}~\bibnamefont {Wagner}}, \bibinfo {author}
  {\bibfnamefont {L.}~\bibnamefont {K\"orber}}, \bibinfo {author}
  {\bibfnamefont {T.}~\bibnamefont {Hula}}, \bibinfo {author} {\bibfnamefont
  {T.}~\bibnamefont {Hache}}, \bibinfo {author} {\bibfnamefont
  {A.}~\bibnamefont {K\'akay}}, \bibinfo {author} {\bibfnamefont {A.~A.}\
  \bibnamefont {Awad}}, \bibinfo {author} {\bibfnamefont {V.}~\bibnamefont
  {Tiberkevich}}, \bibinfo {author} {\bibfnamefont {A.~N.}\ \bibnamefont
  {Slavin}}, \bibinfo {author} {\bibfnamefont {J.}~\bibnamefont {Fassbender}},
  \ and\ \bibinfo {author} {\bibfnamefont {H.}~\bibnamefont {Schultheiss}},\
  }\bibfield  {title} {\enquote {\bibinfo {title} {Excitation of whispering
  gallery magnons in a magnetic vortex},}\ }\href {\doibase
  10.1103/PhysRevLett.122.097202} {\bibfield  {journal} {\bibinfo  {journal}
  {Phys. Rev. Lett.}\ }\textbf {\bibinfo {volume} {122}},\ \bibinfo {pages}
  {097202} (\bibinfo {year} {2019})}\BibitemShut {NoStop}%
\bibitem [{\citenamefont {Wang}\ \emph
  {et~al.}(2022{\natexlab{a}})\citenamefont {Wang}, \citenamefont {Yuan},
  \citenamefont {Cao},\ and\ \citenamefont {Yan}}]{WangPRL2022}%
  \BibitemOpen
  \bibfield  {author} {\bibinfo {author} {\bibfnamefont {Z.}~\bibnamefont
  {Wang}}, \bibinfo {author} {\bibfnamefont {H.~Y.}\ \bibnamefont {Yuan}},
  \bibinfo {author} {\bibfnamefont {Y.}~\bibnamefont {Cao}}, \ and\ \bibinfo
  {author} {\bibfnamefont {P.}~\bibnamefont {Yan}},\ }\bibfield  {title}
  {\enquote {\bibinfo {title} {Twisted magnon frequency comb and penrose
  superradiance},}\ }\href {\doibase 10.1103/PhysRevLett.129.107203} {\bibfield
   {journal} {\bibinfo  {journal} {Phys. Rev. Lett.}\ }\textbf {\bibinfo
  {volume} {129}},\ \bibinfo {pages} {107203} (\bibinfo {year}
  {2022}{\natexlab{a}})}\BibitemShut {NoStop}%
\bibitem [{\citenamefont {Penrose}(2002)}]{PenroseGRG2002}%
  \BibitemOpen
  \bibfield  {author} {\bibinfo {author} {\bibfnamefont {R.}~\bibnamefont
  {Penrose}},\ }\bibfield  {title} {\enquote {\bibinfo {title} {``golden
  oldie'': Gravitational collapse: The role of general relativity},}\ }\href
  {\doibase 10.1023/A:1016578408204} {\bibfield  {journal} {\bibinfo  {journal}
  {Gen. Relativ. Gravit.}\ }\textbf {\bibinfo {volume} {34}},\ \bibinfo {pages}
  {1141--1165} (\bibinfo {year} {2002})}\BibitemShut {NoStop}%
\bibitem [{\citenamefont {Tabuchi}\ \emph {et~al.}(2016)\citenamefont
  {Tabuchi}, \citenamefont {Ishino}, \citenamefont {Noguchi}, \citenamefont
  {Ishikawa}, \citenamefont {Yamazaki}, \citenamefont {Usami},\ and\
  \citenamefont {Nakamura}}]{Tabuchi2016}%
  \BibitemOpen
  \bibfield  {author} {\bibinfo {author} {\bibfnamefont {Y.}~\bibnamefont
  {Tabuchi}}, \bibinfo {author} {\bibfnamefont {S.}~\bibnamefont {Ishino}},
  \bibinfo {author} {\bibfnamefont {A.}~\bibnamefont {Noguchi}}, \bibinfo
  {author} {\bibfnamefont {T.}~\bibnamefont {Ishikawa}}, \bibinfo {author}
  {\bibfnamefont {R.}~\bibnamefont {Yamazaki}}, \bibinfo {author}
  {\bibfnamefont {K.}~\bibnamefont {Usami}}, \ and\ \bibinfo {author}
  {\bibfnamefont {Y.}~\bibnamefont {Nakamura}},\ }\bibfield  {title} {\enquote
  {\bibinfo {title} {Quantum magnonics: The magnon meets the superconducting
  qubit},}\ }\href {\doibase https://doi.org/10.1016/j.crhy.2016.07.009}
  {\bibfield  {journal} {\bibinfo  {journal} {Comptes Rendus Physique}\
  }\textbf {\bibinfo {volume} {17}},\ \bibinfo {pages} {729--739} (\bibinfo
  {year} {2016})}\BibitemShut {NoStop}%
\bibitem [{\citenamefont {Lachance-Quirion}\ \emph {et~al.}(2019)\citenamefont
  {Lachance-Quirion}, \citenamefont {Tabuchi}, \citenamefont {Gloppe},
  \citenamefont {Usami},\ and\ \citenamefont {Nakamura}}]{Lachance2019}%
  \BibitemOpen
  \bibfield  {author} {\bibinfo {author} {\bibfnamefont {D.}~\bibnamefont
  {Lachance-Quirion}}, \bibinfo {author} {\bibfnamefont {Y.}~\bibnamefont
  {Tabuchi}}, \bibinfo {author} {\bibfnamefont {A.}~\bibnamefont {Gloppe}},
  \bibinfo {author} {\bibfnamefont {K.}~\bibnamefont {Usami}}, \ and\ \bibinfo
  {author} {\bibfnamefont {Y.}~\bibnamefont {Nakamura}},\ }\bibfield  {title}
  {\enquote {\bibinfo {title} {Hybrid quantum systems based on magnonics},}\
  }\href {\doibase 10.7567/1882-0786/ab248d} {\bibfield  {journal} {\bibinfo
  {journal} {Appl. Phys. Express}\ }\textbf {\bibinfo {volume} {12}},\ \bibinfo
  {pages} {070101} (\bibinfo {year} {2019})}\BibitemShut {NoStop}%
\bibitem [{\citenamefont {Bunkov}(2020)}]{Bunkov2020}%
  \BibitemOpen
  \bibfield  {author} {\bibinfo {author} {\bibfnamefont {Y.~M.}\ \bibnamefont
  {Bunkov}},\ }\bibfield  {title} {\enquote {\bibinfo {title} {Quantum
  magnonics},}\ }\href {\doibase 10.1134/S1063776120070018} {\bibfield
  {journal} {\bibinfo  {journal} {J. Exp. Theor. Phys.}\ }\textbf {\bibinfo
  {volume} {131}},\ \bibinfo {pages} {18--28} (\bibinfo {year}
  {2020})}\BibitemShut {NoStop}%
\bibitem [{\citenamefont {Zhang}\ \emph
  {et~al.}(2016{\natexlab{b}})\citenamefont {Zhang}, \citenamefont {Zou},
  \citenamefont {Jiang},\ and\ \citenamefont {Tang}}]{ZhangXufeng2016}%
  \BibitemOpen
  \bibfield  {author} {\bibinfo {author} {\bibfnamefont {X.~F.}\ \bibnamefont
  {Zhang}}, \bibinfo {author} {\bibfnamefont {C.~L.}\ \bibnamefont {Zou}},
  \bibinfo {author} {\bibfnamefont {L.}~\bibnamefont {Jiang}}, \ and\ \bibinfo
  {author} {\bibfnamefont {H.~X.}\ \bibnamefont {Tang}},\ }\bibfield  {title}
  {\enquote {\bibinfo {title} {Cavity magnomechanics},}\ }\href
  {https://advances.sciencemag.org/content/2/3/e1501286} {\bibfield  {journal}
  {\bibinfo  {journal} {Sci. Adv.}\ }\textbf {\bibinfo {volume} {2}},\ \bibinfo
  {pages} {e1501286} (\bibinfo {year} {2016}{\natexlab{b}})}\BibitemShut
  {NoStop}%
\bibitem [{\citenamefont {Li}, \citenamefont {Zhu},\ and\ \citenamefont
  {Agarwal}(2018)}]{LiJie2018PRL}%
  \BibitemOpen
  \bibfield  {author} {\bibinfo {author} {\bibfnamefont {J.}~\bibnamefont
  {Li}}, \bibinfo {author} {\bibfnamefont {S.~Y.}\ \bibnamefont {Zhu}}, \ and\
  \bibinfo {author} {\bibfnamefont {G.~S.}\ \bibnamefont {Agarwal}},\
  }\bibfield  {title} {\enquote {\bibinfo {title} {Magnon-photon-phonon
  entanglement in cavity magnomechanics},}\ }\href {\doibase
  10.1103/PhysRevLett.121.203601} {\bibfield  {journal} {\bibinfo  {journal}
  {Phys. Rev. Lett.}\ }\textbf {\bibinfo {volume} {121}},\ \bibinfo {pages}
  {203601} (\bibinfo {year} {2018})}\BibitemShut {NoStop}%
\bibitem [{\citenamefont {Shen}\ \emph
  {et~al.}(2022{\natexlab{b}})\citenamefont {Shen}, \citenamefont {Xu},
  \citenamefont {Zhang}, \citenamefont {Zhang}, \citenamefont {Wang},
  \citenamefont {Chai}, \citenamefont {Zou}, \citenamefont {Guo},\ and\
  \citenamefont {Dong}}]{DongChunhua2022PRL_MagOptomechanics}%
  \BibitemOpen
  \bibfield  {author} {\bibinfo {author} {\bibfnamefont {Z.}~\bibnamefont
  {Shen}}, \bibinfo {author} {\bibfnamefont {G.-T.}\ \bibnamefont {Xu}},
  \bibinfo {author} {\bibfnamefont {M.}~\bibnamefont {Zhang}}, \bibinfo
  {author} {\bibfnamefont {Y.-L.}\ \bibnamefont {Zhang}}, \bibinfo {author}
  {\bibfnamefont {Y.}~\bibnamefont {Wang}}, \bibinfo {author} {\bibfnamefont
  {C.-Z.}\ \bibnamefont {Chai}}, \bibinfo {author} {\bibfnamefont {C.-L.}\
  \bibnamefont {Zou}}, \bibinfo {author} {\bibfnamefont {G.-C.}\ \bibnamefont
  {Guo}}, \ and\ \bibinfo {author} {\bibfnamefont {C.-H.}\ \bibnamefont
  {Dong}},\ }\bibfield  {title} {\enquote {\bibinfo {title} {Coherent coupling
  between phonons, magnons, and photons},}\ }\href {\doibase
  10.1103/PhysRevLett.129.243601} {\bibfield  {journal} {\bibinfo  {journal}
  {Phys. Rev. Lett.}\ }\textbf {\bibinfo {volume} {129}},\ \bibinfo {pages}
  {243601} (\bibinfo {year} {2022}{\natexlab{b}})}\BibitemShut {NoStop}%
\bibitem [{\citenamefont {Soykal}\ and\ \citenamefont
  {Flatt\'e}(2010{\natexlab{b}})}]{Soykal2010}%
  \BibitemOpen
  \bibfield  {author} {\bibinfo {author} {\bibfnamefont {{\"{O}}.~O.}\
  \bibnamefont {Soykal}}\ and\ \bibinfo {author} {\bibfnamefont {M.~E.}\
  \bibnamefont {Flatt\'e}},\ }\bibfield  {title} {\enquote {\bibinfo {title}
  {Strong field interactions between a nanomagnet and a photonic cavity},}\
  }\href {\doibase 10.1103/PhysRevLett.104.077202} {\bibfield  {journal}
  {\bibinfo  {journal} {Phys. Rev. Lett.}\ }\textbf {\bibinfo {volume} {104}},\
  \bibinfo {pages} {077202} (\bibinfo {year} {2010}{\natexlab{b}})}\BibitemShut
  {NoStop}%
\bibitem [{\citenamefont {Huebl}\ \emph {et~al.}(2013)\citenamefont {Huebl},
  \citenamefont {Zollitsch}, \citenamefont {Lotze}, \citenamefont {Hocke},
  \citenamefont {Greifenstein}, \citenamefont {Marx}, \citenamefont {Gross},\
  and\ \citenamefont {Goennenwein}}]{Hubel2013}%
  \BibitemOpen
  \bibfield  {author} {\bibinfo {author} {\bibfnamefont {H.}~\bibnamefont
  {Huebl}}, \bibinfo {author} {\bibfnamefont {C.~W.}\ \bibnamefont
  {Zollitsch}}, \bibinfo {author} {\bibfnamefont {J.}~\bibnamefont {Lotze}},
  \bibinfo {author} {\bibfnamefont {F.}~\bibnamefont {Hocke}}, \bibinfo
  {author} {\bibfnamefont {M.}~\bibnamefont {Greifenstein}}, \bibinfo {author}
  {\bibfnamefont {A.}~\bibnamefont {Marx}}, \bibinfo {author} {\bibfnamefont
  {R.}~\bibnamefont {Gross}}, \ and\ \bibinfo {author} {\bibfnamefont
  {S.~T.~B.}\ \bibnamefont {Goennenwein}},\ }\bibfield  {title} {\enquote
  {\bibinfo {title} {High cooperativity in coupled microwave resonator
  ferrimagnetic insulator hybrids},}\ }\href {\doibase
  10.1103/PhysRevLett.111.127003} {\bibfield  {journal} {\bibinfo  {journal}
  {Phys. Rev. Lett.}\ }\textbf {\bibinfo {volume} {111}},\ \bibinfo {pages}
  {127003} (\bibinfo {year} {2013})}\BibitemShut {NoStop}%
\bibitem [{\citenamefont {Goryachev}\ \emph {et~al.}(2014)\citenamefont
  {Goryachev}, \citenamefont {Farr}, \citenamefont {Creedon}, \citenamefont
  {Fan}, \citenamefont {Kostylev},\ and\ \citenamefont
  {Tobar}}]{Goryachev2014}%
  \BibitemOpen
  \bibfield  {author} {\bibinfo {author} {\bibfnamefont {M.}~\bibnamefont
  {Goryachev}}, \bibinfo {author} {\bibfnamefont {W.~G.}\ \bibnamefont {Farr}},
  \bibinfo {author} {\bibfnamefont {D.~L.}\ \bibnamefont {Creedon}}, \bibinfo
  {author} {\bibfnamefont {Y.}~\bibnamefont {Fan}}, \bibinfo {author}
  {\bibfnamefont {M.}~\bibnamefont {Kostylev}}, \ and\ \bibinfo {author}
  {\bibfnamefont {M.~E.}\ \bibnamefont {Tobar}},\ }\bibfield  {title} {\enquote
  {\bibinfo {title} {High-cooperativity cavity qed with magnons at microwave
  frequencies},}\ }\href {\doibase 10.1103/PhysRevApplied.2.054002} {\bibfield
  {journal} {\bibinfo  {journal} {Phys. Rev. Appl.}\ }\textbf {\bibinfo
  {volume} {2}},\ \bibinfo {pages} {054002} (\bibinfo {year}
  {2014})}\BibitemShut {NoStop}%
\bibitem [{\citenamefont {Zhang}\ \emph
  {et~al.}(2014{\natexlab{b}})\citenamefont {Zhang}, \citenamefont {Zou},
  \citenamefont {Jiang},\ and\ \citenamefont {Tang}}]{ZhangXufeng2014}%
  \BibitemOpen
  \bibfield  {author} {\bibinfo {author} {\bibfnamefont {X.}~\bibnamefont
  {Zhang}}, \bibinfo {author} {\bibfnamefont {C.~L.}\ \bibnamefont {Zou}},
  \bibinfo {author} {\bibfnamefont {L.}~\bibnamefont {Jiang}}, \ and\ \bibinfo
  {author} {\bibfnamefont {H.~X.}\ \bibnamefont {Tang}},\ }\bibfield  {title}
  {\enquote {\bibinfo {title} {Strongly coupled magnons and cavity microwave
  photons},}\ }\href {\doibase 10.1103/PhysRevLett.113.156401} {\bibfield
  {journal} {\bibinfo  {journal} {Phys. Rev. Lett.}\ }\textbf {\bibinfo
  {volume} {113}},\ \bibinfo {pages} {156401} (\bibinfo {year}
  {2014}{\natexlab{b}})}\BibitemShut {NoStop}%
\bibitem [{\citenamefont {Tabuchi}\ \emph
  {et~al.}(2014{\natexlab{b}})\citenamefont {Tabuchi}, \citenamefont {Ishino},
  \citenamefont {Ishikawa}, \citenamefont {Yamazaki}, \citenamefont {Usami},\
  and\ \citenamefont {Nakamura}}]{Tabuchi2014}%
  \BibitemOpen
  \bibfield  {author} {\bibinfo {author} {\bibfnamefont {Y.}~\bibnamefont
  {Tabuchi}}, \bibinfo {author} {\bibfnamefont {S.}~\bibnamefont {Ishino}},
  \bibinfo {author} {\bibfnamefont {T.}~\bibnamefont {Ishikawa}}, \bibinfo
  {author} {\bibfnamefont {R.}~\bibnamefont {Yamazaki}}, \bibinfo {author}
  {\bibfnamefont {K.}~\bibnamefont {Usami}}, \ and\ \bibinfo {author}
  {\bibfnamefont {Y.}~\bibnamefont {Nakamura}},\ }\bibfield  {title} {\enquote
  {\bibinfo {title} {Hybridizing ferromagnetic magnons and microwave photons in
  the quantum limit},}\ }\href {\doibase 10.1103/PhysRevLett.113.083603}
  {\bibfield  {journal} {\bibinfo  {journal} {Phys. Rev. Lett.}\ }\textbf
  {\bibinfo {volume} {113}},\ \bibinfo {pages} {083603} (\bibinfo {year}
  {2014}{\natexlab{b}})}\BibitemShut {NoStop}%
\bibitem [{\citenamefont {Mergenthaler}\ \emph {et~al.}(2017)\citenamefont
  {Mergenthaler}, \citenamefont {Liu}, \citenamefont {Le~Roy}, \citenamefont
  {Ares}, \citenamefont {Thompson}, \citenamefont {Bogani}, \citenamefont
  {Luis}, \citenamefont {Blundell}, \citenamefont {Lancaster}, \citenamefont
  {Ardavan}, \citenamefont {Briggs}, \citenamefont {Leek},\ and\ \citenamefont
  {Laird}}]{Mergenthaler2017PRL_MagMicroPhoton}%
  \BibitemOpen
  \bibfield  {author} {\bibinfo {author} {\bibfnamefont {M.}~\bibnamefont
  {Mergenthaler}}, \bibinfo {author} {\bibfnamefont {J.}~\bibnamefont {Liu}},
  \bibinfo {author} {\bibfnamefont {J.~J.}\ \bibnamefont {Le~Roy}}, \bibinfo
  {author} {\bibfnamefont {N.}~\bibnamefont {Ares}}, \bibinfo {author}
  {\bibfnamefont {A.~L.}\ \bibnamefont {Thompson}}, \bibinfo {author}
  {\bibfnamefont {L.}~\bibnamefont {Bogani}}, \bibinfo {author} {\bibfnamefont
  {F.}~\bibnamefont {Luis}}, \bibinfo {author} {\bibfnamefont {S.~J.}\
  \bibnamefont {Blundell}}, \bibinfo {author} {\bibfnamefont {T.}~\bibnamefont
  {Lancaster}}, \bibinfo {author} {\bibfnamefont {A.}~\bibnamefont {Ardavan}},
  \bibinfo {author} {\bibfnamefont {G.~A.~D.}\ \bibnamefont {Briggs}}, \bibinfo
  {author} {\bibfnamefont {P.~J.}\ \bibnamefont {Leek}}, \ and\ \bibinfo
  {author} {\bibfnamefont {E.~A.}\ \bibnamefont {Laird}},\ }\bibfield  {title}
  {\enquote {\bibinfo {title} {Strong coupling of microwave photons to
  antiferromagnetic fluctuations in an organic magnet},}\ }\href {\doibase
  10.1103/PhysRevLett.119.147701} {\bibfield  {journal} {\bibinfo  {journal}
  {Phys. Rev. Lett.}\ }\textbf {\bibinfo {volume} {119}},\ \bibinfo {pages}
  {147701} (\bibinfo {year} {2017})}\BibitemShut {NoStop}%
\bibitem [{\citenamefont {Bourhill}\ \emph {et~al.}(2016)\citenamefont
  {Bourhill}, \citenamefont {Kostylev}, \citenamefont {Goryachev},
  \citenamefont {Creedon},\ and\ \citenamefont
  {Tobar}}]{Bourhill2016PRB_MagMicroPhoton}%
  \BibitemOpen
  \bibfield  {author} {\bibinfo {author} {\bibfnamefont {J.}~\bibnamefont
  {Bourhill}}, \bibinfo {author} {\bibfnamefont {N.}~\bibnamefont {Kostylev}},
  \bibinfo {author} {\bibfnamefont {M.}~\bibnamefont {Goryachev}}, \bibinfo
  {author} {\bibfnamefont {D.~L.}\ \bibnamefont {Creedon}}, \ and\ \bibinfo
  {author} {\bibfnamefont {M.~E.}\ \bibnamefont {Tobar}},\ }\bibfield  {title}
  {\enquote {\bibinfo {title} {Ultrahigh cooperativity interactions between
  magnons and resonant photons in a yig sphere},}\ }\href {\doibase
  10.1103/PhysRevB.93.144420} {\bibfield  {journal} {\bibinfo  {journal} {Phys.
  Rev. B}\ }\textbf {\bibinfo {volume} {93}},\ \bibinfo {pages} {144420}
  (\bibinfo {year} {2016})}\BibitemShut {NoStop}%
\bibitem [{\citenamefont {Kostylev}, \citenamefont {Goryachev},\ and\
  \citenamefont {Tobar}(2016)}]{Kostylev2016APL_MagMicroPhoton_SuperStrong}%
  \BibitemOpen
  \bibfield  {author} {\bibinfo {author} {\bibfnamefont {N.}~\bibnamefont
  {Kostylev}}, \bibinfo {author} {\bibfnamefont {M.}~\bibnamefont {Goryachev}},
  \ and\ \bibinfo {author} {\bibfnamefont {M.~E.}\ \bibnamefont {Tobar}},\
  }\bibfield  {title} {\enquote {\bibinfo {title} {Superstrong coupling of a
  microwave cavity to yttrium iron garnet magnons},}\ }\href {\doibase
  10.1063/1.4941730} {\bibfield  {journal} {\bibinfo  {journal} {Appl. Phys.
  Lett.}\ }\textbf {\bibinfo {volume} {108}},\ \bibinfo {pages} {062402}
  (\bibinfo {year} {2016})}\BibitemShut {NoStop}%
\bibitem [{\citenamefont {Haigh}\ \emph
  {et~al.}(2015{\natexlab{a}})\citenamefont {Haigh}, \citenamefont
  {Langenfeld}, \citenamefont {Lambert}, \citenamefont {Baumberg},
  \citenamefont {Ramsay}, \citenamefont {Nunnenkamp},\ and\ \citenamefont
  {Ferguson}}]{Haigh2015PRA_MO}%
  \BibitemOpen
  \bibfield  {author} {\bibinfo {author} {\bibfnamefont {J.~A.}\ \bibnamefont
  {Haigh}}, \bibinfo {author} {\bibfnamefont {S.}~\bibnamefont {Langenfeld}},
  \bibinfo {author} {\bibfnamefont {N.~J.}\ \bibnamefont {Lambert}}, \bibinfo
  {author} {\bibfnamefont {J.~J.}\ \bibnamefont {Baumberg}}, \bibinfo {author}
  {\bibfnamefont {A.~J.}\ \bibnamefont {Ramsay}}, \bibinfo {author}
  {\bibfnamefont {A.}~\bibnamefont {Nunnenkamp}}, \ and\ \bibinfo {author}
  {\bibfnamefont {A.~J.}\ \bibnamefont {Ferguson}},\ }\bibfield  {title}
  {\enquote {\bibinfo {title} {Magneto-optical coupling in
  whispering-gallery-mode resonators},}\ }\href {\doibase
  10.1103/PhysRevA.92.063845} {\bibfield  {journal} {\bibinfo  {journal} {Phys.
  Rev. A}\ }\textbf {\bibinfo {volume} {92}},\ \bibinfo {pages} {063845}
  (\bibinfo {year} {2015}{\natexlab{a}})}\BibitemShut {NoStop}%
\bibitem [{\citenamefont {Osada}\ \emph {et~al.}(2016)\citenamefont {Osada},
  \citenamefont {Hisatomi}, \citenamefont {Noguchi}, \citenamefont {Tabuchi},
  \citenamefont {Yamazaki}, \citenamefont {Usami}, \citenamefont {Sadgrove},
  \citenamefont {Yalla}, \citenamefont {Nomura},\ and\ \citenamefont
  {Nakamura}}]{Osada2016PRL_MO}%
  \BibitemOpen
  \bibfield  {author} {\bibinfo {author} {\bibfnamefont {A.}~\bibnamefont
  {Osada}}, \bibinfo {author} {\bibfnamefont {R.}~\bibnamefont {Hisatomi}},
  \bibinfo {author} {\bibfnamefont {A.}~\bibnamefont {Noguchi}}, \bibinfo
  {author} {\bibfnamefont {Y.}~\bibnamefont {Tabuchi}}, \bibinfo {author}
  {\bibfnamefont {R.}~\bibnamefont {Yamazaki}}, \bibinfo {author}
  {\bibfnamefont {K.}~\bibnamefont {Usami}}, \bibinfo {author} {\bibfnamefont
  {M.}~\bibnamefont {Sadgrove}}, \bibinfo {author} {\bibfnamefont
  {R.}~\bibnamefont {Yalla}}, \bibinfo {author} {\bibfnamefont
  {M.}~\bibnamefont {Nomura}}, \ and\ \bibinfo {author} {\bibfnamefont
  {Y.}~\bibnamefont {Nakamura}},\ }\bibfield  {title} {\enquote {\bibinfo
  {title} {Cavity optomagnonics with spin-orbit coupled photons},}\ }\href
  {\doibase 10.1103/PhysRevLett.116.223601} {\bibfield  {journal} {\bibinfo
  {journal} {Phys. Rev. Lett.}\ }\textbf {\bibinfo {volume} {116}},\ \bibinfo
  {pages} {223601} (\bibinfo {year} {2016})}\BibitemShut {NoStop}%
\bibitem [{\citenamefont {Zhang}\ \emph
  {et~al.}(2016{\natexlab{c}})\citenamefont {Zhang}, \citenamefont {Zhu},
  \citenamefont {Zou},\ and\ \citenamefont {Tang}}]{ZhangXufeng2016PRL_MO}%
  \BibitemOpen
  \bibfield  {author} {\bibinfo {author} {\bibfnamefont {X.}~\bibnamefont
  {Zhang}}, \bibinfo {author} {\bibfnamefont {N.}~\bibnamefont {Zhu}}, \bibinfo
  {author} {\bibfnamefont {C.-L.}\ \bibnamefont {Zou}}, \ and\ \bibinfo
  {author} {\bibfnamefont {H.~X.}\ \bibnamefont {Tang}},\ }\bibfield  {title}
  {\enquote {\bibinfo {title} {Optomagnonic whispering gallery
  microresonators},}\ }\href {\doibase 10.1103/PhysRevLett.117.123605}
  {\bibfield  {journal} {\bibinfo  {journal} {Phys. Rev. Lett.}\ }\textbf
  {\bibinfo {volume} {117}},\ \bibinfo {pages} {123605} (\bibinfo {year}
  {2016}{\natexlab{c}})}\BibitemShut {NoStop}%
\bibitem [{\citenamefont {Viola~Kusminskiy}, \citenamefont {Tang},\ and\
  \citenamefont {Marquardt}(2016)}]{Silvia2016PRA_MO}%
  \BibitemOpen
  \bibfield  {author} {\bibinfo {author} {\bibfnamefont {S.}~\bibnamefont
  {Viola~Kusminskiy}}, \bibinfo {author} {\bibfnamefont {H.~X.}\ \bibnamefont
  {Tang}}, \ and\ \bibinfo {author} {\bibfnamefont {F.}~\bibnamefont
  {Marquardt}},\ }\bibfield  {title} {\enquote {\bibinfo {title} {Coupled
  spin-light dynamics in cavity optomagnonics},}\ }\href {\doibase
  10.1103/PhysRevA.94.033821} {\bibfield  {journal} {\bibinfo  {journal} {Phys.
  Rev. A}\ }\textbf {\bibinfo {volume} {94}},\ \bibinfo {pages} {033821}
  (\bibinfo {year} {2016})}\BibitemShut {NoStop}%
\bibitem [{\citenamefont {Haigh}\ \emph {et~al.}(2016)\citenamefont {Haigh},
  \citenamefont {Nunnenkamp}, \citenamefont {Ramsay},\ and\ \citenamefont
  {Ferguson}}]{Haigh2016PRL_MO}%
  \BibitemOpen
  \bibfield  {author} {\bibinfo {author} {\bibfnamefont {J.~A.}\ \bibnamefont
  {Haigh}}, \bibinfo {author} {\bibfnamefont {A.}~\bibnamefont {Nunnenkamp}},
  \bibinfo {author} {\bibfnamefont {A.~J.}\ \bibnamefont {Ramsay}}, \ and\
  \bibinfo {author} {\bibfnamefont {A.~J.}\ \bibnamefont {Ferguson}},\
  }\bibfield  {title} {\enquote {\bibinfo {title} {Triple-resonant brillouin
  light scattering in magneto-optical cavities},}\ }\href {\doibase
  10.1103/PhysRevLett.117.133602} {\bibfield  {journal} {\bibinfo  {journal}
  {Phys. Rev. Lett.}\ }\textbf {\bibinfo {volume} {117}},\ \bibinfo {pages}
  {133602} (\bibinfo {year} {2016})}\BibitemShut {NoStop}%
\bibitem [{\citenamefont {Osada}\ \emph {et~al.}(2018)\citenamefont {Osada},
  \citenamefont {Gloppe}, \citenamefont {Hisatomi}, \citenamefont {Noguchi},
  \citenamefont {Yamazaki}, \citenamefont {Nomura}, \citenamefont {Nakamura},\
  and\ \citenamefont {Usami}}]{Osada2018PRL_MO}%
  \BibitemOpen
  \bibfield  {author} {\bibinfo {author} {\bibfnamefont {A.}~\bibnamefont
  {Osada}}, \bibinfo {author} {\bibfnamefont {A.}~\bibnamefont {Gloppe}},
  \bibinfo {author} {\bibfnamefont {R.}~\bibnamefont {Hisatomi}}, \bibinfo
  {author} {\bibfnamefont {A.}~\bibnamefont {Noguchi}}, \bibinfo {author}
  {\bibfnamefont {R.}~\bibnamefont {Yamazaki}}, \bibinfo {author}
  {\bibfnamefont {M.}~\bibnamefont {Nomura}}, \bibinfo {author} {\bibfnamefont
  {Y.}~\bibnamefont {Nakamura}}, \ and\ \bibinfo {author} {\bibfnamefont
  {K.}~\bibnamefont {Usami}},\ }\bibfield  {title} {\enquote {\bibinfo {title}
  {Brillouin light scattering by magnetic quasivortices in cavity
  optomagnonics},}\ }\href {\doibase 10.1103/PhysRevLett.120.133602} {\bibfield
   {journal} {\bibinfo  {journal} {Phys. Rev. Lett.}\ }\textbf {\bibinfo
  {volume} {120}},\ \bibinfo {pages} {133602} (\bibinfo {year}
  {2018})}\BibitemShut {NoStop}%
\bibitem [{\citenamefont {Parvini}, \citenamefont {Bittencourt},\ and\
  \citenamefont {Kusminskiy}(2020)}]{Parvini2019PRR_MO_antiferro}%
  \BibitemOpen
  \bibfield  {author} {\bibinfo {author} {\bibfnamefont {T.~S.}\ \bibnamefont
  {Parvini}}, \bibinfo {author} {\bibfnamefont {V.~A. S.~V.}\ \bibnamefont
  {Bittencourt}}, \ and\ \bibinfo {author} {\bibfnamefont {S.~V.}\ \bibnamefont
  {Kusminskiy}},\ }\bibfield  {title} {\enquote {\bibinfo {title}
  {Antiferromagnetic cavity optomagnonics},}\ }\href {\doibase
  10.1103/PhysRevResearch.2.022027} {\bibfield  {journal} {\bibinfo  {journal}
  {Phys. Rev. Res.}\ }\textbf {\bibinfo {volume} {2}},\ \bibinfo {pages}
  {022027} (\bibinfo {year} {2020})}\BibitemShut {NoStop}%
\bibitem [{\citenamefont {Tabuchi}\ \emph
  {et~al.}(2015{\natexlab{b}})\citenamefont {Tabuchi}, \citenamefont {Ishino},
  \citenamefont {Noguchi}, \citenamefont {Ishikawa}, \citenamefont {Yamazaki},
  \citenamefont {Usami},\ and\ \citenamefont {Nakamura}}]{Tabuchi2015}%
  \BibitemOpen
  \bibfield  {author} {\bibinfo {author} {\bibfnamefont {Y.}~\bibnamefont
  {Tabuchi}}, \bibinfo {author} {\bibfnamefont {S.}~\bibnamefont {Ishino}},
  \bibinfo {author} {\bibfnamefont {A.}~\bibnamefont {Noguchi}}, \bibinfo
  {author} {\bibfnamefont {T.}~\bibnamefont {Ishikawa}}, \bibinfo {author}
  {\bibfnamefont {R.}~\bibnamefont {Yamazaki}}, \bibinfo {author}
  {\bibfnamefont {K.}~\bibnamefont {Usami}}, \ and\ \bibinfo {author}
  {\bibfnamefont {Y.}~\bibnamefont {Nakamura}},\ }\bibfield  {title} {\enquote
  {\bibinfo {title} {Coherent coupling between a ferromagnetic magnon and a
  superconducting qubit},}\ }\href {\doibase 10.1126/science.aaa3693}
  {\bibfield  {journal} {\bibinfo  {journal} {Science}\ }\textbf {\bibinfo
  {volume} {349}},\ \bibinfo {pages} {405} (\bibinfo {year}
  {2015}{\natexlab{b}})}\BibitemShut {NoStop}%
\bibitem [{\citenamefont {Lachance-Quirion}\ \emph
  {et~al.}(2017{\natexlab{b}})\citenamefont {Lachance-Quirion}, \citenamefont
  {Tabuchi}, \citenamefont {Ishino}, \citenamefont {Noguchi}, \citenamefont
  {Ishikawa}, \citenamefont {Yamazaki},\ and\ \citenamefont
  {Nakamura}}]{Lachance-Quirione2017NumberState}%
  \BibitemOpen
  \bibfield  {author} {\bibinfo {author} {\bibfnamefont {D.}~\bibnamefont
  {Lachance-Quirion}}, \bibinfo {author} {\bibfnamefont {Y.}~\bibnamefont
  {Tabuchi}}, \bibinfo {author} {\bibfnamefont {S.}~\bibnamefont {Ishino}},
  \bibinfo {author} {\bibfnamefont {A.}~\bibnamefont {Noguchi}}, \bibinfo
  {author} {\bibfnamefont {T.}~\bibnamefont {Ishikawa}}, \bibinfo {author}
  {\bibfnamefont {R.}~\bibnamefont {Yamazaki}}, \ and\ \bibinfo {author}
  {\bibfnamefont {Y.}~\bibnamefont {Nakamura}},\ }\bibfield  {title} {\enquote
  {\bibinfo {title} {Resolving quanta of collective spin excitations in a
  millimeter-sized ferromagnet},}\ }\href {\doibase 10.1126/sciadv.1603150}
  {\bibfield  {journal} {\bibinfo  {journal} {Sci. Adv.}\ }\textbf {\bibinfo
  {volume} {3}},\ \bibinfo {pages} {e1603150} (\bibinfo {year}
  {2017}{\natexlab{b}})}\BibitemShut {NoStop}%
\bibitem [{\citenamefont {Lachance-Quirion}\ \emph
  {et~al.}(2020{\natexlab{b}})\citenamefont {Lachance-Quirion}, \citenamefont
  {Wolski}, \citenamefont {Tabuchi}, \citenamefont {Kono}, \citenamefont
  {Usami},\ and\ \citenamefont {Nakamura}}]{Lachance-Quirion2020SingleMagnon}%
  \BibitemOpen
  \bibfield  {author} {\bibinfo {author} {\bibfnamefont {D.}~\bibnamefont
  {Lachance-Quirion}}, \bibinfo {author} {\bibfnamefont {S.~P.}\ \bibnamefont
  {Wolski}}, \bibinfo {author} {\bibfnamefont {Y.}~\bibnamefont {Tabuchi}},
  \bibinfo {author} {\bibfnamefont {S.}~\bibnamefont {Kono}}, \bibinfo {author}
  {\bibfnamefont {K.}~\bibnamefont {Usami}}, \ and\ \bibinfo {author}
  {\bibfnamefont {Y.}~\bibnamefont {Nakamura}},\ }\bibfield  {title} {\enquote
  {\bibinfo {title} {Entanglement-based single-shot detection of a single
  magnon with a superconducting qubit},}\ }\href {\doibase
  10.1126/science.aaz9236} {\bibfield  {journal} {\bibinfo  {journal}
  {Science}\ }\textbf {\bibinfo {volume} {367}},\ \bibinfo {pages} {425}
  (\bibinfo {year} {2020}{\natexlab{b}})}\BibitemShut {NoStop}%
\bibitem [{\citenamefont {Yuan}\ \emph
  {et~al.}(2020{\natexlab{a}})\citenamefont {Yuan}, \citenamefont {Zheng},
  \citenamefont {Ficek}, \citenamefont {He},\ and\ \citenamefont
  {Yung}}]{Yuan2020PRB}%
  \BibitemOpen
  \bibfield  {author} {\bibinfo {author} {\bibfnamefont {H.~Y.}\ \bibnamefont
  {Yuan}}, \bibinfo {author} {\bibfnamefont {S.}~\bibnamefont {Zheng}},
  \bibinfo {author} {\bibfnamefont {Z.}~\bibnamefont {Ficek}}, \bibinfo
  {author} {\bibfnamefont {Q.~Y.}\ \bibnamefont {He}}, \ and\ \bibinfo {author}
  {\bibfnamefont {M.~H.}\ \bibnamefont {Yung}},\ }\bibfield  {title} {\enquote
  {\bibinfo {title} {Enhancement of magnon-magnon entanglement inside a
  cavity},}\ }\href {\doibase 10.1103/PhysRevB.101.014419} {\bibfield
  {journal} {\bibinfo  {journal} {Phys. Rev. B}\ }\textbf {\bibinfo {volume}
  {101}},\ \bibinfo {pages} {014419} (\bibinfo {year}
  {2020}{\natexlab{a}})}\BibitemShut {NoStop}%
\bibitem [{\citenamefont {Zheng}\ \emph {et~al.}(2020)\citenamefont {Zheng},
  \citenamefont {Sun}, \citenamefont {Yuan}, \citenamefont {Ficek},
  \citenamefont {Gong},\ and\ \citenamefont {He}}]{ZhengShasha2020SciChina}%
  \BibitemOpen
  \bibfield  {author} {\bibinfo {author} {\bibfnamefont {S.-S.}\ \bibnamefont
  {Zheng}}, \bibinfo {author} {\bibfnamefont {F.-X.}\ \bibnamefont {Sun}},
  \bibinfo {author} {\bibfnamefont {H.-Y.}\ \bibnamefont {Yuan}}, \bibinfo
  {author} {\bibfnamefont {Z.}~\bibnamefont {Ficek}}, \bibinfo {author}
  {\bibfnamefont {Q.-H.}\ \bibnamefont {Gong}}, \ and\ \bibinfo {author}
  {\bibfnamefont {Q.-Y.}\ \bibnamefont {He}},\ }\bibfield  {title} {\enquote
  {\bibinfo {title} {Enhanced entanglement and asymmetric epr steering between
  magnons},}\ }\href {\doibase 10.1007/s11433-020-1587-5} {\bibfield  {journal}
  {\bibinfo  {journal} {Sci. China Phys. Mech. Astron.}\ }\textbf {\bibinfo
  {volume} {64}},\ \bibinfo {pages} {210311} (\bibinfo {year}
  {2020})}\BibitemShut {NoStop}%
\bibitem [{\citenamefont {Bai}\ \emph {et~al.}(2015)\citenamefont {Bai},
  \citenamefont {Harder}, \citenamefont {Chen}, \citenamefont {Fan},
  \citenamefont {Xiao},\ and\ \citenamefont {Hu}}]{BaiLihui2015SpinPumping}%
  \BibitemOpen
  \bibfield  {author} {\bibinfo {author} {\bibfnamefont {L.}~\bibnamefont
  {Bai}}, \bibinfo {author} {\bibfnamefont {M.}~\bibnamefont {Harder}},
  \bibinfo {author} {\bibfnamefont {Y.~P.}\ \bibnamefont {Chen}}, \bibinfo
  {author} {\bibfnamefont {X.}~\bibnamefont {Fan}}, \bibinfo {author}
  {\bibfnamefont {J.~Q.}\ \bibnamefont {Xiao}}, \ and\ \bibinfo {author}
  {\bibfnamefont {C.~M.}\ \bibnamefont {Hu}},\ }\bibfield  {title} {\enquote
  {\bibinfo {title} {Spin pumping in electrodynamically coupled magnon-photon
  systems},}\ }\href {\doibase 10.1103/PhysRevLett.114.227201} {\bibfield
  {journal} {\bibinfo  {journal} {Phys. Rev. Lett.}\ }\textbf {\bibinfo
  {volume} {114}},\ \bibinfo {pages} {227201} (\bibinfo {year}
  {2015})}\BibitemShut {NoStop}%
\bibitem [{\citenamefont {Chumak}\ \emph
  {et~al.}(2015{\natexlab{b}})\citenamefont {Chumak}, \citenamefont
  {Vasyuchka}, \citenamefont {Serga},\ and\ \citenamefont
  {Hillebrands}}]{Chumak2015}%
  \BibitemOpen
  \bibfield  {author} {\bibinfo {author} {\bibfnamefont {A.~V.}\ \bibnamefont
  {Chumak}}, \bibinfo {author} {\bibfnamefont {V.~I.}\ \bibnamefont
  {Vasyuchka}}, \bibinfo {author} {\bibfnamefont {A.~A.}\ \bibnamefont
  {Serga}}, \ and\ \bibinfo {author} {\bibfnamefont {B.}~\bibnamefont
  {Hillebrands}},\ }\bibfield  {title} {\enquote {\bibinfo {title} {Magnon
  spintronics},}\ }\href {\doibase 10.1038/nphys3347} {\bibfield  {journal}
  {\bibinfo  {journal} {Nat. Phys.}\ }\textbf {\bibinfo {volume} {11}},\
  \bibinfo {pages} {453} (\bibinfo {year} {2015}{\natexlab{b}})}\BibitemShut
  {NoStop}%
\bibitem [{\citenamefont {Zhang}\ \emph {et~al.}(2015)\citenamefont {Zhang},
  \citenamefont {Zou}, \citenamefont {Zhu}, \citenamefont {Marquardt},
  \citenamefont {Jiang},\ and\ \citenamefont
  {Tang}}]{ZhangXufeng2015NC_Memory}%
  \BibitemOpen
  \bibfield  {author} {\bibinfo {author} {\bibfnamefont {X.}~\bibnamefont
  {Zhang}}, \bibinfo {author} {\bibfnamefont {C.-L.}\ \bibnamefont {Zou}},
  \bibinfo {author} {\bibfnamefont {N.}~\bibnamefont {Zhu}}, \bibinfo {author}
  {\bibfnamefont {F.}~\bibnamefont {Marquardt}}, \bibinfo {author}
  {\bibfnamefont {L.}~\bibnamefont {Jiang}}, \ and\ \bibinfo {author}
  {\bibfnamefont {H.~X.}\ \bibnamefont {Tang}},\ }\href {\doibase
  10.1038/ncomms9914} {\bibfield  {journal} {\bibinfo  {journal} {Nat.
  Commun.}\ }\textbf {\bibinfo {volume} {6}},\ \bibinfo {pages} {8914}
  (\bibinfo {year} {2015})}\BibitemShut {NoStop}%
\bibitem [{\citenamefont {Kosub}\ \emph {et~al.}(2017)\citenamefont {Kosub},
  \citenamefont {Kopte}, \citenamefont {H{\"u}hne}, \citenamefont {Appel},
  \citenamefont {Shields}, \citenamefont {Maletinsky}, \citenamefont
  {H{\"u}bner}, \citenamefont {Liedke}, \citenamefont {Fassbender},
  \citenamefont {Schmidt},\ and\ \citenamefont {Makarov}}]{Kosub2017NC_Memory}%
  \BibitemOpen
  \bibfield  {author} {\bibinfo {author} {\bibfnamefont {T.}~\bibnamefont
  {Kosub}}, \bibinfo {author} {\bibfnamefont {M.}~\bibnamefont {Kopte}},
  \bibinfo {author} {\bibfnamefont {R.}~\bibnamefont {H{\"u}hne}}, \bibinfo
  {author} {\bibfnamefont {P.}~\bibnamefont {Appel}}, \bibinfo {author}
  {\bibfnamefont {B.}~\bibnamefont {Shields}}, \bibinfo {author} {\bibfnamefont
  {P.}~\bibnamefont {Maletinsky}}, \bibinfo {author} {\bibfnamefont
  {R.}~\bibnamefont {H{\"u}bner}}, \bibinfo {author} {\bibfnamefont {M.~O.}\
  \bibnamefont {Liedke}}, \bibinfo {author} {\bibfnamefont {J.}~\bibnamefont
  {Fassbender}}, \bibinfo {author} {\bibfnamefont {O.~G.}\ \bibnamefont
  {Schmidt}}, \ and\ \bibinfo {author} {\bibfnamefont {D.}~\bibnamefont
  {Makarov}},\ }\bibfield  {title} {\enquote {\bibinfo {title} {Purely
  antiferromagnetic magnetoelectric random access memory},}\ }\href {\doibase
  10.1038/ncomms13985} {\bibfield  {journal} {\bibinfo  {journal} {Nat.
  Commun.}\ }\textbf {\bibinfo {volume} {8}},\ \bibinfo {pages} {13985}
  (\bibinfo {year} {2017})}\BibitemShut {NoStop}%
\bibitem [{\citenamefont {Hisatomi}\ \emph {et~al.}(2016)\citenamefont
  {Hisatomi}, \citenamefont {Osada}, \citenamefont {Tabuchi}, \citenamefont
  {Ishikawa}, \citenamefont {Noguchi}, \citenamefont {Yamazaki}, \citenamefont
  {Usami},\ and\ \citenamefont {Nakamura}}]{Hisatomi2016PRB_MOTrans}%
  \BibitemOpen
  \bibfield  {author} {\bibinfo {author} {\bibfnamefont {R.}~\bibnamefont
  {Hisatomi}}, \bibinfo {author} {\bibfnamefont {A.}~\bibnamefont {Osada}},
  \bibinfo {author} {\bibfnamefont {Y.}~\bibnamefont {Tabuchi}}, \bibinfo
  {author} {\bibfnamefont {T.}~\bibnamefont {Ishikawa}}, \bibinfo {author}
  {\bibfnamefont {A.}~\bibnamefont {Noguchi}}, \bibinfo {author} {\bibfnamefont
  {R.}~\bibnamefont {Yamazaki}}, \bibinfo {author} {\bibfnamefont
  {K.}~\bibnamefont {Usami}}, \ and\ \bibinfo {author} {\bibfnamefont
  {Y.}~\bibnamefont {Nakamura}},\ }\bibfield  {title} {\enquote {\bibinfo
  {title} {Bidirectional conversion between microwave and light via
  ferromagnetic magnons},}\ }\href {\doibase 10.1103/PhysRevB.93.174427}
  {\bibfield  {journal} {\bibinfo  {journal} {Phys. Rev. B}\ }\textbf {\bibinfo
  {volume} {93}},\ \bibinfo {pages} {174427} (\bibinfo {year}
  {2016})}\BibitemShut {NoStop}%
\bibitem [{\citenamefont {Chai}\ \emph {et~al.}(2022)\citenamefont {Chai},
  \citenamefont {Shen}, \citenamefont {Zhang}, \citenamefont {Zhao},
  \citenamefont {Guo}, \citenamefont {Zou},\ and\ \citenamefont
  {Dong}}]{ChaiChengzhe2022PhotonRes_MOTrans}%
  \BibitemOpen
  \bibfield  {author} {\bibinfo {author} {\bibfnamefont {C.-Z.}\ \bibnamefont
  {Chai}}, \bibinfo {author} {\bibfnamefont {Z.}~\bibnamefont {Shen}}, \bibinfo
  {author} {\bibfnamefont {Y.-L.}\ \bibnamefont {Zhang}}, \bibinfo {author}
  {\bibfnamefont {H.-Q.}\ \bibnamefont {Zhao}}, \bibinfo {author}
  {\bibfnamefont {G.-C.}\ \bibnamefont {Guo}}, \bibinfo {author} {\bibfnamefont
  {C.-L.}\ \bibnamefont {Zou}}, \ and\ \bibinfo {author} {\bibfnamefont
  {C.-H.}\ \bibnamefont {Dong}},\ }\bibfield  {title} {\enquote {\bibinfo
  {title} {Single-sideband microwave-to-optical conversion in high-q
  ferrimagnetic microspheres},}\ }\href {\doibase 10.1364/PRJ.446226}
  {\bibfield  {journal} {\bibinfo  {journal} {Photon. Res.}\ }\textbf {\bibinfo
  {volume} {10}},\ \bibinfo {pages} {820} (\bibinfo {year} {2022})}\BibitemShut
  {NoStop}%
\bibitem [{\citenamefont {Lounis}\ and\ \citenamefont
  {Orrit}(2005)}]{Lounis2005RPP_SinglePhoton}%
  \BibitemOpen
  \bibfield  {author} {\bibinfo {author} {\bibfnamefont {B.}~\bibnamefont
  {Lounis}}\ and\ \bibinfo {author} {\bibfnamefont {M.}~\bibnamefont {Orrit}},\
  }\bibfield  {title} {\enquote {\bibinfo {title} {Single-photon sources},}\
  }\href {\doibase 10.1088/0034-4885/68/5/R04} {\bibfield  {journal} {\bibinfo
  {journal} {Rep. Prog. Phys.}\ }\textbf {\bibinfo {volume} {68}},\ \bibinfo
  {pages} {1129} (\bibinfo {year} {2005})}\BibitemShut {NoStop}%
\bibitem [{\citenamefont {Milburn}\ and\ \citenamefont
  {Basiri-Esfahani}(2015)}]{Milburn2015Review_SinglePhoton}%
  \BibitemOpen
  \bibfield  {author} {\bibinfo {author} {\bibfnamefont {G.~J.}\ \bibnamefont
  {Milburn}}\ and\ \bibinfo {author} {\bibfnamefont {S.}~\bibnamefont
  {Basiri-Esfahani}},\ }\bibfield  {title} {\enquote {\bibinfo {title} {Quantum
  optics with one or two photons},}\ }\href {\doibase 10.1098/rspa.2015.0208}
  {\bibfield  {journal} {\bibinfo  {journal} {Proc. Math. Phys. Eng. Sci.}\
  }\textbf {\bibinfo {volume} {471}},\ \bibinfo {pages} {20150208} (\bibinfo
  {year} {2015})}\BibitemShut {NoStop}%
\bibitem [{\citenamefont {Kounalakis}, \citenamefont {Bauer},\ and\
  \citenamefont {Blanter}(2022)}]{Kounalakis2022PRL_CatState}%
  \BibitemOpen
  \bibfield  {author} {\bibinfo {author} {\bibfnamefont {M.}~\bibnamefont
  {Kounalakis}}, \bibinfo {author} {\bibfnamefont {G.~E.~W.}\ \bibnamefont
  {Bauer}}, \ and\ \bibinfo {author} {\bibfnamefont {Y.~M.}\ \bibnamefont
  {Blanter}},\ }\bibfield  {title} {\enquote {\bibinfo {title} {Analog quantum
  control of magnonic cat states on a chip by a superconducting qubit},}\
  }\href {\doibase 10.1103/PhysRevLett.129.037205} {\bibfield  {journal}
  {\bibinfo  {journal} {Phys. Rev. Lett.}\ }\textbf {\bibinfo {volume} {129}},\
  \bibinfo {pages} {037205} (\bibinfo {year} {2022})}\BibitemShut {NoStop}%
\bibitem [{\citenamefont {Galland}\ \emph {et~al.}(2014)\citenamefont
  {Galland}, \citenamefont {Sangouard}, \citenamefont {Piro}, \citenamefont
  {Gisin},\ and\ \citenamefont {Kippenberg}}]{Galland2014PRL_PhononFock}%
  \BibitemOpen
  \bibfield  {author} {\bibinfo {author} {\bibfnamefont {C.}~\bibnamefont
  {Galland}}, \bibinfo {author} {\bibfnamefont {N.}~\bibnamefont {Sangouard}},
  \bibinfo {author} {\bibfnamefont {N.}~\bibnamefont {Piro}}, \bibinfo {author}
  {\bibfnamefont {N.}~\bibnamefont {Gisin}}, \ and\ \bibinfo {author}
  {\bibfnamefont {T.~J.}\ \bibnamefont {Kippenberg}},\ }\bibfield  {title}
  {\enquote {\bibinfo {title} {Heralded single-phonon preparation, storage, and
  readout in cavity optomechanics},}\ }\href {\doibase
  10.1103/PhysRevLett.112.143602} {\bibfield  {journal} {\bibinfo  {journal}
  {Phys. Rev. Lett.}\ }\textbf {\bibinfo {volume} {112}},\ \bibinfo {pages}
  {143602} (\bibinfo {year} {2014})}\BibitemShut {NoStop}%
\bibitem [{\citenamefont {Riedinger}\ \emph {et~al.}(2016)\citenamefont
  {Riedinger}, \citenamefont {Hong}, \citenamefont {Norte}, \citenamefont
  {Slater}, \citenamefont {Shang}, \citenamefont {Krause}, \citenamefont
  {Anant}, \citenamefont {Aspelmeyer},\ and\ \citenamefont
  {Gr{\"o}blacher}}]{Riedinger2016Nature_PhononFock}%
  \BibitemOpen
  \bibfield  {author} {\bibinfo {author} {\bibfnamefont {R.}~\bibnamefont
  {Riedinger}}, \bibinfo {author} {\bibfnamefont {S.}~\bibnamefont {Hong}},
  \bibinfo {author} {\bibfnamefont {R.~A.}\ \bibnamefont {Norte}}, \bibinfo
  {author} {\bibfnamefont {J.~A.}\ \bibnamefont {Slater}}, \bibinfo {author}
  {\bibfnamefont {J.}~\bibnamefont {Shang}}, \bibinfo {author} {\bibfnamefont
  {A.~G.}\ \bibnamefont {Krause}}, \bibinfo {author} {\bibfnamefont
  {V.}~\bibnamefont {Anant}}, \bibinfo {author} {\bibfnamefont
  {M.}~\bibnamefont {Aspelmeyer}}, \ and\ \bibinfo {author} {\bibfnamefont
  {S.}~\bibnamefont {Gr{\"o}blacher}},\ }\bibfield  {title} {\enquote {\bibinfo
  {title} {Non-classical correlations between single photons and phonons from a
  mechanical oscillator},}\ }\href {\doibase 10.1038/nature16536} {\bibfield
  {journal} {\bibinfo  {journal} {Nature}\ }\textbf {\bibinfo {volume} {530}},\
  \bibinfo {pages} {313} (\bibinfo {year} {2016})}\BibitemShut {NoStop}%
\bibitem [{\citenamefont {Hong}\ \emph {et~al.}(2017)\citenamefont {Hong},
  \citenamefont {Riedinger}, \citenamefont {Marinkovi{\'c}}, \citenamefont
  {Wallucks}, \citenamefont {Hofer}, \citenamefont {Norte}, \citenamefont
  {Aspelmeyer},\ and\ \citenamefont
  {Gr{\"o}blacher}}]{Hong2017Science_PhononFock}%
  \BibitemOpen
  \bibfield  {author} {\bibinfo {author} {\bibfnamefont {S.}~\bibnamefont
  {Hong}}, \bibinfo {author} {\bibfnamefont {R.}~\bibnamefont {Riedinger}},
  \bibinfo {author} {\bibfnamefont {I.}~\bibnamefont {Marinkovi{\'c}}},
  \bibinfo {author} {\bibfnamefont {A.}~\bibnamefont {Wallucks}}, \bibinfo
  {author} {\bibfnamefont {S.~G.}\ \bibnamefont {Hofer}}, \bibinfo {author}
  {\bibfnamefont {R.~A.}\ \bibnamefont {Norte}}, \bibinfo {author}
  {\bibfnamefont {M.}~\bibnamefont {Aspelmeyer}}, \ and\ \bibinfo {author}
  {\bibfnamefont {S.}~\bibnamefont {Gr{\"o}blacher}},\ }\bibfield  {title}
  {\enquote {\bibinfo {title} {Hanbury brown and twiss interferometry of single
  phonons from an optomechanical resonator},}\ }\href {\doibase
  10.1126/science.aan7939} {\bibfield  {journal} {\bibinfo  {journal}
  {Science}\ }\textbf {\bibinfo {volume} {358}},\ \bibinfo {pages} {203}
  (\bibinfo {year} {2017})}\BibitemShut {NoStop}%
\bibitem [{\citenamefont {Bittencourt}, \citenamefont {Feulner},\ and\
  \citenamefont {Kusminskiy}(2019)}]{Bittencourt2019PRA_MagnonFockState}%
  \BibitemOpen
  \bibfield  {author} {\bibinfo {author} {\bibfnamefont {V.~A. S.~V.}\
  \bibnamefont {Bittencourt}}, \bibinfo {author} {\bibfnamefont
  {V.}~\bibnamefont {Feulner}}, \ and\ \bibinfo {author} {\bibfnamefont
  {S.~V.}\ \bibnamefont {Kusminskiy}},\ }\bibfield  {title} {\enquote {\bibinfo
  {title} {Magnon heralding in cavity optomagnonics},}\ }\href {\doibase
  10.1103/PhysRevA.100.013810} {\bibfield  {journal} {\bibinfo  {journal}
  {Phys. Rev. A}\ }\textbf {\bibinfo {volume} {100}},\ \bibinfo {pages}
  {013810} (\bibinfo {year} {2019})}\BibitemShut {NoStop}%
\bibitem [{\citenamefont {Liu}, \citenamefont {Xiong},\ and\ \citenamefont
  {Wu}(2019)}]{LiuZengXing2019PRB_MagnonBlockade}%
  \BibitemOpen
  \bibfield  {author} {\bibinfo {author} {\bibfnamefont {Z.-X.}\ \bibnamefont
  {Liu}}, \bibinfo {author} {\bibfnamefont {H.}~\bibnamefont {Xiong}}, \ and\
  \bibinfo {author} {\bibfnamefont {Y.}~\bibnamefont {Wu}},\ }\bibfield
  {title} {\enquote {\bibinfo {title} {Magnon blockade in a hybrid
  ferromagnet-superconductor quantum system},}\ }\href {\doibase
  10.1103/PhysRevB.100.134421} {\bibfield  {journal} {\bibinfo  {journal}
  {Phys. Rev. B}\ }\textbf {\bibinfo {volume} {100}},\ \bibinfo {pages}
  {134421} (\bibinfo {year} {2019})}\BibitemShut {NoStop}%
\bibitem [{\citenamefont {Xie}, \citenamefont {Ma},\ and\ \citenamefont
  {Li}(2020)}]{XieJiKun2020PRA_MagnonBlockade}%
  \BibitemOpen
  \bibfield  {author} {\bibinfo {author} {\bibfnamefont {J.-k.}\ \bibnamefont
  {Xie}}, \bibinfo {author} {\bibfnamefont {S.-l.}\ \bibnamefont {Ma}}, \ and\
  \bibinfo {author} {\bibfnamefont {F.-l.}\ \bibnamefont {Li}},\ }\bibfield
  {title} {\enquote {\bibinfo {title} {Quantum-interference-enhanced magnon
  blockade in an yttrium-iron-garnet sphere coupled to superconducting
  circuits},}\ }\href {\doibase 10.1103/PhysRevA.101.042331} {\bibfield
  {journal} {\bibinfo  {journal} {Phys. Rev. A}\ }\textbf {\bibinfo {volume}
  {101}},\ \bibinfo {pages} {042331} (\bibinfo {year} {2020})}\BibitemShut
  {NoStop}%
\bibitem [{\citenamefont {jun Xu}\ \emph {et~al.}(2021)\citenamefont {jun Xu},
  \citenamefont {le~Yang}, \citenamefont {Lin},\ and\ \citenamefont
  {Song}}]{XuYejun2021JOSAB_MagnonBlockade}%
  \BibitemOpen
  \bibfield  {author} {\bibinfo {author} {\bibfnamefont {Y.}~\bibnamefont {jun
  Xu}}, \bibinfo {author} {\bibfnamefont {T.}~\bibnamefont {le~Yang}}, \bibinfo
  {author} {\bibfnamefont {L.}~\bibnamefont {Lin}}, \ and\ \bibinfo {author}
  {\bibfnamefont {J.}~\bibnamefont {Song}},\ }\bibfield  {title} {\enquote
  {\bibinfo {title} {Conventional and unconventional magnon blockades in a
  qubit-magnon hybrid quantum system},}\ }\href {\doibase 10.1364/JOSAB.414600}
  {\bibfield  {journal} {\bibinfo  {journal} {J. Opt. Soc. Am. B}\ }\textbf
  {\bibinfo {volume} {38}},\ \bibinfo {pages} {876} (\bibinfo {year}
  {2021})}\BibitemShut {NoStop}%
\bibitem [{\citenamefont {Yuan}\ and\ \citenamefont
  {Duine}(2020)}]{YuanHY2020PRB_Antibunching}%
  \BibitemOpen
  \bibfield  {author} {\bibinfo {author} {\bibfnamefont {H.~Y.}\ \bibnamefont
  {Yuan}}\ and\ \bibinfo {author} {\bibfnamefont {R.~A.}\ \bibnamefont
  {Duine}},\ }\bibfield  {title} {\enquote {\bibinfo {title} {Magnon
  antibunching in a nanomagnet},}\ }\href {\doibase
  10.1103/PhysRevB.102.100402} {\bibfield  {journal} {\bibinfo  {journal}
  {Phys. Rev. B}\ }\textbf {\bibinfo {volume} {102}},\ \bibinfo {pages}
  {100402(R)} (\bibinfo {year} {2020})}\BibitemShut {NoStop}%
\bibitem [{\citenamefont {Paul}(1982)}]{Paul1982RMP_PhotonAntibunching}%
  \BibitemOpen
  \bibfield  {author} {\bibinfo {author} {\bibfnamefont {H.}~\bibnamefont
  {Paul}},\ }\bibfield  {title} {\enquote {\bibinfo {title} {Photon
  antibunching},}\ }\href {\doibase 10.1103/RevModPhys.54.1061} {\bibfield
  {journal} {\bibinfo  {journal} {Rev. Mod. Phys.}\ }\textbf {\bibinfo {volume}
  {54}},\ \bibinfo {pages} {1061} (\bibinfo {year} {1982})}\BibitemShut
  {NoStop}%
\bibitem [{\citenamefont {Liew}\ and\ \citenamefont
  {Savona}(2010)}]{Liew2010PRL_SinglePhoton}%
  \BibitemOpen
  \bibfield  {author} {\bibinfo {author} {\bibfnamefont {T.~C.~H.}\
  \bibnamefont {Liew}}\ and\ \bibinfo {author} {\bibfnamefont {V.}~\bibnamefont
  {Savona}},\ }\bibfield  {title} {\enquote {\bibinfo {title} {Single photons
  from coupled quantum modes},}\ }\href {\doibase
  10.1103/PhysRevLett.104.183601} {\bibfield  {journal} {\bibinfo  {journal}
  {Phys. Rev. Lett.}\ }\textbf {\bibinfo {volume} {104}},\ \bibinfo {pages}
  {183601} (\bibinfo {year} {2010})}\BibitemShut {NoStop}%
\bibitem [{\citenamefont {Rabl}(2011)}]{Rabl2011PRL_PhotonBlockade}%
  \BibitemOpen
  \bibfield  {author} {\bibinfo {author} {\bibfnamefont {P.}~\bibnamefont
  {Rabl}},\ }\bibfield  {title} {\enquote {\bibinfo {title} {Photon blockade
  effect in optomechanical systems},}\ }\href {\doibase
  10.1103/PhysRevLett.107.063601} {\bibfield  {journal} {\bibinfo  {journal}
  {Phys. Rev. Lett.}\ }\textbf {\bibinfo {volume} {107}},\ \bibinfo {pages}
  {063601} (\bibinfo {year} {2011})}\BibitemShut {NoStop}%
\bibitem [{\citenamefont {Hoffman}\ \emph {et~al.}(2011)\citenamefont
  {Hoffman}, \citenamefont {Srinivasan}, \citenamefont {Schmidt}, \citenamefont
  {Spietz}, \citenamefont {Aumentado}, \citenamefont {T\"ureci},\ and\
  \citenamefont {Houck}}]{Hoffman2011PRL_PhotonBlockade}%
  \BibitemOpen
  \bibfield  {author} {\bibinfo {author} {\bibfnamefont {A.~J.}\ \bibnamefont
  {Hoffman}}, \bibinfo {author} {\bibfnamefont {S.~J.}\ \bibnamefont
  {Srinivasan}}, \bibinfo {author} {\bibfnamefont {S.}~\bibnamefont {Schmidt}},
  \bibinfo {author} {\bibfnamefont {L.}~\bibnamefont {Spietz}}, \bibinfo
  {author} {\bibfnamefont {J.}~\bibnamefont {Aumentado}}, \bibinfo {author}
  {\bibfnamefont {H.~E.}\ \bibnamefont {T\"ureci}}, \ and\ \bibinfo {author}
  {\bibfnamefont {A.~A.}\ \bibnamefont {Houck}},\ }\bibfield  {title} {\enquote
  {\bibinfo {title} {Dispersive photon blockade in a superconducting
  circuit},}\ }\href {\doibase 10.1103/PhysRevLett.107.053602} {\bibfield
  {journal} {\bibinfo  {journal} {Phys. Rev. Lett.}\ }\textbf {\bibinfo
  {volume} {107}},\ \bibinfo {pages} {053602} (\bibinfo {year}
  {2011})}\BibitemShut {NoStop}%
\bibitem [{\citenamefont {Schr{\"o}dinger}(1935)}]{Schrodinger1935_CatState}%
  \BibitemOpen
  \bibfield  {author} {\bibinfo {author} {\bibfnamefont {E.}~\bibnamefont
  {Schr{\"o}dinger}},\ }\bibfield  {title} {\enquote {\bibinfo {title} {Die
  gegenw{\"a}rtige situation in der quantenmechanik},}\ }\href
  {https://doi.org/10.1007/BF01491914} {\bibfield  {journal} {\bibinfo
  {journal} {Naturwissenschaften}\ }\textbf {\bibinfo {volume} {23}},\ \bibinfo
  {pages} {823} (\bibinfo {year} {1935})}\BibitemShut {NoStop}%
\bibitem [{\citenamefont {Wineland}(2013)}]{Wineland2013RMP_NobelLecture}%
  \BibitemOpen
  \bibfield  {author} {\bibinfo {author} {\bibfnamefont {D.~J.}\ \bibnamefont
  {Wineland}},\ }\bibfield  {title} {\enquote {\bibinfo {title} {Nobel lecture:
  Superposition, entanglement, and raising schr{\"o}dinger’s cat},}\ }\href
  {\doibase 10.1103/RevModPhys.85.1103} {\bibfield  {journal} {\bibinfo
  {journal} {Rev. Mod. Phys.}\ }\textbf {\bibinfo {volume} {85}},\ \bibinfo
  {pages} {1103} (\bibinfo {year} {2013})}\BibitemShut {NoStop}%
\bibitem [{\citenamefont {Arndt}\ and\ \citenamefont
  {Hornberger}(2014)}]{Arndt2014NatPhys_FoundamentalTest}%
  \BibitemOpen
  \bibfield  {author} {\bibinfo {author} {\bibfnamefont {M.}~\bibnamefont
  {Arndt}}\ and\ \bibinfo {author} {\bibfnamefont {K.}~\bibnamefont
  {Hornberger}},\ }\bibfield  {title} {\enquote {\bibinfo {title} {Testing the
  limits of quantum mechanical superpositions},}\ }\href {\doibase
  10.1038/nphys2863} {\bibfield  {journal} {\bibinfo  {journal} {Nat. Phys.}\
  }\textbf {\bibinfo {volume} {10}},\ \bibinfo {pages} {271} (\bibinfo {year}
  {2014})}\BibitemShut {NoStop}%
\bibitem [{\citenamefont {Cochrane}, \citenamefont {Milburn},\ and\
  \citenamefont {Munro}(1999)}]{Cochrane1999PRA_Cat_BosonicCode}%
  \BibitemOpen
  \bibfield  {author} {\bibinfo {author} {\bibfnamefont {P.~T.}\ \bibnamefont
  {Cochrane}}, \bibinfo {author} {\bibfnamefont {G.~J.}\ \bibnamefont
  {Milburn}}, \ and\ \bibinfo {author} {\bibfnamefont {W.~J.}\ \bibnamefont
  {Munro}},\ }\bibfield  {title} {\enquote {\bibinfo {title} {Macroscopically
  distinct quantum-superposition states as a bosonic code for amplitude
  damping},}\ }\href {\doibase 10.1103/PhysRevA.59.2631} {\bibfield  {journal}
  {\bibinfo  {journal} {Phys. Rev. A}\ }\textbf {\bibinfo {volume} {59}},\
  \bibinfo {pages} {2631} (\bibinfo {year} {1999})}\BibitemShut {NoStop}%
\bibitem [{\citenamefont {Lund}, \citenamefont {Ralph},\ and\ \citenamefont
  {Haselgrove}(2008)}]{Lund2008PRL_Cat_FaultTolerantComput}%
  \BibitemOpen
  \bibfield  {author} {\bibinfo {author} {\bibfnamefont {A.~P.}\ \bibnamefont
  {Lund}}, \bibinfo {author} {\bibfnamefont {T.~C.}\ \bibnamefont {Ralph}}, \
  and\ \bibinfo {author} {\bibfnamefont {H.~L.}\ \bibnamefont {Haselgrove}},\
  }\bibfield  {title} {\enquote {\bibinfo {title} {Fault-tolerant linear
  optical quantum computing with small-amplitude coherent states},}\ }\href
  {\doibase 10.1103/PhysRevLett.100.030503} {\bibfield  {journal} {\bibinfo
  {journal} {Phys. Rev. Lett.}\ }\textbf {\bibinfo {volume} {100}},\ \bibinfo
  {pages} {030503} (\bibinfo {year} {2008})}\BibitemShut {NoStop}%
\bibitem [{\citenamefont {Neergaard-Nielsen}\ \emph {et~al.}(2010)\citenamefont
  {Neergaard-Nielsen}, \citenamefont {Takeuchi}, \citenamefont {Wakui},
  \citenamefont {Takahashi}, \citenamefont {Hayasaka}, \citenamefont
  {Takeoka},\ and\ \citenamefont
  {Sasaki}}]{Neergaard-Nielsen2010PRL_Cat_QuanCompu}%
  \BibitemOpen
  \bibfield  {author} {\bibinfo {author} {\bibfnamefont {J.~S.}\ \bibnamefont
  {Neergaard-Nielsen}}, \bibinfo {author} {\bibfnamefont {M.}~\bibnamefont
  {Takeuchi}}, \bibinfo {author} {\bibfnamefont {K.}~\bibnamefont {Wakui}},
  \bibinfo {author} {\bibfnamefont {H.}~\bibnamefont {Takahashi}}, \bibinfo
  {author} {\bibfnamefont {K.}~\bibnamefont {Hayasaka}}, \bibinfo {author}
  {\bibfnamefont {M.}~\bibnamefont {Takeoka}}, \ and\ \bibinfo {author}
  {\bibfnamefont {M.}~\bibnamefont {Sasaki}},\ }\bibfield  {title} {\enquote
  {\bibinfo {title} {Optical continuous-variable qubit},}\ }\href {\doibase
  10.1103/PhysRevLett.105.053602} {\bibfield  {journal} {\bibinfo  {journal}
  {Phys. Rev. Lett.}\ }\textbf {\bibinfo {volume} {105}},\ \bibinfo {pages}
  {053602} (\bibinfo {year} {2010})}\BibitemShut {NoStop}%
\bibitem [{\citenamefont {Ofek}\ \emph {et~al.}(2016)\citenamefont {Ofek},
  \citenamefont {Petrenko}, \citenamefont {Heeres}, \citenamefont {Reinhold},
  \citenamefont {Leghtas}, \citenamefont {Vlastakis}, \citenamefont {Liu},
  \citenamefont {Frunzio}, \citenamefont {Girvin}, \citenamefont {Jiang},
  \citenamefont {Mirrahimi}, \citenamefont {Devoret},\ and\ \citenamefont
  {Schoelkopf}}]{Ofek2016Nature_Cat_ErrorCorrelation}%
  \BibitemOpen
  \bibfield  {author} {\bibinfo {author} {\bibfnamefont {N.}~\bibnamefont
  {Ofek}}, \bibinfo {author} {\bibfnamefont {A.}~\bibnamefont {Petrenko}},
  \bibinfo {author} {\bibfnamefont {R.}~\bibnamefont {Heeres}}, \bibinfo
  {author} {\bibfnamefont {P.}~\bibnamefont {Reinhold}}, \bibinfo {author}
  {\bibfnamefont {Z.}~\bibnamefont {Leghtas}}, \bibinfo {author} {\bibfnamefont
  {B.}~\bibnamefont {Vlastakis}}, \bibinfo {author} {\bibfnamefont
  {Y.}~\bibnamefont {Liu}}, \bibinfo {author} {\bibfnamefont {L.}~\bibnamefont
  {Frunzio}}, \bibinfo {author} {\bibfnamefont {S.~M.}\ \bibnamefont {Girvin}},
  \bibinfo {author} {\bibfnamefont {L.}~\bibnamefont {Jiang}}, \bibinfo
  {author} {\bibfnamefont {M.}~\bibnamefont {Mirrahimi}}, \bibinfo {author}
  {\bibfnamefont {M.~H.}\ \bibnamefont {Devoret}}, \ and\ \bibinfo {author}
  {\bibfnamefont {R.~J.}\ \bibnamefont {Schoelkopf}},\ }\bibfield  {title}
  {\enquote {\bibinfo {title} {Extending the lifetime of a quantum bit with
  error correction in superconducting circuits},}\ }\href {\doibase
  10.1038/nature18949} {\bibfield  {journal} {\bibinfo  {journal} {Nature}\
  }\textbf {\bibinfo {volume} {536}},\ \bibinfo {pages} {441} (\bibinfo {year}
  {2016})}\BibitemShut {NoStop}%
\bibitem [{\citenamefont {Karimipour}, \citenamefont {Bahraminasab},\ and\
  \citenamefont {Bagherinezhad}(2002)}]{Karimipour2002PRA_Cat_QSS}%
  \BibitemOpen
  \bibfield  {author} {\bibinfo {author} {\bibfnamefont {V.}~\bibnamefont
  {Karimipour}}, \bibinfo {author} {\bibfnamefont {A.}~\bibnamefont
  {Bahraminasab}}, \ and\ \bibinfo {author} {\bibfnamefont {S.}~\bibnamefont
  {Bagherinezhad}},\ }\bibfield  {title} {\enquote {\bibinfo {title}
  {Entanglement swapping of generalized cat states and secret sharing},}\
  }\href {\doibase 10.1103/PhysRevA.65.042320} {\bibfield  {journal} {\bibinfo
  {journal} {Phys. Rev. A}\ }\textbf {\bibinfo {volume} {65}},\ \bibinfo
  {pages} {042320} (\bibinfo {year} {2002})}\BibitemShut {NoStop}%
\bibitem [{\citenamefont {Brask}\ \emph {et~al.}(2010)\citenamefont {Brask},
  \citenamefont {Rigas}, \citenamefont {Polzik}, \citenamefont {Andersen},\
  and\ \citenamefont {S\o{}rensen}}]{Brask2010PRL_Cat_QuanRepeater}%
  \BibitemOpen
  \bibfield  {author} {\bibinfo {author} {\bibfnamefont {J.~B.}\ \bibnamefont
  {Brask}}, \bibinfo {author} {\bibfnamefont {I.}~\bibnamefont {Rigas}},
  \bibinfo {author} {\bibfnamefont {E.~S.}\ \bibnamefont {Polzik}}, \bibinfo
  {author} {\bibfnamefont {U.~L.}\ \bibnamefont {Andersen}}, \ and\ \bibinfo
  {author} {\bibfnamefont {A.~S.}\ \bibnamefont {S\o{}rensen}},\ }\bibfield
  {title} {\enquote {\bibinfo {title} {Hybrid long-distance entanglement
  distribution protocol},}\ }\href {\doibase 10.1103/PhysRevLett.105.160501}
  {\bibfield  {journal} {\bibinfo  {journal} {Phys. Rev. Lett.}\ }\textbf
  {\bibinfo {volume} {105}},\ \bibinfo {pages} {160501} (\bibinfo {year}
  {2010})}\BibitemShut {NoStop}%
\bibitem [{\citenamefont {Sangouard}\ \emph {et~al.}(2010)\citenamefont
  {Sangouard}, \citenamefont {Simon}, \citenamefont {Gisin}, \citenamefont
  {Laurat}, \citenamefont {Tualle-Brouri},\ and\ \citenamefont
  {Grangier}}]{Sangouard2010JOSAB_Cat_QuanRepeater}%
  \BibitemOpen
  \bibfield  {author} {\bibinfo {author} {\bibfnamefont {N.}~\bibnamefont
  {Sangouard}}, \bibinfo {author} {\bibfnamefont {C.}~\bibnamefont {Simon}},
  \bibinfo {author} {\bibfnamefont {N.}~\bibnamefont {Gisin}}, \bibinfo
  {author} {\bibfnamefont {J.}~\bibnamefont {Laurat}}, \bibinfo {author}
  {\bibfnamefont {R.}~\bibnamefont {Tualle-Brouri}}, \ and\ \bibinfo {author}
  {\bibfnamefont {P.}~\bibnamefont {Grangier}},\ }\bibfield  {title} {\enquote
  {\bibinfo {title} {Quantum repeaters with entangled coherent states},}\
  }\href {\doibase 10.1364/JOSAB.27.00A137} {\bibfield  {journal} {\bibinfo
  {journal} {J. Opt. Soc. Am. B}\ }\textbf {\bibinfo {volume} {27}},\ \bibinfo
  {pages} {A137--A145} (\bibinfo {year} {2010})}\BibitemShut {NoStop}%
\bibitem [{\citenamefont {Phien}\ and\ \citenamefont
  {An}(2008)}]{Phien2008PLA_Cat_Teleportation}%
  \BibitemOpen
  \bibfield  {author} {\bibinfo {author} {\bibfnamefont {H.~N.}\ \bibnamefont
  {Phien}}\ and\ \bibinfo {author} {\bibfnamefont {N.~B.}\ \bibnamefont {An}},\
  }\bibfield  {title} {\enquote {\bibinfo {title} {Quantum teleportation of an
  arbitrary two-mode coherent state using only linear optics elements},}\
  }\href {\doibase https://doi.org/10.1016/j.physleta.2007.12.069} {\bibfield
  {journal} {\bibinfo  {journal} {Phys. Lett. A}\ }\textbf {\bibinfo {volume}
  {372}},\ \bibinfo {pages} {2825} (\bibinfo {year} {2008})}\BibitemShut
  {NoStop}%
\bibitem [{\citenamefont {Joo}, \citenamefont {Munro},\ and\ \citenamefont
  {Spiller}(2011)}]{Joo2011PRL_Cat_Metrology}%
  \BibitemOpen
  \bibfield  {author} {\bibinfo {author} {\bibfnamefont {J.}~\bibnamefont
  {Joo}}, \bibinfo {author} {\bibfnamefont {W.~J.}\ \bibnamefont {Munro}}, \
  and\ \bibinfo {author} {\bibfnamefont {T.~P.}\ \bibnamefont {Spiller}},\
  }\bibfield  {title} {\enquote {\bibinfo {title} {Quantum metrology with
  entangled coherent states},}\ }\href {\doibase
  10.1103/PhysRevLett.107.083601} {\bibfield  {journal} {\bibinfo  {journal}
  {Phys. Rev. Lett.}\ }\textbf {\bibinfo {volume} {107}},\ \bibinfo {pages}
  {083601} (\bibinfo {year} {2011})}\BibitemShut {NoStop}%
\bibitem [{\citenamefont {Yurke}\ and\ \citenamefont
  {Stoler}(1986)}]{Yurke1986PRL_CatState}%
  \BibitemOpen
  \bibfield  {author} {\bibinfo {author} {\bibfnamefont {B.}~\bibnamefont
  {Yurke}}\ and\ \bibinfo {author} {\bibfnamefont {D.}~\bibnamefont {Stoler}},\
  }\bibfield  {title} {\enquote {\bibinfo {title} {Generating quantum
  mechanical superpositions of macroscopically distinguishable states via
  amplitude dispersion},}\ }\href {\doibase 10.1103/PhysRevLett.57.13}
  {\bibfield  {journal} {\bibinfo  {journal} {Phys. Rev. Lett.}\ }\textbf
  {\bibinfo {volume} {57}},\ \bibinfo {pages} {13--16} (\bibinfo {year}
  {1986})}\BibitemShut {NoStop}%
\bibitem [{\citenamefont {Glancy}\ and\ \citenamefont
  {de~Vasconcelos}(2008)}]{Glancy2008JOSAB_Cat_Kerr}%
  \BibitemOpen
  \bibfield  {author} {\bibinfo {author} {\bibfnamefont {S.}~\bibnamefont
  {Glancy}}\ and\ \bibinfo {author} {\bibfnamefont {H.~M.}\ \bibnamefont
  {de~Vasconcelos}},\ }\bibfield  {title} {\enquote {\bibinfo {title} {Methods
  for producing optical coherent state superpositions},}\ }\href {\doibase
  10.1364/JOSAB.25.000712} {\bibfield  {journal} {\bibinfo  {journal} {J. Opt.
  Soc. Am. B}\ }\textbf {\bibinfo {volume} {25}},\ \bibinfo {pages} {712--733}
  (\bibinfo {year} {2008})}\BibitemShut {NoStop}%
\bibitem [{\citenamefont {Song}, \citenamefont {Caves},\ and\ \citenamefont
  {Yurke}(1990)}]{SongShang1990PRA_Cat_BackactionEvading}%
  \BibitemOpen
  \bibfield  {author} {\bibinfo {author} {\bibfnamefont {S.}~\bibnamefont
  {Song}}, \bibinfo {author} {\bibfnamefont {C.~M.}\ \bibnamefont {Caves}}, \
  and\ \bibinfo {author} {\bibfnamefont {B.}~\bibnamefont {Yurke}},\ }\bibfield
   {title} {\enquote {\bibinfo {title} {Generation of superpositions of
  classically distinguishable quantum states from optical back-action
  evasion},}\ }\href {\doibase 10.1103/PhysRevA.41.5261} {\bibfield  {journal}
  {\bibinfo  {journal} {Phys. Rev. A}\ }\textbf {\bibinfo {volume} {41}},\
  \bibinfo {pages} {5261} (\bibinfo {year} {1990})}\BibitemShut {NoStop}%
\bibitem [{\citenamefont {Yurke}, \citenamefont {Schleich},\ and\ \citenamefont
  {Walls}(1990)}]{Yurke1990PRA_Cat_BackactionEvading}%
  \BibitemOpen
  \bibfield  {author} {\bibinfo {author} {\bibfnamefont {B.}~\bibnamefont
  {Yurke}}, \bibinfo {author} {\bibfnamefont {W.}~\bibnamefont {Schleich}}, \
  and\ \bibinfo {author} {\bibfnamefont {D.~F.}\ \bibnamefont {Walls}},\
  }\bibfield  {title} {\enquote {\bibinfo {title} {Quantum superpositions
  generated by quantum nondemolition measurements},}\ }\href {\doibase
  10.1103/PhysRevA.42.1703} {\bibfield  {journal} {\bibinfo  {journal} {Phys.
  Rev. A}\ }\textbf {\bibinfo {volume} {42}},\ \bibinfo {pages} {1703}
  (\bibinfo {year} {1990})}\BibitemShut {NoStop}%
\bibitem [{\citenamefont {Dakna}\ \emph
  {et~al.}(1997{\natexlab{a}})\citenamefont {Dakna}, \citenamefont {Anhut},
  \citenamefont {Opatrn\'y}, \citenamefont {Kn\"oll},\ and\ \citenamefont
  {Welsch}}]{Dakna1997PRA_CatTheory_PhoSubstraction}%
  \BibitemOpen
  \bibfield  {author} {\bibinfo {author} {\bibfnamefont {M.}~\bibnamefont
  {Dakna}}, \bibinfo {author} {\bibfnamefont {T.}~\bibnamefont {Anhut}},
  \bibinfo {author} {\bibfnamefont {T.}~\bibnamefont {Opatrn\'y}}, \bibinfo
  {author} {\bibfnamefont {L.}~\bibnamefont {Kn\"oll}}, \ and\ \bibinfo
  {author} {\bibfnamefont {D.-G.}\ \bibnamefont {Welsch}},\ }\bibfield  {title}
  {\enquote {\bibinfo {title} {Generating schr\"odinger-cat-like states by
  means of conditional measurements on a beam splitter},}\ }\href {\doibase
  10.1103/PhysRevA.55.3184} {\bibfield  {journal} {\bibinfo  {journal} {Phys.
  Rev. A}\ }\textbf {\bibinfo {volume} {55}},\ \bibinfo {pages} {3184}
  (\bibinfo {year} {1997}{\natexlab{a}})}\BibitemShut {NoStop}%
\bibitem [{\citenamefont {Lund}\ \emph {et~al.}(2004)\citenamefont {Lund},
  \citenamefont {Jeong}, \citenamefont {Ralph},\ and\ \citenamefont
  {Kim}}]{Lund2004PRA_CatTheory_PhoSubstraction}%
  \BibitemOpen
  \bibfield  {author} {\bibinfo {author} {\bibfnamefont {A.~P.}\ \bibnamefont
  {Lund}}, \bibinfo {author} {\bibfnamefont {H.}~\bibnamefont {Jeong}},
  \bibinfo {author} {\bibfnamefont {T.~C.}\ \bibnamefont {Ralph}}, \ and\
  \bibinfo {author} {\bibfnamefont {M.~S.}\ \bibnamefont {Kim}},\ }\bibfield
  {title} {\enquote {\bibinfo {title} {Conditional production of superpositions
  of coherent states with inefficient photon detection},}\ }\href {\doibase
  10.1103/PhysRevA.70.020101} {\bibfield  {journal} {\bibinfo  {journal} {Phys.
  Rev. A}\ }\textbf {\bibinfo {volume} {70}},\ \bibinfo {pages} {020101}
  (\bibinfo {year} {2004})}\BibitemShut {NoStop}%
\bibitem [{\citenamefont {Ourjoumtsev}\ \emph
  {et~al.}(2006{\natexlab{a}})\citenamefont {Ourjoumtsev}, \citenamefont
  {Tualle-Brouri}, \citenamefont {Laurat},\ and\ \citenamefont
  {Grangier}}]{AlexeiOurjoumtsev2006Science_CatExperiment}%
  \BibitemOpen
  \bibfield  {author} {\bibinfo {author} {\bibfnamefont {A.}~\bibnamefont
  {Ourjoumtsev}}, \bibinfo {author} {\bibfnamefont {R.}~\bibnamefont
  {Tualle-Brouri}}, \bibinfo {author} {\bibfnamefont {J.}~\bibnamefont
  {Laurat}}, \ and\ \bibinfo {author} {\bibfnamefont {P.}~\bibnamefont
  {Grangier}},\ }\bibfield  {title} {\enquote {\bibinfo {title} {Generating
  optical schr{\"o}dinger kittens for quantum information processing},}\ }\href
  {\doibase 10.1126/science.1122858} {\bibfield  {journal} {\bibinfo  {journal}
  {Science}\ }\textbf {\bibinfo {volume} {312}},\ \bibinfo {pages} {83--86}
  (\bibinfo {year} {2006}{\natexlab{a}})}\BibitemShut {NoStop}%
\bibitem [{\citenamefont {Neergaard-Nielsen}\ \emph {et~al.}(2006)\citenamefont
  {Neergaard-Nielsen}, \citenamefont {Nielsen}, \citenamefont {Hettich},
  \citenamefont {M\o{}lmer},\ and\ \citenamefont
  {Polzik}}]{Neergaard-Nielsen2006PRL_CatExperiment}%
  \BibitemOpen
  \bibfield  {author} {\bibinfo {author} {\bibfnamefont {J.~S.}\ \bibnamefont
  {Neergaard-Nielsen}}, \bibinfo {author} {\bibfnamefont {B.~M.}\ \bibnamefont
  {Nielsen}}, \bibinfo {author} {\bibfnamefont {C.}~\bibnamefont {Hettich}},
  \bibinfo {author} {\bibfnamefont {K.}~\bibnamefont {M\o{}lmer}}, \ and\
  \bibinfo {author} {\bibfnamefont {E.~S.}\ \bibnamefont {Polzik}},\ }\bibfield
   {title} {\enquote {\bibinfo {title} {Generation of a superposition of odd
  photon number states for quantum information networks},}\ }\href {\doibase
  10.1103/PhysRevLett.97.083604} {\bibfield  {journal} {\bibinfo  {journal}
  {Phys. Rev. Lett.}\ }\textbf {\bibinfo {volume} {97}},\ \bibinfo {pages}
  {083604} (\bibinfo {year} {2006})}\BibitemShut {NoStop}%
\bibitem [{\citenamefont {Wakui}\ \emph {et~al.}(2007)\citenamefont {Wakui},
  \citenamefont {Takahashi}, \citenamefont {Furusawa},\ and\ \citenamefont
  {Sasaki}}]{Wakui2007OE_CatExperiment}%
  \BibitemOpen
  \bibfield  {author} {\bibinfo {author} {\bibfnamefont {K.}~\bibnamefont
  {Wakui}}, \bibinfo {author} {\bibfnamefont {H.}~\bibnamefont {Takahashi}},
  \bibinfo {author} {\bibfnamefont {A.}~\bibnamefont {Furusawa}}, \ and\
  \bibinfo {author} {\bibfnamefont {M.}~\bibnamefont {Sasaki}},\ }\bibfield
  {title} {\enquote {\bibinfo {title} {Photon subtracted squeezed states
  generated with periodically poled ktiopo4},}\ }\href {\doibase
  10.1364/OE.15.003568} {\bibfield  {journal} {\bibinfo  {journal} {Opt.
  Express}\ }\textbf {\bibinfo {volume} {15}},\ \bibinfo {pages} {3568}
  (\bibinfo {year} {2007})}\BibitemShut {NoStop}%
\bibitem [{\citenamefont {Takahashi}\ \emph {et~al.}(2008)\citenamefont
  {Takahashi}, \citenamefont {Wakui}, \citenamefont {Suzuki}, \citenamefont
  {Takeoka}, \citenamefont {Hayasaka}, \citenamefont {Furusawa},\ and\
  \citenamefont {Sasaki}}]{Takahashi2008PRL_CatExperiment}%
  \BibitemOpen
  \bibfield  {author} {\bibinfo {author} {\bibfnamefont {H.}~\bibnamefont
  {Takahashi}}, \bibinfo {author} {\bibfnamefont {K.}~\bibnamefont {Wakui}},
  \bibinfo {author} {\bibfnamefont {S.}~\bibnamefont {Suzuki}}, \bibinfo
  {author} {\bibfnamefont {M.}~\bibnamefont {Takeoka}}, \bibinfo {author}
  {\bibfnamefont {K.}~\bibnamefont {Hayasaka}}, \bibinfo {author}
  {\bibfnamefont {A.}~\bibnamefont {Furusawa}}, \ and\ \bibinfo {author}
  {\bibfnamefont {M.}~\bibnamefont {Sasaki}},\ }\bibfield  {title} {\enquote
  {\bibinfo {title} {Generation of large-amplitude coherent-state superposition
  via ancilla-assisted photon subtraction},}\ }\href {\doibase
  10.1103/PhysRevLett.101.233605} {\bibfield  {journal} {\bibinfo  {journal}
  {Phys. Rev. Lett.}\ }\textbf {\bibinfo {volume} {101}},\ \bibinfo {pages}
  {233605} (\bibinfo {year} {2008})}\BibitemShut {NoStop}%
\bibitem [{\citenamefont {Gerrits}\ \emph {et~al.}(2010)\citenamefont
  {Gerrits}, \citenamefont {Glancy}, \citenamefont {Clement}, \citenamefont
  {Calkins}, \citenamefont {Lita}, \citenamefont {Miller}, \citenamefont
  {Migdall}, \citenamefont {Nam}, \citenamefont {Mirin},\ and\ \citenamefont
  {Knill}}]{Gerrits2010PRA_CatExp_Pluse}%
  \BibitemOpen
  \bibfield  {author} {\bibinfo {author} {\bibfnamefont {T.}~\bibnamefont
  {Gerrits}}, \bibinfo {author} {\bibfnamefont {S.}~\bibnamefont {Glancy}},
  \bibinfo {author} {\bibfnamefont {T.~S.}\ \bibnamefont {Clement}}, \bibinfo
  {author} {\bibfnamefont {B.}~\bibnamefont {Calkins}}, \bibinfo {author}
  {\bibfnamefont {A.~E.}\ \bibnamefont {Lita}}, \bibinfo {author}
  {\bibfnamefont {A.~J.}\ \bibnamefont {Miller}}, \bibinfo {author}
  {\bibfnamefont {A.~L.}\ \bibnamefont {Migdall}}, \bibinfo {author}
  {\bibfnamefont {S.~W.}\ \bibnamefont {Nam}}, \bibinfo {author} {\bibfnamefont
  {R.~P.}\ \bibnamefont {Mirin}}, \ and\ \bibinfo {author} {\bibfnamefont
  {E.}~\bibnamefont {Knill}},\ }\bibfield  {title} {\enquote {\bibinfo {title}
  {Generation of optical coherent-state superpositions by number-resolved
  photon subtraction from the squeezed vacuum},}\ }\href {\doibase
  10.1103/PhysRevA.82.031802} {\bibfield  {journal} {\bibinfo  {journal} {Phys.
  Rev. A}\ }\textbf {\bibinfo {volume} {82}},\ \bibinfo {pages} {031802}
  (\bibinfo {year} {2010})}\BibitemShut {NoStop}%
\bibitem [{\citenamefont {Sharma}\ \emph {et~al.}(2021)\citenamefont {Sharma},
  \citenamefont {Bittencourt}, \citenamefont {Karenowska},\ and\ \citenamefont
  {Kusminskiy}}]{Sharma2021PRB_CatState}%
  \BibitemOpen
  \bibfield  {author} {\bibinfo {author} {\bibfnamefont {S.}~\bibnamefont
  {Sharma}}, \bibinfo {author} {\bibfnamefont {V.~A. S.~V.}\ \bibnamefont
  {Bittencourt}}, \bibinfo {author} {\bibfnamefont {A.~D.}\ \bibnamefont
  {Karenowska}}, \ and\ \bibinfo {author} {\bibfnamefont {S.~V.}\ \bibnamefont
  {Kusminskiy}},\ }\bibfield  {title} {\enquote {\bibinfo {title} {Spin cat
  states in ferromagnetic insulators},}\ }\href {\doibase
  10.1103/PhysRevB.103.L100403} {\bibfield  {journal} {\bibinfo  {journal}
  {Phys. Rev. B}\ }\textbf {\bibinfo {volume} {103}},\ \bibinfo {pages}
  {L100403} (\bibinfo {year} {2021})}\BibitemShut {NoStop}%
\bibitem [{\citenamefont {Sun}\ \emph {et~al.}(2021)\citenamefont {Sun},
  \citenamefont {Zheng}, \citenamefont {Xiao}, \citenamefont {Gong},
  \citenamefont {He},\ and\ \citenamefont {Xia}}]{SunFX2021PRL_CatState}%
  \BibitemOpen
  \bibfield  {author} {\bibinfo {author} {\bibfnamefont {F.-X.}\ \bibnamefont
  {Sun}}, \bibinfo {author} {\bibfnamefont {S.-S.}\ \bibnamefont {Zheng}},
  \bibinfo {author} {\bibfnamefont {Y.}~\bibnamefont {Xiao}}, \bibinfo {author}
  {\bibfnamefont {Q.}~\bibnamefont {Gong}}, \bibinfo {author} {\bibfnamefont
  {Q.}~\bibnamefont {He}}, \ and\ \bibinfo {author} {\bibfnamefont
  {K.}~\bibnamefont {Xia}},\ }\bibfield  {title} {\enquote {\bibinfo {title}
  {Remote generation of magnon schr\"odinger cat state via magnon-photon
  entanglement},}\ }\href {\doibase 10.1103/PhysRevLett.127.087203} {\bibfield
  {journal} {\bibinfo  {journal} {Phys. Rev. Lett.}\ }\textbf {\bibinfo
  {volume} {127}},\ \bibinfo {pages} {087203} (\bibinfo {year}
  {2021})}\BibitemShut {NoStop}%
\bibitem [{\citenamefont {Huang}\ \emph {et~al.}(2015)\citenamefont {Huang},
  \citenamefont {Le~Jeannic}, \citenamefont {Ruaudel}, \citenamefont {Verma},
  \citenamefont {Shaw}, \citenamefont {Marsili}, \citenamefont {Nam},
  \citenamefont {Wu}, \citenamefont {Zeng}, \citenamefont {Jeong},
  \citenamefont {Filip}, \citenamefont {Morin},\ and\ \citenamefont
  {Laurat}}]{Huang2015PRL_Cat_QuanOptics}%
  \BibitemOpen
  \bibfield  {author} {\bibinfo {author} {\bibfnamefont {K.}~\bibnamefont
  {Huang}}, \bibinfo {author} {\bibfnamefont {H.}~\bibnamefont {Le~Jeannic}},
  \bibinfo {author} {\bibfnamefont {J.}~\bibnamefont {Ruaudel}}, \bibinfo
  {author} {\bibfnamefont {V.~B.}\ \bibnamefont {Verma}}, \bibinfo {author}
  {\bibfnamefont {M.~D.}\ \bibnamefont {Shaw}}, \bibinfo {author}
  {\bibfnamefont {F.}~\bibnamefont {Marsili}}, \bibinfo {author} {\bibfnamefont
  {S.~W.}\ \bibnamefont {Nam}}, \bibinfo {author} {\bibfnamefont
  {E.}~\bibnamefont {Wu}}, \bibinfo {author} {\bibfnamefont {H.}~\bibnamefont
  {Zeng}}, \bibinfo {author} {\bibfnamefont {Y.-C.}\ \bibnamefont {Jeong}},
  \bibinfo {author} {\bibfnamefont {R.}~\bibnamefont {Filip}}, \bibinfo
  {author} {\bibfnamefont {O.}~\bibnamefont {Morin}}, \ and\ \bibinfo {author}
  {\bibfnamefont {J.}~\bibnamefont {Laurat}},\ }\bibfield  {title} {\enquote
  {\bibinfo {title} {Optical synthesis of large-amplitude squeezed
  coherent-state superpositions with minimal resources},}\ }\href {\doibase
  10.1103/PhysRevLett.115.023602} {\bibfield  {journal} {\bibinfo  {journal}
  {Phys. Rev. Lett.}\ }\textbf {\bibinfo {volume} {115}},\ \bibinfo {pages}
  {023602} (\bibinfo {year} {2015})}\BibitemShut {NoStop}%
\bibitem [{\citenamefont {Dakna}\ \emph
  {et~al.}(1997{\natexlab{b}})\citenamefont {Dakna}, \citenamefont {Anhut},
  \citenamefont {Opatrn\'y}, \citenamefont {Kn\"oll},\ and\ \citenamefont
  {Welsch}}]{Dakna1997PRA_Cat_QuanOptics}%
  \BibitemOpen
  \bibfield  {author} {\bibinfo {author} {\bibfnamefont {M.}~\bibnamefont
  {Dakna}}, \bibinfo {author} {\bibfnamefont {T.}~\bibnamefont {Anhut}},
  \bibinfo {author} {\bibfnamefont {T.}~\bibnamefont {Opatrn\'y}}, \bibinfo
  {author} {\bibfnamefont {L.}~\bibnamefont {Kn\"oll}}, \ and\ \bibinfo
  {author} {\bibfnamefont {D.-G.}\ \bibnamefont {Welsch}},\ }\bibfield  {title}
  {\enquote {\bibinfo {title} {Generating schr\"odinger-cat-like states by
  means of conditional measurements on a beam splitter},}\ }\href {\doibase
  10.1103/PhysRevA.55.3184} {\bibfield  {journal} {\bibinfo  {journal} {Phys.
  Rev. A}\ }\textbf {\bibinfo {volume} {55}},\ \bibinfo {pages} {3184}
  (\bibinfo {year} {1997}{\natexlab{b}})}\BibitemShut {NoStop}%
\bibitem [{\citenamefont {Ourjoumtsev}\ \emph
  {et~al.}(2006{\natexlab{b}})\citenamefont {Ourjoumtsev}, \citenamefont
  {Tualle-Brouri}, \citenamefont {Laurat},\ and\ \citenamefont
  {Grangier}}]{Alexei2006Science_Cat_QuanOptics}%
  \BibitemOpen
  \bibfield  {author} {\bibinfo {author} {\bibfnamefont {A.}~\bibnamefont
  {Ourjoumtsev}}, \bibinfo {author} {\bibfnamefont {R.}~\bibnamefont
  {Tualle-Brouri}}, \bibinfo {author} {\bibfnamefont {J.}~\bibnamefont
  {Laurat}}, \ and\ \bibinfo {author} {\bibfnamefont {P.}~\bibnamefont
  {Grangier}},\ }\bibfield  {title} {\enquote {\bibinfo {title} {Generating
  optical schr\"odinger kittens for quantum information processing},}\ }\href
  {\doibase 10.1126/science.1122858} {\bibfield  {journal} {\bibinfo  {journal}
  {Science}\ }\textbf {\bibinfo {volume} {312}},\ \bibinfo {pages} {83}
  (\bibinfo {year} {2006}{\natexlab{b}})}\BibitemShut {NoStop}%
\bibitem [{\citenamefont {Hofer}\ \emph {et~al.}(2011)\citenamefont {Hofer},
  \citenamefont {Wieczorek}, \citenamefont {Aspelmeyer},\ and\ \citenamefont
  {Hammerer}}]{Hofer2011PulseEnt}%
  \BibitemOpen
  \bibfield  {author} {\bibinfo {author} {\bibfnamefont {S.~G.}\ \bibnamefont
  {Hofer}}, \bibinfo {author} {\bibfnamefont {W.}~\bibnamefont {Wieczorek}},
  \bibinfo {author} {\bibfnamefont {M.}~\bibnamefont {Aspelmeyer}}, \ and\
  \bibinfo {author} {\bibfnamefont {K.}~\bibnamefont {Hammerer}},\ }\bibfield
  {title} {\enquote {\bibinfo {title} {Quantum entanglement and teleportation
  in pulsed cavity optomechanics},}\ }\href {\doibase
  10.1103/PhysRevA.84.052327} {\bibfield  {journal} {\bibinfo  {journal} {Phys.
  Rev. A}\ }\textbf {\bibinfo {volume} {84}},\ \bibinfo {pages} {052327}
  (\bibinfo {year} {2011})}\BibitemShut {NoStop}%
\bibitem [{\citenamefont {Palomaki}\ \emph {et~al.}(2013)\citenamefont
  {Palomaki}, \citenamefont {Teufel}, \citenamefont {Simmonds},\ and\
  \citenamefont {Lehnert}}]{Palomaki710}%
  \BibitemOpen
  \bibfield  {author} {\bibinfo {author} {\bibfnamefont {T.~A.}\ \bibnamefont
  {Palomaki}}, \bibinfo {author} {\bibfnamefont {J.~D.}\ \bibnamefont
  {Teufel}}, \bibinfo {author} {\bibfnamefont {R.~W.}\ \bibnamefont
  {Simmonds}}, \ and\ \bibinfo {author} {\bibfnamefont {K.~W.}\ \bibnamefont
  {Lehnert}},\ }\bibfield  {title} {\enquote {\bibinfo {title} {Entangling
  mechanical motion with microwave fields},}\ }\href {\doibase
  10.1126/science.1244563} {\bibfield  {journal} {\bibinfo  {journal}
  {Science}\ }\textbf {\bibinfo {volume} {342}},\ \bibinfo {pages} {710}
  (\bibinfo {year} {2013})}\BibitemShut {NoStop}%
\bibitem [{\citenamefont {Landau}(1965)}]{Landau1935}%
  \BibitemOpen
  \bibfield  {author} {\bibinfo {author} {\bibfnamefont {L.~D.}\ \bibnamefont
  {Landau}},\ }\bibfield  {title} {\enquote {\bibinfo {title} {On the theory of
  the dispersion of magnetic permeability in ferromagnetic bodies},}\ }in\
  \href {\doibase https://doi.org/10.1016/B978-0-08-010586-4.50023-7} {\emph
  {\bibinfo {booktitle} {Collected Papers of L.D. Landau}}},\ \bibinfo {editor}
  {edited by\ \bibinfo {editor} {\bibfnamefont {D.}~\bibnamefont {{TER
  HAAR}}}}\ (\bibinfo  {publisher} {Pergamon},\ \bibinfo {year} {1965})\ pp.\
  \bibinfo {pages} {101--114}\BibitemShut {NoStop}%
\bibitem [{\citenamefont {Gilbert}(2004)}]{Gilbert2004}%
  \BibitemOpen
  \bibfield  {author} {\bibinfo {author} {\bibfnamefont {T.}~\bibnamefont
  {Gilbert}},\ }\bibfield  {title} {\enquote {\bibinfo {title} {A
  phenomenological theory of damping in ferromagnetic materials},}\ }\href
  {\doibase 10.1109/TMAG.2004.836740} {\bibfield  {journal} {\bibinfo
  {journal} {IEEE Transactions on Magnetics}\ }\textbf {\bibinfo {volume}
  {40}},\ \bibinfo {pages} {3443--3449} (\bibinfo {year} {2004})}\BibitemShut
  {NoStop}%
\bibitem [{\citenamefont {Yuan}\ \emph
  {et~al.}(2022{\natexlab{b}})\citenamefont {Yuan}, \citenamefont {Sterk},
  \citenamefont {Kamra},\ and\ \citenamefont {Duine}}]{YuanDephasing2022}%
  \BibitemOpen
  \bibfield  {author} {\bibinfo {author} {\bibfnamefont {H.~Y.}\ \bibnamefont
  {Yuan}}, \bibinfo {author} {\bibfnamefont {W.~P.}\ \bibnamefont {Sterk}},
  \bibinfo {author} {\bibfnamefont {A.}~\bibnamefont {Kamra}}, \ and\ \bibinfo
  {author} {\bibfnamefont {R.~A.}\ \bibnamefont {Duine}},\ }\bibfield  {title}
  {\enquote {\bibinfo {title} {Pure dephasing of magnonic quantum states},}\
  }\href {\doibase 10.1103/PhysRevB.106.L100403} {\bibfield  {journal}
  {\bibinfo  {journal} {Phys. Rev. B}\ }\textbf {\bibinfo {volume} {106}},\
  \bibinfo {pages} {L100403} (\bibinfo {year}
  {2022}{\natexlab{b}})}\BibitemShut {NoStop}%
\bibitem [{\citenamefont {Feynman}, \citenamefont {Leighton},\ and\
  \citenamefont {Sands}(2015)}]{FeynmanBook2015}%
  \BibitemOpen
  \bibfield  {author} {\bibinfo {author} {\bibfnamefont {R.}~\bibnamefont
  {Feynman}}, \bibinfo {author} {\bibfnamefont {R.}~\bibnamefont {Leighton}}, \
  and\ \bibinfo {author} {\bibfnamefont {M.}~\bibnamefont {Sands}},\ }\href
  {https://books.google.es/books?id=d76DBQAAQBAJ} {\emph {\bibinfo {title} {The
  Feynman Lectures on Physics, Vol. I}}}\ (\bibinfo  {publisher} {Basic
  Books},\ \bibinfo {year} {2015})\BibitemShut {NoStop}%
\bibitem [{\citenamefont {Yuan}\ \emph
  {et~al.}(2022{\natexlab{c}})\citenamefont {Yuan}, \citenamefont {Sterk},
  \citenamefont {Kamra},\ and\ \citenamefont {Duine}}]{YuanME2022}%
  \BibitemOpen
  \bibfield  {author} {\bibinfo {author} {\bibfnamefont {H.~Y.}\ \bibnamefont
  {Yuan}}, \bibinfo {author} {\bibfnamefont {W.~P.}\ \bibnamefont {Sterk}},
  \bibinfo {author} {\bibfnamefont {A.}~\bibnamefont {Kamra}}, \ and\ \bibinfo
  {author} {\bibfnamefont {R.~A.}\ \bibnamefont {Duine}},\ }\bibfield  {title}
  {\enquote {\bibinfo {title} {Master equation approach to magnon relaxation
  and dephasing},}\ }\href {\doibase 10.1103/PhysRevB.106.224422} {\bibfield
  {journal} {\bibinfo  {journal} {Phys. Rev. B}\ }\textbf {\bibinfo {volume}
  {106}},\ \bibinfo {pages} {224422} (\bibinfo {year}
  {2022}{\natexlab{c}})}\BibitemShut {NoStop}%
\bibitem [{\citenamefont {Carmichael}(2013)}]{CarmichaelBook2013}%
  \BibitemOpen
  \bibfield  {author} {\bibinfo {author} {\bibfnamefont {H.~J.}\ \bibnamefont
  {Carmichael}},\ }\href
  {https://link.springer.com/book/10.1007/978-3-662-03875-8} {\emph {\bibinfo
  {title} {Statistical Methods in Quantum Optics}}}\ (\bibinfo  {publisher}
  {Springer Berlin, Heidelberg},\ \bibinfo {year} {2013})\BibitemShut {NoStop}%
\bibitem [{\citenamefont {Hornak}\ and\ \citenamefont {of~Technology.
  Center~for Imaging~Science}(1996)}]{HornakBook1996}%
  \BibitemOpen
  \bibfield  {author} {\bibinfo {author} {\bibfnamefont {J.}~\bibnamefont
  {Hornak}}\ and\ \bibinfo {author} {\bibfnamefont {R.~I.}\ \bibnamefont
  {of~Technology. Center~for Imaging~Science}},\ }\href
  {https://books.google.es/books?id=PIBKvgAACAAJ} {\emph {\bibinfo {title} {The
  Basics of NMR}}}\ (\bibinfo  {publisher} {Rochester Institute of Technology,
  Center for Imaging Science},\ \bibinfo {year} {1996})\BibitemShut {NoStop}%
\bibitem [{\citenamefont {Horodecki}\ \emph {et~al.}(2009)\citenamefont
  {Horodecki}, \citenamefont {Horodecki}, \citenamefont {Horodecki},\ and\
  \citenamefont {Horodecki}}]{Horodecki2009RMP_Ent}%
  \BibitemOpen
  \bibfield  {author} {\bibinfo {author} {\bibfnamefont {R.}~\bibnamefont
  {Horodecki}}, \bibinfo {author} {\bibfnamefont {P.}~\bibnamefont
  {Horodecki}}, \bibinfo {author} {\bibfnamefont {M.}~\bibnamefont
  {Horodecki}}, \ and\ \bibinfo {author} {\bibfnamefont {K.}~\bibnamefont
  {Horodecki}},\ }\bibfield  {title} {\enquote {\bibinfo {title} {{Quantum
  entanglement}},}\ }\href {\doibase 10.1103/RevModPhys.81.865} {\bibfield
  {journal} {\bibinfo  {journal} {Rev. Mod. Phys.}\ }\textbf {\bibinfo {volume}
  {81}},\ \bibinfo {pages} {865--942} (\bibinfo {year} {2009})}\BibitemShut
  {NoStop}%
\bibitem [{\citenamefont {Simon}\ \emph {et~al.}(2017)\citenamefont {Simon},
  \citenamefont {Jaeger}, \citenamefont {Sergienko}, \citenamefont {Simon},\
  and\ \citenamefont {Sergienko}}]{Simon2017QuantumCommunication}%
  \BibitemOpen
  \bibfield  {author} {\bibinfo {author} {\bibfnamefont {D.~S.}\ \bibnamefont
  {Simon}}, \bibinfo {author} {\bibfnamefont {G.}~\bibnamefont {Jaeger}},
  \bibinfo {author} {\bibfnamefont {A.~V.}\ \bibnamefont {Sergienko}}, \bibinfo
  {author} {\bibfnamefont {D.~S.}\ \bibnamefont {Simon}}, \ and\ \bibinfo
  {author} {\bibfnamefont {A.~V.}\ \bibnamefont {Sergienko}},\ }\href@noop {}
  {\emph {\bibinfo {title} {Quantum metrology, imaging, and communication}}}\
  (\bibinfo  {publisher} {Springer International Publishing},\ \bibinfo {year}
  {2017})\BibitemShut {NoStop}%
\bibitem [{\citenamefont {Georgescu}, \citenamefont {Ashhab},\ and\
  \citenamefont {Nori}(2014)}]{Georgescu2014QuantumSimulation}%
  \BibitemOpen
  \bibfield  {author} {\bibinfo {author} {\bibfnamefont {I.~M.}\ \bibnamefont
  {Georgescu}}, \bibinfo {author} {\bibfnamefont {S.}~\bibnamefont {Ashhab}}, \
  and\ \bibinfo {author} {\bibfnamefont {F.}~\bibnamefont {Nori}},\ }\bibfield
  {title} {\enquote {\bibinfo {title} {Quantum simulation},}\ }\href {\doibase
  10.1103/RevModPhys.86.153} {\bibfield  {journal} {\bibinfo  {journal} {Rev.
  Mod. Phys.}\ }\textbf {\bibinfo {volume} {86}},\ \bibinfo {pages} {153}
  (\bibinfo {year} {2014})}\BibitemShut {NoStop}%
\bibitem [{\citenamefont {Nielsen}\ and\ \citenamefont
  {Chuang}(2010)}]{Nielsen2010QuantumComputation}%
  \BibitemOpen
  \bibfield  {author} {\bibinfo {author} {\bibfnamefont {M.}~\bibnamefont
  {Nielsen}}\ and\ \bibinfo {author} {\bibfnamefont {I.}~\bibnamefont
  {Chuang}},\ }\href@noop {} {\emph {\bibinfo {title} {Quantum computation and
  quantum information}}}\ (\bibinfo  {publisher} {Cambridge University Press},\
  \bibinfo {year} {2010})\BibitemShut {NoStop}%
\bibitem [{\citenamefont {Degen}, \citenamefont {Reinhard},\ and\ \citenamefont
  {Cappellaro}(2017)}]{Degen2017QuantumSensing}%
  \BibitemOpen
  \bibfield  {author} {\bibinfo {author} {\bibfnamefont {C.~L.}\ \bibnamefont
  {Degen}}, \bibinfo {author} {\bibfnamefont {F.}~\bibnamefont {Reinhard}}, \
  and\ \bibinfo {author} {\bibfnamefont {P.}~\bibnamefont {Cappellaro}},\
  }\bibfield  {title} {\enquote {\bibinfo {title} {Quantum sensing},}\ }\href
  {\doibase 10.1103/RevModPhys.89.035002} {\bibfield  {journal} {\bibinfo
  {journal} {Rev. Mod. Phys.}\ }\textbf {\bibinfo {volume} {89}},\ \bibinfo
  {pages} {035002} (\bibinfo {year} {2017})}\BibitemShut {NoStop}%
\bibitem [{\citenamefont {Pirandola}\ \emph {et~al.}(2018)\citenamefont
  {Pirandola}, \citenamefont {Bardhan}, \citenamefont {Gehring}, \citenamefont
  {Weedbrook},\ and\ \citenamefont {Lloyd}}]{Pirandola2018QuantumSensing}%
  \BibitemOpen
  \bibfield  {author} {\bibinfo {author} {\bibfnamefont {S.}~\bibnamefont
  {Pirandola}}, \bibinfo {author} {\bibfnamefont {B.~R.}\ \bibnamefont
  {Bardhan}}, \bibinfo {author} {\bibfnamefont {T.}~\bibnamefont {Gehring}},
  \bibinfo {author} {\bibfnamefont {C.}~\bibnamefont {Weedbrook}}, \ and\
  \bibinfo {author} {\bibfnamefont {S.}~\bibnamefont {Lloyd}},\ }\bibfield
  {title} {\enquote {\bibinfo {title} {Advances in photonic quantum sensing},}\
  }\href {\doibase 10.1038/s41566-018-0301-6} {\bibfield  {journal} {\bibinfo
  {journal} {Nat. Photon.}\ }\textbf {\bibinfo {volume} {12}},\ \bibinfo
  {pages} {724} (\bibinfo {year} {2018})}\BibitemShut {NoStop}%
\bibitem [{\citenamefont {Schr{\"{o}}dinger}(1935)}]{Schrodinger1935}%
  \BibitemOpen
  \bibfield  {author} {\bibinfo {author} {\bibfnamefont {E.}~\bibnamefont
  {Schr{\"{o}}dinger}},\ }\bibfield  {title} {\enquote {\bibinfo {title}
  {{Probability relations between separated systems}},}\ }\href@noop {}
  {\bibfield  {journal} {\bibinfo  {journal} {Proc. Cambridge Philos. Soc.}\
  }\textbf {\bibinfo {volume} {31}},\ \bibinfo {pages} {555} (\bibinfo {year}
  {1935})}\BibitemShut {NoStop}%
\bibitem [{\citenamefont {Einstein}, \citenamefont {Podolsky},\ and\
  \citenamefont {Rosen}(1935)}]{Einstein1935}%
  \BibitemOpen
  \bibfield  {author} {\bibinfo {author} {\bibfnamefont {A.}~\bibnamefont
  {Einstein}}, \bibinfo {author} {\bibfnamefont {B.}~\bibnamefont {Podolsky}},
  \ and\ \bibinfo {author} {\bibfnamefont {N.}~\bibnamefont {Rosen}},\
  }\bibfield  {title} {\enquote {\bibinfo {title} {{Can quantum-mechanical
  description of physical reality be considered correct?}}}\ }\href {\doibase
  10.1007/s10701-010-9411-9} {\bibfield  {journal} {\bibinfo  {journal} {Phys.
  Rev.}\ }\textbf {\bibinfo {volume} {47}},\ \bibinfo {pages} {777} (\bibinfo
  {year} {1935})}\BibitemShut {NoStop}%
\bibitem [{\citenamefont {Reid}(1989)}]{Reid1989}%
  \BibitemOpen
  \bibfield  {author} {\bibinfo {author} {\bibfnamefont {M.~D.}\ \bibnamefont
  {Reid}},\ }\bibfield  {title} {\enquote {\bibinfo {title} {{Demonstration of
  the Einstein-Podolsky-Rosen paradox using nondegenerate parametric
  amplification}},}\ }\href {\doibase 10.1103/PhysRevA.40.913} {\bibfield
  {journal} {\bibinfo  {journal} {Phys. Rev. A}\ }\textbf {\bibinfo {volume}
  {40}},\ \bibinfo {pages} {913} (\bibinfo {year} {1989})}\BibitemShut
  {NoStop}%
\bibitem [{\citenamefont {Wiseman}, \citenamefont {Jones},\ and\ \citenamefont
  {Doherty}(2007)}]{Wiseman2007}%
  \BibitemOpen
  \bibfield  {author} {\bibinfo {author} {\bibfnamefont {H.~M.}\ \bibnamefont
  {Wiseman}}, \bibinfo {author} {\bibfnamefont {S.~J.}\ \bibnamefont {Jones}},
  \ and\ \bibinfo {author} {\bibfnamefont {A.~C.}\ \bibnamefont {Doherty}},\
  }\bibfield  {title} {\enquote {\bibinfo {title} {{Steering, entanglement,
  nonlocality, and the Einstein-Podolsky-Rosen paradox}},}\ }\href {\doibase
  10.1103/PhysRevLett.98.140402} {\bibfield  {journal} {\bibinfo  {journal}
  {Phys. Rev. Lett.}\ }\textbf {\bibinfo {volume} {98}},\ \bibinfo {pages}
  {140402} (\bibinfo {year} {2007})}\BibitemShut {NoStop}%
\bibitem [{\citenamefont {Jones}, \citenamefont {Wiseman},\ and\ \citenamefont
  {Doherty}(2007)}]{Jones2007}%
  \BibitemOpen
  \bibfield  {author} {\bibinfo {author} {\bibfnamefont {S.~J.}\ \bibnamefont
  {Jones}}, \bibinfo {author} {\bibfnamefont {H.~M.}\ \bibnamefont {Wiseman}},
  \ and\ \bibinfo {author} {\bibfnamefont {A.~C.}\ \bibnamefont {Doherty}},\
  }\bibfield  {title} {\enquote {\bibinfo {title} {{Entanglement,
  Einstein-Podolsky-Rosen correlations, Bell nonlocality, and steering}},}\
  }\href {\doibase 10.1103/PhysRevA.76.052116} {\bibfield  {journal} {\bibinfo
  {journal} {Phys. Rev. A}\ }\textbf {\bibinfo {volume} {76}},\ \bibinfo
  {pages} {052116} (\bibinfo {year} {2007})}\BibitemShut {NoStop}%
\bibitem [{\citenamefont {Cavalcanti}\ \emph {et~al.}(2009)\citenamefont
  {Cavalcanti}, \citenamefont {Jones}, \citenamefont {Wiseman},\ and\
  \citenamefont {Reid}}]{Cavalcanti2009}%
  \BibitemOpen
  \bibfield  {author} {\bibinfo {author} {\bibfnamefont {E.~G.}\ \bibnamefont
  {Cavalcanti}}, \bibinfo {author} {\bibfnamefont {S.~J.}\ \bibnamefont
  {Jones}}, \bibinfo {author} {\bibfnamefont {H.~M.}\ \bibnamefont {Wiseman}},
  \ and\ \bibinfo {author} {\bibfnamefont {M.~D.}\ \bibnamefont {Reid}},\
  }\bibfield  {title} {\enquote {\bibinfo {title} {{Experimental criteria for
  steering and the Einstein-Podolsky-Rosen paradox}},}\ }\href {\doibase
  10.1103/PhysRevA.80.032112} {\bibfield  {journal} {\bibinfo  {journal} {Phys.
  Rev. A}\ }\textbf {\bibinfo {volume} {80}},\ \bibinfo {pages} {032112}
  (\bibinfo {year} {2009})}\BibitemShut {NoStop}%
\bibitem [{\citenamefont {Uola}\ \emph {et~al.}(2020)\citenamefont {Uola},
  \citenamefont {Costa}, \citenamefont {Nguyen},\ and\ \citenamefont
  {G\"uhne}}]{Uola2020RMP_SteeringReview}%
  \BibitemOpen
  \bibfield  {author} {\bibinfo {author} {\bibfnamefont {R.}~\bibnamefont
  {Uola}}, \bibinfo {author} {\bibfnamefont {A.~C.~S.}\ \bibnamefont {Costa}},
  \bibinfo {author} {\bibfnamefont {H.~C.}\ \bibnamefont {Nguyen}}, \ and\
  \bibinfo {author} {\bibfnamefont {O.}~\bibnamefont {G\"uhne}},\ }\bibfield
  {title} {\enquote {\bibinfo {title} {Quantum steering},}\ }\href {\doibase
  10.1103/RevModPhys.92.015001} {\bibfield  {journal} {\bibinfo  {journal}
  {Rev. Mod. Phys.}\ }\textbf {\bibinfo {volume} {92}},\ \bibinfo {pages}
  {015001} (\bibinfo {year} {2020})}\BibitemShut {NoStop}%
\bibitem [{\citenamefont {Reid}\ \emph {et~al.}(2009)\citenamefont {Reid},
  \citenamefont {Drummond}, \citenamefont {Bowen}, \citenamefont {Cavalcanti},
  \citenamefont {Lam}, \citenamefont {Bachor}, \citenamefont {Andersen},\ and\
  \citenamefont {Leuchs}}]{Reid2009}%
  \BibitemOpen
  \bibfield  {author} {\bibinfo {author} {\bibfnamefont {M.~D.}\ \bibnamefont
  {Reid}}, \bibinfo {author} {\bibfnamefont {P.~D.}\ \bibnamefont {Drummond}},
  \bibinfo {author} {\bibfnamefont {W.~P.}\ \bibnamefont {Bowen}}, \bibinfo
  {author} {\bibfnamefont {E.~G.}\ \bibnamefont {Cavalcanti}}, \bibinfo
  {author} {\bibfnamefont {P.~K.}\ \bibnamefont {Lam}}, \bibinfo {author}
  {\bibfnamefont {H.~A.}\ \bibnamefont {Bachor}}, \bibinfo {author}
  {\bibfnamefont {U.~L.}\ \bibnamefont {Andersen}}, \ and\ \bibinfo {author}
  {\bibfnamefont {G.}~\bibnamefont {Leuchs}},\ }\bibfield  {title} {\enquote
  {\bibinfo {title} {Colloquium: The einstein-podolsky-rosen paradox: From
  concepts to applications},}\ }\href {\doibase 10.1103/RevModPhys.81.1727}
  {\bibfield  {journal} {\bibinfo  {journal} {Rev. Mod. Phys.}\ }\textbf
  {\bibinfo {volume} {81}},\ \bibinfo {pages} {1727} (\bibinfo {year}
  {2009})}\BibitemShut {NoStop}%
\bibitem [{\citenamefont {Cavalcanti}\ and\ \citenamefont
  {Skrzypczyk}(2017)}]{PhysRep2017}%
  \BibitemOpen
  \bibfield  {author} {\bibinfo {author} {\bibfnamefont {D.}~\bibnamefont
  {Cavalcanti}}\ and\ \bibinfo {author} {\bibfnamefont {P.}~\bibnamefont
  {Skrzypczyk}},\ }\bibfield  {title} {\enquote {\bibinfo {title} {{Quantum
  steering : a review with focus on semidefinite programming}},}\ }\href
  {\doibase 10.1088/1361-6633/80/2/024001} {\bibfield  {journal} {\bibinfo
  {journal} {Rep. Prog. Phys.}\ }\textbf {\bibinfo {volume} {80}},\ \bibinfo
  {pages} {024001} (\bibinfo {year} {2017})}\BibitemShut {NoStop}%
\bibitem [{\citenamefont {Wagner}\ \emph {et~al.}(2008)\citenamefont {Wagner},
  \citenamefont {Janousek}, \citenamefont {Delaubert}, \citenamefont {Zou},
  \citenamefont {Harb}, \citenamefont {Treps}, \citenamefont {Morizur},
  \citenamefont {Lam},\ and\ \citenamefont {Bachor}}]{Wagner2008}%
  \BibitemOpen
  \bibfield  {author} {\bibinfo {author} {\bibfnamefont {K.}~\bibnamefont
  {Wagner}}, \bibinfo {author} {\bibfnamefont {J.}~\bibnamefont {Janousek}},
  \bibinfo {author} {\bibfnamefont {V.}~\bibnamefont {Delaubert}}, \bibinfo
  {author} {\bibfnamefont {H.}~\bibnamefont {Zou}}, \bibinfo {author}
  {\bibfnamefont {C.}~\bibnamefont {Harb}}, \bibinfo {author} {\bibfnamefont
  {N.}~\bibnamefont {Treps}}, \bibinfo {author} {\bibfnamefont {J.~F.}\
  \bibnamefont {Morizur}}, \bibinfo {author} {\bibfnamefont {P.~K.}\
  \bibnamefont {Lam}}, \ and\ \bibinfo {author} {\bibfnamefont {H.~A.}\
  \bibnamefont {Bachor}},\ }\bibfield  {title} {\enquote {\bibinfo {title}
  {Entangling the spatial properties of laser beams},}\ }\href {\doibase
  10.1126/science.1159663} {\bibfield  {journal} {\bibinfo  {journal}
  {Science}\ }\textbf {\bibinfo {volume} {321}},\ \bibinfo {pages} {541--543}
  (\bibinfo {year} {2008})}\BibitemShut {NoStop}%
\bibitem [{\citenamefont {H{\"{a}}ndchen}\ \emph {et~al.}(2012)\citenamefont
  {H{\"{a}}ndchen}, \citenamefont {Eberle}, \citenamefont {Steinlechner},
  \citenamefont {Samblowski}, \citenamefont {Franz}, \citenamefont {Werner},\
  and\ \citenamefont {Schnabel}}]{Handchen2012}%
  \BibitemOpen
  \bibfield  {author} {\bibinfo {author} {\bibfnamefont {V.}~\bibnamefont
  {H{\"{a}}ndchen}}, \bibinfo {author} {\bibfnamefont {T.}~\bibnamefont
  {Eberle}}, \bibinfo {author} {\bibfnamefont {S.}~\bibnamefont
  {Steinlechner}}, \bibinfo {author} {\bibfnamefont {A.}~\bibnamefont
  {Samblowski}}, \bibinfo {author} {\bibfnamefont {T.}~\bibnamefont {Franz}},
  \bibinfo {author} {\bibfnamefont {R.~F.}\ \bibnamefont {Werner}}, \ and\
  \bibinfo {author} {\bibfnamefont {R.}~\bibnamefont {Schnabel}},\ }\bibfield
  {title} {\enquote {\bibinfo {title} {{Observation of one-way
  Einstein-Podolsky-Rosen steering}},}\ }\href {\doibase
  10.1038/nphoton.2012.202} {\bibfield  {journal} {\bibinfo  {journal} {Nat.
  Photon.}\ }\textbf {\bibinfo {volume} {6}},\ \bibinfo {pages} {596--599}
  (\bibinfo {year} {2012})}\BibitemShut {NoStop}%
\bibitem [{\citenamefont {Olsen}\ and\ \citenamefont
  {Bradley}(2008)}]{Olsen2008PRA}%
  \BibitemOpen
  \bibfield  {author} {\bibinfo {author} {\bibfnamefont {M.~K.}\ \bibnamefont
  {Olsen}}\ and\ \bibinfo {author} {\bibfnamefont {A.~S.}\ \bibnamefont
  {Bradley}},\ }\bibfield  {title} {\enquote {\bibinfo {title} {{Bright
  bichromatic entanglement and quantum dynamics of sum frequency
  generation}},}\ }\href {\doibase 10.1103/PhysRevA.77.023813} {\bibfield
  {journal} {\bibinfo  {journal} {Phys. Rev. A}\ }\textbf {\bibinfo {volume}
  {77}},\ \bibinfo {pages} {023813} (\bibinfo {year} {2008})}\BibitemShut
  {NoStop}%
\bibitem [{\citenamefont {Midgley}, \citenamefont {Ferris},\ and\ \citenamefont
  {Olsen}(2010)}]{Midgley2010PRA}%
  \BibitemOpen
  \bibfield  {author} {\bibinfo {author} {\bibfnamefont {S.~L.~W.}\
  \bibnamefont {Midgley}}, \bibinfo {author} {\bibfnamefont {A.~J.}\
  \bibnamefont {Ferris}}, \ and\ \bibinfo {author} {\bibfnamefont {M.~K.}\
  \bibnamefont {Olsen}},\ }\bibfield  {title} {\enquote {\bibinfo {title}
  {{Asymmetric gaussian steering: When Alice and Bob disagree}},}\ }\href
  {\doibase 10.1103/PhysRevA.81.022101} {\bibfield  {journal} {\bibinfo
  {journal} {Phys. Rev. A}\ }\textbf {\bibinfo {volume} {81}},\ \bibinfo
  {pages} {022101} (\bibinfo {year} {2010})}\BibitemShut {NoStop}%
\bibitem [{\citenamefont {Olsen}(2013)}]{Olsen2013PRA}%
  \BibitemOpen
  \bibfield  {author} {\bibinfo {author} {\bibfnamefont {M.~K.}\ \bibnamefont
  {Olsen}},\ }\bibfield  {title} {\enquote {\bibinfo {title} {{Asymmetric
  Gaussian harmonic steering in second-harmonic generation}},}\ }\href
  {\doibase 10.1103/PhysRevA.88.051802} {\bibfield  {journal} {\bibinfo
  {journal} {Phys. Rev. A}\ }\textbf {\bibinfo {volume} {88}},\ \bibinfo
  {pages} {051802(R)} (\bibinfo {year} {2013})}\BibitemShut {NoStop}%
\bibitem [{\citenamefont {Schneeloch}\ \emph {et~al.}(2013)\citenamefont
  {Schneeloch}, \citenamefont {Broadbent}, \citenamefont {Walborn},
  \citenamefont {Cavalcanti},\ and\ \citenamefont
  {Howell}}]{Schneeloch2013PRA}%
  \BibitemOpen
  \bibfield  {author} {\bibinfo {author} {\bibfnamefont {J.}~\bibnamefont
  {Schneeloch}}, \bibinfo {author} {\bibfnamefont {C.~J.}\ \bibnamefont
  {Broadbent}}, \bibinfo {author} {\bibfnamefont {S.~P.}\ \bibnamefont
  {Walborn}}, \bibinfo {author} {\bibfnamefont {E.~G.}\ \bibnamefont
  {Cavalcanti}}, \ and\ \bibinfo {author} {\bibfnamefont {J.~C.}\ \bibnamefont
  {Howell}},\ }\bibfield  {title} {\enquote {\bibinfo {title}
  {{Einstein-Podolsky-Rosen steering inequalities from entropic uncertainty
  relations}},}\ }\href {\doibase 10.1103/PhysRevA.87.062103} {\bibfield
  {journal} {\bibinfo  {journal} {Phys. Rev. A}\ }\textbf {\bibinfo {volume}
  {87}},\ \bibinfo {pages} {062103} (\bibinfo {year} {2013})}\BibitemShut
  {NoStop}%
\bibitem [{\citenamefont {He}\ and\ \citenamefont
  {Reid}(2013{\natexlab{a}})}]{HeQY2013PRA_PulseOpto_Ent}%
  \BibitemOpen
  \bibfield  {author} {\bibinfo {author} {\bibfnamefont {Q.~Y.}\ \bibnamefont
  {He}}\ and\ \bibinfo {author} {\bibfnamefont {M.~D.}\ \bibnamefont {Reid}},\
  }\bibfield  {title} {\enquote {\bibinfo {title} {Einstein-podolsky-rosen
  paradox and quantum steering in pulsed optomechanics},}\ }\href {\doibase
  10.1103/PhysRevA.88.052121} {\bibfield  {journal} {\bibinfo  {journal} {Phys.
  Rev. A}\ }\textbf {\bibinfo {volume} {88}},\ \bibinfo {pages} {052121}
  (\bibinfo {year} {2013}{\natexlab{a}})}\BibitemShut {NoStop}%
\bibitem [{\citenamefont {He}\ and\ \citenamefont {Ficek}(2014)}]{HeQY2014PRA}%
  \BibitemOpen
  \bibfield  {author} {\bibinfo {author} {\bibfnamefont {Q.~Y.}\ \bibnamefont
  {He}}\ and\ \bibinfo {author} {\bibfnamefont {Z.}~\bibnamefont {Ficek}},\
  }\bibfield  {title} {\enquote {\bibinfo {title} {{Einstein-Podolsky-Rosen
  paradox and quantum steering in a three-mode optomechanical system}},}\
  }\href {\doibase 10.1103/PhysRevA.89.022332} {\bibfield  {journal} {\bibinfo
  {journal} {Phys. Rev. A}\ }\textbf {\bibinfo {volume} {89}},\ \bibinfo
  {pages} {022332} (\bibinfo {year} {2014})}\BibitemShut {NoStop}%
\bibitem [{\citenamefont {Evans}\ and\ \citenamefont
  {Wiseman}(2014)}]{Evans2014PRA}%
  \BibitemOpen
  \bibfield  {author} {\bibinfo {author} {\bibfnamefont {D.~A.}\ \bibnamefont
  {Evans}}\ and\ \bibinfo {author} {\bibfnamefont {H.~M.}\ \bibnamefont
  {Wiseman}},\ }\bibfield  {title} {\enquote {\bibinfo {title} {{Optimal
  measurements for tests of Einstein-Podolsky-Rosen steering with no detection
  loophole using two-qubit Werner states}},}\ }\href {\doibase
  10.1103/PhysRevA.90.012114} {\bibfield  {journal} {\bibinfo  {journal} {Phys.
  Rev. A}\ }\textbf {\bibinfo {volume} {90}},\ \bibinfo {pages} {012114}
  (\bibinfo {year} {2014})}\BibitemShut {NoStop}%
\bibitem [{\citenamefont {Bowles}\ \emph {et~al.}(2014)\citenamefont {Bowles},
  \citenamefont {V{\'{e}}rtesi}, \citenamefont {Quintino},\ and\ \citenamefont
  {Brunner}}]{Bowles2014PRL}%
  \BibitemOpen
  \bibfield  {author} {\bibinfo {author} {\bibfnamefont {J.}~\bibnamefont
  {Bowles}}, \bibinfo {author} {\bibfnamefont {T.}~\bibnamefont
  {V{\'{e}}rtesi}}, \bibinfo {author} {\bibfnamefont {M.~T.}\ \bibnamefont
  {Quintino}}, \ and\ \bibinfo {author} {\bibfnamefont {N.}~\bibnamefont
  {Brunner}},\ }\bibfield  {title} {\enquote {\bibinfo {title} {{One-way
  Einstein-Podolsky-Rosen Steering}},}\ }\href {\doibase
  10.1103/PhysRevLett.112.200402} {\bibfield  {journal} {\bibinfo  {journal}
  {Phys. Rev. Lett.}\ }\textbf {\bibinfo {volume} {112}},\ \bibinfo {pages}
  {200402} (\bibinfo {year} {2014})}\BibitemShut {NoStop}%
\bibitem [{\citenamefont {Skrzypczyk}, \citenamefont {Navascu{\'{e}}s},\ and\
  \citenamefont {Cavalcanti}(2014)}]{Skrzypczyk2014PRL}%
  \BibitemOpen
  \bibfield  {author} {\bibinfo {author} {\bibfnamefont {P.}~\bibnamefont
  {Skrzypczyk}}, \bibinfo {author} {\bibfnamefont {M.}~\bibnamefont
  {Navascu{\'{e}}s}}, \ and\ \bibinfo {author} {\bibfnamefont {D.}~\bibnamefont
  {Cavalcanti}},\ }\bibfield  {title} {\enquote {\bibinfo {title} {{Quantifying
  Einstein-Podolsky-Rosen Steering}},}\ }\href {\doibase
  10.1103/PhysRevLett.112.180404} {\bibfield  {journal} {\bibinfo  {journal}
  {Phys. Rev. Lett.}\ }\textbf {\bibinfo {volume} {112}},\ \bibinfo {pages}
  {180404} (\bibinfo {year} {2014})}\BibitemShut {NoStop}%
\bibitem [{\citenamefont {Wang}\ \emph {et~al.}(2014)\citenamefont {Wang},
  \citenamefont {Gong}, \citenamefont {Ficek},\ and\ \citenamefont
  {He}}]{WangM2014PRA}%
  \BibitemOpen
  \bibfield  {author} {\bibinfo {author} {\bibfnamefont {M.}~\bibnamefont
  {Wang}}, \bibinfo {author} {\bibfnamefont {Q.~H.}\ \bibnamefont {Gong}},
  \bibinfo {author} {\bibfnamefont {Z.}~\bibnamefont {Ficek}}, \ and\ \bibinfo
  {author} {\bibfnamefont {Q.~Y.}\ \bibnamefont {He}},\ }\bibfield  {title}
  {\enquote {\bibinfo {title} {{Role of thermal noise in tripartite quantum
  steering}},}\ }\href {\doibase 10.1103/PhysRevA.90.023801} {\bibfield
  {journal} {\bibinfo  {journal} {Phys. Rev. A}\ }\textbf {\bibinfo {volume}
  {90}},\ \bibinfo {pages} {023801} (\bibinfo {year} {2014})}\BibitemShut
  {NoStop}%
\bibitem [{\citenamefont {Rosales-Z\'{a}rate}\ \emph
  {et~al.}(2015)\citenamefont {Rosales-Z\'{a}rate}, \citenamefont {Teh},
  \citenamefont {Kiesewetter}, \citenamefont {Brolis}, \citenamefont {Ng},\
  and\ \citenamefont {Reid}}]{Reidjosab2015}%
  \BibitemOpen
  \bibfield  {author} {\bibinfo {author} {\bibfnamefont {L.}~\bibnamefont
  {Rosales-Z\'{a}rate}}, \bibinfo {author} {\bibfnamefont {R.~Y.}\ \bibnamefont
  {Teh}}, \bibinfo {author} {\bibfnamefont {S.}~\bibnamefont {Kiesewetter}},
  \bibinfo {author} {\bibfnamefont {A.}~\bibnamefont {Brolis}}, \bibinfo
  {author} {\bibfnamefont {K.}~\bibnamefont {Ng}}, \ and\ \bibinfo {author}
  {\bibfnamefont {M.~D.}\ \bibnamefont {Reid}},\ }\bibfield  {title} {\enquote
  {\bibinfo {title} {Decoherence of einstein-podolsky-rosen steering},}\ }\href
  {\doibase 10.1364/JOSAB.32.000A82} {\bibfield  {journal} {\bibinfo  {journal}
  {J. Opt. Soc. Am. B}\ }\textbf {\bibinfo {volume} {32}},\ \bibinfo {pages}
  {A82--A91} (\bibinfo {year} {2015})}\BibitemShut {NoStop}%
\bibitem [{\citenamefont {He}, \citenamefont {Gong},\ and\ \citenamefont
  {Reid}(2015)}]{HeQY2015PRL}%
  \BibitemOpen
  \bibfield  {author} {\bibinfo {author} {\bibfnamefont {Q.~Y.}\ \bibnamefont
  {He}}, \bibinfo {author} {\bibfnamefont {Q.~H.}\ \bibnamefont {Gong}}, \ and\
  \bibinfo {author} {\bibfnamefont {M.~D.}\ \bibnamefont {Reid}},\ }\bibfield
  {title} {\enquote {\bibinfo {title} {{Classifying Directional Gaussian
  Entanglement, Einstein-Podolsky-Rosen Steering, and Discord}},}\ }\href
  {\doibase 10.1103/PhysRevLett.114.060402} {\bibfield  {journal} {\bibinfo
  {journal} {Phys. Rev. Lett.}\ }\textbf {\bibinfo {volume} {114}},\ \bibinfo
  {pages} {060402} (\bibinfo {year} {2015})}\BibitemShut {NoStop}%
\bibitem [{\citenamefont {Tan}, \citenamefont {Zhang},\ and\ \citenamefont
  {Li}(2015)}]{TanHT2015PRA}%
  \BibitemOpen
  \bibfield  {author} {\bibinfo {author} {\bibfnamefont {H.~T.}\ \bibnamefont
  {Tan}}, \bibinfo {author} {\bibfnamefont {X.~C.}\ \bibnamefont {Zhang}}, \
  and\ \bibinfo {author} {\bibfnamefont {G.~X.}\ \bibnamefont {Li}},\
  }\bibfield  {title} {\enquote {\bibinfo {title} {{Steady-state one-way
  Einstein-Podolsky-Rosen steering in optomechanical interfaces}},}\ }\href
  {\doibase 10.1103/PhysRevA.91.032121} {\bibfield  {journal} {\bibinfo
  {journal} {Phys. Rev. A}\ }\textbf {\bibinfo {volume} {91}},\ \bibinfo
  {pages} {032121} (\bibinfo {year} {2015})}\BibitemShut {NoStop}%
\bibitem [{\citenamefont {Quintino}\ \emph {et~al.}(2015)\citenamefont
  {Quintino}, \citenamefont {V\'ertesi}, \citenamefont {Cavalcanti},
  \citenamefont {Augusiak}, \citenamefont {Demianowicz}, \citenamefont
  {Ac\'{i}n},\ and\ \citenamefont {Brunner}}]{Marco2015PRA}%
  \BibitemOpen
  \bibfield  {author} {\bibinfo {author} {\bibfnamefont {M.~T.}\ \bibnamefont
  {Quintino}}, \bibinfo {author} {\bibfnamefont {T.}~\bibnamefont {V\'ertesi}},
  \bibinfo {author} {\bibfnamefont {D.}~\bibnamefont {Cavalcanti}}, \bibinfo
  {author} {\bibfnamefont {R.}~\bibnamefont {Augusiak}}, \bibinfo {author}
  {\bibfnamefont {M.}~\bibnamefont {Demianowicz}}, \bibinfo {author}
  {\bibfnamefont {A.}~\bibnamefont {Ac\'{i}n}}, \ and\ \bibinfo {author}
  {\bibfnamefont {N.}~\bibnamefont {Brunner}},\ }\bibfield  {title} {\enquote
  {\bibinfo {title} {{Inequivalence of entanglement , steering , and Bell
  nonlocality for general measurements}},}\ }\href {\doibase
  10.1103/PhysRevA.92.032107} {\bibfield  {journal} {\bibinfo  {journal} {Phys.
  Rev. A}\ }\textbf {\bibinfo {volume} {92}},\ \bibinfo {pages} {032107}
  (\bibinfo {year} {2015})}\BibitemShut {NoStop}%
\bibitem [{\citenamefont {Bowles}\ \emph {et~al.}(2016)\citenamefont {Bowles},
  \citenamefont {Hirsch}, \citenamefont {Quintino},\ and\ \citenamefont
  {Brunner}}]{Bowles2016PRA}%
  \BibitemOpen
  \bibfield  {author} {\bibinfo {author} {\bibfnamefont {J.}~\bibnamefont
  {Bowles}}, \bibinfo {author} {\bibfnamefont {F.}~\bibnamefont {Hirsch}},
  \bibinfo {author} {\bibfnamefont {M.~T.}\ \bibnamefont {Quintino}}, \ and\
  \bibinfo {author} {\bibfnamefont {N.}~\bibnamefont {Brunner}},\ }\bibfield
  {title} {\enquote {\bibinfo {title} {{Sufficient criterion for guaranteeing
  that a two-qubit state is unsteerable}},}\ }\href {\doibase
  10.1103/PhysRevA.93.022121} {\bibfield  {journal} {\bibinfo  {journal} {Phys.
  Rev. A}\ }\textbf {\bibinfo {volume} {93}},\ \bibinfo {pages} {022121}
  (\bibinfo {year} {2016})}\BibitemShut {NoStop}%
\bibitem [{\citenamefont {Olsen}(2017)}]{Olsen2017PRL}%
  \BibitemOpen
  \bibfield  {author} {\bibinfo {author} {\bibfnamefont {M.~K.}\ \bibnamefont
  {Olsen}},\ }\bibfield  {title} {\enquote {\bibinfo {title} {{Controlled
  Asymmetry of Einstein-Podolsky-Rosen Steering with an Injected Nondegenerate
  Optical Parametric Oscillator}},}\ }\href {\doibase
  10.1103/PhysRevLett.119.160501} {\bibfield  {journal} {\bibinfo  {journal}
  {Phys. Rev. Lett.}\ }\textbf {\bibinfo {volume} {119}},\ \bibinfo {pages}
  {160501} (\bibinfo {year} {2017})}\BibitemShut {NoStop}%
\bibitem [{\citenamefont {Baker}, \citenamefont {Wollmann},\ and\ \citenamefont
  {Pryde}(2018)}]{Baker2018JOpt}%
  \BibitemOpen
  \bibfield  {author} {\bibinfo {author} {\bibfnamefont {T.~J.}\ \bibnamefont
  {Baker}}, \bibinfo {author} {\bibfnamefont {S.}~\bibnamefont {Wollmann}}, \
  and\ \bibinfo {author} {\bibfnamefont {G.~J.}\ \bibnamefont {Pryde}},\
  }\bibfield  {title} {\enquote {\bibinfo {title} {{Necessary condition for
  steerability of arbitrary two-qubit states with loss}},}\ }\href {\doibase
  10.1088/2040-8986/aaaa3c} {\bibfield  {journal} {\bibinfo  {journal} {J.
  Opt.}\ }\textbf {\bibinfo {volume} {20}},\ \bibinfo {pages} {034008}
  (\bibinfo {year} {2018})}\BibitemShut {NoStop}%
\bibitem [{\citenamefont {Zheng}\ \emph {et~al.}(2019)\citenamefont {Zheng},
  \citenamefont {Sun}, \citenamefont {Lai}, \citenamefont {Gong},\ and\
  \citenamefont {He}}]{ZhengShasha2019PRA}%
  \BibitemOpen
  \bibfield  {author} {\bibinfo {author} {\bibfnamefont {S.}~\bibnamefont
  {Zheng}}, \bibinfo {author} {\bibfnamefont {F.}~\bibnamefont {Sun}}, \bibinfo
  {author} {\bibfnamefont {Y.}~\bibnamefont {Lai}}, \bibinfo {author}
  {\bibfnamefont {Q.}~\bibnamefont {Gong}}, \ and\ \bibinfo {author}
  {\bibfnamefont {Q.}~\bibnamefont {He}},\ }\bibfield  {title} {\enquote
  {\bibinfo {title} {Manipulation and enhancement of asymmetric steering via
  interference effects induced by closed-loop coupling},}\ }\href {\doibase
  10.1103/PhysRevA.99.022335} {\bibfield  {journal} {\bibinfo  {journal} {Phys.
  Rev. A}\ }\textbf {\bibinfo {volume} {99}},\ \bibinfo {pages} {022335}
  (\bibinfo {year} {2019})}\BibitemShut {NoStop}%
\bibitem [{\citenamefont {Armstrong}\ \emph
  {et~al.}(2015{\natexlab{a}})\citenamefont {Armstrong}, \citenamefont {Wang},
  \citenamefont {Teh}, \citenamefont {Gong}, \citenamefont {He}, \citenamefont
  {Janousek}, \citenamefont {Bachor}, \citenamefont {Reid},\ and\ \citenamefont
  {Lam}}]{Armstrong2015}%
  \BibitemOpen
  \bibfield  {author} {\bibinfo {author} {\bibfnamefont {S.}~\bibnamefont
  {Armstrong}}, \bibinfo {author} {\bibfnamefont {M.}~\bibnamefont {Wang}},
  \bibinfo {author} {\bibfnamefont {R.~Y.}\ \bibnamefont {Teh}}, \bibinfo
  {author} {\bibfnamefont {Q.~H.}\ \bibnamefont {Gong}}, \bibinfo {author}
  {\bibfnamefont {Q.~Y.}\ \bibnamefont {He}}, \bibinfo {author} {\bibfnamefont
  {J.}~\bibnamefont {Janousek}}, \bibinfo {author} {\bibfnamefont {H.~A.}\
  \bibnamefont {Bachor}}, \bibinfo {author} {\bibfnamefont {M.~D.}\
  \bibnamefont {Reid}}, \ and\ \bibinfo {author} {\bibfnamefont {P.~K.}\
  \bibnamefont {Lam}},\ }\bibfield  {title} {\enquote {\bibinfo {title}
  {{Multipartite Einstein-Podolsky-Rosen steering and genuine tripartite
  entanglement with optical networks}},}\ }\href {\doibase 10.1038/nphys3202}
  {\bibfield  {journal} {\bibinfo  {journal} {Nat. Phys.}\ }\textbf {\bibinfo
  {volume} {11}},\ \bibinfo {pages} {167--172} (\bibinfo {year}
  {2015}{\natexlab{a}})}\BibitemShut {NoStop}%
\bibitem [{\citenamefont {Qin}\ \emph {et~al.}(2017)\citenamefont {Qin},
  \citenamefont {Deng}, \citenamefont {Tian}, \citenamefont {Wang},
  \citenamefont {Su}, \citenamefont {Xie},\ and\ \citenamefont
  {Peng}}]{QinZZ2017PRA}%
  \BibitemOpen
  \bibfield  {author} {\bibinfo {author} {\bibfnamefont {Z.~Z.}\ \bibnamefont
  {Qin}}, \bibinfo {author} {\bibfnamefont {X.~W.}\ \bibnamefont {Deng}},
  \bibinfo {author} {\bibfnamefont {C.~X.}\ \bibnamefont {Tian}}, \bibinfo
  {author} {\bibfnamefont {M.~H.}\ \bibnamefont {Wang}}, \bibinfo {author}
  {\bibfnamefont {X.~L.}\ \bibnamefont {Su}}, \bibinfo {author} {\bibfnamefont
  {C.~D.}\ \bibnamefont {Xie}}, \ and\ \bibinfo {author} {\bibfnamefont
  {K.~C.}\ \bibnamefont {Peng}},\ }\bibfield  {title} {\enquote {\bibinfo
  {title} {{Manipulating the direction of Einstein-Podolsky-Rosen steering}},}\
  }\href {\doibase 10.1103/PhysRevA.95.052114} {\bibfield  {journal} {\bibinfo
  {journal} {Phys. Rev. A}\ }\textbf {\bibinfo {volume} {95}},\ \bibinfo
  {pages} {052114} (\bibinfo {year} {2017})}\BibitemShut {NoStop}%
\bibitem [{\citenamefont {Sun}\ \emph {et~al.}(2016)\citenamefont {Sun},
  \citenamefont {Ye}, \citenamefont {Xu}, \citenamefont {Xu}, \citenamefont
  {Tang}, \citenamefont {Wu}, \citenamefont {Chen}, \citenamefont {Li},\ and\
  \citenamefont {Guo}}]{SunKai2016PRL}%
  \BibitemOpen
  \bibfield  {author} {\bibinfo {author} {\bibfnamefont {K.}~\bibnamefont
  {Sun}}, \bibinfo {author} {\bibfnamefont {X.~J.}\ \bibnamefont {Ye}},
  \bibinfo {author} {\bibfnamefont {J.~S.}\ \bibnamefont {Xu}}, \bibinfo
  {author} {\bibfnamefont {X.~Y.}\ \bibnamefont {Xu}}, \bibinfo {author}
  {\bibfnamefont {J.~S.}\ \bibnamefont {Tang}}, \bibinfo {author}
  {\bibfnamefont {Y.~C.}\ \bibnamefont {Wu}}, \bibinfo {author} {\bibfnamefont
  {J.~L.}\ \bibnamefont {Chen}}, \bibinfo {author} {\bibfnamefont {C.~F.}\
  \bibnamefont {Li}}, \ and\ \bibinfo {author} {\bibfnamefont {G.~C.}\
  \bibnamefont {Guo}},\ }\bibfield  {title} {\enquote {\bibinfo {title}
  {{Experimental Quantification of Asymmetric Einstein-Podolsky-Rosen
  Steering}},}\ }\href {\doibase 10.1103/PhysRevLett.116.160404} {\bibfield
  {journal} {\bibinfo  {journal} {Phys. Rev. Lett.}\ }\textbf {\bibinfo
  {volume} {116}},\ \bibinfo {pages} {160404} (\bibinfo {year}
  {2016})}\BibitemShut {NoStop}%
\bibitem [{\citenamefont {Wollmann}\ \emph {et~al.}(2016)\citenamefont
  {Wollmann}, \citenamefont {Walk}, \citenamefont {Bennet}, \citenamefont
  {Wiseman},\ and\ \citenamefont {Pryde}}]{Wollmann2016}%
  \BibitemOpen
  \bibfield  {author} {\bibinfo {author} {\bibfnamefont {S.}~\bibnamefont
  {Wollmann}}, \bibinfo {author} {\bibfnamefont {N.}~\bibnamefont {Walk}},
  \bibinfo {author} {\bibfnamefont {A.~J.}\ \bibnamefont {Bennet}}, \bibinfo
  {author} {\bibfnamefont {H.~M.}\ \bibnamefont {Wiseman}}, \ and\ \bibinfo
  {author} {\bibfnamefont {G.~J.}\ \bibnamefont {Pryde}},\ }\bibfield  {title}
  {\enquote {\bibinfo {title} {{Observation of Genuine One-Way
  Einstein-Podolsky-Rosen Steering}},}\ }\href {\doibase
  10.1103/PhysRevLett.116.160403} {\bibfield  {journal} {\bibinfo  {journal}
  {Phys. Rev. Lett.}\ }\textbf {\bibinfo {volume} {116}},\ \bibinfo {pages}
  {160403} (\bibinfo {year} {2016})}\BibitemShut {NoStop}%
\bibitem [{\citenamefont {Xiao}\ \emph {et~al.}(2017)\citenamefont {Xiao},
  \citenamefont {Ye}, \citenamefont {Sun}, \citenamefont {Xu}, \citenamefont
  {Li},\ and\ \citenamefont {Guo}}]{XiaoYa2017PRL}%
  \BibitemOpen
  \bibfield  {author} {\bibinfo {author} {\bibfnamefont {Y.}~\bibnamefont
  {Xiao}}, \bibinfo {author} {\bibfnamefont {X.~J.}\ \bibnamefont {Ye}},
  \bibinfo {author} {\bibfnamefont {K.}~\bibnamefont {Sun}}, \bibinfo {author}
  {\bibfnamefont {J.~S.}\ \bibnamefont {Xu}}, \bibinfo {author} {\bibfnamefont
  {C.~F.}\ \bibnamefont {Li}}, \ and\ \bibinfo {author} {\bibfnamefont {G.~C.}\
  \bibnamefont {Guo}},\ }\bibfield  {title} {\enquote {\bibinfo {title}
  {{Demonstration of Multisetting One-Way Einstein-Podolsky-Rosen Steering in
  Two-Qubit Systems}},}\ }\href {\doibase 10.1103/PhysRevLett.118.140404}
  {\bibfield  {journal} {\bibinfo  {journal} {Phys. Rev. Lett.}\ }\textbf
  {\bibinfo {volume} {118}},\ \bibinfo {pages} {140404} (\bibinfo {year}
  {2017})}\BibitemShut {NoStop}%
\bibitem [{\citenamefont {Tischler}\ \emph {et~al.}(2018)\citenamefont
  {Tischler}, \citenamefont {Ghafari}, \citenamefont {Baker}, \citenamefont
  {Slussarenko}, \citenamefont {Patel}, \citenamefont {Weston}, \citenamefont
  {Wollmann}, \citenamefont {Shalm}, \citenamefont {Verma}, \citenamefont
  {Nam}, \citenamefont {Nguyen}, \citenamefont {Wiseman},\ and\ \citenamefont
  {Pryde}}]{Tischler2018PRL}%
  \BibitemOpen
  \bibfield  {author} {\bibinfo {author} {\bibfnamefont {N.}~\bibnamefont
  {Tischler}}, \bibinfo {author} {\bibfnamefont {F.}~\bibnamefont {Ghafari}},
  \bibinfo {author} {\bibfnamefont {T.~J.}\ \bibnamefont {Baker}}, \bibinfo
  {author} {\bibfnamefont {S.}~\bibnamefont {Slussarenko}}, \bibinfo {author}
  {\bibfnamefont {R.~B.}\ \bibnamefont {Patel}}, \bibinfo {author}
  {\bibfnamefont {M.~M.}\ \bibnamefont {Weston}}, \bibinfo {author}
  {\bibfnamefont {S.}~\bibnamefont {Wollmann}}, \bibinfo {author}
  {\bibfnamefont {L.~K.}\ \bibnamefont {Shalm}}, \bibinfo {author}
  {\bibfnamefont {V.~B.}\ \bibnamefont {Verma}}, \bibinfo {author}
  {\bibfnamefont {S.~W.}\ \bibnamefont {Nam}}, \bibinfo {author} {\bibfnamefont
  {H.~C.}\ \bibnamefont {Nguyen}}, \bibinfo {author} {\bibfnamefont {H.~M.}\
  \bibnamefont {Wiseman}}, \ and\ \bibinfo {author} {\bibfnamefont {G.~J.}\
  \bibnamefont {Pryde}},\ }\bibfield  {title} {\enquote {\bibinfo {title}
  {{Conclusive experimental demonstration of one-way Einstein-Podolsky-Rosen
  steering}},}\ }\href {\doibase 10.1103/PhysRevLett.121.100401} {\bibfield
  {journal} {\bibinfo  {journal} {Phys. Rev. Lett.}\ }\textbf {\bibinfo
  {volume} {121}},\ \bibinfo {pages} {100401} (\bibinfo {year}
  {2018})}\BibitemShut {NoStop}%
\bibitem [{\citenamefont {Fadel}\ \emph {et~al.}(2018)\citenamefont {Fadel},
  \citenamefont {Zibold}, \citenamefont {D\'{e}camps},\ and\ \citenamefont
  {Treutlein}}]{Fadel2018Science}%
  \BibitemOpen
  \bibfield  {author} {\bibinfo {author} {\bibfnamefont {M.}~\bibnamefont
  {Fadel}}, \bibinfo {author} {\bibfnamefont {T.}~\bibnamefont {Zibold}},
  \bibinfo {author} {\bibfnamefont {B.}~\bibnamefont {D\'{e}camps}}, \ and\
  \bibinfo {author} {\bibfnamefont {P.}~\bibnamefont {Treutlein}},\ }\bibfield
  {title} {\enquote {\bibinfo {title} {{Spatial entanglement patterns and
  Einstein-Podolsky-Rosen steering in Bose-Einstein condensates}},}\ }\href
  {\doibase 10.1126/science.aao1850} {\bibfield  {journal} {\bibinfo  {journal}
  {Science}\ }\textbf {\bibinfo {volume} {360}},\ \bibinfo {pages} {409}
  (\bibinfo {year} {2018})}\BibitemShut {NoStop}%
\bibitem [{\citenamefont {Opanchuk}, \citenamefont {Arnaud},\ and\
  \citenamefont {Reid}(2014)}]{Opanchuk2014}%
  \BibitemOpen
  \bibfield  {author} {\bibinfo {author} {\bibfnamefont {B.}~\bibnamefont
  {Opanchuk}}, \bibinfo {author} {\bibfnamefont {L.}~\bibnamefont {Arnaud}}, \
  and\ \bibinfo {author} {\bibfnamefont {M.~D.}\ \bibnamefont {Reid}},\
  }\bibfield  {title} {\enquote {\bibinfo {title} {Detecting faked
  continuous-variable entanglement using one-sided device-independent
  entanglement witnesses},}\ }\href {\doibase 10.1103/PhysRevA.89.062101}
  {\bibfield  {journal} {\bibinfo  {journal} {Phys. Rev. A}\ }\textbf {\bibinfo
  {volume} {89}},\ \bibinfo {pages} {062101} (\bibinfo {year}
  {2014})}\BibitemShut {NoStop}%
\bibitem [{\citenamefont {He}\ \emph {et~al.}(2015)\citenamefont {He},
  \citenamefont {Rosales-Z{\'{a}}rate}, \citenamefont {Adesso},\ and\
  \citenamefont {Reid}}]{He2015PRL_Teleportation}%
  \BibitemOpen
  \bibfield  {author} {\bibinfo {author} {\bibfnamefont {Q.~Y.}\ \bibnamefont
  {He}}, \bibinfo {author} {\bibfnamefont {L.}~\bibnamefont
  {Rosales-Z{\'{a}}rate}}, \bibinfo {author} {\bibfnamefont {G.}~\bibnamefont
  {Adesso}}, \ and\ \bibinfo {author} {\bibfnamefont {M.~D.}\ \bibnamefont
  {Reid}},\ }\bibfield  {title} {\enquote {\bibinfo {title} {{Secure Continuous
  Variable Teleportation and Einstein-Podolsky-Rosen Steering}},}\ }\href
  {\doibase 10.1103/PhysRevLett.115.180502} {\bibfield  {journal} {\bibinfo
  {journal} {Phys. Rev. Lett.}\ }\textbf {\bibinfo {volume} {115}},\ \bibinfo
  {pages} {180502} (\bibinfo {year} {2015})}\BibitemShut {NoStop}%
\bibitem [{\citenamefont {Reid}(2013)}]{Reid2013PRA}%
  \BibitemOpen
  \bibfield  {author} {\bibinfo {author} {\bibfnamefont {M.~D.}\ \bibnamefont
  {Reid}},\ }\bibfield  {title} {\enquote {\bibinfo {title} {{Signifying
  quantum benchmarks for qubit teleportation and secure quantum communication
  using Einstein-Podolsky-Rosen steering inequalities}},}\ }\href {\doibase
  10.1103/PhysRevA.88.062338} {\bibfield  {journal} {\bibinfo  {journal} {Phys.
  Rev. A}\ }\textbf {\bibinfo {volume} {88}},\ \bibinfo {pages} {062338}
  (\bibinfo {year} {2013})}\BibitemShut {NoStop}%
\bibitem [{\citenamefont {Chiu}\ \emph {et~al.}(2016)\citenamefont {Chiu},
  \citenamefont {Lambert}, \citenamefont {Liao}, \citenamefont {Nori},\ and\
  \citenamefont {Li}}]{CMLi2016NPJ_QI}%
  \BibitemOpen
  \bibfield  {author} {\bibinfo {author} {\bibfnamefont {C.-Y.}\ \bibnamefont
  {Chiu}}, \bibinfo {author} {\bibfnamefont {N.}~\bibnamefont {Lambert}},
  \bibinfo {author} {\bibfnamefont {T.-L.}\ \bibnamefont {Liao}}, \bibinfo
  {author} {\bibfnamefont {F.}~\bibnamefont {Nori}}, \ and\ \bibinfo {author}
  {\bibfnamefont {C.-M.}\ \bibnamefont {Li}},\ }\bibfield  {title} {\enquote
  {\bibinfo {title} {{No-cloning of quantum steering}},}\ }\href {\doibase
  10.1038/npjqi.2016.20} {\bibfield  {journal} {\bibinfo  {journal} {npj
  Quantum Information}\ }\textbf {\bibinfo {volume} {2}},\ \bibinfo {pages}
  {16020} (\bibinfo {year} {2016})}\BibitemShut {NoStop}%
\bibitem [{\citenamefont {Branciard}\ \emph {et~al.}(2012)\citenamefont
  {Branciard}, \citenamefont {Cavalcanti}, \citenamefont {Walborn},
  \citenamefont {Scarani},\ and\ \citenamefont {Wiseman}}]{Branciard2012QKD}%
  \BibitemOpen
  \bibfield  {author} {\bibinfo {author} {\bibfnamefont {C.}~\bibnamefont
  {Branciard}}, \bibinfo {author} {\bibfnamefont {E.~G.}\ \bibnamefont
  {Cavalcanti}}, \bibinfo {author} {\bibfnamefont {S.~P.}\ \bibnamefont
  {Walborn}}, \bibinfo {author} {\bibfnamefont {V.}~\bibnamefont {Scarani}}, \
  and\ \bibinfo {author} {\bibfnamefont {H.~M.}\ \bibnamefont {Wiseman}},\
  }\bibfield  {title} {\enquote {\bibinfo {title} {{One-sided
  device-independent quantum key distribution: Security, feasibility, and the
  connection with steering}},}\ }\href {\doibase 10.1103/PhysRevA.85.010301}
  {\bibfield  {journal} {\bibinfo  {journal} {Phys. Rev. A}\ }\textbf {\bibinfo
  {volume} {85}},\ \bibinfo {pages} {010301(R)} (\bibinfo {year}
  {2012})}\BibitemShut {NoStop}%
\bibitem [{\citenamefont {Gehring}\ \emph {et~al.}(2015)\citenamefont
  {Gehring}, \citenamefont {H{\"{a}}ndchen}, \citenamefont {Duhme},
  \citenamefont {Furrer}, \citenamefont {Franz}, \citenamefont {Pacher},
  \citenamefont {Werner},\ and\ \citenamefont {Schnabel}}]{Gehring2015QKD}%
  \BibitemOpen
  \bibfield  {author} {\bibinfo {author} {\bibfnamefont {T.}~\bibnamefont
  {Gehring}}, \bibinfo {author} {\bibfnamefont {V.}~\bibnamefont
  {H{\"{a}}ndchen}}, \bibinfo {author} {\bibfnamefont {J.}~\bibnamefont
  {Duhme}}, \bibinfo {author} {\bibfnamefont {F.}~\bibnamefont {Furrer}},
  \bibinfo {author} {\bibfnamefont {T.}~\bibnamefont {Franz}}, \bibinfo
  {author} {\bibfnamefont {C.}~\bibnamefont {Pacher}}, \bibinfo {author}
  {\bibfnamefont {R.~F.}\ \bibnamefont {Werner}}, \ and\ \bibinfo {author}
  {\bibfnamefont {R.}~\bibnamefont {Schnabel}},\ }\bibfield  {title} {\enquote
  {\bibinfo {title} {{Implementation of continuous-variable quantum key
  distribution with composable and one-sided-device-independent security
  against coherent attacks}},}\ }\href {\doibase 10.1038/ncomms9795} {\bibfield
   {journal} {\bibinfo  {journal} {Nat. Commun.}\ }\textbf {\bibinfo {volume}
  {6}},\ \bibinfo {pages} {8795} (\bibinfo {year} {2015})}\BibitemShut
  {NoStop}%
\bibitem [{\citenamefont {Walk}\ \emph {et~al.}(2016)\citenamefont {Walk},
  \citenamefont {Hosseni}, \citenamefont {Geng}, \citenamefont {Thearle},
  \citenamefont {Haw}, \citenamefont {Armstrong}, \citenamefont {Assad},
  \citenamefont {Janousek}, \citenamefont {Ralph}, \citenamefont {Symul},
  \citenamefont {Wiseman},\ and\ \citenamefont {Lam}}]{Walk2016QKD}%
  \BibitemOpen
  \bibfield  {author} {\bibinfo {author} {\bibfnamefont {N.}~\bibnamefont
  {Walk}}, \bibinfo {author} {\bibfnamefont {S.}~\bibnamefont {Hosseni}},
  \bibinfo {author} {\bibfnamefont {J.}~\bibnamefont {Geng}}, \bibinfo {author}
  {\bibfnamefont {O.}~\bibnamefont {Thearle}}, \bibinfo {author} {\bibfnamefont
  {J.~Y.}\ \bibnamefont {Haw}}, \bibinfo {author} {\bibfnamefont
  {S.}~\bibnamefont {Armstrong}}, \bibinfo {author} {\bibfnamefont {S.~M.}\
  \bibnamefont {Assad}}, \bibinfo {author} {\bibfnamefont {J.}~\bibnamefont
  {Janousek}}, \bibinfo {author} {\bibfnamefont {T.~C.}\ \bibnamefont {Ralph}},
  \bibinfo {author} {\bibfnamefont {T.}~\bibnamefont {Symul}}, \bibinfo
  {author} {\bibfnamefont {H.~M.}\ \bibnamefont {Wiseman}}, \ and\ \bibinfo
  {author} {\bibfnamefont {P.~K.}\ \bibnamefont {Lam}},\ }\bibfield  {title}
  {\enquote {\bibinfo {title} {{Experimental demonstration of Gaussian
  protocols for one-sided device-independent quantum key distribution}},}\
  }\href {\doibase 10.1364/OPTICA.3.000634} {\bibfield  {journal} {\bibinfo
  {journal} {Optica}\ }\textbf {\bibinfo {volume} {3}},\ \bibinfo {pages} {634}
  (\bibinfo {year} {2016})}\BibitemShut {NoStop}%
\bibitem [{\citenamefont {Armstrong}\ \emph
  {et~al.}(2015{\natexlab{b}})\citenamefont {Armstrong}, \citenamefont {Wang},
  \citenamefont {Teh}, \citenamefont {Gong}, \citenamefont {He}, \citenamefont
  {Janousek}, \citenamefont {Bachor}, \citenamefont {Reid},\ and\ \citenamefont
  {Lam}}]{Armstrong2015QSS}%
  \BibitemOpen
  \bibfield  {author} {\bibinfo {author} {\bibfnamefont {S.}~\bibnamefont
  {Armstrong}}, \bibinfo {author} {\bibfnamefont {M.}~\bibnamefont {Wang}},
  \bibinfo {author} {\bibfnamefont {R.~Y.}\ \bibnamefont {Teh}}, \bibinfo
  {author} {\bibfnamefont {Q.~H.}\ \bibnamefont {Gong}}, \bibinfo {author}
  {\bibfnamefont {Q.~Y.}\ \bibnamefont {He}}, \bibinfo {author} {\bibfnamefont
  {J.}~\bibnamefont {Janousek}}, \bibinfo {author} {\bibfnamefont {H.~A.}\
  \bibnamefont {Bachor}}, \bibinfo {author} {\bibfnamefont {M.~D.}\
  \bibnamefont {Reid}}, \ and\ \bibinfo {author} {\bibfnamefont {P.~K.}\
  \bibnamefont {Lam}},\ }\bibfield  {title} {\enquote {\bibinfo {title}
  {{Multipartite Einstein-Podolsky-Rosen steering and genuine tripartite
  entanglement with optical networks}},}\ }\href {\doibase 10.1038/nphys3202}
  {\bibfield  {journal} {\bibinfo  {journal} {Nat. Phys.}\ }\textbf {\bibinfo
  {volume} {11}},\ \bibinfo {pages} {167} (\bibinfo {year}
  {2015}{\natexlab{b}})}\BibitemShut {NoStop}%
\bibitem [{\citenamefont {Xiang}\ \emph
  {et~al.}(2017{\natexlab{a}})\citenamefont {Xiang}, \citenamefont {Kogias},
  \citenamefont {Adesso},\ and\ \citenamefont {He}}]{Xiang2017QSS}%
  \BibitemOpen
  \bibfield  {author} {\bibinfo {author} {\bibfnamefont {Y.}~\bibnamefont
  {Xiang}}, \bibinfo {author} {\bibfnamefont {I.}~\bibnamefont {Kogias}},
  \bibinfo {author} {\bibfnamefont {G.}~\bibnamefont {Adesso}}, \ and\ \bibinfo
  {author} {\bibfnamefont {Q.~Y.}\ \bibnamefont {He}},\ }\bibfield  {title}
  {\enquote {\bibinfo {title} {Multipartite gaussian steering: Monogamy
  constraints and quantum cryptography applications},}\ }\href {\doibase
  10.1103/PhysRevA.95.010101} {\bibfield  {journal} {\bibinfo  {journal} {Phys.
  Rev. A}\ }\textbf {\bibinfo {volume} {95}},\ \bibinfo {pages} {010101(R)}
  (\bibinfo {year} {2017}{\natexlab{a}})}\BibitemShut {NoStop}%
\bibitem [{\citenamefont {Kogias}\ \emph {et~al.}(2017)\citenamefont {Kogias},
  \citenamefont {Xiang}, \citenamefont {He},\ and\ \citenamefont
  {Adesso}}]{Kogias2017QSS}%
  \BibitemOpen
  \bibfield  {author} {\bibinfo {author} {\bibfnamefont {I.}~\bibnamefont
  {Kogias}}, \bibinfo {author} {\bibfnamefont {Y.}~\bibnamefont {Xiang}},
  \bibinfo {author} {\bibfnamefont {Q.~Y.}\ \bibnamefont {He}}, \ and\ \bibinfo
  {author} {\bibfnamefont {G.}~\bibnamefont {Adesso}},\ }\bibfield  {title}
  {\enquote {\bibinfo {title} {{Unconditional security of entanglement-based
  continuous-variable quantum secret sharing}},}\ }\href {\doibase
  10.1103/PhysRevA.95.012315} {\bibfield  {journal} {\bibinfo  {journal} {Phys.
  Rev. A}\ }\textbf {\bibinfo {volume} {95}},\ \bibinfo {pages} {012315}
  (\bibinfo {year} {2017})}\BibitemShut {NoStop}%
\bibitem [{\citenamefont {Li}\ \emph {et~al.}(2015)\citenamefont {Li},
  \citenamefont {Chen}, \citenamefont {Chen}, \citenamefont {Zhang},
  \citenamefont {Chen},\ and\ \citenamefont {Pan}}]{Li2015OneWayComputation}%
  \BibitemOpen
  \bibfield  {author} {\bibinfo {author} {\bibfnamefont {C.~M.}\ \bibnamefont
  {Li}}, \bibinfo {author} {\bibfnamefont {K.}~\bibnamefont {Chen}}, \bibinfo
  {author} {\bibfnamefont {Y.~N.}\ \bibnamefont {Chen}}, \bibinfo {author}
  {\bibfnamefont {Q.}~\bibnamefont {Zhang}}, \bibinfo {author} {\bibfnamefont
  {Y.~A.}\ \bibnamefont {Chen}}, \ and\ \bibinfo {author} {\bibfnamefont
  {J.~W.}\ \bibnamefont {Pan}},\ }\bibfield  {title} {\enquote {\bibinfo
  {title} {{Genuine High-Order Einstein-Podolsky-Rosen Steering}},}\ }\href
  {\doibase 10.1103/PhysRevLett.115.010402} {\bibfield  {journal} {\bibinfo
  {journal} {Phys. Rev. Lett.}\ }\textbf {\bibinfo {volume} {115}},\ \bibinfo
  {pages} {010402} (\bibinfo {year} {2015})}\BibitemShut {NoStop}%
\bibitem [{\citenamefont {Skrzypczyk}\ and\ \citenamefont
  {Cavalcanti}(2018)}]{Skrzypczyk2018PRL_RandomnessGeneration}%
  \BibitemOpen
  \bibfield  {author} {\bibinfo {author} {\bibfnamefont {P.}~\bibnamefont
  {Skrzypczyk}}\ and\ \bibinfo {author} {\bibfnamefont {D.}~\bibnamefont
  {Cavalcanti}},\ }\bibfield  {title} {\enquote {\bibinfo {title} {Maximal
  randomness generation from steering inequality violations using qudits},}\
  }\href {\doibase 10.1103/PhysRevLett.120.260401} {\bibfield  {journal}
  {\bibinfo  {journal} {Phys. Rev. Lett.}\ }\textbf {\bibinfo {volume} {120}},\
  \bibinfo {pages} {260401} (\bibinfo {year} {2018})}\BibitemShut {NoStop}%
\bibitem [{\citenamefont {Guo}\ \emph {et~al.}(2019)\citenamefont {Guo},
  \citenamefont {Cheng}, \citenamefont {Hu}, \citenamefont {Liu}, \citenamefont
  {Huang}, \citenamefont {Huang}, \citenamefont {Li}, \citenamefont {Guo},\
  and\ \citenamefont {Cavalcanti}}]{GuoYu2019PRL_RandomnessGeneration}%
  \BibitemOpen
  \bibfield  {author} {\bibinfo {author} {\bibfnamefont {Y.}~\bibnamefont
  {Guo}}, \bibinfo {author} {\bibfnamefont {S.}~\bibnamefont {Cheng}}, \bibinfo
  {author} {\bibfnamefont {X.}~\bibnamefont {Hu}}, \bibinfo {author}
  {\bibfnamefont {B.-H.}\ \bibnamefont {Liu}}, \bibinfo {author} {\bibfnamefont
  {E.-M.}\ \bibnamefont {Huang}}, \bibinfo {author} {\bibfnamefont {Y.-F.}\
  \bibnamefont {Huang}}, \bibinfo {author} {\bibfnamefont {C.-F.}\ \bibnamefont
  {Li}}, \bibinfo {author} {\bibfnamefont {G.-C.}\ \bibnamefont {Guo}}, \ and\
  \bibinfo {author} {\bibfnamefont {E.~G.}\ \bibnamefont {Cavalcanti}},\
  }\bibfield  {title} {\enquote {\bibinfo {title} {Experimental
  measurement-device-independent quantum steering and randomness generation
  beyond qubits},}\ }\href {\doibase 10.1103/PhysRevLett.123.170402} {\bibfield
   {journal} {\bibinfo  {journal} {Phys. Rev. Lett.}\ }\textbf {\bibinfo
  {volume} {123}},\ \bibinfo {pages} {170402} (\bibinfo {year}
  {2019})}\BibitemShut {NoStop}%
\bibitem [{\citenamefont {Piani}\ and\ \citenamefont
  {Watrous}(2015)}]{Piani2015PRL}%
  \BibitemOpen
  \bibfield  {author} {\bibinfo {author} {\bibfnamefont {M.}~\bibnamefont
  {Piani}}\ and\ \bibinfo {author} {\bibfnamefont {J.}~\bibnamefont
  {Watrous}},\ }\bibfield  {title} {\enquote {\bibinfo {title} {{Necessary and
  sufficient quantum information characterization of Einstein-Podolsky-Rosen
  steering}},}\ }\href {\doibase 10.1103/PhysRevLett.114.060404} {\bibfield
  {journal} {\bibinfo  {journal} {Phys. Rev. Lett.}\ }\textbf {\bibinfo
  {volume} {114}},\ \bibinfo {pages} {060604} (\bibinfo {year}
  {2015})}\BibitemShut {NoStop}%
\bibitem [{\citenamefont {{Holevo}}(1975)}]{Holevo1975Gaussian}%
  \BibitemOpen
  \bibfield  {author} {\bibinfo {author} {\bibfnamefont {A.~S.}\ \bibnamefont
  {{Holevo}}},\ }\bibfield  {title} {\enquote {\bibinfo {title} {Some
  statistical problems for quantum gaussian states},}\ }\href {\doibase
  10.1109/TIT.1975.1055441} {\bibfield  {journal} {\bibinfo  {journal} {IEEE
  Trans. Inf. Theory}\ }\textbf {\bibinfo {volume} {21}},\ \bibinfo {pages}
  {533} (\bibinfo {year} {1975})}\BibitemShut {NoStop}%
\bibitem [{\citenamefont {{Holevo}}(2011)}]{Holevo2011Gaussian}%
  \BibitemOpen
  \bibfield  {author} {\bibinfo {author} {\bibfnamefont {A.~S.}\ \bibnamefont
  {{Holevo}}},\ }\href@noop {} {\emph {\bibinfo {title} {Probabilistic and
  statistical aspects of quantum theory}}}\ (\bibinfo  {publisher} {Edizioni
  della Normale},\ \bibinfo {year} {2011})\BibitemShut {NoStop}%
\bibitem [{\citenamefont {Vidal}\ and\ \citenamefont
  {Werner}(2002)}]{Vidal2002_LogNegativity}%
  \BibitemOpen
  \bibfield  {author} {\bibinfo {author} {\bibfnamefont {G.}~\bibnamefont
  {Vidal}}\ and\ \bibinfo {author} {\bibfnamefont {R.~F.}\ \bibnamefont
  {Werner}},\ }\bibfield  {title} {\enquote {\bibinfo {title} {{Computable
  measure of entanglement}},}\ }\href {\doibase 10.1103/PhysRevA.65.032314}
  {\bibfield  {journal} {\bibinfo  {journal} {Phys. Rev. A}\ }\textbf {\bibinfo
  {volume} {65}},\ \bibinfo {pages} {032314} (\bibinfo {year}
  {2002})}\BibitemShut {NoStop}%
\bibitem [{\citenamefont {Adesso}, \citenamefont {Serafini},\ and\
  \citenamefont {Illuminati}(2004)}]{Adesso2004_LogNegativity}%
  \BibitemOpen
  \bibfield  {author} {\bibinfo {author} {\bibfnamefont {G.}~\bibnamefont
  {Adesso}}, \bibinfo {author} {\bibfnamefont {A.}~\bibnamefont {Serafini}}, \
  and\ \bibinfo {author} {\bibfnamefont {F.}~\bibnamefont {Illuminati}},\
  }\bibfield  {title} {\enquote {\bibinfo {title} {{Extremal entanglement and
  mixedness in continuous variable systems}},}\ }\href {\doibase
  10.1103/PhysRevA.70.022318} {\bibfield  {journal} {\bibinfo  {journal} {Phys.
  Rev. A}\ }\textbf {\bibinfo {volume} {70}},\ \bibinfo {pages} {022318}
  (\bibinfo {year} {2004})}\BibitemShut {NoStop}%
\bibitem [{\citenamefont {Kogias}\ \emph {et~al.}(2015)\citenamefont {Kogias},
  \citenamefont {Lee}, \citenamefont {Ragy},\ and\ \citenamefont
  {Adesso}}]{Kogias2015_BipartiteSteering}%
  \BibitemOpen
  \bibfield  {author} {\bibinfo {author} {\bibfnamefont {I.}~\bibnamefont
  {Kogias}}, \bibinfo {author} {\bibfnamefont {A.~R.}\ \bibnamefont {Lee}},
  \bibinfo {author} {\bibfnamefont {S.}~\bibnamefont {Ragy}}, \ and\ \bibinfo
  {author} {\bibfnamefont {G.}~\bibnamefont {Adesso}},\ }\bibfield  {title}
  {\enquote {\bibinfo {title} {{Quantification of Gaussian quantum
  steering}},}\ }\href {\doibase 10.1103/PhysRevLett.114.060403} {\bibfield
  {journal} {\bibinfo  {journal} {Phys. Rev. Lett.}\ }\textbf {\bibinfo
  {volume} {114}},\ \bibinfo {pages} {060403} (\bibinfo {year}
  {2015})}\BibitemShut {NoStop}%
\bibitem [{\citenamefont {Duan}\ \emph {et~al.}(2000)\citenamefont {Duan},
  \citenamefont {Giedke}, \citenamefont {Cirac},\ and\ \citenamefont
  {Zoller}}]{DuanLuming2020_Ent_Inequality}%
  \BibitemOpen
  \bibfield  {author} {\bibinfo {author} {\bibfnamefont {L.-M.}\ \bibnamefont
  {Duan}}, \bibinfo {author} {\bibfnamefont {G.}~\bibnamefont {Giedke}},
  \bibinfo {author} {\bibfnamefont {J.~I.}\ \bibnamefont {Cirac}}, \ and\
  \bibinfo {author} {\bibfnamefont {P.}~\bibnamefont {Zoller}},\ }\bibfield
  {title} {\enquote {\bibinfo {title} {Inseparability criterion for continuous
  variable systems},}\ }\href {\doibase 10.1103/PhysRevLett.84.2722} {\bibfield
   {journal} {\bibinfo  {journal} {Phys. Rev. Lett.}\ }\textbf {\bibinfo
  {volume} {84}},\ \bibinfo {pages} {2722--2725} (\bibinfo {year}
  {2000})}\BibitemShut {NoStop}%
\bibitem [{\citenamefont {Adesso}\ and\ \citenamefont
  {Illuminati}(2006)}]{Gerardo2006NJP_TripartiteEnt}%
  \BibitemOpen
  \bibfield  {author} {\bibinfo {author} {\bibfnamefont {G.}~\bibnamefont
  {Adesso}}\ and\ \bibinfo {author} {\bibfnamefont {F.}~\bibnamefont
  {Illuminati}},\ }\bibfield  {title} {\enquote {\bibinfo {title} {Continuous
  variable tangle, monogamy inequality, and entanglement sharing in gaussian
  states of continuous variable systems},}\ }\href {\doibase
  10.1088/1367-2630/8/1/015} {\bibfield  {journal} {\bibinfo  {journal} {New J.
  Phys.}\ }\textbf {\bibinfo {volume} {8}},\ \bibinfo {pages} {15} (\bibinfo
  {year} {2006})}\BibitemShut {NoStop}%
\bibitem [{\citenamefont {Adesso}\ and\ \citenamefont
  {Illuminati}(2007)}]{Gerardo2007JPA_EntReview}%
  \BibitemOpen
  \bibfield  {author} {\bibinfo {author} {\bibfnamefont {G.}~\bibnamefont
  {Adesso}}\ and\ \bibinfo {author} {\bibfnamefont {F.}~\bibnamefont
  {Illuminati}},\ }\bibfield  {title} {\enquote {\bibinfo {title} {Entanglement
  in continuous-variable systems: recent advances and current perspectives},}\
  }\href {\doibase 10.1088/1751-8113/40/28/S01} {\bibfield  {journal} {\bibinfo
   {journal} {J. Phys. A}\ }\textbf {\bibinfo {volume} {40}},\ \bibinfo {pages}
  {7821} (\bibinfo {year} {2007})}\BibitemShut {NoStop}%
\bibitem [{\citenamefont {Xiang}\ \emph
  {et~al.}(2017{\natexlab{b}})\citenamefont {Xiang}, \citenamefont {Kogias},
  \citenamefont {Adesso},\ and\ \citenamefont
  {He}}]{XiangYu2017PRA_SteeringMonogamy}%
  \BibitemOpen
  \bibfield  {author} {\bibinfo {author} {\bibfnamefont {Y.}~\bibnamefont
  {Xiang}}, \bibinfo {author} {\bibfnamefont {I.}~\bibnamefont {Kogias}},
  \bibinfo {author} {\bibfnamefont {G.}~\bibnamefont {Adesso}}, \ and\ \bibinfo
  {author} {\bibfnamefont {Q.}~\bibnamefont {He}},\ }\bibfield  {title}
  {\enquote {\bibinfo {title} {Multipartite gaussian steering: Monogamy
  constraints and quantum cryptography applications},}\ }\href {\doibase
  10.1103/PhysRevA.95.010101} {\bibfield  {journal} {\bibinfo  {journal} {Phys.
  Rev. A}\ }\textbf {\bibinfo {volume} {95}},\ \bibinfo {pages} {010101}
  (\bibinfo {year} {2017}{\natexlab{b}})}\BibitemShut {NoStop}%
\bibitem [{\citenamefont {He}\ and\ \citenamefont
  {Reid}(2013{\natexlab{b}})}]{HeQY2013PRL_MultipartiteSteering}%
  \BibitemOpen
  \bibfield  {author} {\bibinfo {author} {\bibfnamefont {Q.~Y.}\ \bibnamefont
  {He}}\ and\ \bibinfo {author} {\bibfnamefont {M.~D.}\ \bibnamefont {Reid}},\
  }\bibfield  {title} {\enquote {\bibinfo {title} {Genuine multipartite
  einstein-podolsky-rosen steering},}\ }\href {\doibase
  10.1103/PhysRevLett.111.250403} {\bibfield  {journal} {\bibinfo  {journal}
  {Phys. Rev. Lett.}\ }\textbf {\bibinfo {volume} {111}},\ \bibinfo {pages}
  {250403} (\bibinfo {year} {2013}{\natexlab{b}})}\BibitemShut {NoStop}%
\bibitem [{\citenamefont {Zurek}(2003)}]{Zurek2003review_Decoherence}%
  \BibitemOpen
  \bibfield  {author} {\bibinfo {author} {\bibfnamefont {W.~H.}\ \bibnamefont
  {Zurek}},\ }\bibfield  {title} {\enquote {\bibinfo {title} {Decoherence,
  einselection, and the quantum origins of the classical},}\ }\href {\doibase
  10.1103/RevModPhys.75.715} {\bibfield  {journal} {\bibinfo  {journal} {Rev.
  Mod. Phys.}\ }\textbf {\bibinfo {volume} {75}},\ \bibinfo {pages} {715}
  (\bibinfo {year} {2003})}\BibitemShut {NoStop}%
\bibitem [{\citenamefont {Fr\"owis}\ \emph {et~al.}(2018)\citenamefont
  {Fr\"owis}, \citenamefont {Sekatski}, \citenamefont {D\"ur}, \citenamefont
  {Gisin},\ and\ \citenamefont {Sangouard}}]{Florian2018RMP_MacroState}%
  \BibitemOpen
  \bibfield  {author} {\bibinfo {author} {\bibfnamefont {F.}~\bibnamefont
  {Fr\"owis}}, \bibinfo {author} {\bibfnamefont {P.}~\bibnamefont {Sekatski}},
  \bibinfo {author} {\bibfnamefont {W.}~\bibnamefont {D\"ur}}, \bibinfo
  {author} {\bibfnamefont {N.}~\bibnamefont {Gisin}}, \ and\ \bibinfo {author}
  {\bibfnamefont {N.}~\bibnamefont {Sangouard}},\ }\bibfield  {title} {\enquote
  {\bibinfo {title} {Macroscopic quantum states: Measures, fragility, and
  implementations},}\ }\href {\doibase 10.1103/RevModPhys.90.025004} {\bibfield
   {journal} {\bibinfo  {journal} {Rev. Mod. Phys.}\ }\textbf {\bibinfo
  {volume} {90}},\ \bibinfo {pages} {025004} (\bibinfo {year}
  {2018})}\BibitemShut {NoStop}%
\bibitem [{\citenamefont {Bassi}\ \emph {et~al.}(2013)\citenamefont {Bassi},
  \citenamefont {Lochan}, \citenamefont {Satin}, \citenamefont {Singh},\ and\
  \citenamefont {Ulbricht}}]{Angelo2013RMP_WaveCollapse}%
  \BibitemOpen
  \bibfield  {author} {\bibinfo {author} {\bibfnamefont {A.}~\bibnamefont
  {Bassi}}, \bibinfo {author} {\bibfnamefont {K.}~\bibnamefont {Lochan}},
  \bibinfo {author} {\bibfnamefont {S.}~\bibnamefont {Satin}}, \bibinfo
  {author} {\bibfnamefont {T.~P.}\ \bibnamefont {Singh}}, \ and\ \bibinfo
  {author} {\bibfnamefont {H.}~\bibnamefont {Ulbricht}},\ }\bibfield  {title}
  {\enquote {\bibinfo {title} {Models of wave-function collapse, underlying
  theories, and experimental tests},}\ }\href {\doibase
  10.1103/RevModPhys.85.471} {\bibfield  {journal} {\bibinfo  {journal} {Rev.
  Mod. Phys.}\ }\textbf {\bibinfo {volume} {85}},\ \bibinfo {pages} {471}
  (\bibinfo {year} {2013})}\BibitemShut {NoStop}%
\bibitem [{\citenamefont {Zhang}, \citenamefont {Zhang},\ and\ \citenamefont
  {Li}(2017)}]{ZhangJing2017PRA_WaveCollaspe}%
  \BibitemOpen
  \bibfield  {author} {\bibinfo {author} {\bibfnamefont {J.}~\bibnamefont
  {Zhang}}, \bibinfo {author} {\bibfnamefont {T.}~\bibnamefont {Zhang}}, \ and\
  \bibinfo {author} {\bibfnamefont {J.}~\bibnamefont {Li}},\ }\bibfield
  {title} {\enquote {\bibinfo {title} {Probing spontaneous wave-function
  collapse with entangled levitating nanospheres},}\ }\href {\doibase
  10.1103/PhysRevA.95.012141} {\bibfield  {journal} {\bibinfo  {journal} {Phys.
  Rev. A}\ }\textbf {\bibinfo {volume} {95}},\ \bibinfo {pages} {012141}
  (\bibinfo {year} {2017})}\BibitemShut {NoStop}%
\bibitem [{\citenamefont {Kamra}\ \emph {et~al.}(2019)\citenamefont {Kamra},
  \citenamefont {Thingstad}, \citenamefont {Rastelli}, \citenamefont {Duine},
  \citenamefont {Brataas}, \citenamefont {Belzig},\ and\ \citenamefont
  {Sudb\o{}}}]{Kamra2019PRB_AntiferroEnt}%
  \BibitemOpen
  \bibfield  {author} {\bibinfo {author} {\bibfnamefont {A.}~\bibnamefont
  {Kamra}}, \bibinfo {author} {\bibfnamefont {E.}~\bibnamefont {Thingstad}},
  \bibinfo {author} {\bibfnamefont {G.}~\bibnamefont {Rastelli}}, \bibinfo
  {author} {\bibfnamefont {R.~A.}\ \bibnamefont {Duine}}, \bibinfo {author}
  {\bibfnamefont {A.}~\bibnamefont {Brataas}}, \bibinfo {author} {\bibfnamefont
  {W.}~\bibnamefont {Belzig}}, \ and\ \bibinfo {author} {\bibfnamefont
  {A.}~\bibnamefont {Sudb\o{}}},\ }\bibfield  {title} {\enquote {\bibinfo
  {title} {Antiferromagnetic magnons as highly squeezed fock states underlying
  quantum correlations},}\ }\href {\doibase 10.1103/PhysRevB.100.174407}
  {\bibfield  {journal} {\bibinfo  {journal} {Phys. Rev. B}\ }\textbf {\bibinfo
  {volume} {100}},\ \bibinfo {pages} {174407} (\bibinfo {year}
  {2019})}\BibitemShut {NoStop}%
\bibitem [{\citenamefont {Li}\ and\ \citenamefont
  {Zhu}(2019)}]{LiJie2019Magon-MagnonEnt}%
  \BibitemOpen
  \bibfield  {author} {\bibinfo {author} {\bibfnamefont {J.}~\bibnamefont
  {Li}}\ and\ \bibinfo {author} {\bibfnamefont {S.~Y.}\ \bibnamefont {Zhu}},\
  }\bibfield  {title} {\enquote {\bibinfo {title} {Entangling two magnon modes
  via magnetostrictive interaction},}\ }\href {\doibase
  10.1088/1367-2630/ab3508} {\bibfield  {journal} {\bibinfo  {journal} {New J.
  Phys.}\ }\textbf {\bibinfo {volume} {21}},\ \bibinfo {pages} {085001}
  (\bibinfo {year} {2019})}\BibitemShut {NoStop}%
\bibitem [{\citenamefont {Zhang}, \citenamefont {Scully},\ and\ \citenamefont
  {Agarwal}(2019{\natexlab{b}})}]{ZhangZhedong2019Magnon-MagnonEnt}%
  \BibitemOpen
  \bibfield  {author} {\bibinfo {author} {\bibfnamefont {Z.}~\bibnamefont
  {Zhang}}, \bibinfo {author} {\bibfnamefont {M.~O.}\ \bibnamefont {Scully}}, \
  and\ \bibinfo {author} {\bibfnamefont {G.~S.}\ \bibnamefont {Agarwal}},\
  }\bibfield  {title} {\enquote {\bibinfo {title} {Quantum entanglement between
  two magnon modes via kerr nonlinearity driven far from equilibrium},}\ }\href
  {\doibase 10.1103/PhysRevResearch.1.023021} {\bibfield  {journal} {\bibinfo
  {journal} {Phys. Rev. Res.}\ }\textbf {\bibinfo {volume} {1}},\ \bibinfo
  {pages} {023021} (\bibinfo {year} {2019}{\natexlab{b}})}\BibitemShut
  {NoStop}%
\bibitem [{\citenamefont {Nair}\ and\ \citenamefont
  {Agarwal}(2020)}]{Nair2020Magnon-MagnonEnt}%
  \BibitemOpen
  \bibfield  {author} {\bibinfo {author} {\bibfnamefont {J.~M.~P.}\
  \bibnamefont {Nair}}\ and\ \bibinfo {author} {\bibfnamefont {G.~S.}\
  \bibnamefont {Agarwal}},\ }\bibfield  {title} {\enquote {\bibinfo {title}
  {Deterministic quantum entanglement between macroscopic ferrite samples},}\
  }\href {\doibase 10.1063/5.0015195} {\bibfield  {journal} {\bibinfo
  {journal} {Appl. Phys. Lett.}\ }\textbf {\bibinfo {volume} {117}},\ \bibinfo
  {pages} {084001} (\bibinfo {year} {2020})}\BibitemShut {NoStop}%
\bibitem [{\citenamefont {Yuan}\ \emph
  {et~al.}(2020{\natexlab{b}})\citenamefont {Yuan}, \citenamefont {Yan},
  \citenamefont {Zheng}, \citenamefont {He}, \citenamefont {Xia},\ and\
  \citenamefont {Yung}}]{Yuan2020PRL_Bell}%
  \BibitemOpen
  \bibfield  {author} {\bibinfo {author} {\bibfnamefont {H.~Y.}\ \bibnamefont
  {Yuan}}, \bibinfo {author} {\bibfnamefont {P.}~\bibnamefont {Yan}}, \bibinfo
  {author} {\bibfnamefont {S.}~\bibnamefont {Zheng}}, \bibinfo {author}
  {\bibfnamefont {Q.~Y.}\ \bibnamefont {He}}, \bibinfo {author} {\bibfnamefont
  {K.}~\bibnamefont {Xia}}, \ and\ \bibinfo {author} {\bibfnamefont {M.-H.}\
  \bibnamefont {Yung}},\ }\bibfield  {title} {\enquote {\bibinfo {title}
  {Steady {Bell} state generation via magnon-photon coupling},}\ }\href
  {\doibase 10.1103/PhysRevLett.124.053602} {\bibfield  {journal} {\bibinfo
  {journal} {Phys. Rev. Lett.}\ }\textbf {\bibinfo {volume} {124}},\ \bibinfo
  {pages} {053602} (\bibinfo {year} {2020}{\natexlab{b}})}\BibitemShut
  {NoStop}%
\bibitem [{\citenamefont {Harder}\ \emph
  {et~al.}(2018{\natexlab{b}})\citenamefont {Harder}, \citenamefont {Yang},
  \citenamefont {Yao}, \citenamefont {Yu}, \citenamefont {Rao}, \citenamefont
  {Gui}, \citenamefont {Stamps},\ and\ \citenamefont
  {Hu}}]{Harder2018PRL_LevelAttraction}%
  \BibitemOpen
  \bibfield  {author} {\bibinfo {author} {\bibfnamefont {M.}~\bibnamefont
  {Harder}}, \bibinfo {author} {\bibfnamefont {Y.}~\bibnamefont {Yang}},
  \bibinfo {author} {\bibfnamefont {B.~M.}\ \bibnamefont {Yao}}, \bibinfo
  {author} {\bibfnamefont {C.~H.}\ \bibnamefont {Yu}}, \bibinfo {author}
  {\bibfnamefont {J.~W.}\ \bibnamefont {Rao}}, \bibinfo {author} {\bibfnamefont
  {Y.~S.}\ \bibnamefont {Gui}}, \bibinfo {author} {\bibfnamefont {R.~L.}\
  \bibnamefont {Stamps}}, \ and\ \bibinfo {author} {\bibfnamefont {C.~M.}\
  \bibnamefont {Hu}},\ }\bibfield  {title} {\enquote {\bibinfo {title} {Level
  attraction due to dissipative magnon-photon coupling},}\ }\href {\doibase
  10.1103/PhysRevLett.121.137203} {\bibfield  {journal} {\bibinfo  {journal}
  {Phys. Rev. Lett.}\ }\textbf {\bibinfo {volume} {121}},\ \bibinfo {pages}
  {137203} (\bibinfo {year} {2018}{\natexlab{b}})}\BibitemShut {NoStop}%
\bibitem [{\citenamefont {Bhoi}\ \emph {et~al.}(2019)\citenamefont {Bhoi},
  \citenamefont {Kim}, \citenamefont {Jang}, \citenamefont {Kim}, \citenamefont
  {Yang}, \citenamefont {Cho},\ and\ \citenamefont
  {Kim}}]{Bhoi2019PRB_LevelAttraction}%
  \BibitemOpen
  \bibfield  {author} {\bibinfo {author} {\bibfnamefont {B.}~\bibnamefont
  {Bhoi}}, \bibinfo {author} {\bibfnamefont {B.}~\bibnamefont {Kim}}, \bibinfo
  {author} {\bibfnamefont {S.-H.}\ \bibnamefont {Jang}}, \bibinfo {author}
  {\bibfnamefont {J.}~\bibnamefont {Kim}}, \bibinfo {author} {\bibfnamefont
  {J.}~\bibnamefont {Yang}}, \bibinfo {author} {\bibfnamefont {Y.-J.}\
  \bibnamefont {Cho}}, \ and\ \bibinfo {author} {\bibfnamefont {S.-K.}\
  \bibnamefont {Kim}},\ }\bibfield  {title} {\enquote {\bibinfo {title}
  {Abnormal anticrossing effect in photon-magnon coupling},}\ }\href {\doibase
  10.1103/PhysRevB.99.134426} {\bibfield  {journal} {\bibinfo  {journal} {Phys.
  Rev. B}\ }\textbf {\bibinfo {volume} {99}},\ \bibinfo {pages} {134426}
  (\bibinfo {year} {2019})}\BibitemShut {NoStop}%
\bibitem [{\citenamefont {Boventer}\ \emph {et~al.}(2020)\citenamefont
  {Boventer}, \citenamefont {D\"orflinger}, \citenamefont {Wolz}, \citenamefont
  {Mac\^edo}, \citenamefont {Lebrun}, \citenamefont {Kl\"aui},\ and\
  \citenamefont {Weides}}]{Boventer2020PRR_LevelAttraction}%
  \BibitemOpen
  \bibfield  {author} {\bibinfo {author} {\bibfnamefont {I.}~\bibnamefont
  {Boventer}}, \bibinfo {author} {\bibfnamefont {C.}~\bibnamefont
  {D\"orflinger}}, \bibinfo {author} {\bibfnamefont {T.}~\bibnamefont {Wolz}},
  \bibinfo {author} {\bibfnamefont {R.}~\bibnamefont {Mac\^edo}}, \bibinfo
  {author} {\bibfnamefont {R.}~\bibnamefont {Lebrun}}, \bibinfo {author}
  {\bibfnamefont {M.}~\bibnamefont {Kl\"aui}}, \ and\ \bibinfo {author}
  {\bibfnamefont {M.}~\bibnamefont {Weides}},\ }\bibfield  {title} {\enquote
  {\bibinfo {title} {Control of the coupling strength and linewidth of a cavity
  magnon-polariton},}\ }\href {\doibase 10.1103/PhysRevResearch.2.013154}
  {\bibfield  {journal} {\bibinfo  {journal} {Phys. Rev. Res.}\ }\textbf
  {\bibinfo {volume} {2}},\ \bibinfo {pages} {013154} (\bibinfo {year}
  {2020})}\BibitemShut {NoStop}%
\bibitem [{\citenamefont {Wang}\ \emph
  {et~al.}(2022{\natexlab{b}})\citenamefont {Wang}, \citenamefont {Gou},
  \citenamefont {Xu},\ and\ \citenamefont
  {Gong}}]{WangFei2022PRA_MagnonAtomEnt}%
  \BibitemOpen
  \bibfield  {author} {\bibinfo {author} {\bibfnamefont {F.}~\bibnamefont
  {Wang}}, \bibinfo {author} {\bibfnamefont {C.}~\bibnamefont {Gou}}, \bibinfo
  {author} {\bibfnamefont {J.}~\bibnamefont {Xu}}, \ and\ \bibinfo {author}
  {\bibfnamefont {C.}~\bibnamefont {Gong}},\ }\bibfield  {title} {\enquote
  {\bibinfo {title} {Hybrid magnon-atom entanglement and magnon blockade via
  quantum interference},}\ }\href {\doibase 10.1103/PhysRevA.106.013705}
  {\bibfield  {journal} {\bibinfo  {journal} {Phys. Rev. A}\ }\textbf {\bibinfo
  {volume} {106}},\ \bibinfo {pages} {013705} (\bibinfo {year}
  {2022}{\natexlab{b}})}\BibitemShut {NoStop}%
\bibitem [{\citenamefont {Kong}\ \emph {et~al.}(2021)\citenamefont {Kong},
  \citenamefont {Hu}, \citenamefont {Hu},\ and\ \citenamefont
  {Xu}}]{KongDeyi2021PRB_MagnonAtomEnt}%
  \BibitemOpen
  \bibfield  {author} {\bibinfo {author} {\bibfnamefont {D.}~\bibnamefont
  {Kong}}, \bibinfo {author} {\bibfnamefont {X.}~\bibnamefont {Hu}}, \bibinfo
  {author} {\bibfnamefont {L.}~\bibnamefont {Hu}}, \ and\ \bibinfo {author}
  {\bibfnamefont {J.}~\bibnamefont {Xu}},\ }\bibfield  {title} {\enquote
  {\bibinfo {title} {Magnon-atom interaction via dispersive cavities: Magnon
  entanglement},}\ }\href {\doibase 10.1103/PhysRevB.103.224416} {\bibfield
  {journal} {\bibinfo  {journal} {Phys. Rev. B}\ }\textbf {\bibinfo {volume}
  {103}},\ \bibinfo {pages} {224416} (\bibinfo {year} {2021})}\BibitemShut
  {NoStop}%
\bibitem [{\citenamefont {Amazioug}, \citenamefont {Teklu},\ and\ \citenamefont
  {Asjad}(2022)}]{Muhammad2022TriEntEnhancement_Feedback}%
  \BibitemOpen
  \bibfield  {author} {\bibinfo {author} {\bibfnamefont {M.}~\bibnamefont
  {Amazioug}}, \bibinfo {author} {\bibfnamefont {B.}~\bibnamefont {Teklu}}, \
  and\ \bibinfo {author} {\bibfnamefont {M.}~\bibnamefont {Asjad}},\ }\bibfield
   {title} {\enquote {\bibinfo {title} {Enhancement of magnon-photon-phonon
  entanglement in a cavity magnomechanics with coherent feedback loop},}\
  }\href {https://arxiv.org/abs/2211.17052} {\bibfield  {journal} {\bibinfo
  {journal} {arXiv: 2211.17052}\ } (\bibinfo {year} {2022})}\BibitemShut
  {NoStop}%
\bibitem [{\citenamefont {Tan}(2019)}]{TanHuatang2019PRR_Steering}%
  \BibitemOpen
  \bibfield  {author} {\bibinfo {author} {\bibfnamefont {H.}~\bibnamefont
  {Tan}},\ }\bibfield  {title} {\enquote {\bibinfo {title} {Genuine
  photon-magnon-phonon einstein-podolsky-rosen steerable nonlocality in a
  continuously-monitored cavity magnomechanical system},}\ }\href {\doibase
  10.1103/PhysRevResearch.1.033161} {\bibfield  {journal} {\bibinfo  {journal}
  {Phys. Rev. Res.}\ }\textbf {\bibinfo {volume} {1}},\ \bibinfo {pages}
  {033161} (\bibinfo {year} {2019})}\BibitemShut {NoStop}%
\bibitem [{\citenamefont {Yu}, \citenamefont {Shen},\ and\ \citenamefont
  {Li}(2020)}]{YuMei2020EntMagnon}%
  \BibitemOpen
  \bibfield  {author} {\bibinfo {author} {\bibfnamefont {M.}~\bibnamefont
  {Yu}}, \bibinfo {author} {\bibfnamefont {H.}~\bibnamefont {Shen}}, \ and\
  \bibinfo {author} {\bibfnamefont {J.}~\bibnamefont {Li}},\ }\bibfield
  {title} {\enquote {\bibinfo {title} {Magnetostrictively induced stationary
  entanglement between two microwave fields},}\ }\href {\doibase
  10.1103/PhysRevLett.124.213604} {\bibfield  {journal} {\bibinfo  {journal}
  {Phys. Rev. Lett.}\ }\textbf {\bibinfo {volume} {124}},\ \bibinfo {pages}
  {213604} (\bibinfo {year} {2020})}\BibitemShut {NoStop}%
\bibitem [{\citenamefont {Yang}\ \emph {et~al.}(2020)\citenamefont {Yang},
  \citenamefont {Liu}, \citenamefont {Jin}, \citenamefont {Zhu}, \citenamefont
  {Zhu}, \citenamefont {Liu}, \citenamefont {Ming},\ and\ \citenamefont
  {Yang}}]{YangZhibo2020OE_BiTripartite_KerrNonlinearity}%
  \BibitemOpen
  \bibfield  {author} {\bibinfo {author} {\bibfnamefont {Z.-B.}\ \bibnamefont
  {Yang}}, \bibinfo {author} {\bibfnamefont {J.-S.}\ \bibnamefont {Liu}},
  \bibinfo {author} {\bibfnamefont {H.}~\bibnamefont {Jin}}, \bibinfo {author}
  {\bibfnamefont {Q.-H.}\ \bibnamefont {Zhu}}, \bibinfo {author} {\bibfnamefont
  {A.-D.}\ \bibnamefont {Zhu}}, \bibinfo {author} {\bibfnamefont {H.-Y.}\
  \bibnamefont {Liu}}, \bibinfo {author} {\bibfnamefont {Y.}~\bibnamefont
  {Ming}}, \ and\ \bibinfo {author} {\bibfnamefont {R.-C.}\ \bibnamefont
  {Yang}},\ }\bibfield  {title} {\enquote {\bibinfo {title} {Entanglement
  enhanced by kerr nonlinearity in a cavity-optomagnonics system},}\ }\href
  {\doibase 10.1364/OE.404522} {\bibfield  {journal} {\bibinfo  {journal} {Opt.
  Express}\ }\textbf {\bibinfo {volume} {28}},\ \bibinfo {pages} {31862}
  (\bibinfo {year} {2020})}\BibitemShut {NoStop}%
\bibitem [{\citenamefont {Ying}, \citenamefont {Shuang-Yuan},\ and\
  \citenamefont {Jing-Ping}(2020)}]{ZhouYing2020SqueezingDrive_Bi-MultiEnt}%
  \BibitemOpen
  \bibfield  {author} {\bibinfo {author} {\bibfnamefont {Z.}~\bibnamefont
  {Ying}}, \bibinfo {author} {\bibfnamefont {X.}~\bibnamefont {Shuang-Yuan}}, \
  and\ \bibinfo {author} {\bibfnamefont {X.}~\bibnamefont {Jing-Ping}},\
  }\bibfield  {title} {\enquote {\bibinfo {title} {Bipartite and tripartite
  entanglement caused by squeezed drive in magnetic-cavity quantum
  electrodynamics system},}\ }\href
  {https://wulixb.iphy.ac.cn/cn/article/id/67f6902d-1b40-4a0c-abcc-93257e8fcbb7}
  {\bibfield  {journal} {\bibinfo  {journal} {Acta Phys. Sin.}\ }\textbf
  {\bibinfo {volume} {69}},\ \bibinfo {pages} {220301} (\bibinfo {year}
  {2020})}\BibitemShut {NoStop}%
\bibitem [{\citenamefont {Li}\ and\ \citenamefont
  {Gröblacher}(2021)}]{JieLi2021QuanSciTec_TwoModeSqueezingDrive}%
  \BibitemOpen
  \bibfield  {author} {\bibinfo {author} {\bibfnamefont {J.}~\bibnamefont
  {Li}}\ and\ \bibinfo {author} {\bibfnamefont {S.}~\bibnamefont
  {Gröblacher}},\ }\bibfield  {title} {\enquote {\bibinfo {title} {Entangling
  the vibrational modes of two massive ferromagnetic spheres using cavity
  magnomechanics},}\ }\href {\doibase 10.1088/2058-9565/abd982} {\bibfield
  {journal} {\bibinfo  {journal} {Quantum Sci. Technol.}\ }\textbf {\bibinfo
  {volume} {6}},\ \bibinfo {pages} {024005} (\bibinfo {year}
  {2021})}\BibitemShut {NoStop}%
\bibitem [{\citenamefont {Vitali}\ \emph {et~al.}(2007)\citenamefont {Vitali},
  \citenamefont {Gigan}, \citenamefont {Ferreira}, \citenamefont {B\"ohm},
  \citenamefont {Tombesi}, \citenamefont {Guerreiro}, \citenamefont {Vedral},
  \citenamefont {Zeilinger},\ and\ \citenamefont
  {Aspelmeyer}}]{Vitali2007PRL_Optomechanics}%
  \BibitemOpen
  \bibfield  {author} {\bibinfo {author} {\bibfnamefont {D.}~\bibnamefont
  {Vitali}}, \bibinfo {author} {\bibfnamefont {S.}~\bibnamefont {Gigan}},
  \bibinfo {author} {\bibfnamefont {A.}~\bibnamefont {Ferreira}}, \bibinfo
  {author} {\bibfnamefont {H.~R.}\ \bibnamefont {B\"ohm}}, \bibinfo {author}
  {\bibfnamefont {P.}~\bibnamefont {Tombesi}}, \bibinfo {author} {\bibfnamefont
  {A.}~\bibnamefont {Guerreiro}}, \bibinfo {author} {\bibfnamefont
  {V.}~\bibnamefont {Vedral}}, \bibinfo {author} {\bibfnamefont
  {A.}~\bibnamefont {Zeilinger}}, \ and\ \bibinfo {author} {\bibfnamefont
  {M.}~\bibnamefont {Aspelmeyer}},\ }\bibfield  {title} {\enquote {\bibinfo
  {title} {Optomechanical entanglement between a movable mirror and a cavity
  field},}\ }\href {\doibase 10.1103/PhysRevLett.98.030405} {\bibfield
  {journal} {\bibinfo  {journal} {Phys. Rev. Lett.}\ }\textbf {\bibinfo
  {volume} {98}},\ \bibinfo {pages} {030405} (\bibinfo {year}
  {2007})}\BibitemShut {NoStop}%
\bibitem [{\citenamefont {Genes}\ \emph {et~al.}(2008)\citenamefont {Genes},
  \citenamefont {Mari}, \citenamefont {Tombesi},\ and\ \citenamefont
  {Vitali}}]{Genes2008Ent_Optomechanics}%
  \BibitemOpen
  \bibfield  {author} {\bibinfo {author} {\bibfnamefont {C.}~\bibnamefont
  {Genes}}, \bibinfo {author} {\bibfnamefont {A.}~\bibnamefont {Mari}},
  \bibinfo {author} {\bibfnamefont {P.}~\bibnamefont {Tombesi}}, \ and\
  \bibinfo {author} {\bibfnamefont {D.}~\bibnamefont {Vitali}},\ }\bibfield
  {title} {\enquote {\bibinfo {title} {Robust entanglement of a micromechanical
  resonator with output optical fields},}\ }\href {\doibase
  10.1103/PhysRevA.78.032316} {\bibfield  {journal} {\bibinfo  {journal} {Phys.
  Rev. A}\ }\textbf {\bibinfo {volume} {78}},\ \bibinfo {pages} {032316}
  (\bibinfo {year} {2008})}\BibitemShut {NoStop}%
\bibitem [{\citenamefont {Wang}\ \emph
  {et~al.}(2016{\natexlab{b}})\citenamefont {Wang}, \citenamefont {Zhang},
  \citenamefont {Zhang}, \citenamefont {Luo}, \citenamefont {Xiong},
  \citenamefont {Wang}, \citenamefont {Li}, \citenamefont {Hu},\ and\
  \citenamefont {You}}]{WangYipu2016PRB_KerrExp}%
  \BibitemOpen
  \bibfield  {author} {\bibinfo {author} {\bibfnamefont {Y.-P.}\ \bibnamefont
  {Wang}}, \bibinfo {author} {\bibfnamefont {G.-Q.}\ \bibnamefont {Zhang}},
  \bibinfo {author} {\bibfnamefont {D.}~\bibnamefont {Zhang}}, \bibinfo
  {author} {\bibfnamefont {X.-Q.}\ \bibnamefont {Luo}}, \bibinfo {author}
  {\bibfnamefont {W.}~\bibnamefont {Xiong}}, \bibinfo {author} {\bibfnamefont
  {S.-P.}\ \bibnamefont {Wang}}, \bibinfo {author} {\bibfnamefont {T.-F.}\
  \bibnamefont {Li}}, \bibinfo {author} {\bibfnamefont {C.-M.}\ \bibnamefont
  {Hu}}, \ and\ \bibinfo {author} {\bibfnamefont {J.~Q.}\ \bibnamefont {You}},\
  }\bibfield  {title} {\enquote {\bibinfo {title} {Magnon kerr effect in a
  strongly coupled cavity-magnon system},}\ }\href {\doibase
  10.1103/PhysRevB.94.224410} {\bibfield  {journal} {\bibinfo  {journal} {Phys.
  Rev. B}\ }\textbf {\bibinfo {volume} {94}},\ \bibinfo {pages} {224410}
  (\bibinfo {year} {2016}{\natexlab{b}})}\BibitemShut {NoStop}%
\bibitem [{\citenamefont {Wang}\ \emph
  {et~al.}(2018{\natexlab{b}})\citenamefont {Wang}, \citenamefont {Zhang},
  \citenamefont {Zhang}, \citenamefont {Li}, \citenamefont {Hu},\ and\
  \citenamefont {You}}]{WangYipu2018PRL_KerrExpBistability}%
  \BibitemOpen
  \bibfield  {author} {\bibinfo {author} {\bibfnamefont {Y.-P.}\ \bibnamefont
  {Wang}}, \bibinfo {author} {\bibfnamefont {G.-Q.}\ \bibnamefont {Zhang}},
  \bibinfo {author} {\bibfnamefont {D.}~\bibnamefont {Zhang}}, \bibinfo
  {author} {\bibfnamefont {T.-F.}\ \bibnamefont {Li}}, \bibinfo {author}
  {\bibfnamefont {C.-M.}\ \bibnamefont {Hu}}, \ and\ \bibinfo {author}
  {\bibfnamefont {J.~Q.}\ \bibnamefont {You}},\ }\bibfield  {title} {\enquote
  {\bibinfo {title} {Bistability of cavity magnon polaritons},}\ }\href
  {\doibase 10.1103/PhysRevLett.120.057202} {\bibfield  {journal} {\bibinfo
  {journal} {Phys. Rev. Lett.}\ }\textbf {\bibinfo {volume} {120}},\ \bibinfo
  {pages} {057202} (\bibinfo {year} {2018}{\natexlab{b}})}\BibitemShut
  {NoStop}%
\bibitem [{\citenamefont {Jiang}, \citenamefont {Tang},\ and\ \citenamefont
  {Li}(2022)}]{JiangXi2022Photonics_KerrNonclassical}%
  \BibitemOpen
  \bibfield  {author} {\bibinfo {author} {\bibfnamefont {X.}~\bibnamefont
  {Jiang}}, \bibinfo {author} {\bibfnamefont {S.}~\bibnamefont {Tang}}, \ and\
  \bibinfo {author} {\bibfnamefont {S.}~\bibnamefont {Li}},\ }\bibfield
  {title} {\enquote {\bibinfo {title} {Kerr-nonlinearity-triggered
  nonclassicality of magnons in a photon-magnon coupling system},}\ }\href
  {\doibase 10.3390/photonics9100681} {\bibfield  {journal} {\bibinfo
  {journal} {Photonics}\ }\textbf {\bibinfo {volume} {9}},\ \bibinfo {pages}
  {681} (\bibinfo {year} {2022})}\BibitemShut {NoStop}%
\bibitem [{\citenamefont {Kong}, \citenamefont {Xiong},\ and\ \citenamefont
  {Wu}(2019{\natexlab{b}})}]{KongCui2019PRAppl_KerrNonreciprocity}%
  \BibitemOpen
  \bibfield  {author} {\bibinfo {author} {\bibfnamefont {C.}~\bibnamefont
  {Kong}}, \bibinfo {author} {\bibfnamefont {H.}~\bibnamefont {Xiong}}, \ and\
  \bibinfo {author} {\bibfnamefont {Y.}~\bibnamefont {Wu}},\ }\bibfield
  {title} {\enquote {\bibinfo {title} {Magnon-induced nonreciprocity based on
  the magnon kerr effect},}\ }\href {\doibase 10.1103/PhysRevApplied.12.034001}
  {\bibfield  {journal} {\bibinfo  {journal} {Phys. Rev. Appl.}\ }\textbf
  {\bibinfo {volume} {12}},\ \bibinfo {pages} {034001} (\bibinfo {year}
  {2019}{\natexlab{b}})}\BibitemShut {NoStop}%
\bibitem [{\citenamefont {Xie}\ \emph {et~al.}(2023)\citenamefont {Xie},
  \citenamefont {He}, \citenamefont {Liao}, \citenamefont {Chen},\ and\
  \citenamefont {Lin}}]{XieHong2023OE_MO_Ent}%
  \BibitemOpen
  \bibfield  {author} {\bibinfo {author} {\bibfnamefont {H.}~\bibnamefont
  {Xie}}, \bibinfo {author} {\bibfnamefont {L.-W.}\ \bibnamefont {He}},
  \bibinfo {author} {\bibfnamefont {C.-G.}\ \bibnamefont {Liao}}, \bibinfo
  {author} {\bibfnamefont {Z.-H.}\ \bibnamefont {Chen}}, \ and\ \bibinfo
  {author} {\bibfnamefont {X.-M.}\ \bibnamefont {Lin}},\ }\bibfield  {title}
  {\enquote {\bibinfo {title} {Generation of robust optical entanglement in
  cavity optomagnonics},}\ }\href {\doibase 10.1364/OE.478963} {\bibfield
  {journal} {\bibinfo  {journal} {Opt. Express}\ }\textbf {\bibinfo {volume}
  {31}},\ \bibinfo {pages} {7994} (\bibinfo {year} {2023})}\BibitemShut
  {NoStop}%
\bibitem [{\citenamefont {Zhong}\ \emph {et~al.}(2013)\citenamefont {Zhong},
  \citenamefont {Menzel}, \citenamefont {Candia}, \citenamefont {Eder},
  \citenamefont {Ihmig}, \citenamefont {Baust}, \citenamefont {Haeberlein},
  \citenamefont {Hoffmann}, \citenamefont {Inomata}, \citenamefont {Yamamoto},
  \citenamefont {Nakamura}, \citenamefont {Solano}, \citenamefont {Deppe},
  \citenamefont {Marx},\ and\ \citenamefont
  {Gross}}]{ZhongL2013NJP_OneModeSqueezing_JPA}%
  \BibitemOpen
  \bibfield  {author} {\bibinfo {author} {\bibfnamefont {L.}~\bibnamefont
  {Zhong}}, \bibinfo {author} {\bibfnamefont {E.~P.}\ \bibnamefont {Menzel}},
  \bibinfo {author} {\bibfnamefont {R.~D.}\ \bibnamefont {Candia}}, \bibinfo
  {author} {\bibfnamefont {P.}~\bibnamefont {Eder}}, \bibinfo {author}
  {\bibfnamefont {M.}~\bibnamefont {Ihmig}}, \bibinfo {author} {\bibfnamefont
  {A.}~\bibnamefont {Baust}}, \bibinfo {author} {\bibfnamefont
  {M.}~\bibnamefont {Haeberlein}}, \bibinfo {author} {\bibfnamefont
  {E.}~\bibnamefont {Hoffmann}}, \bibinfo {author} {\bibfnamefont
  {K.}~\bibnamefont {Inomata}}, \bibinfo {author} {\bibfnamefont
  {T.}~\bibnamefont {Yamamoto}}, \bibinfo {author} {\bibfnamefont
  {Y.}~\bibnamefont {Nakamura}}, \bibinfo {author} {\bibfnamefont
  {E.}~\bibnamefont {Solano}}, \bibinfo {author} {\bibfnamefont
  {F.}~\bibnamefont {Deppe}}, \bibinfo {author} {\bibfnamefont
  {A.}~\bibnamefont {Marx}}, \ and\ \bibinfo {author} {\bibfnamefont
  {R.}~\bibnamefont {Gross}},\ }\bibfield  {title} {\enquote {\bibinfo {title}
  {Squeezing with a flux-driven josephson parametric amplifier},}\ }\href
  {\doibase 10.1088/1367-2630/15/12/125013} {\bibfield  {journal} {\bibinfo
  {journal} {New J. Phys.}\ }\textbf {\bibinfo {volume} {15}},\ \bibinfo
  {pages} {125013} (\bibinfo {year} {2013})}\BibitemShut {NoStop}%
\bibitem [{\citenamefont {Fedorov}\ \emph {et~al.}(2016)\citenamefont
  {Fedorov}, \citenamefont {Zhong}, \citenamefont {Pogorzalek}, \citenamefont
  {Eder}, \citenamefont {Fischer}, \citenamefont {Goetz}, \citenamefont {Xie},
  \citenamefont {Wulschner}, \citenamefont {Inomata}, \citenamefont {Yamamoto},
  \citenamefont {Nakamura}, \citenamefont {Di~Candia}, \citenamefont
  {Las~Heras}, \citenamefont {Sanz}, \citenamefont {Solano}, \citenamefont
  {Menzel}, \citenamefont {Deppe}, \citenamefont {Marx},\ and\ \citenamefont
  {Gross}}]{Fedorov2016PRL_OneModeSqueezing_JPA}%
  \BibitemOpen
  \bibfield  {author} {\bibinfo {author} {\bibfnamefont {K.~G.}\ \bibnamefont
  {Fedorov}}, \bibinfo {author} {\bibfnamefont {L.}~\bibnamefont {Zhong}},
  \bibinfo {author} {\bibfnamefont {S.}~\bibnamefont {Pogorzalek}}, \bibinfo
  {author} {\bibfnamefont {P.}~\bibnamefont {Eder}}, \bibinfo {author}
  {\bibfnamefont {M.}~\bibnamefont {Fischer}}, \bibinfo {author} {\bibfnamefont
  {J.}~\bibnamefont {Goetz}}, \bibinfo {author} {\bibfnamefont
  {E.}~\bibnamefont {Xie}}, \bibinfo {author} {\bibfnamefont {F.}~\bibnamefont
  {Wulschner}}, \bibinfo {author} {\bibfnamefont {K.}~\bibnamefont {Inomata}},
  \bibinfo {author} {\bibfnamefont {T.}~\bibnamefont {Yamamoto}}, \bibinfo
  {author} {\bibfnamefont {Y.}~\bibnamefont {Nakamura}}, \bibinfo {author}
  {\bibfnamefont {R.}~\bibnamefont {Di~Candia}}, \bibinfo {author}
  {\bibfnamefont {U.}~\bibnamefont {Las~Heras}}, \bibinfo {author}
  {\bibfnamefont {M.}~\bibnamefont {Sanz}}, \bibinfo {author} {\bibfnamefont
  {E.}~\bibnamefont {Solano}}, \bibinfo {author} {\bibfnamefont {E.~P.}\
  \bibnamefont {Menzel}}, \bibinfo {author} {\bibfnamefont {F.}~\bibnamefont
  {Deppe}}, \bibinfo {author} {\bibfnamefont {A.}~\bibnamefont {Marx}}, \ and\
  \bibinfo {author} {\bibfnamefont {R.}~\bibnamefont {Gross}},\ }\bibfield
  {title} {\enquote {\bibinfo {title} {Displacement of propagating squeezed
  microwave states},}\ }\href {\doibase 10.1103/PhysRevLett.117.020502}
  {\bibfield  {journal} {\bibinfo  {journal} {Phys. Rev. Lett.}\ }\textbf
  {\bibinfo {volume} {117}},\ \bibinfo {pages} {020502} (\bibinfo {year}
  {2016})}\BibitemShut {NoStop}%
\bibitem [{\citenamefont {Bienfait}\ \emph {et~al.}(2017)\citenamefont
  {Bienfait}, \citenamefont {Campagne-Ibarcq}, \citenamefont {Kiilerich},
  \citenamefont {Zhou}, \citenamefont {Probst}, \citenamefont {Pla},
  \citenamefont {Schenkel}, \citenamefont {Vion}, \citenamefont {Esteve},
  \citenamefont {Morton}, \citenamefont {Moelmer},\ and\ \citenamefont
  {Bertet}}]{Bienfait2017PRX_OneModeSqueezing_JPA}%
  \BibitemOpen
  \bibfield  {author} {\bibinfo {author} {\bibfnamefont {A.}~\bibnamefont
  {Bienfait}}, \bibinfo {author} {\bibfnamefont {P.}~\bibnamefont
  {Campagne-Ibarcq}}, \bibinfo {author} {\bibfnamefont {A.~H.}\ \bibnamefont
  {Kiilerich}}, \bibinfo {author} {\bibfnamefont {X.}~\bibnamefont {Zhou}},
  \bibinfo {author} {\bibfnamefont {S.}~\bibnamefont {Probst}}, \bibinfo
  {author} {\bibfnamefont {J.~J.}\ \bibnamefont {Pla}}, \bibinfo {author}
  {\bibfnamefont {T.}~\bibnamefont {Schenkel}}, \bibinfo {author}
  {\bibfnamefont {D.}~\bibnamefont {Vion}}, \bibinfo {author} {\bibfnamefont
  {D.}~\bibnamefont {Esteve}}, \bibinfo {author} {\bibfnamefont {J.~J.~L.}\
  \bibnamefont {Morton}}, \bibinfo {author} {\bibfnamefont {K.}~\bibnamefont
  {Moelmer}}, \ and\ \bibinfo {author} {\bibfnamefont {P.}~\bibnamefont
  {Bertet}},\ }\bibfield  {title} {\enquote {\bibinfo {title} {Magnetic
  resonance with squeezed microwaves},}\ }\href {\doibase
  10.1103/PhysRevX.7.041011} {\bibfield  {journal} {\bibinfo  {journal} {Phys.
  Rev. X}\ }\textbf {\bibinfo {volume} {7}},\ \bibinfo {pages} {041011}
  (\bibinfo {year} {2017})}\BibitemShut {NoStop}%
\bibitem [{\citenamefont {Malnou}\ \emph {et~al.}(2018)\citenamefont {Malnou},
  \citenamefont {Palken}, \citenamefont {Vale}, \citenamefont {Hilton},\ and\
  \citenamefont {Lehnert}}]{Malnou2018PRAppl_OneModeSqueezing_JPA}%
  \BibitemOpen
  \bibfield  {author} {\bibinfo {author} {\bibfnamefont {M.}~\bibnamefont
  {Malnou}}, \bibinfo {author} {\bibfnamefont {D.~A.}\ \bibnamefont {Palken}},
  \bibinfo {author} {\bibfnamefont {L.~R.}\ \bibnamefont {Vale}}, \bibinfo
  {author} {\bibfnamefont {G.~C.}\ \bibnamefont {Hilton}}, \ and\ \bibinfo
  {author} {\bibfnamefont {K.~W.}\ \bibnamefont {Lehnert}},\ }\bibfield
  {title} {\enquote {\bibinfo {title} {Optimal operation of a josephson
  parametric amplifier for vacuum squeezing},}\ }\href {\doibase
  10.1103/PhysRevApplied.9.044023} {\bibfield  {journal} {\bibinfo  {journal}
  {Phys. Rev. Appl.}\ }\textbf {\bibinfo {volume} {9}},\ \bibinfo {pages}
  {044023} (\bibinfo {year} {2018})}\BibitemShut {NoStop}%
\bibitem [{\citenamefont {Zhang}\ \emph {et~al.}(2022)\citenamefont {Zhang},
  \citenamefont {Wang}, \citenamefont {Han}, \citenamefont {Zhang},\ and\
  \citenamefont {Wang}}]{ZhangWei2022OE_SqueezingDrive_Bi-EntSteering}%
  \BibitemOpen
  \bibfield  {author} {\bibinfo {author} {\bibfnamefont {W.}~\bibnamefont
  {Zhang}}, \bibinfo {author} {\bibfnamefont {T.}~\bibnamefont {Wang}},
  \bibinfo {author} {\bibfnamefont {X.}~\bibnamefont {Han}}, \bibinfo {author}
  {\bibfnamefont {S.}~\bibnamefont {Zhang}}, \ and\ \bibinfo {author}
  {\bibfnamefont {H.-F.}\ \bibnamefont {Wang}},\ }\bibfield  {title} {\enquote
  {\bibinfo {title} {Quantum entanglement and one-way steering in a cavity
  magnomechanical system via a squeezed vacuum field},}\ }\href {\doibase
  10.1364/OE.453787} {\bibfield  {journal} {\bibinfo  {journal} {Opt. Express}\
  }\textbf {\bibinfo {volume} {30}},\ \bibinfo {pages} {10969} (\bibinfo {year}
  {2022})}\BibitemShut {NoStop}%
\bibitem [{\citenamefont {Eichler}\ \emph {et~al.}(2011)\citenamefont
  {Eichler}, \citenamefont {Bozyigit}, \citenamefont {Lang}, \citenamefont
  {Baur}, \citenamefont {Steffen}, \citenamefont {Fink}, \citenamefont
  {Filipp},\ and\ \citenamefont
  {Wallraff}}]{Eichler2011PRL_TwoModeSqueezing_JPA}%
  \BibitemOpen
  \bibfield  {author} {\bibinfo {author} {\bibfnamefont {C.}~\bibnamefont
  {Eichler}}, \bibinfo {author} {\bibfnamefont {D.}~\bibnamefont {Bozyigit}},
  \bibinfo {author} {\bibfnamefont {C.}~\bibnamefont {Lang}}, \bibinfo {author}
  {\bibfnamefont {M.}~\bibnamefont {Baur}}, \bibinfo {author} {\bibfnamefont
  {L.}~\bibnamefont {Steffen}}, \bibinfo {author} {\bibfnamefont {J.~M.}\
  \bibnamefont {Fink}}, \bibinfo {author} {\bibfnamefont {S.}~\bibnamefont
  {Filipp}}, \ and\ \bibinfo {author} {\bibfnamefont {A.}~\bibnamefont
  {Wallraff}},\ }\bibfield  {title} {\enquote {\bibinfo {title} {Observation of
  two-mode squeezing in the microwave frequency domain},}\ }\href {\doibase
  10.1103/PhysRevLett.107.113601} {\bibfield  {journal} {\bibinfo  {journal}
  {Phys. Rev. Lett.}\ }\textbf {\bibinfo {volume} {107}},\ \bibinfo {pages}
  {113601} (\bibinfo {year} {2011})}\BibitemShut {NoStop}%
\bibitem [{\citenamefont {Flurin}\ \emph {et~al.}(2012)\citenamefont {Flurin},
  \citenamefont {Roch}, \citenamefont {Mallet}, \citenamefont {Devoret},\ and\
  \citenamefont {Huard}}]{Flurin2012PRL_TwoModeSqueezing_JPM}%
  \BibitemOpen
  \bibfield  {author} {\bibinfo {author} {\bibfnamefont {E.}~\bibnamefont
  {Flurin}}, \bibinfo {author} {\bibfnamefont {N.}~\bibnamefont {Roch}},
  \bibinfo {author} {\bibfnamefont {F.}~\bibnamefont {Mallet}}, \bibinfo
  {author} {\bibfnamefont {M.~H.}\ \bibnamefont {Devoret}}, \ and\ \bibinfo
  {author} {\bibfnamefont {B.}~\bibnamefont {Huard}},\ }\bibfield  {title}
  {\enquote {\bibinfo {title} {Generating entangled microwave radiation over
  two transmission lines},}\ }\href {\doibase 10.1103/PhysRevLett.109.183901}
  {\bibfield  {journal} {\bibinfo  {journal} {Phys. Rev. Lett.}\ }\textbf
  {\bibinfo {volume} {109}},\ \bibinfo {pages} {183901} (\bibinfo {year}
  {2012})}\BibitemShut {NoStop}%
\bibitem [{\citenamefont {Yu}, \citenamefont {Zhu},\ and\ \citenamefont
  {Li}(2020)}]{YuMei2020JPB_MagMagEnt}%
  \BibitemOpen
  \bibfield  {author} {\bibinfo {author} {\bibfnamefont {M.}~\bibnamefont
  {Yu}}, \bibinfo {author} {\bibfnamefont {S.-Y.}\ \bibnamefont {Zhu}}, \ and\
  \bibinfo {author} {\bibfnamefont {J.}~\bibnamefont {Li}},\ }\bibfield
  {title} {\enquote {\bibinfo {title} {Macroscopic entanglement of two magnon
  modes via quantum correlated microwave fields},}\ }\href {\doibase
  10.1088/1361-6455/ab68b5} {\bibfield  {journal} {\bibinfo  {journal} {J.
  Phys. B}\ }\textbf {\bibinfo {volume} {53}},\ \bibinfo {pages} {065402}
  (\bibinfo {year} {2020})}\BibitemShut {NoStop}%
\bibitem [{\citenamefont {Schoelkopf}\ and\ \citenamefont
  {Girvin}(2008)}]{Schoelkopf2008Wiring}%
  \BibitemOpen
  \bibfield  {author} {\bibinfo {author} {\bibfnamefont {R.}~\bibnamefont
  {Schoelkopf}}\ and\ \bibinfo {author} {\bibfnamefont {S.}~\bibnamefont
  {Girvin}},\ }\bibfield  {title} {\enquote {\bibinfo {title} {Wiring up
  quantum systems},}\ }\href {\doibase 10.1038/451664a} {\bibfield  {journal}
  {\bibinfo  {journal} {Nature}\ }\textbf {\bibinfo {volume} {451}},\ \bibinfo
  {pages} {664} (\bibinfo {year} {2008})}\BibitemShut {NoStop}%
\bibitem [{\citenamefont {Xiang}\ \emph {et~al.}(2013)\citenamefont {Xiang},
  \citenamefont {Ashhab}, \citenamefont {You},\ and\ \citenamefont
  {Nori}}]{Xiang2013Hybrid}%
  \BibitemOpen
  \bibfield  {author} {\bibinfo {author} {\bibfnamefont {Z.-L.}\ \bibnamefont
  {Xiang}}, \bibinfo {author} {\bibfnamefont {S.}~\bibnamefont {Ashhab}},
  \bibinfo {author} {\bibfnamefont {J.~Q.}\ \bibnamefont {You}}, \ and\
  \bibinfo {author} {\bibfnamefont {F.}~\bibnamefont {Nori}},\ }\bibfield
  {title} {\enquote {\bibinfo {title} {Hybrid quantum circuits: Superconducting
  circuits interacting with other quantum systems},}\ }\href {\doibase
  10.1103/RevModPhys.85.623} {\bibfield  {journal} {\bibinfo  {journal} {Rev.
  Mod. Phys.}\ }\textbf {\bibinfo {volume} {85}},\ \bibinfo {pages} {623}
  (\bibinfo {year} {2013})}\BibitemShut {NoStop}%
\bibitem [{\citenamefont {Clarke}\ and\ \citenamefont
  {Wilhelm}(2008)}]{Clarke2008SuperconductingQubits}%
  \BibitemOpen
  \bibfield  {author} {\bibinfo {author} {\bibfnamefont {J.}~\bibnamefont
  {Clarke}}\ and\ \bibinfo {author} {\bibfnamefont {F.~K.}\ \bibnamefont
  {Wilhelm}},\ }\bibfield  {title} {\enquote {\bibinfo {title} {Superconducting
  quantum bits},}\ }\href {\doibase 10.1038/nature07128} {\bibfield  {journal}
  {\bibinfo  {journal} {Nature}\ }\textbf {\bibinfo {volume} {453}},\ \bibinfo
  {pages} {1031} (\bibinfo {year} {2008})}\BibitemShut {NoStop}%
\bibitem [{\citenamefont {O'brien}, \citenamefont {Furusawa},\ and\
  \citenamefont {Vu{\v{c}}kovi{\'c}}(2009)}]{Obrien2009PhotonicQuantum}%
  \BibitemOpen
  \bibfield  {author} {\bibinfo {author} {\bibfnamefont {J.~L.}\ \bibnamefont
  {O'brien}}, \bibinfo {author} {\bibfnamefont {A.}~\bibnamefont {Furusawa}}, \
  and\ \bibinfo {author} {\bibfnamefont {J.}~\bibnamefont
  {Vu{\v{c}}kovi{\'c}}},\ }\bibfield  {title} {\enquote {\bibinfo {title}
  {Photonic quantum technologies},}\ }\href {\doibase 10.1038/nphoton.2009.229}
  {\bibfield  {journal} {\bibinfo  {journal} {Nat. Photon.}\ }\textbf {\bibinfo
  {volume} {3}},\ \bibinfo {pages} {687} (\bibinfo {year} {2009})}\BibitemShut
  {NoStop}%
\bibitem [{\citenamefont {Lvovsky}, \citenamefont {Sanders},\ and\
  \citenamefont {Tittel}(2009)}]{Lvovsky2009Optical}%
  \BibitemOpen
  \bibfield  {author} {\bibinfo {author} {\bibfnamefont {A.~I.}\ \bibnamefont
  {Lvovsky}}, \bibinfo {author} {\bibfnamefont {B.~C.}\ \bibnamefont
  {Sanders}}, \ and\ \bibinfo {author} {\bibfnamefont {W.}~\bibnamefont
  {Tittel}},\ }\bibfield  {title} {\enquote {\bibinfo {title} {Optical quantum
  memory},}\ }\href {\doibase 10.1038/nphoton.2009.231} {\bibfield  {journal}
  {\bibinfo  {journal} {Nat. photon.}\ }\textbf {\bibinfo {volume} {3}},\
  \bibinfo {pages} {706} (\bibinfo {year} {2009})}\BibitemShut {NoStop}%
\bibitem [{\citenamefont {Kok}\ \emph {et~al.}(2007)\citenamefont {Kok},
  \citenamefont {Munro}, \citenamefont {Nemoto}, \citenamefont {Ralph},
  \citenamefont {Dowling},\ and\ \citenamefont
  {Milburn}}]{Kok2007LinearOptical}%
  \BibitemOpen
  \bibfield  {author} {\bibinfo {author} {\bibfnamefont {P.}~\bibnamefont
  {Kok}}, \bibinfo {author} {\bibfnamefont {W.~J.}\ \bibnamefont {Munro}},
  \bibinfo {author} {\bibfnamefont {K.}~\bibnamefont {Nemoto}}, \bibinfo
  {author} {\bibfnamefont {T.~C.}\ \bibnamefont {Ralph}}, \bibinfo {author}
  {\bibfnamefont {J.~P.}\ \bibnamefont {Dowling}}, \ and\ \bibinfo {author}
  {\bibfnamefont {G.~J.}\ \bibnamefont {Milburn}},\ }\bibfield  {title}
  {\enquote {\bibinfo {title} {Linear optical quantum computing with photonic
  qubits},}\ }\href {\doibase 10.1103/RevModPhys.79.135} {\bibfield  {journal}
  {\bibinfo  {journal} {Rev. Mod. Phys.}\ }\textbf {\bibinfo {volume} {79}},\
  \bibinfo {pages} {135} (\bibinfo {year} {2007})}\BibitemShut {NoStop}%
\bibitem [{\citenamefont {Bochmann}\ \emph {et~al.}(2013)\citenamefont
  {Bochmann}, \citenamefont {Vainsencher}, \citenamefont {Awschalom},\ and\
  \citenamefont {Cleland}}]{Bochmann2013NatPhys_MOTrans_Optomechanics}%
  \BibitemOpen
  \bibfield  {author} {\bibinfo {author} {\bibfnamefont {J.}~\bibnamefont
  {Bochmann}}, \bibinfo {author} {\bibfnamefont {A.}~\bibnamefont
  {Vainsencher}}, \bibinfo {author} {\bibfnamefont {D.~D.}\ \bibnamefont
  {Awschalom}}, \ and\ \bibinfo {author} {\bibfnamefont {A.~N.}\ \bibnamefont
  {Cleland}},\ }\bibfield  {title} {\enquote {\bibinfo {title} {Nanomechanical
  coupling between microwave and optical photons},}\ }\href {\doibase
  10.1038/nphys2748} {\bibfield  {journal} {\bibinfo  {journal} {Nat. Phys.}\
  }\textbf {\bibinfo {volume} {9}},\ \bibinfo {pages} {712} (\bibinfo {year}
  {2013})}\BibitemShut {NoStop}%
\bibitem [{\citenamefont {Andrews}\ \emph {et~al.}(2014)\citenamefont
  {Andrews}, \citenamefont {Peterson}, \citenamefont {Purdy}, \citenamefont
  {Cicak}, \citenamefont {Simmonds}, \citenamefont {Regal},\ and\ \citenamefont
  {Lehnert}}]{Andrews2014NatPhys_MOTrans_Optomechanics}%
  \BibitemOpen
  \bibfield  {author} {\bibinfo {author} {\bibfnamefont {R.~W.}\ \bibnamefont
  {Andrews}}, \bibinfo {author} {\bibfnamefont {R.~W.}\ \bibnamefont
  {Peterson}}, \bibinfo {author} {\bibfnamefont {T.~P.}\ \bibnamefont {Purdy}},
  \bibinfo {author} {\bibfnamefont {K.}~\bibnamefont {Cicak}}, \bibinfo
  {author} {\bibfnamefont {R.~W.}\ \bibnamefont {Simmonds}}, \bibinfo {author}
  {\bibfnamefont {C.~A.}\ \bibnamefont {Regal}}, \ and\ \bibinfo {author}
  {\bibfnamefont {K.~W.}\ \bibnamefont {Lehnert}},\ }\bibfield  {title}
  {\enquote {\bibinfo {title} {Bidirectional and efficient conversion between
  microwave and optical light},}\ }\href {\doibase 10.1038/nphys2911}
  {\bibfield  {journal} {\bibinfo  {journal} {Nat. Phys.}\ }\textbf {\bibinfo
  {volume} {10}},\ \bibinfo {pages} {321} (\bibinfo {year} {2014})}\BibitemShut
  {NoStop}%
\bibitem [{\citenamefont {Fan}\ \emph {et~al.}(2015)\citenamefont {Fan},
  \citenamefont {Fong}, \citenamefont {Poot},\ and\ \citenamefont
  {Tang}}]{FanLinran2015NatCommun_MOTrans_Optomechanics}%
  \BibitemOpen
  \bibfield  {author} {\bibinfo {author} {\bibfnamefont {L.}~\bibnamefont
  {Fan}}, \bibinfo {author} {\bibfnamefont {K.~Y.}\ \bibnamefont {Fong}},
  \bibinfo {author} {\bibfnamefont {M.}~\bibnamefont {Poot}}, \ and\ \bibinfo
  {author} {\bibfnamefont {H.~X.}\ \bibnamefont {Tang}},\ }\bibfield  {title}
  {\enquote {\bibinfo {title} {Cascaded optical transparency in
  multimode-cavity optomechanical systems},}\ }\href {\doibase
  10.1038/ncomms6850} {\bibfield  {journal} {\bibinfo  {journal} {Nat.
  Commun.}\ }\textbf {\bibinfo {volume} {6}},\ \bibinfo {pages} {5850}
  (\bibinfo {year} {2015})}\BibitemShut {NoStop}%
\bibitem [{\citenamefont {Lecocq}\ \emph {et~al.}(2016)\citenamefont {Lecocq},
  \citenamefont {Clark}, \citenamefont {Simmonds}, \citenamefont {Aumentado},\
  and\ \citenamefont {Teufel}}]{Lecocq2016PRL_MOTrans_Optomechanics}%
  \BibitemOpen
  \bibfield  {author} {\bibinfo {author} {\bibfnamefont {F.}~\bibnamefont
  {Lecocq}}, \bibinfo {author} {\bibfnamefont {J.~B.}\ \bibnamefont {Clark}},
  \bibinfo {author} {\bibfnamefont {R.~W.}\ \bibnamefont {Simmonds}}, \bibinfo
  {author} {\bibfnamefont {J.}~\bibnamefont {Aumentado}}, \ and\ \bibinfo
  {author} {\bibfnamefont {J.~D.}\ \bibnamefont {Teufel}},\ }\bibfield  {title}
  {\enquote {\bibinfo {title} {Mechanically mediated microwave frequency
  conversion in the quantum regime},}\ }\href {\doibase
  10.1103/PhysRevLett.116.043601} {\bibfield  {journal} {\bibinfo  {journal}
  {Phys. Rev. Lett.}\ }\textbf {\bibinfo {volume} {116}},\ \bibinfo {pages}
  {043601} (\bibinfo {year} {2016})}\BibitemShut {NoStop}%
\bibitem [{\citenamefont {Balram}\ \emph {et~al.}(2016)\citenamefont {Balram},
  \citenamefont {Davan{\c c}o}, \citenamefont {Song},\ and\ \citenamefont
  {Srinivasan}}]{Balram2016NatPhoton_MOTrans_Optomechanics}%
  \BibitemOpen
  \bibfield  {author} {\bibinfo {author} {\bibfnamefont {K.~C.}\ \bibnamefont
  {Balram}}, \bibinfo {author} {\bibfnamefont {M.~I.}\ \bibnamefont {Davan{\c
  c}o}}, \bibinfo {author} {\bibfnamefont {J.~D.}\ \bibnamefont {Song}}, \ and\
  \bibinfo {author} {\bibfnamefont {K.}~\bibnamefont {Srinivasan}},\ }\bibfield
   {title} {\enquote {\bibinfo {title} {Coherent coupling between
  radiofrequency, optical and acoustic waves in piezo-optomechanical
  circuits},}\ }\href {\doibase 10.1038/nphoton.2016.46} {\bibfield  {journal}
  {\bibinfo  {journal} {Nat. Photon.}\ }\textbf {\bibinfo {volume} {10}},\
  \bibinfo {pages} {346} (\bibinfo {year} {2016})}\BibitemShut {NoStop}%
\bibitem [{\citenamefont {Fan}\ \emph {et~al.}(2016)\citenamefont {Fan},
  \citenamefont {Zou}, \citenamefont {Poot}, \citenamefont {Cheng},
  \citenamefont {Guo}, \citenamefont {Han},\ and\ \citenamefont
  {Tang}}]{FanLinRan2016NatPhoton_MOTrans_Optomechanics}%
  \BibitemOpen
  \bibfield  {author} {\bibinfo {author} {\bibfnamefont {L.}~\bibnamefont
  {Fan}}, \bibinfo {author} {\bibfnamefont {C.-L.}\ \bibnamefont {Zou}},
  \bibinfo {author} {\bibfnamefont {M.}~\bibnamefont {Poot}}, \bibinfo {author}
  {\bibfnamefont {R.}~\bibnamefont {Cheng}}, \bibinfo {author} {\bibfnamefont
  {X.}~\bibnamefont {Guo}}, \bibinfo {author} {\bibfnamefont {X.}~\bibnamefont
  {Han}}, \ and\ \bibinfo {author} {\bibfnamefont {H.~X.}\ \bibnamefont
  {Tang}},\ }\bibfield  {title} {\enquote {\bibinfo {title} {Integrated
  optomechanical single-photon frequency shifter},}\ }\href {\doibase
  10.1038/nphoton.2016.206} {\bibfield  {journal} {\bibinfo  {journal} {Nat.
  Photon.}\ }\textbf {\bibinfo {volume} {10}},\ \bibinfo {pages} {766}
  (\bibinfo {year} {2016})}\BibitemShut {NoStop}%
\bibitem [{\citenamefont {Shao}\ \emph {et~al.}(2019)\citenamefont {Shao},
  \citenamefont {Yu}, \citenamefont {Maity}, \citenamefont {Sinclair},
  \citenamefont {Zheng}, \citenamefont {Chia}, \citenamefont {Shams-Ansari},
  \citenamefont {Wang}, \citenamefont {Zhang}, \citenamefont {Lai},\ and\
  \citenamefont {Lon\v{c}ar}}]{ShaoLinbo2019Optica_MOTrans_Optomechanics}%
  \BibitemOpen
  \bibfield  {author} {\bibinfo {author} {\bibfnamefont {L.}~\bibnamefont
  {Shao}}, \bibinfo {author} {\bibfnamefont {M.}~\bibnamefont {Yu}}, \bibinfo
  {author} {\bibfnamefont {S.}~\bibnamefont {Maity}}, \bibinfo {author}
  {\bibfnamefont {N.}~\bibnamefont {Sinclair}}, \bibinfo {author}
  {\bibfnamefont {L.}~\bibnamefont {Zheng}}, \bibinfo {author} {\bibfnamefont
  {C.}~\bibnamefont {Chia}}, \bibinfo {author} {\bibfnamefont {A.}~\bibnamefont
  {Shams-Ansari}}, \bibinfo {author} {\bibfnamefont {C.}~\bibnamefont {Wang}},
  \bibinfo {author} {\bibfnamefont {M.}~\bibnamefont {Zhang}}, \bibinfo
  {author} {\bibfnamefont {K.}~\bibnamefont {Lai}}, \ and\ \bibinfo {author}
  {\bibfnamefont {M.}~\bibnamefont {Lon\v{c}ar}},\ }\bibfield  {title}
  {\enquote {\bibinfo {title} {Microwave-to-optical conversion using lithium
  niobate thin-film acoustic resonators},}\ }\href {\doibase
  10.1364/OPTICA.6.001498} {\bibfield  {journal} {\bibinfo  {journal} {Optica}\
  }\textbf {\bibinfo {volume} {6}},\ \bibinfo {pages} {1498} (\bibinfo {year}
  {2019})}\BibitemShut {NoStop}%
\bibitem [{\citenamefont {Han}\ \emph {et~al.}(2020)\citenamefont {Han},
  \citenamefont {Fu}, \citenamefont {Zhong}, \citenamefont {Zou}, \citenamefont
  {Xu}, \citenamefont {Sayem}, \citenamefont {Xu}, \citenamefont {Wang},
  \citenamefont {Cheng}, \citenamefont {Jiang},\ and\ \citenamefont
  {Tang}}]{XuHan2020NatCommun_MOTrans_Optomechanics}%
  \BibitemOpen
  \bibfield  {author} {\bibinfo {author} {\bibfnamefont {X.}~\bibnamefont
  {Han}}, \bibinfo {author} {\bibfnamefont {W.}~\bibnamefont {Fu}}, \bibinfo
  {author} {\bibfnamefont {C.}~\bibnamefont {Zhong}}, \bibinfo {author}
  {\bibfnamefont {C.-L.}\ \bibnamefont {Zou}}, \bibinfo {author} {\bibfnamefont
  {Y.}~\bibnamefont {Xu}}, \bibinfo {author} {\bibfnamefont {A.~A.}\
  \bibnamefont {Sayem}}, \bibinfo {author} {\bibfnamefont {M.}~\bibnamefont
  {Xu}}, \bibinfo {author} {\bibfnamefont {S.}~\bibnamefont {Wang}}, \bibinfo
  {author} {\bibfnamefont {R.}~\bibnamefont {Cheng}}, \bibinfo {author}
  {\bibfnamefont {L.}~\bibnamefont {Jiang}}, \ and\ \bibinfo {author}
  {\bibfnamefont {H.~X.}\ \bibnamefont {Tang}},\ }\bibfield  {title} {\enquote
  {\bibinfo {title} {Cavity piezo-mechanics for superconducting-nanophotonic
  quantum interface},}\ }\href {\doibase 10.1038/s41467-020-17053-3} {\bibfield
   {journal} {\bibinfo  {journal} {Nat. Commun.}\ }\textbf {\bibinfo {volume}
  {11}},\ \bibinfo {pages} {3237} (\bibinfo {year} {2020})}\BibitemShut
  {NoStop}%
\bibitem [{\citenamefont {Tsang}(2010)}]{Tsang2010PRA_MOTrans_EO}%
  \BibitemOpen
  \bibfield  {author} {\bibinfo {author} {\bibfnamefont {M.}~\bibnamefont
  {Tsang}},\ }\bibfield  {title} {\enquote {\bibinfo {title} {Cavity quantum
  electro-optics},}\ }\href {\doibase 10.1103/PhysRevA.81.063837} {\bibfield
  {journal} {\bibinfo  {journal} {Phys. Rev. A}\ }\textbf {\bibinfo {volume}
  {81}},\ \bibinfo {pages} {063837} (\bibinfo {year} {2010})}\BibitemShut
  {NoStop}%
\bibitem [{\citenamefont {Tsang}(2011)}]{Tsang2011PRA2_MOTrans_EO}%
  \BibitemOpen
  \bibfield  {author} {\bibinfo {author} {\bibfnamefont {M.}~\bibnamefont
  {Tsang}},\ }\bibfield  {title} {\enquote {\bibinfo {title} {Cavity quantum
  electro-optics. ii. input-output relations between traveling optical and
  microwave fields},}\ }\href {\doibase 10.1103/PhysRevA.84.043845} {\bibfield
  {journal} {\bibinfo  {journal} {Phys. Rev. A}\ }\textbf {\bibinfo {volume}
  {84}},\ \bibinfo {pages} {043845} (\bibinfo {year} {2011})}\BibitemShut
  {NoStop}%
\bibitem [{\citenamefont {Rueda}\ \emph {et~al.}(2016)\citenamefont {Rueda},
  \citenamefont {Sedlmeir}, \citenamefont {Collodo}, \citenamefont {Vogl},
  \citenamefont {Stiller}, \citenamefont {Schunk}, \citenamefont {Strekalov},
  \citenamefont {Marquardt}, \citenamefont {Fink}, \citenamefont {Painter},
  \citenamefont {Leuchs},\ and\ \citenamefont
  {Schwefel}}]{Rueda2016Optica_MOTrans_EO}%
  \BibitemOpen
  \bibfield  {author} {\bibinfo {author} {\bibfnamefont {A.}~\bibnamefont
  {Rueda}}, \bibinfo {author} {\bibfnamefont {F.}~\bibnamefont {Sedlmeir}},
  \bibinfo {author} {\bibfnamefont {M.~C.}\ \bibnamefont {Collodo}}, \bibinfo
  {author} {\bibfnamefont {U.}~\bibnamefont {Vogl}}, \bibinfo {author}
  {\bibfnamefont {B.}~\bibnamefont {Stiller}}, \bibinfo {author} {\bibfnamefont
  {G.}~\bibnamefont {Schunk}}, \bibinfo {author} {\bibfnamefont {D.~V.}\
  \bibnamefont {Strekalov}}, \bibinfo {author} {\bibfnamefont {C.}~\bibnamefont
  {Marquardt}}, \bibinfo {author} {\bibfnamefont {J.~M.}\ \bibnamefont {Fink}},
  \bibinfo {author} {\bibfnamefont {O.}~\bibnamefont {Painter}}, \bibinfo
  {author} {\bibfnamefont {G.}~\bibnamefont {Leuchs}}, \ and\ \bibinfo {author}
  {\bibfnamefont {H.~G.~L.}\ \bibnamefont {Schwefel}},\ }\bibfield  {title}
  {\enquote {\bibinfo {title} {Efficient microwave to optical photon
  conversion: an electro-optical realization},}\ }\href {\doibase
  10.1364/OPTICA.3.000597} {\bibfield  {journal} {\bibinfo  {journal} {Optica}\
  }\textbf {\bibinfo {volume} {3}},\ \bibinfo {pages} {597} (\bibinfo {year}
  {2016})}\BibitemShut {NoStop}%
\bibitem [{\citenamefont {Javerzac-Galy}\ \emph {et~al.}(2016)\citenamefont
  {Javerzac-Galy}, \citenamefont {Plekhanov}, \citenamefont {Bernier},
  \citenamefont {Toth}, \citenamefont {Feofanov},\ and\ \citenamefont
  {Kippenberg}}]{Javerzac-Galy2016PRA_MOTrans_EO}%
  \BibitemOpen
  \bibfield  {author} {\bibinfo {author} {\bibfnamefont {C.}~\bibnamefont
  {Javerzac-Galy}}, \bibinfo {author} {\bibfnamefont {K.}~\bibnamefont
  {Plekhanov}}, \bibinfo {author} {\bibfnamefont {N.~R.}\ \bibnamefont
  {Bernier}}, \bibinfo {author} {\bibfnamefont {L.~D.}\ \bibnamefont {Toth}},
  \bibinfo {author} {\bibfnamefont {A.~K.}\ \bibnamefont {Feofanov}}, \ and\
  \bibinfo {author} {\bibfnamefont {T.~J.}\ \bibnamefont {Kippenberg}},\
  }\bibfield  {title} {\enquote {\bibinfo {title} {On-chip microwave-to-optical
  quantum coherent converter based on a superconducting resonator coupled to an
  electro-optic microresonator},}\ }\href {\doibase 10.1103/PhysRevA.94.053815}
  {\bibfield  {journal} {\bibinfo  {journal} {Phys. Rev. A}\ }\textbf {\bibinfo
  {volume} {94}},\ \bibinfo {pages} {053815} (\bibinfo {year}
  {2016})}\BibitemShut {NoStop}%
\bibitem [{\citenamefont {Fan}\ \emph {et~al.}(2018)\citenamefont {Fan},
  \citenamefont {Zou}, \citenamefont {Cheng}, \citenamefont {Guo},
  \citenamefont {Han}, \citenamefont {Gong}, \citenamefont {Wang},\ and\
  \citenamefont {Tang}}]{FanLinran2018SciAdv_MOTrans_EO}%
  \BibitemOpen
  \bibfield  {author} {\bibinfo {author} {\bibfnamefont {L.}~\bibnamefont
  {Fan}}, \bibinfo {author} {\bibfnamefont {C.-L.}\ \bibnamefont {Zou}},
  \bibinfo {author} {\bibfnamefont {R.}~\bibnamefont {Cheng}}, \bibinfo
  {author} {\bibfnamefont {X.}~\bibnamefont {Guo}}, \bibinfo {author}
  {\bibfnamefont {X.}~\bibnamefont {Han}}, \bibinfo {author} {\bibfnamefont
  {Z.}~\bibnamefont {Gong}}, \bibinfo {author} {\bibfnamefont {S.}~\bibnamefont
  {Wang}}, \ and\ \bibinfo {author} {\bibfnamefont {H.~X.}\ \bibnamefont
  {Tang}},\ }\bibfield  {title} {\enquote {\bibinfo {title} {Superconducting
  cavity electro-optics: A platform for coherent photon conversion between
  superconducting and photonic circuits},}\ }\href {\doibase
  10.1126/sciadv.aar4994} {\bibfield  {journal} {\bibinfo  {journal} {Sci.
  Adv.}\ }\textbf {\bibinfo {volume} {4}},\ \bibinfo {pages} {eaar4994}
  (\bibinfo {year} {2018})}\BibitemShut {NoStop}%
\bibitem [{\citenamefont {Xu}\ \emph {et~al.}(2021)\citenamefont {Xu},
  \citenamefont {Sayem}, \citenamefont {Fan}, \citenamefont {Zou},
  \citenamefont {Wang}, \citenamefont {Cheng}, \citenamefont {Fu},
  \citenamefont {Yang}, \citenamefont {Xu},\ and\ \citenamefont
  {Tang}}]{XuYuntao2021NatCommun_MOTrans_EO}%
  \BibitemOpen
  \bibfield  {author} {\bibinfo {author} {\bibfnamefont {Y.}~\bibnamefont
  {Xu}}, \bibinfo {author} {\bibfnamefont {A.~A.}\ \bibnamefont {Sayem}},
  \bibinfo {author} {\bibfnamefont {L.}~\bibnamefont {Fan}}, \bibinfo {author}
  {\bibfnamefont {C.-L.}\ \bibnamefont {Zou}}, \bibinfo {author} {\bibfnamefont
  {S.}~\bibnamefont {Wang}}, \bibinfo {author} {\bibfnamefont {R.}~\bibnamefont
  {Cheng}}, \bibinfo {author} {\bibfnamefont {W.}~\bibnamefont {Fu}}, \bibinfo
  {author} {\bibfnamefont {L.}~\bibnamefont {Yang}}, \bibinfo {author}
  {\bibfnamefont {M.}~\bibnamefont {Xu}}, \ and\ \bibinfo {author}
  {\bibfnamefont {H.~X.}\ \bibnamefont {Tang}},\ }\bibfield  {title} {\enquote
  {\bibinfo {title} {Bidirectional interconversion of microwave and light with
  thin-film lithium niobate},}\ }\href {\doibase 10.1038/s41467-021-24809-y}
  {\bibfield  {journal} {\bibinfo  {journal} {Nat. Commun.}\ }\textbf {\bibinfo
  {volume} {12}},\ \bibinfo {pages} {4453} (\bibinfo {year}
  {2021})}\BibitemShut {NoStop}%
\bibitem [{\citenamefont {Hafezi}\ \emph {et~al.}(2012)\citenamefont {Hafezi},
  \citenamefont {Kim}, \citenamefont {Rolston}, \citenamefont {Orozco},
  \citenamefont {Lev},\ and\ \citenamefont
  {Taylor}}]{Hafezi2012PRA_MOTrans_Atom}%
  \BibitemOpen
  \bibfield  {author} {\bibinfo {author} {\bibfnamefont {M.}~\bibnamefont
  {Hafezi}}, \bibinfo {author} {\bibfnamefont {Z.}~\bibnamefont {Kim}},
  \bibinfo {author} {\bibfnamefont {S.~L.}\ \bibnamefont {Rolston}}, \bibinfo
  {author} {\bibfnamefont {L.~A.}\ \bibnamefont {Orozco}}, \bibinfo {author}
  {\bibfnamefont {B.~L.}\ \bibnamefont {Lev}}, \ and\ \bibinfo {author}
  {\bibfnamefont {J.~M.}\ \bibnamefont {Taylor}},\ }\bibfield  {title}
  {\enquote {\bibinfo {title} {Atomic interface between microwave and optical
  photons},}\ }\href {\doibase 10.1103/PhysRevA.85.020302} {\bibfield
  {journal} {\bibinfo  {journal} {Phys. Rev. A}\ }\textbf {\bibinfo {volume}
  {85}},\ \bibinfo {pages} {020302} (\bibinfo {year} {2012})}\BibitemShut
  {NoStop}%
\bibitem [{\citenamefont {Gard}\ \emph {et~al.}(2017)\citenamefont {Gard},
  \citenamefont {Jacobs}, \citenamefont {McDermott},\ and\ \citenamefont
  {Saffman}}]{Gard2017PRA_MOTrans_Atom}%
  \BibitemOpen
  \bibfield  {author} {\bibinfo {author} {\bibfnamefont {B.~T.}\ \bibnamefont
  {Gard}}, \bibinfo {author} {\bibfnamefont {K.}~\bibnamefont {Jacobs}},
  \bibinfo {author} {\bibfnamefont {R.}~\bibnamefont {McDermott}}, \ and\
  \bibinfo {author} {\bibfnamefont {M.}~\bibnamefont {Saffman}},\ }\bibfield
  {title} {\enquote {\bibinfo {title} {Microwave-to-optical frequency
  conversion using a cesium atom coupled to a superconducting resonator},}\
  }\href {\doibase 10.1103/PhysRevA.96.013833} {\bibfield  {journal} {\bibinfo
  {journal} {Phys. Rev. A}\ }\textbf {\bibinfo {volume} {96}},\ \bibinfo
  {pages} {013833} (\bibinfo {year} {2017})}\BibitemShut {NoStop}%
\bibitem [{\citenamefont {Han}\ \emph {et~al.}(2018)\citenamefont {Han},
  \citenamefont {Vogt}, \citenamefont {Gross}, \citenamefont {Jaksch},
  \citenamefont {Kiffner},\ and\ \citenamefont
  {Li}}]{HanJingshan2018PRL_MOTrans_Atom}%
  \BibitemOpen
  \bibfield  {author} {\bibinfo {author} {\bibfnamefont {J.}~\bibnamefont
  {Han}}, \bibinfo {author} {\bibfnamefont {T.}~\bibnamefont {Vogt}}, \bibinfo
  {author} {\bibfnamefont {C.}~\bibnamefont {Gross}}, \bibinfo {author}
  {\bibfnamefont {D.}~\bibnamefont {Jaksch}}, \bibinfo {author} {\bibfnamefont
  {M.}~\bibnamefont {Kiffner}}, \ and\ \bibinfo {author} {\bibfnamefont
  {W.}~\bibnamefont {Li}},\ }\bibfield  {title} {\enquote {\bibinfo {title}
  {Coherent microwave-to-optical conversion via six-wave mixing in rydberg
  atoms},}\ }\href {\doibase 10.1103/PhysRevLett.120.093201} {\bibfield
  {journal} {\bibinfo  {journal} {Phys. Rev. Lett.}\ }\textbf {\bibinfo
  {volume} {120}},\ \bibinfo {pages} {093201} (\bibinfo {year}
  {2018})}\BibitemShut {NoStop}%
\bibitem [{\citenamefont {Zhu}\ \emph {et~al.}(2020)\citenamefont {Zhu},
  \citenamefont {Zhang}, \citenamefont {Han}, \citenamefont {Zou},
  \citenamefont {Zhong}, \citenamefont {Wang}, \citenamefont {Jiang},\ and\
  \citenamefont {Tang}}]{ZhuNa2020Optica_MOTrans}%
  \BibitemOpen
  \bibfield  {author} {\bibinfo {author} {\bibfnamefont {N.}~\bibnamefont
  {Zhu}}, \bibinfo {author} {\bibfnamefont {X.}~\bibnamefont {Zhang}}, \bibinfo
  {author} {\bibfnamefont {X.}~\bibnamefont {Han}}, \bibinfo {author}
  {\bibfnamefont {C.-L.}\ \bibnamefont {Zou}}, \bibinfo {author} {\bibfnamefont
  {C.}~\bibnamefont {Zhong}}, \bibinfo {author} {\bibfnamefont {C.-H.}\
  \bibnamefont {Wang}}, \bibinfo {author} {\bibfnamefont {L.}~\bibnamefont
  {Jiang}}, \ and\ \bibinfo {author} {\bibfnamefont {H.~X.}\ \bibnamefont
  {Tang}},\ }\bibfield  {title} {\enquote {\bibinfo {title} {Waveguide cavity
  optomagnonics for microwave-to-optics conversion},}\ }\href {\doibase
  10.1364/OPTICA.397967} {\bibfield  {journal} {\bibinfo  {journal} {Optica}\
  }\textbf {\bibinfo {volume} {7}},\ \bibinfo {pages} {1291} (\bibinfo {year}
  {2020})}\BibitemShut {NoStop}%
\bibitem [{\citenamefont {Bagci}\ \emph {et~al.}(2014)\citenamefont {Bagci},
  \citenamefont {Simonsen}, \citenamefont {Schmid}, \citenamefont {Villanueva},
  \citenamefont {Zeuthen}, \citenamefont {Appel}, \citenamefont {Taylor},
  \citenamefont {S{\o}rensen}, \citenamefont {Usami}, \citenamefont
  {Schliesser},\ and\ \citenamefont
  {Polzik}}]{Bagci2014Nature_MicOptConvertor}%
  \BibitemOpen
  \bibfield  {author} {\bibinfo {author} {\bibfnamefont {T.}~\bibnamefont
  {Bagci}}, \bibinfo {author} {\bibfnamefont {A.}~\bibnamefont {Simonsen}},
  \bibinfo {author} {\bibfnamefont {S.}~\bibnamefont {Schmid}}, \bibinfo
  {author} {\bibfnamefont {L.~G.}\ \bibnamefont {Villanueva}}, \bibinfo
  {author} {\bibfnamefont {E.}~\bibnamefont {Zeuthen}}, \bibinfo {author}
  {\bibfnamefont {J.}~\bibnamefont {Appel}}, \bibinfo {author} {\bibfnamefont
  {J.~M.}\ \bibnamefont {Taylor}}, \bibinfo {author} {\bibfnamefont
  {A.}~\bibnamefont {S{\o}rensen}}, \bibinfo {author} {\bibfnamefont
  {K.}~\bibnamefont {Usami}}, \bibinfo {author} {\bibfnamefont
  {A.}~\bibnamefont {Schliesser}}, \ and\ \bibinfo {author} {\bibfnamefont
  {E.~S.}\ \bibnamefont {Polzik}},\ }\bibfield  {title} {\enquote {\bibinfo
  {title} {Optical detection of radio waves through a nanomechanical
  transducer},}\ }\href {\doibase 10.1038/nature13029} {\bibfield  {journal}
  {\bibinfo  {journal} {Nature}\ }\textbf {\bibinfo {volume} {507}},\ \bibinfo
  {pages} {81} (\bibinfo {year} {2014})}\BibitemShut {NoStop}%
\bibitem [{\citenamefont {Wolski}\ \emph {et~al.}(2020)\citenamefont {Wolski},
  \citenamefont {Lachance-Quirion}, \citenamefont {Tabuchi}, \citenamefont
  {Kono}, \citenamefont {Noguchi}, \citenamefont {Usami},\ and\ \citenamefont
  {Nakamura}}]{Wolski2020QuantumSensing}%
  \BibitemOpen
  \bibfield  {author} {\bibinfo {author} {\bibfnamefont {S.~P.}\ \bibnamefont
  {Wolski}}, \bibinfo {author} {\bibfnamefont {D.}~\bibnamefont
  {Lachance-Quirion}}, \bibinfo {author} {\bibfnamefont {Y.}~\bibnamefont
  {Tabuchi}}, \bibinfo {author} {\bibfnamefont {S.}~\bibnamefont {Kono}},
  \bibinfo {author} {\bibfnamefont {A.}~\bibnamefont {Noguchi}}, \bibinfo
  {author} {\bibfnamefont {K.}~\bibnamefont {Usami}}, \ and\ \bibinfo {author}
  {\bibfnamefont {Y.}~\bibnamefont {Nakamura}},\ }\bibfield  {title} {\enquote
  {\bibinfo {title} {Dissipation-based quantum sensing of magnons with a
  superconducting qubit},}\ }\href {\doibase 10.1103/PhysRevLett.125.117701}
  {\bibfield  {journal} {\bibinfo  {journal} {Phys. Rev. Lett.}\ }\textbf
  {\bibinfo {volume} {125}},\ \bibinfo {pages} {117701} (\bibinfo {year}
  {2020})}\BibitemShut {NoStop}%
\bibitem [{\citenamefont {Lloyd}(2008)}]{Lloyd2008QuantumIllumination}%
  \BibitemOpen
  \bibfield  {author} {\bibinfo {author} {\bibfnamefont {S.}~\bibnamefont
  {Lloyd}},\ }\bibfield  {title} {\enquote {\bibinfo {title} {Enhanced
  sensitivity of photodetection via quantum illumination},}\ }\href {\doibase
  10.1126/science.1160627} {\bibfield  {journal} {\bibinfo  {journal}
  {Science}\ }\textbf {\bibinfo {volume} {321}},\ \bibinfo {pages} {1463}
  (\bibinfo {year} {2008})}\BibitemShut {NoStop}%
\bibitem [{\citenamefont {Tan}\ \emph {et~al.}(2008)\citenamefont {Tan},
  \citenamefont {Erkmen}, \citenamefont {Giovannetti}, \citenamefont {Guha},
  \citenamefont {Lloyd}, \citenamefont {Maccone}, \citenamefont {Pirandola},\
  and\ \citenamefont {Shapiro}}]{Tan2008QuantumIllumination}%
  \BibitemOpen
  \bibfield  {author} {\bibinfo {author} {\bibfnamefont {S.-H.}\ \bibnamefont
  {Tan}}, \bibinfo {author} {\bibfnamefont {B.~I.}\ \bibnamefont {Erkmen}},
  \bibinfo {author} {\bibfnamefont {V.}~\bibnamefont {Giovannetti}}, \bibinfo
  {author} {\bibfnamefont {S.}~\bibnamefont {Guha}}, \bibinfo {author}
  {\bibfnamefont {S.}~\bibnamefont {Lloyd}}, \bibinfo {author} {\bibfnamefont
  {L.}~\bibnamefont {Maccone}}, \bibinfo {author} {\bibfnamefont
  {S.}~\bibnamefont {Pirandola}}, \ and\ \bibinfo {author} {\bibfnamefont
  {J.~H.}\ \bibnamefont {Shapiro}},\ }\bibfield  {title} {\enquote {\bibinfo
  {title} {Quantum illumination with gaussian states},}\ }\href {\doibase
  10.1103/PhysRevLett.101.253601} {\bibfield  {journal} {\bibinfo  {journal}
  {Phys. Rev. Lett.}\ }\textbf {\bibinfo {volume} {101}},\ \bibinfo {pages}
  {253601} (\bibinfo {year} {2008})}\BibitemShut {NoStop}%
\bibitem [{\citenamefont {Barzanjeh}\ \emph {et~al.}(2015)\citenamefont
  {Barzanjeh}, \citenamefont {Guha}, \citenamefont {Weedbrook}, \citenamefont
  {Vitali}, \citenamefont {Shapiro},\ and\ \citenamefont
  {Pirandola}}]{Barzanjeh2015QuantumIllumination}%
  \BibitemOpen
  \bibfield  {author} {\bibinfo {author} {\bibfnamefont {S.}~\bibnamefont
  {Barzanjeh}}, \bibinfo {author} {\bibfnamefont {S.}~\bibnamefont {Guha}},
  \bibinfo {author} {\bibfnamefont {C.}~\bibnamefont {Weedbrook}}, \bibinfo
  {author} {\bibfnamefont {D.}~\bibnamefont {Vitali}}, \bibinfo {author}
  {\bibfnamefont {J.~H.}\ \bibnamefont {Shapiro}}, \ and\ \bibinfo {author}
  {\bibfnamefont {S.}~\bibnamefont {Pirandola}},\ }\bibfield  {title} {\enquote
  {\bibinfo {title} {Microwave quantum illumination},}\ }\href {\doibase
  10.1103/PhysRevLett.114.080503} {\bibfield  {journal} {\bibinfo  {journal}
  {Phys. Rev. Lett.}\ }\textbf {\bibinfo {volume} {114}},\ \bibinfo {pages}
  {080503} (\bibinfo {year} {2015})}\BibitemShut {NoStop}%
\bibitem [{\citenamefont {Cai}\ \emph {et~al.}(2021)\citenamefont {Cai},
  \citenamefont {Liao}, \citenamefont {Shen}, \citenamefont {Guo},\ and\
  \citenamefont {Zhou}}]{Cai2021QuantumIllumination}%
  \BibitemOpen
  \bibfield  {author} {\bibinfo {author} {\bibfnamefont {Q.}~\bibnamefont
  {Cai}}, \bibinfo {author} {\bibfnamefont {J.}~\bibnamefont {Liao}}, \bibinfo
  {author} {\bibfnamefont {B.}~\bibnamefont {Shen}}, \bibinfo {author}
  {\bibfnamefont {G.}~\bibnamefont {Guo}}, \ and\ \bibinfo {author}
  {\bibfnamefont {Q.}~\bibnamefont {Zhou}},\ }\bibfield  {title} {\enquote
  {\bibinfo {title} {Microwave quantum illumination via cavity magnonics},}\
  }\href {\doibase 10.1103/PhysRevA.103.052419} {\bibfield  {journal} {\bibinfo
   {journal} {Phys. Rev. A}\ }\textbf {\bibinfo {volume} {103}},\ \bibinfo
  {pages} {052419} (\bibinfo {year} {2021})}\BibitemShut {NoStop}%
\bibitem [{\citenamefont {Zhang}\ \emph {et~al.}(2017)\citenamefont {Zhang},
  \citenamefont {Luo}, \citenamefont {Wang}, \citenamefont {Li},\ and\
  \citenamefont {You}}]{Zhang2017Observation}%
  \BibitemOpen
  \bibfield  {author} {\bibinfo {author} {\bibfnamefont {D.}~\bibnamefont
  {Zhang}}, \bibinfo {author} {\bibfnamefont {X.-Q.}\ \bibnamefont {Luo}},
  \bibinfo {author} {\bibfnamefont {Y.-P.}\ \bibnamefont {Wang}}, \bibinfo
  {author} {\bibfnamefont {T.-F.}\ \bibnamefont {Li}}, \ and\ \bibinfo {author}
  {\bibfnamefont {J.~Q.}\ \bibnamefont {You}},\ }\bibfield  {title} {\enquote
  {\bibinfo {title} {Observation of the exceptional point in cavity
  magnon-polaritons},}\ }\href {\doibase 10.1038/s41467-017-01634-w} {\bibfield
   {journal} {\bibinfo  {journal} {Nat. Commun.}\ }\textbf {\bibinfo {volume}
  {8}},\ \bibinfo {pages} {1368} (\bibinfo {year} {2017})}\BibitemShut
  {NoStop}%
\bibitem [{\citenamefont {Zhang}\ and\ \citenamefont
  {You}(2019)}]{Zhang2019HigherOrder}%
  \BibitemOpen
  \bibfield  {author} {\bibinfo {author} {\bibfnamefont {G.-Q.}\ \bibnamefont
  {Zhang}}\ and\ \bibinfo {author} {\bibfnamefont {J.~Q.}\ \bibnamefont
  {You}},\ }\bibfield  {title} {\enquote {\bibinfo {title} {Higher-order
  exceptional point in a cavity magnonics system},}\ }\href {\doibase
  10.1103/PhysRevB.99.054404} {\bibfield  {journal} {\bibinfo  {journal} {Phys.
  Rev. B}\ }\textbf {\bibinfo {volume} {99}},\ \bibinfo {pages} {054404}
  (\bibinfo {year} {2019})}\BibitemShut {NoStop}%
\bibitem [{\citenamefont {Wiersig}(2014)}]{Jan2014Enhancing}%
  \BibitemOpen
  \bibfield  {author} {\bibinfo {author} {\bibfnamefont {J.}~\bibnamefont
  {Wiersig}},\ }\bibfield  {title} {\enquote {\bibinfo {title} {Enhancing the
  sensitivity of frequency and energy splitting detection by using exceptional
  points: Application to microcavity sensors for single-particle detection},}\
  }\href {\doibase 10.1103/PhysRevLett.112.203901} {\bibfield  {journal}
  {\bibinfo  {journal} {Phys. Rev. Lett.}\ }\textbf {\bibinfo {volume} {112}},\
  \bibinfo {pages} {203901} (\bibinfo {year} {2014})}\BibitemShut {NoStop}%
\bibitem [{\citenamefont {Liu}\ \emph {et~al.}(2016)\citenamefont {Liu},
  \citenamefont {Zhang}, \citenamefont {\"Ozdemir}, \citenamefont {Peng},
  \citenamefont {Jing}, \citenamefont {L\"u}, \citenamefont {Li}, \citenamefont
  {Yang}, \citenamefont {Nori},\ and\ \citenamefont {Liu}}]{Liu2016Metrology}%
  \BibitemOpen
  \bibfield  {author} {\bibinfo {author} {\bibfnamefont {Z.-P.}\ \bibnamefont
  {Liu}}, \bibinfo {author} {\bibfnamefont {J.}~\bibnamefont {Zhang}}, \bibinfo
  {author} {\bibfnamefont {i.~m. c.~K.}\ \bibnamefont {\"Ozdemir}}, \bibinfo
  {author} {\bibfnamefont {B.}~\bibnamefont {Peng}}, \bibinfo {author}
  {\bibfnamefont {H.}~\bibnamefont {Jing}}, \bibinfo {author} {\bibfnamefont
  {X.-Y.}\ \bibnamefont {L\"u}}, \bibinfo {author} {\bibfnamefont {C.-W.}\
  \bibnamefont {Li}}, \bibinfo {author} {\bibfnamefont {L.}~\bibnamefont
  {Yang}}, \bibinfo {author} {\bibfnamefont {F.}~\bibnamefont {Nori}}, \ and\
  \bibinfo {author} {\bibfnamefont {Y.-x.}\ \bibnamefont {Liu}},\ }\bibfield
  {title} {\enquote {\bibinfo {title} {Metrology with $\mathcal{PT}$-symmetric
  cavities: Enhanced sensitivity near the $\mathcal{PT}$-phase transition},}\
  }\href {\doibase 10.1103/PhysRevLett.117.110802} {\bibfield  {journal}
  {\bibinfo  {journal} {Phys. Rev. Lett.}\ }\textbf {\bibinfo {volume} {117}},\
  \bibinfo {pages} {110802} (\bibinfo {year} {2016})}\BibitemShut {NoStop}%
\bibitem [{\citenamefont {Chen}\ \emph
  {et~al.}(2017{\natexlab{b}})\citenamefont {Chen}, \citenamefont
  {Kaya~{\"O}zdemir}, \citenamefont {Zhao}, \citenamefont {Wiersig},\ and\
  \citenamefont {Yang}}]{Chen2017Exceptional}%
  \BibitemOpen
  \bibfield  {author} {\bibinfo {author} {\bibfnamefont {W.}~\bibnamefont
  {Chen}}, \bibinfo {author} {\bibfnamefont {{\c{S}}.}~\bibnamefont
  {Kaya~{\"O}zdemir}}, \bibinfo {author} {\bibfnamefont {G.}~\bibnamefont
  {Zhao}}, \bibinfo {author} {\bibfnamefont {J.}~\bibnamefont {Wiersig}}, \
  and\ \bibinfo {author} {\bibfnamefont {L.}~\bibnamefont {Yang}},\ }\bibfield
  {title} {\enquote {\bibinfo {title} {Exceptional points enhance sensing in an
  optical microcavity},}\ }\href {\doibase 10.1038/nature23281} {\bibfield
  {journal} {\bibinfo  {journal} {Nature}\ }\textbf {\bibinfo {volume} {548}},\
  \bibinfo {pages} {192} (\bibinfo {year} {2017}{\natexlab{b}})}\BibitemShut
  {NoStop}%
\bibitem [{\citenamefont {Cao}\ and\ \citenamefont
  {Yan}(2019)}]{Cao2019EPsensitivity}%
  \BibitemOpen
  \bibfield  {author} {\bibinfo {author} {\bibfnamefont {Y.}~\bibnamefont
  {Cao}}\ and\ \bibinfo {author} {\bibfnamefont {P.}~\bibnamefont {Yan}},\
  }\bibfield  {title} {\enquote {\bibinfo {title} {Exceptional magnetic
  sensitivity of $\mathcal{P}\mathcal{T}$-symmetric cavity magnon
  polaritons},}\ }\href {\doibase 10.1103/PhysRevB.99.214415} {\bibfield
  {journal} {\bibinfo  {journal} {Phys. Rev. B}\ }\textbf {\bibinfo {volume}
  {99}},\ \bibinfo {pages} {214415} (\bibinfo {year} {2019})}\BibitemShut
  {NoStop}%
\bibitem [{\citenamefont {Wang}, \citenamefont {Guo},\ and\ \citenamefont
  {Berakdar}(2021)}]{Wang2021Enhanced}%
  \BibitemOpen
  \bibfield  {author} {\bibinfo {author} {\bibfnamefont {X.-g.}\ \bibnamefont
  {Wang}}, \bibinfo {author} {\bibfnamefont {G.-h.}\ \bibnamefont {Guo}}, \
  and\ \bibinfo {author} {\bibfnamefont {J.}~\bibnamefont {Berakdar}},\
  }\bibfield  {title} {\enquote {\bibinfo {title} {Enhanced sensitivity at
  magnetic high-order exceptional points and topological energy transfer in
  magnonic planar waveguides},}\ }\href {\doibase
  10.1103/PhysRevApplied.15.034050} {\bibfield  {journal} {\bibinfo  {journal}
  {Phys. Rev. Appl.}\ }\textbf {\bibinfo {volume} {15}},\ \bibinfo {pages}
  {034050} (\bibinfo {year} {2021})}\BibitemShut {NoStop}%
\bibitem [{\citenamefont {Zhang}, \citenamefont {Wang},\ and\ \citenamefont
  {Xiong}(2023)}]{Zhang2023PRB_detection}%
  \BibitemOpen
  \bibfield  {author} {\bibinfo {author} {\bibfnamefont {G.-Q.}\ \bibnamefont
  {Zhang}}, \bibinfo {author} {\bibfnamefont {Y.}~\bibnamefont {Wang}}, \ and\
  \bibinfo {author} {\bibfnamefont {W.}~\bibnamefont {Xiong}},\ }\bibfield
  {title} {\enquote {\bibinfo {title} {Detection sensitivity enhancement of
  magnon kerr nonlinearity in cavity magnonics induced by coherent perfect
  absorption},}\ }\href {\doibase 10.1103/PhysRevB.107.064417} {\bibfield
  {journal} {\bibinfo  {journal} {Phys. Rev. B}\ }\textbf {\bibinfo {volume}
  {107}},\ \bibinfo {pages} {064417} (\bibinfo {year} {2023})}\BibitemShut
  {NoStop}%
\bibitem [{\citenamefont {Stancil}\ and\ \citenamefont
  {Prabhakar}(2009)}]{Stancil2009SpinWave}%
  \BibitemOpen
  \bibfield  {author} {\bibinfo {author} {\bibfnamefont {D.~D.}\ \bibnamefont
  {Stancil}}\ and\ \bibinfo {author} {\bibfnamefont {A.}~\bibnamefont
  {Prabhakar}},\ }\href@noop {} {\emph {\bibinfo {title} {Spin waves}}},\
  Vol.~\bibinfo {volume} {5}\ (\bibinfo  {publisher} {Springer},\ \bibinfo
  {year} {2009})\BibitemShut {NoStop}%
\bibitem [{\citenamefont {Chong}\ \emph {et~al.}(2010)\citenamefont {Chong},
  \citenamefont {Ge}, \citenamefont {Cao},\ and\ \citenamefont
  {Stone}}]{Chong2010CPA}%
  \BibitemOpen
  \bibfield  {author} {\bibinfo {author} {\bibfnamefont {Y.~D.}\ \bibnamefont
  {Chong}}, \bibinfo {author} {\bibfnamefont {L.}~\bibnamefont {Ge}}, \bibinfo
  {author} {\bibfnamefont {H.}~\bibnamefont {Cao}}, \ and\ \bibinfo {author}
  {\bibfnamefont {A.~D.}\ \bibnamefont {Stone}},\ }\bibfield  {title} {\enquote
  {\bibinfo {title} {Coherent perfect absorbers: Time-reversed lasers},}\
  }\href {\doibase 10.1103/PhysRevLett.105.053901} {\bibfield  {journal}
  {\bibinfo  {journal} {Phys. Rev. Lett.}\ }\textbf {\bibinfo {volume} {105}},\
  \bibinfo {pages} {053901} (\bibinfo {year} {2010})}\BibitemShut {NoStop}%
\bibitem [{\citenamefont {Wan}\ \emph {et~al.}(2011)\citenamefont {Wan},
  \citenamefont {Chong}, \citenamefont {Ge}, \citenamefont {Noh}, \citenamefont
  {Stone},\ and\ \citenamefont {Cao}}]{Wan2011CPA}%
  \BibitemOpen
  \bibfield  {author} {\bibinfo {author} {\bibfnamefont {W.}~\bibnamefont
  {Wan}}, \bibinfo {author} {\bibfnamefont {Y.}~\bibnamefont {Chong}}, \bibinfo
  {author} {\bibfnamefont {L.}~\bibnamefont {Ge}}, \bibinfo {author}
  {\bibfnamefont {H.}~\bibnamefont {Noh}}, \bibinfo {author} {\bibfnamefont
  {A.~D.}\ \bibnamefont {Stone}}, \ and\ \bibinfo {author} {\bibfnamefont
  {H.}~\bibnamefont {Cao}},\ }\bibfield  {title} {\enquote {\bibinfo {title}
  {Time-reversed lasing and interferometric control of absorption},}\ }\href
  {\doibase 10.1126/science.1200735} {\bibfield  {journal} {\bibinfo  {journal}
  {Science}\ }\textbf {\bibinfo {volume} {331}},\ \bibinfo {pages} {889}
  (\bibinfo {year} {2011})}\BibitemShut {NoStop}%
\bibitem [{\citenamefont {Haigh}\ \emph
  {et~al.}(2015{\natexlab{b}})\citenamefont {Haigh}, \citenamefont {Lambert},
  \citenamefont {Doherty},\ and\ \citenamefont
  {Ferguson}}]{Haigh2015Dispersive}%
  \BibitemOpen
  \bibfield  {author} {\bibinfo {author} {\bibfnamefont {J.~A.}\ \bibnamefont
  {Haigh}}, \bibinfo {author} {\bibfnamefont {N.~J.}\ \bibnamefont {Lambert}},
  \bibinfo {author} {\bibfnamefont {A.~C.}\ \bibnamefont {Doherty}}, \ and\
  \bibinfo {author} {\bibfnamefont {A.~J.}\ \bibnamefont {Ferguson}},\
  }\bibfield  {title} {\enquote {\bibinfo {title} {Dispersive readout of
  ferromagnetic resonance for strongly coupled magnons and microwave
  photons},}\ }\href {\doibase 10.1103/PhysRevB.91.104410} {\bibfield
  {journal} {\bibinfo  {journal} {Phys. Rev. B}\ }\textbf {\bibinfo {volume}
  {91}},\ \bibinfo {pages} {104410} (\bibinfo {year}
  {2015}{\natexlab{b}})}\BibitemShut {NoStop}%
\bibitem [{\citenamefont {Yao}\ \emph {et~al.}(2017)\citenamefont {Yao},
  \citenamefont {Gui}, \citenamefont {Rao}, \citenamefont {Kaur}, \citenamefont
  {Chen}, \citenamefont {Lu}, \citenamefont {Xiao}, \citenamefont {Guo},
  \citenamefont {Marzlin},\ and\ \citenamefont {Hu}}]{Yao2017Cooperative}%
  \BibitemOpen
  \bibfield  {author} {\bibinfo {author} {\bibfnamefont {B.}~\bibnamefont
  {Yao}}, \bibinfo {author} {\bibfnamefont {Y.}~\bibnamefont {Gui}}, \bibinfo
  {author} {\bibfnamefont {J.}~\bibnamefont {Rao}}, \bibinfo {author}
  {\bibfnamefont {S.}~\bibnamefont {Kaur}}, \bibinfo {author} {\bibfnamefont
  {X.}~\bibnamefont {Chen}}, \bibinfo {author} {\bibfnamefont {W.}~\bibnamefont
  {Lu}}, \bibinfo {author} {\bibfnamefont {Y.}~\bibnamefont {Xiao}}, \bibinfo
  {author} {\bibfnamefont {H.}~\bibnamefont {Guo}}, \bibinfo {author}
  {\bibfnamefont {K.-P.}\ \bibnamefont {Marzlin}}, \ and\ \bibinfo {author}
  {\bibfnamefont {C.-M.}\ \bibnamefont {Hu}},\ }\bibfield  {title} {\enquote
  {\bibinfo {title} {Cooperative polariton dynamics in feedback-coupled
  cavities},}\ }\href {\doibase 10.1038/s41467-017-01796-7} {\bibfield
  {journal} {\bibinfo  {journal} {Nat. Commun.}\ }\textbf {\bibinfo {volume}
  {8}},\ \bibinfo {pages} {1437} (\bibinfo {year} {2017})}\BibitemShut
  {NoStop}%
\bibitem [{\citenamefont {Muralidhar}\ \emph {et~al.}(2021)\citenamefont
  {Muralidhar}, \citenamefont {Khymyn}, \citenamefont {Awad}, \citenamefont
  {Alem\'an}, \citenamefont {Hanstorp},\ and\ \citenamefont
  {\AA{}kerman}}]{MuralidharPRL2021}%
  \BibitemOpen
  \bibfield  {author} {\bibinfo {author} {\bibfnamefont {S.}~\bibnamefont
  {Muralidhar}}, \bibinfo {author} {\bibfnamefont {R.}~\bibnamefont {Khymyn}},
  \bibinfo {author} {\bibfnamefont {A.~A.}\ \bibnamefont {Awad}}, \bibinfo
  {author} {\bibfnamefont {A.}~\bibnamefont {Alem\'an}}, \bibinfo {author}
  {\bibfnamefont {D.}~\bibnamefont {Hanstorp}}, \ and\ \bibinfo {author}
  {\bibfnamefont {J.}~\bibnamefont {\AA{}kerman}},\ }\bibfield  {title}
  {\enquote {\bibinfo {title} {Femtosecond laser pulse driven caustic spin wave
  beams},}\ }\href {\doibase 10.1103/PhysRevLett.126.037204} {\bibfield
  {journal} {\bibinfo  {journal} {Phys. Rev. Lett.}\ }\textbf {\bibinfo
  {volume} {126}},\ \bibinfo {pages} {037204} (\bibinfo {year}
  {2021})}\BibitemShut {NoStop}%
\bibitem [{\citenamefont {Harms}, \citenamefont {Yuan},\ and\ \citenamefont
  {Duine}(2022)}]{HarmsarXiv2022}%
  \BibitemOpen
  \bibfield  {author} {\bibinfo {author} {\bibfnamefont {J.~S.}\ \bibnamefont
  {Harms}}, \bibinfo {author} {\bibfnamefont {H.~Y.}\ \bibnamefont {Yuan}}, \
  and\ \bibinfo {author} {\bibfnamefont {R.~A.}\ \bibnamefont {Duine}},\ }\href
  {\doibase 10.48550/ARXIV.2210.16698} {\enquote {\bibinfo {title}
  {Antimagnonics},}\ } (\bibinfo {year} {2022})\BibitemShut {NoStop}%
\bibitem [{\citenamefont {Jin}\ \emph {et~al.}(2023)\citenamefont {Jin},
  \citenamefont {Yao}, \citenamefont {Wang}, \citenamefont {Yuan},
  \citenamefont {Zeng}, \citenamefont {Cao},\ and\ \citenamefont
  {Yan}}]{JinarXiv2023}%
  \BibitemOpen
  \bibfield  {author} {\bibinfo {author} {\bibfnamefont {Z.}~\bibnamefont
  {Jin}}, \bibinfo {author} {\bibfnamefont {X.}~\bibnamefont {Yao}}, \bibinfo
  {author} {\bibfnamefont {Z.}~\bibnamefont {Wang}}, \bibinfo {author}
  {\bibfnamefont {H.~Y.}\ \bibnamefont {Yuan}}, \bibinfo {author}
  {\bibfnamefont {Z.}~\bibnamefont {Zeng}}, \bibinfo {author} {\bibfnamefont
  {Y.}~\bibnamefont {Cao}}, \ and\ \bibinfo {author} {\bibfnamefont
  {P.}~\bibnamefont {Yan}},\ }\href {\doibase 10.48550/ARXIV.2301.03211}
  {\enquote {\bibinfo {title} {Nonlinear topological magnon spin hall
  effect},}\ } (\bibinfo {year} {2023})\BibitemShut {NoStop}%
\end{thebibliography}%
\end{document}